\documentclass[12pt]{article}

\usepackage{latexsym,amsmath,amscd,amssymb,graphics}
\usepackage{enumerate,framed}

\usepackage{graphicx}

\usepackage[square,authoryear]{natbib}
\usepackage{url}

\usepackage[all]{xy}

\makeatletter

\@addtoreset{figure}{section}
\def\thefigure{\thesection.\@arabic\c@figure}
\def\fps@figure{h, t}
\@addtoreset{table}{bsection}
\def\thetable{\thesection.\@arabic\c@table}
\def\fps@table{h, t}
\@addtoreset{equation}{section}

\makeatother

\begin{document}

\newtheorem{theorem}{Theorem}[section]
\newtheorem{definition}[theorem]{Definition}
\newtheorem{lemma}[theorem]{Lemma}
\newtheorem{remark}[theorem]{Remark}
\newtheorem{proposition}[theorem]{Proposition}
\newtheorem{corollary}[theorem]{Corollary}
\newtheorem{example}[theorem]{Example}

\def\below#1#2{\mathrel{\mathop{#1}\limits_{#2}}}



\title{The Geometric Structure of Complex Fluids}
\author{Fran\c{c}ois Gay-Balmaz$^{1}$ and Tudor S. Ratiu$^{1}$}
\addtocounter{footnote}{1} \footnotetext{Section de
Math\'ematiques and Bernoulli Center, \'Ecole Polytechnique F\'ed\'erale de
Lausanne.
CH--1015 Lausanne. Switzerland.
\texttt{Francois.Gay-Balmaz@epfl.ch, Tudor.Ratiu@epfl.ch}
\addtocounter{footnote}{1} }

\date{}
\maketitle

\makeatother

\maketitle




\begin{abstract} This paper develops the theory of affine Euler-Poincar\'e and
affine Lie-Poisson reductions and applies these processes to various examples
of
complex fluids, including Yang-Mills and Hall magnetohydrodynamics for fluids
and superfluids, spin glasses, microfluids, and liquid crystals. As a
consequence of the Lagrangian approach, the variational formulation of the
equations is determined. On the Hamiltonian side, the associated Poisson
brackets are obtained by reduction of a canonical cotangent bundle. A
Kelvin-Noether circulation theorem is presented and is applied to these
examples.
\end{abstract}

\tableofcontents


\section{Introduction}\label{Introduction}

The equations of motion of an \textit{adiabatic compressible fluid} are given
by
\begin{equation}\label{ICAF}
\left\lbrace
\begin{array}{ll}
\vspace{0.2cm}\displaystyle\frac{\partial \mathbf{u}}{\partial
t}+\nabla_\mathbf{u}\mathbf{u}=\frac{1}{\rho}\operatorname{grad}p,\\
\vspace{0.2cm}\displaystyle\frac{\partial \rho}{\partial
t}+\operatorname{div}(\rho \mathbf{u})=0,\quad\frac{\partial S}{\partial
t}+\operatorname{div}(S\mathbf{u})=0,
\end{array} \right.
\end{equation}
where $\rho$ is the mass density, $S$ is the entropy density, and $p$ is
the pressure. It was shown in \cite{MoGr1980} that this system, as well as its
magnetohydrodynamic extension,
admit a non-canonical Poisson formulation, that is, equation \eqref{ICAF} can be
written as
\[
\dot f=\{f,h\},
\]
relative to a Hamiltonian function $h$. It is of great (mathematical and
physical) interest to obtain these Poisson brackets by a reduction procedure
from a canonical Hamiltonian formulation on a cotangent bundle. In
\cite{MaRaWe1984}, the non-canonical Poisson bracket associated to \eqref{ICAF}
is obtained via Lie-Poisson
reduction for a semidirect product group involving the diffeomorphisms group of
the fluid container and the space of the advected quantities $\rho$ and $S$.
The
Lagrangian formulation of these equations is given in \cite{HoMaRa1998}. 

In the same spirit, the non-canonical Hamiltonian structure for
\textit{adiabatic
Yang-Mills charged fluids} discovered in \cite{GiHoKu1983} is obtained by
reduction from a canonical formulation in \cite{GBRa2008}, by using a
Kaluza-Klein point of view involving the automorphism group of the principal
bundle of the theory. This paper gives also the Euler-Poincar\'e formulation of
these equations.

Non-canonical Hamiltonian structures for a wide class of non-dissipative fluid
models were derived in \cite{HoKu1984}, \cite{Ho1987}, \cite{HoKu1987},
\cite{HoKu1988}, and \cite{Ho2001}. These examples include \textit{Yang-Mills
magnetohydrodynamics\,}, \textit{spin glasses\,}, and various models of
\textit{superfluids\,}, and involve Lie-Poisson brackets \textit{with
cocycles}.
Remarkably, from a mathematical point of view, the Hamiltonian structures of
many of these models are identical. This Hamiltonian structure together with
the
corresponding variational principles are studied in more detail, with an
application to liquid crystals, in \cite{Ho2002}. We will refer to all these
models as \textit{complex fluids}.

In this paper we show the remarkable property that these Lie-Poisson brackets
with cocycles can also be obtained by Poisson reduction from a canonical
Hamiltonian structure.
The cocycle in the Hamiltonian structure appears only after reduction and it is
due to the presence of an affine term added to the cotangent lifted action. The
associated reduction process is naturally called \textit{affine Lie-Poisson
reduction\/}.

An important example of such an affine action is given by the usual action of
the automorphism group of a principal bundle on the connection forms. As a
result, we obtain, in a natural way, covariant differentials and covariant
divergences in the expression of the Poisson brackets and of the reduced
equations. These gauge theory aspects in the case of complex fluids are
mathematically and physically interesting since they represent a bridge to
other
possible gauge theories in physics.

The new affine Lie-Poisson reduction principle presented in this paper unifies
all previous approaches 
in the problem of complex fluid dynamics. Eringen and Holm have extensively
studied the roles of 
auxiliary variables that were Lie algebra valued 1-forms frozen into the fluid
flow. In particular, Holm 
observed that
these variables contribute generalized 2-cocycles in the Lie-Poisson structures
of ideal complex fluids. 
We implement Holm's observation as a fundamental principle which,
for physical interactions in complex fluids, is similar to the principle of
spontaneous symmetry
breaking in the coupling of charges and fields in electromagnetism. This new
geometrically formulated 
principle applies also to Lie algebra valued charges (e.g., spins, Yang-Mills
charges, etc). The present 
paper reformulates the theory of complex fluids in the framework of the
fundamental principles of condensed-matter physics. It is clear that other
geometric applications in 
condensed-matter physics are forthcoming, such as micropolar elasticity, for
example.

The paper also formulates a parallel theory on the Lagrangian side by extending
the Euler-Poincar\'e 
framework for semidirect
products developed in \cite{HoMaRa1998} to the case of an affine representation
of the configuration 
space Lie group on the vector space whose dual are the convected variables. The
resulting \textit{affine
Euler-Poincar\'e reduction} is natural in two senses. Firstly, in the case of
complex fluids and at the reduced level, it coincides with that given in
\cite{Ho2002}. Secondly, in the hyperregular case and through the Legendre
transformation, it is compatible with the affine Lie-Poisson reduction for
semidirect products discussed previously. In addition, the Euler-Poincar\'e
formulation immediately 
leads to analogs of the Kelvin circulation theorem for complex fluids.

\medskip  \noindent \textbf{Organization of the paper.}
The paper is organized as follows. The first third, comprising 
\S\ref{Lagrangian_formulation}-\S\ref{Kelvin_Noether}, presents only theoretical
results. These are 
developed and applied to a wide range of examples in the last section
\S\ref{applications} forming the 
bulk of the paper.

The theoretical part begins by recalling at the end of this introduction some
needed facts about the 
Lagrangian and Hamiltonian reductions for semidirect products. It is well-known
that these processes 
form the basic framework for the geometric formulation of various models of
simple fluids. In 
\S\ref{Lagrangian_formulation} we present the theory of affine Euler-Poincar\'e
reduction for a general 
Lie group $G$ acting by affine representation on a dual vector space $V^*$. This
theory is specialized, 
in \S\ref{Lagrangian_PCF}, to the case of general complex fluids. More
precisely, we describe concretely 
the group $G$ and the cocycle needed in the affine representation, in order to
obtain by reduction the 
general equations for complex fluids. In order to carry out the Hamiltonian side
of the theory, we need 
to recall and state some results concerning the reduction of a canonical
symplectic structure with a 
magnetic term. This is the subject of \S\ref{AMLPR}, whose principal result
states that, under some 
conditions, reducing a canonical symplectic form relative to a cotangent lift
with an affine term is 
equivalent to reducing a magnetic symplectic form relative to the cotangent
lift. This observation is 
used in \S\ref{AHSPT} to obtain the theory of affine Lie-Poisson reduction for a
general Lie group $G$ 
acting by affine representation on a dual vector space $V^*$. In particular, we
compute the associated 
momentum map and Poisson bracket as well as the symplectic reduced spaces. This
theory is also 
shown to be a particular case of the process of reduction by stages for
nonequivariant momentum 
maps. In \S\ref{Hamiltonian_PCF} we specialize these results to the case of the
group and the cocycle 
involved in the description of complex fluid dynamics. In particular, we compute
the associated Poisson 
bracket and momentum map.  This part constitutes the Hamiltonian side of the
theory developed in 
\S\ref{Lagrangian_PCF}. The Kelvin-Noether theorem is a version of the classical
Noether conservation 
law that holds for
solutions of the Euler-Poincar\'e equations. For example, an application of
this theorem to the compressible fluid gives the Kelvin circulation theorem. The
generalization of this 
result to the case of affine Euler-Poincar\'e equations is the subject of
\S\ref{Kelvin_Noether}.

The rest of the paper is devoted to applications dealing with spin systems and
complex fluids, all of 
them contained in \S\ref{applications}. In \S\ref{subsec:spin_chains} we start
with the example of spin 
systems, since it illustrates the applicability of our theory in a very simple
situation that exhibits, 
nevertheless, some of the key difficulties of more complicated fluid models. We
then treat, in 
\S\ref{YM_MHD}-\S\ref{HVBK}, the examples of Yang-Mills and Hall
magnetohydrodynamics for fluids 
and superfluids as well as the HVBK dynamics for ${}^4$He with vortices. In each
case, we formulate in 
detail the Lagrangian and Hamiltonian reduction processes as well as the
associated Poisson bracket 
and circulation theorems. In order to compare fluids and superfluids from a
Hamiltonian point of view, 
we present in \S\ref{fluids_vs_superlfuids} a summary of the models treated so
far. In 
\S\ref{VD_spin_glasses} we study the example of the Volovik-Dotsenko spin
glasses and try to 
understand, from a Hamiltonian point of view, the passage from a given spin
system to its 
hydrodynamic analogue. This process allows us to understand mathematically the
link between the two 
approaches to spin glasses appearing in the current literature. All the
Hamiltonian structures obtained 
so far by reduction of the canonical structure, coincide with the ones obtained
previously and by 
various different methods in a series of papers by Holm and Kupershmidt.

In the case of microfluids, treated in \S\ref{MF}, we determine the Lagrangian
and Hamiltonian 
structure of three models proposed by Eringen, namely, the micropolar, the
microstretch, and the 
micromorphic models, corresponding to three groups associated to the internal
structure of the fluid 
particles. Since our theory applies to any group, we can obtain new models of
microlfuids such as the 
anisotropic micropolar or anisotropic microstretch models. In the case of
microstretch and micropolar 
fluids, we show that the internal degree of freedom can be modeled by the group
of invertible or unit 
quaternions, respectively. In \S\ref{LC}, we determine and compare the
Lagrangian and Hamiltonian 
structures of three models of liquid crystals dynamics. We also treat the case
of Eringen's polymeric 
liquid crystals and give some information concerning the new model of
anisotropic micropolar liquid 
crystals.

We close this introduction by recalling some needed facts about 
Euler-Poincar\'e and Lie-Poisson reduction for semidirect products.

\medskip 

\noindent \textbf{Notations for semidirect products.} In the Euler-Poincar\'e
reduction for semidirect products (see \cite{HoMaRa1998}) one is given a Lie group $G $ and a \textit{right\/} representation $\rho : G\rightarrow
\operatorname{Aut}(V)$ of $G$ on the vector space $V$.  As a set, the semidirect product $S=G\,\circledS\,V$
is
the Cartesian product $S=G\times V$ whose group multiplication is given by
\[
(g_1,v_1)(g_2,v_2)=(g_1g_2,v_2+\rho_{g_2}(v_1)).
\]
The Lie algebra of $S$ is the semidirect product Lie algebra,
$\mathfrak{s}=\mathfrak{g}\,\circledS\,V$, whose bracket has the expression
\[
\operatorname{ad}_{(\xi_1,v_1)}(\xi_2,v_2)=[(\xi_1,v_1),(\xi_2,v_2)]=([\xi_1,\xi_2],v_1\xi_2-v_2\xi_1),
\]
where $v\xi$ denotes the induced action of $\mathfrak{g}$ on $V$, that is,
\[
v\xi:=\left.\frac{d}{dt}\right|_{t=0}\rho_{\operatorname{exp}(t\xi)}(v)\in
V.
\]
From the expression for the Lie bracket, it follows that for
$(\xi,v)\in\mathfrak{s}$ and $(\mu,a)\in\mathfrak{s}^*$ we have
\[
\operatorname{ad}^*_{(\xi,v)}(\mu,a)=(\operatorname{ad}^*_\xi\mu+v\diamond
a,a\xi),
\]
where $a\xi\in V^*$ and $v\diamond a\in\mathfrak{g}^*$ are given by
\[
a\xi:=\left.\frac{d}{dt}\right|_{t=0}\rho^*_{\operatorname{exp}(-t\xi)}(a)\quad\text{and}\quad
\langle v\diamond a,\xi\rangle_\mathfrak{g}:=-\langle a\xi,v\rangle_V,
\]
and where $\left\langle\cdot , \cdot \right\rangle_ \mathfrak{g}: \mathfrak{g}
^\ast \times \mathfrak{g}\rightarrow \mathbb{R}$ and $\left\langle \cdot ,
\cdot
\right\rangle_V: V ^\ast \times V \rightarrow \mathbb{R}$ are the duality
parings.

\noindent \textbf{Lagrangian semidirect product theory.}
\begin{itemize}
\item Assume that we have a function $L:TG\times V^*\rightarrow\mathbb{R}$
which
is right $G$-invariant.
\item In particular, if $a_0\in V^*$, define the Lagrangian
$L_{a_0}:TG\rightarrow\mathbb{R}$ by $L_{a_0}(v_g):=L(v_g,a_0)$. Then $L_{a_0}$
is right invariant under the lift to $TG$ of the right action of $G_{a_0}$ on
$G$, where $G_{a_0}$ is the isotropy group of $a_0$.
\item Right $G$-invariance of $L$ permits us to define $l:\mathfrak{g}\times
V^*\rightarrow\mathbb{R}$ by
\[
l(T_gR_{g^{-1}}(v_g),\rho^*_{g}(a_0))=L(v_g,a_0).
\]
\item For a curve $g(t)\in G$, let $\xi(t):=TR_{g(t)^{-1}}(\dot{g}(t))$ and
define the curve $a(t)$ as the unique solution of the following linear
differential equation with time dependent coefficients
\[
\dot{a}(t)=-a(t)\xi(t),
\]
with initial condition $a(0)=a_0$. The solution can be written as
$a(t)=\rho^*_{g(t)}(a_0)$.
\end{itemize}

\begin{theorem}\label{EPSD} With the preceding notations, the following are
equivalent:
\begin{enumerate}
\item[\bf{(i)}] With $a_0$ held fixed, Hamilton's variational principle
\[
\delta\int_{t_1}^{t_2}L_{a_0}(g(t),\dot{g}(t))dt=0,
\]
holds, for variations $\delta g(t)$ of $g(t)$ vanishing at the endpoints.
\item[\bf{(ii)}] $g(t)$ satisfies the Euler-Lagrange equations for $L_{a_0}$ on
$G$.
\item[\bf{(iii)}] The constrained variational principle
\[
\delta\int_{t_1}^{t_2}l(\xi(t),a(t))dt=0,
\]
holds on $\mathfrak{g}\times V^*$, upon using variations of the form
\[
\delta\xi=\frac{\partial\eta}{\partial t}-[\xi,\eta],\quad \delta a=-a\eta,
\]
where $\eta(t)\in\mathfrak{g}$ vanishes at the endpoints.
\item[\bf{(iv)}] The Euler-Poincar\'e equations hold on $\mathfrak{g}\times V^*$:
\begin{equation}\label{EP}
\frac{\partial}{\partial t}\frac{\delta
l}{\delta\xi}=-\operatorname{ad}^*_\xi\frac{\delta l}{\delta\xi}+\frac{\delta
l}{\delta a}\diamond a.
\end{equation}
\end{enumerate}
\end{theorem}

It is worth noting that there is a remarkable symmetry breaking from $L : TG \times V ^\ast \rightarrow \mathbb{R}$ to $L_{a_0}: TG \rightarrow \mathbb{R}$. Whereas the function $L$ is right $G $-invariant, that is,  $L(v_h g, \rho_{g ^{-1}}^\ast(a)) = L(v _h, a)$ for all $v _h \in TG$, $a \in V ^\ast$, $g \in G $, the Lagrangian $L_{a_0} $ is only $G_{a_0} $-invariant, that is, $L_{a_0}(v _h k) =
L_{a_0}(v _h)$ for all  $v _h \in TG$, $k \in G_{a_0}$. Note, however, that $L_{a_0} $ is a Lagrangian function on $TG $, whereas $L $ is not since it is not defined on a tangent bundle.  The relationship between these two different invariance properties are dealt with in the previous theorem and the evolution $a(t) = \rho^\ast_{g (t)} a _0$ or, equivalently, the equation $\dot{a}(t) = - a (t)\xi(t)$, $a (0) = a _0$, that needs to be added to \eqref{EP} in order to get a complete system of equations for all the unknowns, is exactly due to this symmetry breaking from $G $ to $G_{a_0} $. We shall observe the same phenomenon throughout the paper when working out the affine reduction theorem on both the Lagrangian and Hamiltonian side. 

\medskip  

\noindent \textbf{Hamiltonian semidirect product theory.} Let $S :=
G\,\circledS\,V $ be the semidirect product defined at the beginning of
this section. The lift of right translation of $S $ on $T ^\ast S $
induces a right action on $T ^\ast G \times V ^\ast$. Consider a Hamiltonian
function $H:
T ^\ast G \times V ^\ast \rightarrow \mathbb{R}$ right
invariant under the $S $-action on $T ^\ast G \times V ^\ast$. In particular,
the
function $H_{a_0}: = H|_{T ^\ast G\times \{a_0\}}: T ^\ast G
\rightarrow \mathbb{R}$ is invariant under the induced action
of the isotropy subgroup $G_{a_0}: = \{g \in G \mid \rho_g^\ast a_0 = a_0\}$
for
any $a_0 \in V ^\ast$. The following theorem is an easy consequence of the
semidirect product reduction theorem (see \cite{MaRaWe1984}) and the reduction
by stages method (see \cite{MaMiOrPeRa2007}).

\begin{theorem}\label{LPSD}
For $\alpha(t)\in T^*_{g(t)}G$ and
$\mu(t):=T^*R_{g(t)}(\alpha(t))\in\mathfrak{g}^*$, the following are
equivalent:
\begin{enumerate}
\item[\bf{(i)}] $\alpha(t)$ satisfies Hamilton's equations for
$H_{a_0}$ on $T^*G$.
\item[\bf{(ii)}] The Lie-Poisson equation holds on $\mathfrak{s}^*$:
\[
\frac{\partial}{\partial t}(\mu,a)=-\operatorname{ad}^*_{\left(\frac{\delta
h}{\delta\mu},\frac{\delta h}{\delta a}\right)}(\mu,a)
=- \left(\operatorname{ad}^*_{\frac {\delta h}{ \delta \mu}}\mu+\frac {\delta
h}{ \delta a}\diamond
a,a\frac {\delta h}{ \delta \mu}\right),\quad a(0)=a_0
\]
where $\mathfrak{s}$ is the semidirect product Lie algebra
$\mathfrak{s}=\mathfrak{g}\,\circledS\, V$. The associated Poisson bracket is
the Lie-Poisson bracket on the semidirect product Lie algebra $\mathfrak{s}^*$,
that is,
\[
\{f,g\}(\mu,a)=\left\langle\mu,\left[\frac{\delta
f}{\delta\mu},\frac{\delta
g}{\delta\mu}\right]\right\rangle+\left\langle a,\frac{\delta f}{\delta
a}\frac{\delta g}{\delta\mu}-\frac{\delta g}{\delta a}\frac{\delta
f}{\delta\mu}\right\rangle.
\]
\end{enumerate}
As on the Lagrangian side, the evolution of the advected quantities is given by
$a(t)=\rho^*_{g(t)}(a_0)$.
\end{theorem}

For example, one can start with a Lagrangian $L_{a_0}$ as at the beginning of
this
section, suppose that the Legendre transformation $\mathbb{F}L_{a_0}$ is
invertible and form the corresponding Hamiltonian
$H_{a_0}=E_{a_0}\circ\mathbb{F}L_{a_0}^{-1}$, where $E_{a_0}$ is the energy of
$L _{a_0}$. Then the function $H: T^\ast G \times V ^\ast
\rightarrow \mathbb{R}$ so defined is $S $-invariant and one can apply this
theorem. At the level of the reduced space, to a reduced Lagrangian
$l:\mathfrak{g}\times V^*\rightarrow\mathbb{R}$ we associate the reduced
Hamiltonian $h:\mathfrak{g}^*\,\circledS\,V^*\rightarrow\mathbb{R}$ given by
\[
h(\mu,a):=\langle\mu,\xi\rangle-l(\xi,a),\quad\mu=\frac{\delta l}{\delta\xi}.
\]
Since
\[
\frac{\delta h}{\delta\mu}=\xi\quad\text{and}\quad\frac{\delta h}{\delta
a}=-\frac{\delta l}{\delta a},
\]
we see that the Lie-Poisson equations for $h$ on $\mathfrak{s}^*$ are
equivalent
to the Euler-Poincar\'e equations \eqref{EP} for $l$ together with the
advection
equation $\dot{a}+a\xi=0$.

\medskip  

\noindent \textbf{Links with the reduction by stages.} Consider the semidirect product
Lie group $S=G\,\circledS\,V$ acting by right translation on its cotangent
bundle $T^*S$. An equivariant momentum map relative to the canonical symplectic
form is given by
\[
\mathbf{J}_R(\alpha_f,(u,a))=T^*L_{(f,u)}(\alpha_f,(u,a))=(T^*_eL_f(\alpha_f)+u\diamond
a,a).
\]
Since $V$ is a closed normal subgroup of $S$, it also acts on $T^*S$ and has a
momentum map $\mathbf{J}_V : T^*S\to V^*$ given by
\[
\mathbf{J}_V(\alpha_f,(u,a))=a.
\]
Reducing $T^*S$ by $V$ at the value $a$ we get the first reduced space
$(T^*S)_a=\mathbf{J}_V^{-1}(a)/V$. The isotropy subgroup $G_a$, consisting of
elements of $G$ that leave the point $a$ fixed, acts freely and properly on
$(T^*S)_a$ and has an induced equivariant momentum map $\mathbf{J}_a :
(T^*S)_a\to\mathfrak{g}^*_a$, where $\mathfrak{g}_a$ is the Lie algebra of
$G_a$. Reducing $(T^*S)_a$ at the point $\mu_a:=\mu|\mathfrak{g}_a$, we get the
second reduced space
$\left((T^*S)_a\right)_{\mu_a}=\mathbf{J}_a^{-1}(\mu_a)/(G_a)_{\mu_a}$.

Using the Semidirect Product Reduction (\cite{MaRaWe1984}) or the Reduction by
Stages Theorem (\cite{MaMiOrPeRa2007}), the two-stage reduced space
$\left((T^*S)_a\right)_{\mu_a}$  is symplectically diffeomorphic to the reduced
space $(T^*S)_{(\mu,a)}=\mathbf{J}_R^{-1}(\mu,a)/G_{(\mu,a)}$ obtained by
reducing $T^*S$ by the whole group $S$ at the point $(\mu,a)\in\mathfrak{s}^*$.

The first symplectic reduced space $((T^*S)_a,\Omega_a)$ is
symplectically diffeomorphic to the canonical symplectic manifold
$(T^*G,\Omega)$ and the second reduced space
$\left(\left((T^*S)_a\right)_{\mu_a},(\Omega_a)_{\mu_a}\right)$ is
symplectically diffeomorphic to the coadjoint orbit
$\left(\mathcal{O}_{(\mu,a)},\omega_{(\mu,a)}\right)$ together with its orbit
symplectic form. Note also that we can consider the right $G $-invariant
Hamiltonian $H:T^*G\times V^*\to\mathbb{R}$ as being the Poisson reduction of a
$S$-invariant Hamiltonian $\overline{H}:T^*S\to\mathbb{R}$ by the normal
subgroup $\{e\} \times V $ since $(T ^\ast S)/(\{e\} \times V)  \cong T ^\ast G
\times V^\ast$. Here the $G $-action on $T ^\ast G \times V ^\ast$ is induced
by
the lift of right translation of  $S $. Theorem \ref{LPSD} is then a trivial
consequence of these observations.


\section{Affine Lagrangian Semidirect Product
Theory}\label{Lagrangian_formulation}

Consider the \textit{right\/} contragredient representation $\rho^*_{g^{-1}}$
of
$G$ on the vector space $V^*$. We can form an affine \textit{right\/}
representation $\theta_g(a)=\rho^*_{g^{-1}}(a)+c(g)$, where
$c\in\mathcal{F}(G,V^*)$ is a contragredient representation valued right group
one-cocycle, that is, it verifies the property
$c(fg)=\rho_{g^{-1}}^*(c(f))+c(g)$ for all $f, g \in G$. This implies
that
$c(e)=0$ and $c(g^{-1})=-\rho^*_g(c(g))$. Note that
\[
\left.\frac{d}{dt}\right|_{t=0}\theta_{\operatorname{exp}(t\xi)}(a)=a\xi+\mathbf{d}c(\xi).
\]
and
\[
\langle a\xi+\mathbf{d}c(\xi),v\rangle_V=\langle \mathbf{d}c^T(v)-v\diamond
a,\xi\rangle_\mathfrak{g},
\]
where $\mathbf{d}c :\mathfrak{g}\rightarrow V^*$ is defined by
$\mathbf{d}c(\xi):=T_ec(\xi)$, and $\mathbf{d}c^T:V\rightarrow\mathfrak{g}^*$
is
defined by
\[
\langle \mathbf{d}c^T(v),\xi\rangle_\mathfrak{g}:=\langle
\mathbf{d}c(\xi),v\rangle_V.
\]
\begin{itemize}
\item Assume that we have a function $L:TG\times V^*\rightarrow\mathbb{R}$
which
is right $G$-invariant under the affine action 
\[
(v_h,a)\mapsto
(T_hR_g(v_h),\theta_g(a))=(T_hR_g(v_h),\rho^*_{g^{-1}}(a)+c(g)).
\]
\item In particular, if $a_0\in V^*$, define the Lagrangian
$L_{a_0}:TG\rightarrow\mathbb{R}$ by $L_{a_0}(v_g):=L(v_g,a_0)$. Then $L_{a_0}$
is right invariant under the lift to $TG$ of the right action of $G_{a_0}^c$ on
$G$, where $G_{a_0}^c$ is the isotropy group of $a_0$ with respect to the
affine
action $\theta$.
\item Right $G$-invariance of $L$ permits us to define $l:\mathfrak{g}\times
V^*\rightarrow\mathbb{R}$ by
\[
l(T_gR_{g^{-1}}(v_g),\theta_{g^{-1}}(a_0))=L(v_g,a_0).
\]
\item For a curve $g(t)\in G$, let $\xi(t):=TR_{g(t)^{-1}}(\dot{g}(t))$ and
define the curve $a(t)$ as the unique solution of the following affine
differential equation with time dependent coefficients
\[
\dot{a}(t)=-a(t)\xi(t)-\mathbf{d}c(\xi(t)),
\]
with initial condition $a(0)=a_0$. The solution can be written as
$a(t)=\theta_{g(t)^{-1}}(a_0)$.
\end{itemize}

\begin{theorem}\label{AEPSD} With the preceding notations, the following are
equivalent:
\begin{enumerate}
\item[\bf{(i)}] With $a_0$ held fixed, Hamilton's variational principle
\begin{equation}\label{Hamilton_principle}
\delta\int_{t_1}^{t_2}L_{a_0}(g(t),\dot{g}(t))dt=0,
\end{equation}
holds, for variations $\delta g(t)$ of $g(t)$ vanishing at the endpoints.
\item[\bf{(ii)}] $g(t)$ satisfies the Euler-Lagrange equations for $L_{a_0}$ on
$G$.
\item[\bf{(iii)}] The constrained variational principle
\begin{equation}\label{Euler-Poincare_principle}
\delta\int_{t_1}^{t_2}l(\xi(t),a(t))dt=0,
\end{equation}
holds on $\mathfrak{g}\times V^*$, upon using variations of the form
\[
\delta\xi=\frac{\partial\eta}{\partial t}-[\xi,\eta],\quad \delta
a=-a\eta-\mathbf{d}c(\eta),
\]
where $\eta(t)\in\mathfrak{g}$ vanishes at the endpoints.
\item[\bf{(iv)}] The affine Euler-Poincar\'e equations hold on
$\mathfrak{g}\times
V^*$:
\begin{equation}\label{AEP}
\frac{\partial}{\partial t}\frac{\delta
l}{\delta\xi}=-\operatorname{ad}^*_\xi\frac{\delta l}{\delta\xi}+\frac{\delta
l}{\delta a}\diamond a-\mathbf{d}c^T\left(\frac{\delta l}{\delta a}\right).
\end{equation}
\end{enumerate}
\end{theorem}

\noindent \textbf{Proof.} The equivalence of $\textbf{(i)}$ and $\textbf{(ii)}$ is true in general.

Next we show the equivalence of $\textbf{(iii)}$ and $\textbf{(iv)}$. Indeed, using
the definitions, integrating by parts, and taking into account that
$\eta(t_1)=\eta(t_2)=0$, we compute the variation of the integral to be
\begin{align*}
\delta\int_{t_1}^{t_2}l(\xi(t),&a(t))dt=\int_{t_1}^{t_2}\left(\left\langle\frac{\delta
l}{\delta\xi},\delta\xi\right\rangle+\left\langle\delta a,\frac{\delta
l}{\delta
a}\right\rangle\right)dt\\
&=\int_{t_1}^{t_2}\left(\left\langle\frac{\delta
l}{\delta\xi},\dot\eta-\operatorname{ad}_\xi\eta\right\rangle-\left\langle
a\eta+\mathbf{d}c(\eta),\frac{\delta l}{\delta a}\right\rangle\right)dt\\
&=\int_{t_1}^{t_2}\left(\left\langle-\frac{d}{dt}\frac{\delta
l}{\delta\xi}-\operatorname{ad}^*_\xi\frac{\delta
l}{\delta\xi},\eta\right\rangle-\left\langle-\frac{\delta l}{\delta a}\diamond
a+\mathbf{d}c^T\left(\frac{\delta l}{\delta
a}\right),\eta\right\rangle\right)dt\\
&=\int_{t_1}^{t_2}\left(\left\langle-\frac{d}{dt}\frac{\delta
l}{\delta\xi}-\operatorname{ad}^*_\xi\frac{\delta l}{\delta\xi}+\frac{\delta
l}{\delta a}\diamond a-\mathbf{d}c^T\left(\frac{\delta l}{\delta
a}\right),\eta\right\rangle\right)dt
\end{align*}
and so the result follows.

Finally we show that $\textbf{(i)}$ and $\textbf{(iii)}$ are equivalent. First note
that the $G$-invariance of $L:TG\times V^*\rightarrow\mathbb{R}$ and the
definition of $a(t)=\theta_{g(t)^{-1}}(a_0)$ imply that the integrands in
\eqref{Hamilton_principle} and \eqref{Euler-Poincare_principle} are equal. It
is
known that all variations $\delta g(t)\in TG$ of $g(t)$ with fixed endpoints
induce and are induced by variations $\delta\xi(t)\in\mathfrak{g}$ of $\xi(t)$
of the form $\delta\xi=\dot\eta-[\xi,\eta]$ with $\eta(t)$ vanishing at the
endpoints; the relation between $\delta g(t)$ and $\eta(t)$ is given by
$\eta(t)=TR_{g(t)^{-1}}(\delta g(t))$. See \cite{BlKrMaRa1996} for details.

Thus, if $\textbf{(i)}$ holds, we define $\eta(t)=TR_{g(t)^{-1}}(\delta g(t))$
for
a variation $\delta g(t)$ with fixed endpoints. Then if we let
$\delta\xi(t)=TR_{g(t)^{-1}}(\dot g(t))$, we have
$\delta\xi=\dot\eta-[\xi,\eta]$. In addition, the variation of
$a(t)=\theta_{g(t)^{-1}}(a_0)$ is $\delta
a(t)=-a(t)\eta(t)-\mathbf{d}c(\eta(t))$. Conversely, if
$\delta\xi=\dot\eta-[\xi,\eta]$ with $\eta(t)$ vanishing at the endpoints, we
define $\delta g(t)=TR_{g(t)}(\eta(t))$. This $\delta g(t)$ is the general
variation of $g(t)$ vanishing at the endpoints. From $\delta
a(t)=-a(t)\eta(t)-\mathbf{d}c(\eta(t))$ it follows that the variation of
$\theta_{g(t)}(a(t))=a_0$ vanishes, which is consistent with the dependence of
$L_{a_0}$ only on $g(t),\dot g(t).\qquad\blacksquare$


\section{Lagrangian Approach to Continuum Theories of Perfect Complex
Fluids}\label{Lagrangian_PCF}

Recall that in the case of the motion of a fluid on an orientable manifold
$\mathcal{D}$, the configuration space is the group
$G=\operatorname{Diff}(\mathcal{D})$ of all diffeomorphisms of $\mathcal{D}$.
In
the case of incompressible fluids, one chooses the subgroup
$\operatorname{Diff}_{\rm vol}(\mathcal{D})$ of all volume preserving
diffeomorphisms, with respect to a fixed volume form on $\mathcal{D}$. Besides
the diffeomorphism group, the other basic object is the vector space $V^*$ of
advected quantities on which $G$ acts by representations. Typical advected
quantities are for example the \textit{mass density}, the \textit{specific
entropy} or the \textit{magnetic field}. One can obtain the fluid equations by
choosing the appropriate Lagrangian and Hamiltonian functions and by applying
the semidirect Euler-Poincar\'e or Lie-Poisson reduction processes (Theorem
\ref{EPSD} and \ref{LPSD}), see \cite{MaRaWe1984} and \cite{HoMaRa1998}.

The goal of this section is to extend these formulations to the case of complex
fluids. At the reduced level, the Euler-Poincar\'e equations for complex fluids
are given in \cite{Ho2002} (equations (3.23), (3.24), (3.32) and (3.33)). The
two key observations we make regarding these equations are the following.
First,
the two equations (3.23) and (3.24) suggest that the configuration manifold
$\operatorname{Diff}(\mathcal{D})$ has to be enlarged to a bigger group $G$ in
order to contains variables involving the Lie group $\mathcal{O}$ of order
parameters. Second the two advection equations (3.32) and (3.33) suggest that
there is a new advected quantity on which the group $G$ acts by affine
representation. Making use of these two observations, we construct below the
appropriate configuration space and the appropriate affine action for the
dynamics of complex fluids. By using the general process of affine
Euler-Poincar\'e reduction developed before 
(Theorem \ref{AEPSD}), we get (a generalization of) the equations given in \cite{Ho2002}.

Here and in all examples that follow, there are fields different from the
velocity field for which we shall never specify the boundary conditions.
We make the general assumption, valid throughout the paper, that all
integrations by parts have vanishing boundary terms, or that the problem has
periodic boundary conditions (in which case $\mathcal{D}$ is a boundaryless
three dimensional manifold). Of course if one would try to get an analytically
rigorous result, the boundary conditions for all fields need to be carefully
specified.

\medskip  

\noindent\textbf{The configuration manifold.} Consider a finite dimensional Lie group $\mathcal{O}$. In applications
$\mathcal{O}$ will be called the \textit{order parameter Lie group}. Recall
that
in the case of the motion of a fluid on an orientable manifold $\mathcal{D}$,
the configuration space is the group $G=\operatorname{Diff}(\mathcal{D})$ of
all
diffeomorphisms of $\mathcal{D}$. In the case of complex fluids, the basic idea
is to enlarge this group to the semidirect product of groups
$G=\operatorname{Diff}(\mathcal{D})\,\circledS\,\mathcal{F}(\mathcal{D},\mathcal{O})$.
Here $\mathcal{F}(\mathcal{D},\mathcal{O})$ denotes the group of all mappings
$\chi$ defined on $\mathcal{D}$ with values in the Lie group $\mathcal{O}$ of
order parameters. The diffeomorphism group acts on
$\mathcal{F}(\mathcal{D},\mathcal{O})$ via the \textit{right\/} action
\[
(\eta,\chi)\in\operatorname{Diff}(\mathcal{D})\times\mathcal{F}(\mathcal{D},\mathcal{O})\mapsto\chi\circ\eta\in\mathcal{F}(\mathcal{D},\mathcal{O}).
\]
Therefore, the group multiplication is given by
\[
(\eta,\chi)(\varphi,\psi)=(\eta\circ\varphi,(\chi\circ\varphi)\psi).
\]
Recall that the tangent space to $\operatorname{Diff}(\mathcal{D})$ at $\eta$
is
\[
T_\eta\operatorname{Diff}(\mathcal{D})=\{\mathbf{u}_\eta:\mathcal{D}\rightarrow
T\mathcal{D}\mid \mathbf{u}_\eta(x)\in T_{\eta(x)}\mathcal{D}\},
\]
the tangent space to $\mathcal{F}(\mathcal{D},\mathcal{O})$ at $\chi$ is
\[
T_\chi\mathcal{F}(\mathcal{D},\mathcal{O})=\{\nu_\chi:\mathcal{D}\rightarrow
T\mathcal{O}\mid \nu_\chi(x)\in T_{\chi(x)}\mathcal{O}\}.
\]
A direct computation shows that the tangent map of right translation is 
\[
TR_{(\varphi,\psi)}(\mathbf{u}_\eta,\nu_\chi)=(\mathbf{u}_\eta\circ\varphi,TR_\psi(\nu_\chi\circ\varphi)).
\]
For simplicity we fix a volume form $\mu$ on $\mathcal{D}$. Therefore we can
identify the cotangent space $T_\eta^*\operatorname{Diff}(\mathcal{D})$ with a
space of one-forms over $\eta$, that is,
\[
T_\eta^*\operatorname{Diff}(\mathcal{D})=\{\mathbf{m}_\eta:\mathcal{D}\rightarrow
T^*\mathcal{D}\mid\mathbf{m}_\eta(x)\in T^*_{\eta(x)}\mathcal{D}\}.
\]
The cotangent space of $\mathcal{F}(\mathcal{D},\mathcal{O})$ at $\chi$ is
naturally given by
\[
T^*_\chi\mathcal{F}(\mathcal{D},\mathcal{O})=\{\kappa_\chi:\mathcal{D}\rightarrow
T^*\mathcal{O}\mid \kappa_\chi(x) \in  T^*_{\chi(x)} \mathcal{O}\}.
\]
Using these identifications, the cotangent map of right translation is computed
to be
\[
T^*R_{(\varphi,\psi)}(\mathbf{m}_\eta,\kappa_\chi)=J(\varphi^{-1})\left(\mathbf{m}_\eta\circ\varphi^{-1},T^*R_{\psi\circ\varphi^{-1}}(\kappa_\chi\circ\varphi^{-1})\right),
\]
where $J(\varphi ^{-1}) $ is the Jacobian determinant of the diffeomorphism
$\varphi^{-1}$,
and the corresponding cotangent lift, defined by
$R^{T^*}_{(\varphi,\psi)}:=T^*R_{(\varphi,\psi)^{-1}}$, is given by
\[
R^{T^*}_{(\varphi,\psi)}(\mathbf{m}_\eta,\kappa_\chi)=J(\varphi)(\mathbf{m}_\eta\circ\varphi,T^*R_{\psi^{-1}}(\kappa_\chi\circ\varphi)).
\]
The Lie algebra $\mathfrak{g}$ of the semidirect product group is
\[
\mathfrak{g}=\mathfrak{X}(\mathcal{D})\,\circledS\,\mathcal{F}(\mathcal{D},\mathfrak{o}),
\]
and the Lie bracket is computed to be
\[
\operatorname{ad}_{(\mathbf{u},\nu)}(\mathbf{v},\zeta)=(\operatorname{ad}_{\mathbf{u}}\mathbf{v},\operatorname{ad}_{\nu}\zeta+\mathbf{d}\nu\cdot\mathbf{v}-\mathbf{d}\zeta\cdot\mathbf{u}),
\]
where $\operatorname{ad}_{\mathbf{u}}\mathbf{v}=-[\mathbf{u},\mathbf{v}]$,
$\operatorname{ad}_{\nu}\zeta\in\mathcal{F}(\mathcal{D},\mathfrak{o})$ is given
by $\operatorname{ad}_{\nu}\zeta(x):=\operatorname{ad}_{\nu(x)}\zeta(x)$, and
$\mathbf{d}\nu\cdot\mathbf{v}\in\mathcal{F}(\mathcal{D},\mathfrak{o})$ is given
by $\mathbf{d}\nu\cdot\mathbf{v}(x):=\mathbf{d}\nu(x)(\mathbf{v}(x))$.

Using the previous identification of cotangent spaces, the dual Lie algebra
$\mathfrak{g}^*$ can be identified with
\[
\Omega^1(\mathcal{D})\,\circledS\,\mathcal{F}(\mathcal{D},\mathfrak{o}^*)
\]
through the pairing
\[
\langle(\mathbf{m},\kappa),(\mathbf{u},\nu)\rangle=\int_\mathcal{D}\left(\mathbf{m}\cdot\mathbf{u}+\kappa\cdot\nu\right)\mu.
\]
The dual map to $\operatorname{ad}_{(\mathbf{u},\nu)}$ is 
\begin{equation}\label{ad*}
\operatorname{ad}^*_{(\mathbf{u},\nu)}(\mathbf{m},\kappa)=\big({\boldsymbol{\pounds}}_\mathbf{u}\mathbf{m}+(\operatorname{div}\mathbf{u})\mathbf{m}+\kappa\cdot\mathbf{d}\nu,\operatorname{ad}^*_\nu\kappa+\operatorname{div}(\mathbf{u}\kappa)\big).
\end{equation}
This formula needs some explanation. The symbol
$\kappa\cdot\mathbf{d}\nu\in\Omega^1(\mathcal{D})$ denotes the 
one-form defined by 
\[
\kappa\cdot\mathbf{d}\nu(v_x):=\kappa(x)(\mathbf{d}\nu(v_x))
\]
and $\operatorname{ad}^*_\nu\kappa\in\mathcal{F}(\mathcal{D},\mathfrak{o}^*)$
denotes the $\mathfrak{o}^*$-valued mapping defined by
\[
\operatorname{ad}^*_\nu\kappa(x):=\operatorname{ad}^*_{\nu(x)}(\kappa(x)).
\]
The expression $\mathbf{u}\kappa$ denotes the $1$-contravariant tensor field
with values in $\mathfrak{o}^*$ defined by
\[
\mathbf{u}\kappa(\alpha_x):=\alpha_x(\mathbf{u}(x))\kappa(x)\in\mathfrak{o}^*.
\]
Since $\mathbf{u}\kappa $ is a generalization of the notion of a vector field,
we denote by $\mathfrak{X}(\mathcal{D},\mathfrak{o}^*)$ the space of all
$1$-contravariant tensor fields with values in $\mathfrak{o}^*$. In
\eqref{ad*},
$\operatorname{div}(\mathbf{u})$ denotes the divergence of the vector field
$\mathbf{u}$ with respect to the fixed volume form $\mu$. Recall that it is
defined by the condition
\[
(\operatorname{div}\mathbf{u})\mu={\boldsymbol{\pounds}}_\mathbf{u}\mu.
\]
This operator can be naturally extended to the space
$\mathfrak{X}(\mathcal{D},\mathfrak{o}^*)$ as follows. For
$w\in\mathfrak{X}(\mathcal{D},\mathfrak{o}^*)$ we write $w=w_a\varepsilon^a$
where $(\varepsilon^a)$ is a basis of $\mathfrak{o}^*$ and
$w_a\in\mathfrak{X}(\mathcal{D})$. We define
$\operatorname{div}:\mathfrak{X}(\mathcal{D},\mathfrak{o}^*)\rightarrow\mathcal{F}(\mathcal{D},\mathfrak{o}^*)$
by the equality
\[
\operatorname{div} w:=(\operatorname{div}w_a)\varepsilon^a.
\]
Note that if $w=\mathbf{u}\kappa$ then  
\[
\operatorname{div}(\mathbf{u}\kappa)=\textbf{d}\kappa\cdot\mathbf{u}+(\operatorname{div}\mathbf{u})\kappa.
\]

\medskip  

\noindent \textbf{The space of advected quantities.}
In physical applications, the affine representation space $V^*$ of
$G=\operatorname{Diff}(\mathcal{D})\,\circledS\,\mathcal{F}(\mathcal{D},\mathcal{O})$
is a direct product $V_1^*\oplus V^*_2$, where $V_i^*$ are subspaces of the
space of all tensor fields on $\mathcal{D}$ (possibly with values in a vector
space). Moreover:
\begin{itemize}
\item $V_1^*$ is only acted upon by the component
$\operatorname{Diff}(\mathcal{D})$ of $G$.
\item The action of $G$ on $V_2^*$ is affine, with the restriction that the
affine term only depends on the second component
$\mathcal{F}(\mathcal{D},\mathcal{O})$ of $G$.
\end{itemize}

In this way, we obtain the affine representation 
\begin{equation}\label{affine_representation}
(a,\gamma)\in V_1^*\oplus V^*_2\mapsto (a\eta,\gamma(\eta,\chi)+C(\chi))\in
V_1^*\oplus V^*_2,
\end{equation}
where $\gamma(\eta,\chi)$ denotes the representation of $(\eta,\chi)\in G $ on 
$\gamma\in V^*_2$, and
$C\in\mathcal{F}(\mathcal{F}(\mathcal{D},\mathcal{O}),V^*_2)$ satisfies the
identity
\begin{equation}\label{cocyclecondition}
C((\chi\circ\varphi)\psi)=C(\chi)(\varphi,\psi)+C(\psi).
\end{equation}
Note that this is equivalent to say that the representation $\rho$ and the
affine term $c$ of the previous section have the particular form
\[
\rho^*_{(\eta,\chi)^{-1}}(a,\gamma)=(a\eta,\gamma(\eta,\chi))\quad\text{and}\quad
c(\eta,\chi)=(0,C(\chi)).
\]

The infinitesimal action of $(\mathbf{u}, \nu) \in \mathfrak{g}$ on $\gamma\in
V_2^*$ induced by the representation of $G$ on $V_2^*$ is
\begin{align*}
\gamma(\mathbf{u},\mathbf{\nu}):&=\left.\frac{d}{dt}\right|_{t=0}\gamma(\operatorname{exp}(t\mathbf{u}),\operatorname{exp}(t\nu))=\left.\frac{d}{dt}\right|_{t=0}\gamma(\operatorname{exp}(t\mathbf{u}),e)(e,\operatorname{exp}(t\nu))\\
&=\left.\frac{d}{dt}\right|_{t=0}\gamma(\operatorname{exp}(t\mathbf{u}),e)+\left.\frac{d}{dt}\right|_{t=0}\gamma(e,\operatorname{exp}(t\nu))=:\gamma\mathbf{u}+\gamma\nu.
\end{align*}
Therefore, for $(v,w)\in V_1\oplus V_2$ we have
\[
(v,w)\diamond (a,\gamma)=(v\diamond a+w\diamond_1\gamma,w\diamond_2\gamma),
\]
where $\diamond_1$ and $\diamond_2$ are associated to the induced
representations of the first and second component of $G$ on $V^*_2$. On the
right hand side, the diamond operation $\diamond$ is associated to the
representation of $\operatorname{Diff}(\mathcal{D})$ on $V_1^*$. The space
$V_1^*$ is naturally the dual of some space $V_1$ of tensor fields on
$\mathcal{D}$. For example the $(p,q)$ tensor fields are naturally in duality
with the $(q,p)$ tensor fields. For $a\in V_1^*$ and $v\in V_1$, the duality
pairing is given by
\[
\langle a,v\rangle=\int_\mathcal{D}(a\cdot v)\mu,
\]
where $\cdot$ denotes the contraction of tensor fields.

Since the affine cocycle has the particular form $c(\eta,\chi)=(0,C(\chi))$, we
obtain that
\[
\mathbf{d}c^T(v,w)=(0,\mathbf{d}C^T(w)).
\]
For a Lagrangian
$l=l(\mathbf{u},\nu,a,\gamma):[\mathfrak{X}(\mathcal{D})\,\circledS\,\mathcal{F}(\mathcal{D},\mathfrak{o})]\,\circledS\,[V_1^*\oplus
V_2^*]\to\mathbb{R}$, the affine Euler-Poincar\'e equations \eqref{AEP} become
\begin{equation}
\label{AEP_PCF}
\left\lbrace
\begin{array}{ll}
\vspace{0.1cm}\displaystyle\frac{\partial}{\partial t}\frac{\delta
l}{\delta\mathbf{u}}=-{\boldsymbol{\pounds}}_\mathbf{u}\frac{\delta
l}{\delta\mathbf{u}}-(\operatorname{div}\mathbf{u})\frac{\delta
l}{\delta\mathbf{u}}-\frac{\delta l}{\delta\nu}\cdot\mathbf{d}\nu+\frac{\delta
l}{\delta a}\diamond a+\frac{\delta l}{\delta\gamma}\diamond_1\gamma\\
\displaystyle\frac{\partial}{\partial t}\frac{\delta
l}{\delta\nu}=-\operatorname{ad}^*_\nu\frac{\delta
l}{\delta\nu}-\operatorname{div}\left(\mathbf{u}\frac{\delta
l}{\delta\nu}\right)+\frac{\delta
l}{\delta\gamma}\diamond_2\gamma-\mathbf{d}C^T\left(\frac{\delta
l}{\delta\gamma}\right),
\end{array} \right.
\end{equation}
and the advection equations are
\begin{equation}\label{advection_PCF}
\left\lbrace
\begin{array}{ll}

\dot{a}+a\mathbf{u}=0\\
\dot{\gamma}+\gamma\mathbf{u}+\gamma\nu+\mathbf{d}C(\nu)=0.\\
\end{array}\right.
\end{equation}

\noindent \textsc{Remark on the duality pairing.} We remind the reader that, for simplicity, we have assumed that $\mathcal{D}$ is orientable and we have also fixed a volume form $\mu\in \Omega^n(\mathcal{D})$. Thus, using the $L^2$ pairing, the dual space of the Lie algebra $\mathfrak{X}(\mathcal{D})$ of vector fields is identified with the one-forms $\Omega^1(\mathcal{D})$. Therefore, the functional derivative $\delta l/\delta\textbf{u}$ is a one-form on $\mathcal{D}$. 

The natural dual of $\mathfrak{X}( \mathcal{D}) $ is the space $\Omega^1(\mathcal{D})\otimes\mathcal{D}en(\mathcal{D})$ of one-form densities on $\mathcal{D}$. In this general approach, valid also for non-orientable manifolds, the duality pairing of $\alpha\otimes\omega\in \Omega^1(\mathcal{D})\otimes\mathcal{D}en(\mathcal{D})$ with $\mathbf{u}$ reads
\[
\int_\mathcal{D}(\alpha\!\cdot\!\mathbf{u})\omega.
\]
Then the functional derivative $\delta l/\delta\textbf{u}$ is interpreted as a one-form density on $\mathcal{D}$.  The same remark applies for the dual of vector valued tensors on $\mathcal{D}$, in particular for the Lie algebra $\mathcal{F}(\mathcal{D},\mathfrak{o})$.

We chose to work on oriented manifolds and to avoid the use of $\Omega^1(\mathcal{D})\otimes\mathcal{D}en(\mathcal{D})$ because this general case would considerably complicate the writing of various formulas and equations, without adding any essential new information. We emphasize that all results of this paper are also valid for non-orientable manifolds by appealing to the usual exterior differential calculus and the Stokes theorem for twisted $k$-forms on $\mathcal{D}$ (see \cite{AbMaRa1989}, Supplement 7.2A for details). \quad $\blacklozenge$

\medskip  

\noindent \textbf{Basic example.} Take $V_2^*:=\Omega^1(\mathcal{D},\mathfrak{o})$,
the
space of all one-forms on $\mathcal{D}$ with values in $\mathfrak{o}$. This
space is naturally the dual of the space
$V_2=\mathfrak{X}(\mathcal{D},\mathfrak{o}^*)$ of contravariant tensor fields
with values in $\mathfrak{0}^*$, the duality pairing being given, for
$\gamma\in\Omega^1(\mathcal{D},\mathfrak{o})$ and
$w\in\mathfrak{X}(\mathcal{D},\mathfrak{o}^*)$, by
\[
\langle\gamma,w\rangle:=\int_\mathcal{D}(\gamma\cdot w)\mu,
\]
where $\gamma\cdot w$ denotes the contraction of tensors.

We consider for \eqref{affine_representation} the affine representation defined
by
\begin{equation}\label{affine_representation_gamma}
(a,\gamma)\mapsto(a\eta,\operatorname{Ad}_{\chi^{-1}}\eta^*\gamma+\chi^{-1}T\chi),
\end{equation}
where $\operatorname{Ad}_{\chi^{-1}}\eta^*\gamma+\chi^{-1}T\chi$ is the
$\mathfrak{o}$-valued one-form given by
\[
\left(\operatorname{Ad}_{\chi^{-1}}\eta^*\gamma+\chi^{-1}T\chi\right)(v_x):=\operatorname{Ad}_{\chi(x)^{-1}}(\eta^*\gamma(v_x))+\chi(x)^{-1}T_x\chi(v_x),
\]
for $v_x\in T_x\mathcal{D}$. One can check that
$\gamma(\eta,\chi):=\operatorname{Ad}_{\chi^{-1}}\eta^*\gamma$ is a right
representation of $G$ on $V_2^*$ and that $C(\chi)=\chi^{-1}T\chi$ verifies the
condition \eqref{cocyclecondition}. In fact,
\eqref{affine_representation_gamma}
corresponds to the action of the automorphism group of the trivial principal
bundle $\mathcal{O}\times\mathcal{D}$, on the space of connections.

For this example we have
\[
\gamma\mathbf{u}={\boldsymbol{\pounds}}_\mathbf{u}\gamma,\quad\gamma\nu=-\operatorname{ad}_\nu\gamma\quad\text{and}\quad\mathbf{d}C(\nu)=\mathbf{d}\nu,
\]
where $\operatorname{ad}_\nu\gamma\in\Omega^1(M,\mathfrak{o})$ denotes the
one-form given by
\[
\left(\operatorname{ad}_\nu\gamma\right)(v_x):=\operatorname{ad}_{\nu(x)}(\gamma(v_x))=[\nu(x),\gamma(v_x)],
\]
and $\mathbf{d}\nu\in\Omega^1(\mathcal{D},\mathfrak{o})$ is given by
$\mathbf{d}\nu(v_x):=T_x\nu(v_x)\in\mathfrak{o}$.

A direct computation shows that
\begin{align*}
w\diamond_1\gamma&=(\operatorname{div}
w)\cdot\gamma-w\cdot\mathbf{i}_{\_\,}\mathbf{d}\gamma \in \Omega^1(\mathcal{D})
,\\
w\diamond_2\gamma&=-\operatorname{Tr}(\operatorname{ad}^*_\gamma w)
\in\mathcal{F}(\mathcal{D},\mathfrak{o}^*),\\
\mathbf{d}C^T(w)&=-\operatorname{div} w
\in\mathcal{F}(\mathcal{D},\mathfrak{o}^*),
\end{align*}
where $\operatorname{Tr}$ denotes the trace of the $\mathfrak{o}^\ast$-valued
$(1,1)$ tensor
\[
\operatorname{ad}^*_\gamma w: T^*\mathcal{D}\times
T\mathcal{D}\rightarrow\mathfrak{o}^*,\quad(\alpha_x,v_x)\mapsto
\operatorname{ad}^*_{\gamma(v_x)}(w(\alpha_x)).
\]
In coordinates we have $\operatorname{Tr}(\operatorname{ad}^*_\gamma
w)=\operatorname{ad}^*_{\gamma_i}w^i$. Making use of these computations, the
affine Euler-Poincar\'e equations \eqref{AEP_PCF} become
\begin{equation}
\label{EPComplexFluid}
\left\lbrace
\begin{array}{ll}
\vspace{0.1cm}\displaystyle\frac{\partial}{\partial t}\frac{\delta
l}{\delta\mathbf{u}}=-{\boldsymbol{\pounds}}_\mathbf{u}\frac{\delta
l}{\delta\mathbf{u}}-(\operatorname{div}\mathbf{u})\frac{\delta
l}{\delta\mathbf{u}}-\frac{\delta l}{\delta\nu}\cdot\mathbf{d}\nu+\frac{\delta
l}{\delta a}\diamond a\\
\vspace{0.1cm}\displaystyle \qquad \qquad  
+\left(\operatorname{div} \frac{\delta
l}{\delta\gamma}\right)\cdot\gamma-\frac{\delta
l}{\delta\gamma}\cdot\mathbf{i}_{\_\,}\mathbf{d}\gamma\\
\displaystyle\frac{\partial}{\partial t}\frac{\delta
l}{\delta\nu}=-\operatorname{ad}^*_\nu\frac{\delta
l}{\delta\nu}+\operatorname{div}\left(\frac{\delta
l}{\delta\gamma}-\mathbf{u}\frac{\delta
l}{\delta\nu}\right)-\operatorname{Tr}\left(\operatorname{ad}^*_\gamma\frac{\delta
l}{\delta\gamma}\right),
\end{array} \right.
\end{equation}
and the advection equations are
\[
\left\lbrace
\begin{array}{ll}
\dot{a}+a\mathbf{u}=0\\
\dot{\gamma}+{\boldsymbol{\pounds}}_\mathbf{u}\gamma-\operatorname{ad}_\nu\gamma+\mathbf{d}\nu=0.\\
\end{array}\right.
\]
These are, up to sign conventions, the equations for complex fluids as given in
\cite{Ho2002}.

We now rewrite these equations using the covariant differentiation associated
to
a connection.  A one-form $\gamma \in \Omega^1( \mathcal{D}, \mathfrak{o}) $ 
can be considered as a connection one-form on the trivial principal
$\mathcal{O}$-bundle $\mathcal{O} \times \mathcal{D} \rightarrow \mathcal{D}$,
namely, 
\begin{equation}
\label{connection_gamma}
(v_x,\xi_{\mathsf{h}}) \in T_x \mathcal{D} \times T_{\mathsf{h}} \mathcal{O}
\mapsto
\operatorname{Ad}_{\mathsf{h}^{-1}}(\gamma(x)(v_x)+TR_{\mathsf{h}^{-1}}(\xi_{\mathsf{h}}))
\in \mathfrak{o}.
\end{equation}
 The covariant differential associated to this principal connection will be
denoted by $\mathbf{d}^ \gamma$. Therefore, for a function
$\nu\in\mathcal{F}(\mathcal{D},\mathfrak{o})$, we have 
\begin{equation}
\label{gamma_covariant_derivative}
\mathbf{d}^\gamma\nu(\mathbf{v}):=\mathbf{d}\nu(\mathbf{v})+[\gamma(\mathbf{v}),\nu].
\end{equation}
The covariant divergence of $w\in\mathfrak{X}(\mathcal{D},\mathfrak{o}^*)$ is
the function
\begin{equation}
\label{gamma_covariant_divergence}
\operatorname{div}^\gamma
w:=\operatorname{div}w-\operatorname{Tr}(\operatorname{ad}^*_\gamma
w)\in\mathcal{F}(\mathcal{D},\mathfrak{o}^*),
\end{equation}
defined as minus the adjoint of the covariant differential, that is, 
\begin{equation}
\label{divergence_formula}
\int_\mathcal{D}\left(\mathbf{d}^\gamma\nu\cdot
w\right)\mu=-\int_\mathcal{D}\left(\nu\cdot\operatorname{div}^\gamma
w\right)\mu
\end{equation}
for all $\nu \in \mathcal{F}( \mathcal{D}, \mathfrak{o}) $. Note that the Lie
derivative of $\gamma\in\Omega^1(\mathcal{D},\mathfrak{o})$ can be written as
\begin{align}
\label{Covariant_Cartan}
{\boldsymbol{\pounds}}_\mathbf{u}\gamma(\mathbf{v})&=\mathbf{d}(\gamma(\mathbf{u}))(\mathbf{v})+\mathbf{i}_\mathbf{u}\mathbf{d}\gamma(\mathbf{v})\nonumber \\
&=\mathbf{d}^\gamma(\gamma(\mathbf{u}))(\mathbf{v})-[\gamma(\mathbf{v}),\gamma(\mathbf{u})]+\mathbf{d}\gamma^\gamma(\mathbf{u},\mathbf{v})-[\gamma(\mathbf{u}),\gamma(\mathbf{v})] \nonumber \\
&=\mathbf{d}^\gamma(\gamma(\mathbf{u}))(\mathbf{v})+\mathbf{i}_\mathbf{u}B(\mathbf{v}),
\end{align}
where
\[
B:=\mathbf{d}^\gamma\gamma=\mathbf{d}\gamma+[\gamma,\gamma],
\]
is the curvature of the connection induced by $\gamma$.

Note also that, using covariant differentiation, we have
\[
w\diamond_1\gamma=(\operatorname{div}
w)\cdot\gamma-w\cdot\mathbf{i}_{\_\,}\mathbf{d}\gamma=(\operatorname{div}^\gamma
w)\cdot\gamma-w\cdot\mathbf{i}_{\_\,}B.
\]
Therefore, in terms of $\mathbf{d}^\gamma, \operatorname{div}^\gamma$, and
$B=\mathbf{d}^\gamma\gamma$, the equations read
\begin{equation}\label{EPComplexFluid_covariant_form}
\left\lbrace
\begin{array}{ll}
\vspace{0.1cm}\displaystyle\frac{\partial}{\partial t}\frac{\delta
l}{\delta\mathbf{u}}=-{\boldsymbol{\pounds}}_\mathbf{u}\frac{\delta
l}{\delta\mathbf{u}}-(\operatorname{div}\mathbf{u})\frac{\delta
l}{\delta\mathbf{u}}-\frac{\delta l}{\delta\nu}\cdot\mathbf{d}\nu+\frac{\delta
l}{\delta a}\diamond a \\
\vspace{0.1cm}\displaystyle \qquad \qquad 
+\left(\operatorname{div}^\gamma\frac{\delta
l}{\delta\gamma}\right)\cdot\gamma-\frac{\delta
l}{\delta\gamma}\cdot\mathbf{i}_{\_\,}B\\
\displaystyle\frac{\partial}{\partial t}\frac{\delta
l}{\delta\nu}=-\operatorname{ad}^*_\nu\frac{\delta
l}{\delta\nu}-\operatorname{div}\left(\mathbf{u}\frac{\delta
l}{\delta\nu}\right)  +\operatorname{div}^\gamma\frac{\delta l}{\delta\gamma},
\end{array}\right.
\end{equation}
and
\[
\left\lbrace
\begin{array}{ll}
\dot{a}+a\mathbf{u}=0\\
\dot{\gamma}+\mathbf{d}^\gamma(\gamma(\mathbf{u}))+\mathbf{i}_\mathbf{u}B+\mathbf{d}^\gamma\nu=0.\\
\end{array}\right.
\]

\medskip  

\noindent \textbf{The $B$-representation.} In this example we want to reformulate the
reduction process as well as the equations of motion 
\eqref{EPComplexFluid_covariant_form} in terms of another set of variables,
namely, $(\mathbf{u}, \nu, a, B) $, where
$B=\mathbf{d}^\gamma\gamma=\mathbf{d}\gamma+[\gamma,\gamma]\in\Omega^2(\mathcal{D},\mathfrak{o})$
is the \textit{curvature} of $\gamma$, instead of $(\mathbf{u}, \nu,
a,\gamma)$.
As we shall see below, with this choice of variables the action
\eqref{affine_representation_gamma} becomes linear instead of affine. We shall
also assume that the Lagrangian $L $, and hence also $l $, depend on $\gamma$
only through $B$. To do this we shall use the standard Euler-Poincar\'e
reduction for semidirect products (Theorem \ref{EPSD}). 

Indeed, if $\gamma'=\operatorname{Ad}_{\chi^{-1}}\eta^*\gamma+\chi^{-1}T\chi$
then we have
$\mathbf{d}^{\gamma'}\gamma'=\operatorname{Ad}_{\chi^{-1}}\eta^*\mathbf{d}^\gamma\gamma$.
Thus the representation of
$\operatorname{Diff}(\mathcal{D})\,\circledS\,\mathcal{F}(\mathcal{D},\mathcal{O})$
on $V ^*_1 \oplus \Omega^2( \mathcal{D}, \mathfrak{o}) $ is
given by
\[
(a,B)\mapsto (a\eta,\operatorname{Ad}_{\chi^{-1}}\eta^*B).
\]
The associated infinitesimal action of
$(\mathbf{u},\nu)\in\mathfrak{X}(\mathcal{D})\,\circledS\,\mathcal{F}(\mathcal{D},\mathfrak{o})$
is
\[
(a,B)(\mathbf{u},\nu)=(a\mathbf{u},B(\mathbf{u},\nu))=(a\mathbf{u},{\boldsymbol{\pounds}}_\mathbf{u}B-\operatorname{ad}_\nu
B).
\]
The space $\Omega^2(\mathcal{D},\mathfrak{o})$ is, in a natural way, dual to
the
space $\Omega_2(\mathcal{D},\mathfrak{o}^*)$ of $2$-contravariant skew
symmetric
tensor fields with values in $\mathfrak{o}^*$. The duality pairing is given by
contraction and integration with respect to the fixed volume form $\mu$. More
generally, we can consider the space $\Omega_k(\mathcal{D},\mathfrak{o}^*)$ of
$k$-contravariant skew symmetric tensor fields with values in $\mathfrak{o}^*$
and we can define the divergence operators, $\operatorname{div},
\operatorname{div} ^ \gamma
:\Omega_k(\mathcal{D},\mathfrak{o}^*)\rightarrow\Omega_{k-1}(\mathcal{D},\mathfrak{o}^*)$,
to be minus the adjoint of the exterior derivatives $\mathbf{d}$ and
$\mathbf{d}^\gamma$, respectively. For example, $\operatorname{div}^ \gamma$ is
defined on $\Omega_k(\mathcal{D}, \mathfrak{o}^\ast) $ by
\begin{equation}
\label{divergence_connection_formula}
\int _ \mathcal{D} \left(\mathbf{d}^\gamma \alpha \cdot  \omega \right) \mu  =
-
\int_ \mathcal{D} \left(\alpha\cdot \operatorname{div}^ \gamma \omega \right)
\mu,
\end{equation}
where $\alpha\in \Omega^{k-1}( \mathcal{D}, \mathfrak{o}) $ and $\omega \in
\Omega_{k} ( \mathcal{D}, \mathfrak{o}^\ast)$
Note that we have used the notations
$\Omega_1(\mathcal{D},\mathfrak{o}^*)=\mathfrak{X}(\mathcal{D},\mathfrak{o}^*)$
and
$\Omega_0(\mathcal{D},\mathfrak{o}^*)=\mathcal{F}(\mathcal{D},\mathfrak{o}^*)$.

Using the duality pairing defined before, for $(v,b)\in
V_1\,\oplus\,\Omega_2(\mathcal{D},\mathfrak{o}^*)$ and $(a,B)\in
V_1^*\,\oplus\,\Omega^2(\mathcal{D},\mathfrak{o})$, the diamond operation is
given by
\[
(v,b)\diamond(a,B)=(v\diamond a+b\diamond_1 B,b\diamond_2 B),
\]
where
\[
b\diamond_1
B=(\operatorname{div}b)\cdot\mathbf{i}_{\_\,}B-b\cdot\mathbf{i}_{\_\,}\mathbf{d}B\in\Omega^1(\mathcal{D})
\]
and
\[
b\diamond_2
B=-\operatorname{Tr}(\operatorname{ad}^*_Bb)=-\operatorname{ad}^*_{B_{ij}}b^{ij}\in\mathcal{F}(\mathcal{D},
\mathfrak{o}^*).
\]
The Euler-Poincar\'e equation \eqref{EP} are in this case
\begin{equation}\label{EPComplexFluid_B_representation}
\left\lbrace
\begin{array}{ll}
\vspace{0.1cm}\displaystyle\frac{\partial}{\partial t}\frac{\delta
l}{\delta\mathbf{u}}=-{\boldsymbol{\pounds}}_\mathbf{u}\frac{\delta
l}{\delta\mathbf{u}}-(\operatorname{div}\mathbf{u})\frac{\delta
l}{\delta\mathbf{u}}-\frac{\delta l}{\delta\nu}\cdot\mathbf{d}\nu+\frac{\delta
l}{\delta a}\diamond a \\
\vspace{0.2cm} \qquad \qquad \displaystyle
+\left(\operatorname{div}\frac{\delta l}{\delta
B}\right)\cdot\mathbf{i}_{\_\,}B-\frac{\delta l}{\delta
B}\cdot\mathbf{i}_{\_\,}\mathbf{d}B\\
\displaystyle\frac{\partial}{\partial t}\frac{\delta
l}{\delta\nu}=-\operatorname{ad}^*_\nu\frac{\delta l}{\delta\nu}+\frac{\delta
l}{\delta
B}\diamond_2B-\operatorname{Tr}\left(\operatorname{ad}^*_B\frac{\delta
l}{\delta B}\right),
\end{array}\right.
\end{equation}
and the advection equations are
\[
\left\lbrace
\begin{array}{ll}
\dot{a}+a\mathbf{u}=0\\
\dot{B}+{\boldsymbol{\pounds}}_\mathbf{u}B-\operatorname{ad}_\nu B=0.\\
\end{array}\right.
\]

We now show that when $B$ is the curvature of the connection $\gamma$, that is,
$B=\mathbf{d}^\gamma\gamma$, then the affine Euler-Poincar\'e equations
\eqref{EPComplexFluid_covariant_form} imply  the standard Euler Poincar\'e
equations \eqref{EPComplexFluid_B_representation}. Define the map
\[
\Phi:\Omega^1(\mathcal{D},\mathfrak{o})\rightarrow\Omega^2(\mathcal{D},\mathfrak{o})
\quad \text{by} \quad \Phi(\gamma)=\mathbf{d}^\gamma\gamma
\]
and suppose that the Lagrangians $l^1(\mathbf{u},\nu,a,\gamma)$ and 
$l^2(\mathbf{u},\nu,a,B)$ are related by 
\[
l^2(\mathbf{u},\nu,a,\Phi(\gamma))=l^1(\mathbf{u},\nu,a,\gamma).
\]
 We have
$T_\gamma\Phi(\alpha)=\mathbf{d}\alpha+[\gamma,\alpha]+[\alpha,\gamma]=\mathbf{d}^\gamma\alpha$
and
$\Phi(\operatorname{Ad}_{\chi^{-1}}\eta^*\gamma+\chi^{-1}T\chi)=\operatorname{Ad}_{\chi^{-1}}\eta^*\Phi(\gamma)$.
Therefore, by taking the derivative, the infinitesimal action verifies
\[
B\mathbf{u}+B\nu=\mathbf{d}^\gamma(\gamma\mathbf{u}+\gamma\nu+\mathbf{d}\nu),
\]
or, explicitly,
\begin{equation}\label{link_infinitesimal_actions}
{\boldsymbol{\pounds}}_\mathbf{u}B-\operatorname{ad}_\nu
B=\mathbf{d}^\gamma({\boldsymbol{\pounds}}_\mathbf{u}\gamma-\operatorname{ad}_\nu\gamma+\mathbf{d}\nu).
\end{equation}
This proves that the advection equations for $\gamma$ implies the one for $B$.
Using the preceding formula and the definition of the diamond operation we find
that
\[
(\operatorname{div}^\gamma b)\diamond_1\gamma=-b\diamond_1
B \quad\text{and}\quad(\operatorname{div}^\gamma
b)\diamond_2\gamma+\operatorname{div}(\operatorname{div}^\gamma
b)=-b\diamond_2B,
\]
or, explicitly,
\[
\operatorname{div}^\gamma\left((\operatorname{div}^\gamma b)\gamma \right)
-\operatorname{div}^\gamma
b\cdot\mathbf{i}_{\_\,}B=-\operatorname{div}b\cdot\mathbf{i}_{\_\,}B+b\cdot\mathbf{i}_{\_\,}\mathbf{d}B
\]
and
\[
\operatorname{div}^\gamma(\operatorname{div}^\gamma
b)=\operatorname{Tr}(\operatorname{ad}^*_Bb).
\]
Using the equality
\[
-\operatorname{div}^\gamma\left(\frac{\delta l^2}{\delta
B}\right)\circ\Phi=\frac{\delta l^1}{\delta \gamma},
\]
we obtain
\begin{equation}\label{compatible_diamonds}
\frac{\delta l^1}{\delta \gamma}\diamond_1\gamma=\frac{\delta l^2}{\delta
B}\diamond_1B\quad\text{and}\quad\frac{\delta l^1}{\delta
\gamma}\diamond_2\gamma+\operatorname{div}\left(\frac{\delta l^1}{\delta
\gamma}\right)=\frac{\delta l^2}{\delta B}\diamond_2B.
\end{equation}
This proves that the affine Euler-Poincar\'e equations
\eqref{EPComplexFluid_covariant_form} imply the standard Euler-Poincar\'e
equations \eqref{EPComplexFluid_B_representation}.


\section{Affine and Magnetic Lie-Poisson Reduction}\label{AMLPR}

The goal of this section is to carry out a generalization of the standard
process of Lie-Poisson reduction for Lie groups, which is motivated by the
example of complex fluids. The only modification lies in the fact that the Lie
group $G$ acts on its cotangent bundle by a cotangent lift \textit{plus an
affine term}. The principal result of this section states that, under some
conditions, reducing a canonical symplectic form relative to a cotangent lift
with an affine term is equivalent to reduce a magnetic symplectic form relative
to the right-cotangent lift. At the reduced level, we obtain affine Lie-Poisson
brackets and affine coadjoint orbits, whose affine terms depend on the affine
term in the action.

Consider the cotangent lift $R^{T^*}_g$ of the right translation $R_g$ on a Lie
group $G$. Recall that $R^{T^*}_g$ is the right action of $G$ on $T^*G$ given
by
\[
R^{T^*}_g(\alpha_f)=T^*R_{g^{-1}}(\alpha_f).
\]
Consider the map $\Psi_g :T^*G\rightarrow T^*G$ defined by
\begin{equation}\label{affine_cot_lift}
\Psi_g(\alpha_f):=R^{T^*}_g(\alpha_f)+C_g(f),
\end{equation}
where $C:G\times G\rightarrow T^*G$ is a smooth map such that $C_g(f)\in
T^*_{fg}G$, for all $f,g\in G$. The map $\Psi_g$ is seen here as a modification
of the cotangent lift by an affine term $C$. The following lemma gives the
conditions guaranteeing that the map $\Psi_g$ is a right action.

\begin{lemma} Consider the map $\Psi_g$ defined in \eqref{affine_cot_lift}. The
following are equivalent.
\begin{enumerate}
\item[\bf{(i)}] $\Psi_g$ is a right action.
\item[\bf{(ii)}] For all $f,g,h\in G$, the affine term $C$ verifies the property
\begin{equation}\label{cocycle_property}
C_{gh}(f)=C_h(fg)+R^{T^*}_h(C_g(f)).
\end{equation}
\item[\bf{(iii)}] There exists a one-form $\alpha\in\Omega^1(G)$ such that
$C_g(f)=\alpha(fg)-R^{T^*}_g(\alpha(f))$.
\end{enumerate}
\end{lemma}
\textbf{Proof.} The equivalence between $\bf{(i)}$ and $\bf{(ii)}$ is a direct
computation. Suppose that $\bf{(ii)}$ holds. By setting $f=e$ in formula
\eqref{cocycle_property}, we obtain that $C_{gh}(e)=C_h(g)+R^{T^*}_h(C_g(e))$.
Therefore we can define the one-form $\alpha$ by $\alpha(g):=C_g(e)$, and we
have $C_h(g)=\alpha(gh)-R^{T^*}_h(\alpha(g))$. Conversely, suppose that
$\bf{(iii)}$ holds. Then a direct computation shows that \eqref{cocycle_property}
holds.$\qquad\blacksquare$

\medskip

We denote by $\mathcal{C}(G)$ the space of all maps $C : G\times G\rightarrow
T^*G,\;(g,f)\mapsto C_g(f)\in T^*_{fg}G$ verifying the property
\eqref{cocycle_property}.
Remark that given an affine term $C\in\mathcal{C}(G)$, the one-form $\alpha$ in
item $\bf{iii}$, is only determined up to a right-invariant one-form. Denoting
by $\Omega^1_R(G)$ the space of all right-invariant one-forms on $G$, we have
an
isomorphism between $\mathcal{C}(G)$ and $\Omega^1(G)/\Omega^1_R(G)$. This
space
is clearly isomorphic to the space $\Omega^1_0(G)$ of all one-forms $\alpha$ on
$G$ such that $\alpha(e)=0$. From now on, when we say that \textit{the}
one-form
$\alpha$ is associated to $C$, we shall always assume that $\alpha(e) = 0$
which
then guarantees the uniqueness of this one-form.

\medskip

In order to carry out the symplectic reduction associated to an affine action
$\Psi_g$ of the form \eqref{affine_cot_lift}, we make two crucial observations
(see Theorem \ref{affine_LiePoisson_reduction}). 
\begin{itemize}
\item Let $\alpha\in\Omega^1_0(G)$ be the one-form associated to $\Psi_g$ and
consider the associated fiber translation $t_\alpha$ on $T^*G$ defined by
\[
t_\alpha(\beta_f):=\beta_f-\alpha(f).
\]
Then $t_\alpha$, viewed as a map from the canonical cotangent bundle
$(T^*G,\Omega_{\rm can})$ to the magnetic cotangent bundle $(T^*G,\Omega_{\rm
can}-\pi^*_G\mathbf{d}\alpha)$, is a \textit{symplectic map}. Moreover,
$t_\alpha$ is equivariant with respect to the action $\Psi_g$ on
$(T^*G,\Omega_{\rm can})$ and the cotangent lift $R^{T^*}_g$ on
$(T^*G,\Omega-\pi^*_G\mathbf{d}\alpha)$.
\item Suppose that $\mathbf{d}\alpha$ is $G$-invariant. Then the action
$\Psi_g$
is \textit{symplectic} relative to the canonical symplectic form
$\Omega_{can}$.
\end{itemize}
From these observations we conclude that, under some conditions to be specified
later on, reducing the canonical cotangent bundle $(T^*G,\Omega_{\rm can})$
relative to an affine action of the form \eqref{affine_cot_lift} is equivalent
to reduce the magnetic cotangent bundle $(T^*G,\Omega_{\rm
can}-\pi^*_G\mathbf{d}\alpha)$ relative to the cotangent lift of right
translations. It is therefore useful to recall below some facts about the
reduction of magnetic cotangent bundle.

\medskip  

\noindent \textbf{Some facts about magnetic cotangent bundle reduction.}
We first recall the following result (Theorem 7.1.1 in \cite{MaMiOrPeRa2007})
about the existence of momentum maps associated to cotangent bundles with
magnetic terms.

\begin{theorem}\label{Magnetic_term} 
Let $\mathcal{B}$ be a closed two-form on a connected configuration manifold
$Q$. Let $\Phi : G\times Q\rightarrow Q$ be a free and proper right action
leaving the form $\mathcal{B}$ invariant. Consider the cotangent lift
$\Phi^{T^*}_g$ of the $G$-action $\Phi_g$ to the symplectic manifold
$(T^*Q,\Omega_{\rm can}-\pi^*_Q\mathcal{B})$, where $\Omega_{\rm can}$ is the
canonical symplectic form and $\pi_Q : T^*Q\rightarrow Q$ is the cotangent
bundle projection. Suppose that there is a smooth map $\phi : Q\rightarrow
\mathfrak{g}^*$ that satisfies
\[
\mathbf{i}_{\xi_Q}\mathcal{B}=\mathbf{d}\langle\phi,\xi\rangle
\]
for all $\xi\in\mathfrak{g}$. Then the following hold:
\begin{enumerate}
\item[\bf{(i)}] The map
$\mathbf{J}=\mathbf{J}_{\operatorname{can}}-\phi\circ\pi_Q$, where
$\mathbf{J}_{\operatorname{can}}$ is the standard momentum map for the
$G$-action relative to the canonical symplectic form, is a momentum map for the
cotangent lifted action of $G$ on $T^*Q$ with symplectic form $\Omega_{\rm
can}-\pi^*_Q\mathcal{B}$.
\item[\bf{(ii)}] The momentum map is, in general, not equivariant. Its
non-equivariance $\mathfrak{g}^*$-valued group one-cocycle $\sigma :
G\rightarrow\mathfrak{g}^*$ is given by
\begin{equation}\label{nonequivariance_cocycle}
\sigma(g)=-\phi(\Phi_g(q))+\operatorname{Ad}^*_g(\phi(q)),
\end{equation}
with the right hand side independent of $q\in Q$. The cocycle identity is
\[
\sigma(gh)=\operatorname{Ad}^*_h(\sigma(g))+\sigma(h).
\]
\end{enumerate}
\end{theorem}

Note that the $G$-invariance of $\mathcal{B}$ ensures that the action
$\Phi^{T^*}$ is symplectic relative to the symplectic form $\Omega_{\rm
can}-\pi_Q^*\mathcal{B}$. The momentum map $\mathbf{J} :
T^*Q\rightarrow\mathfrak{g}^*$ is equivariant relative to the right affine
action of $G$ on $\mathfrak{g}^*$ given by
\[
\theta_g(\lambda):=\operatorname{Ad}^*_g\lambda+\sigma(g).
\]

We turn now to the particular case when the configuration manifold is a Lie
group $G$ and the action is the cotangent lift of right translation, that is,
$\Phi^{T^*}_g=R^{T^*}_g$. We first recall from \cite{MaMiOrPeRa2007} (Theorem
7.2.1) the magnetic Lie-Poisson reduction theorem.

\begin{theorem} \label{Magnetic_Lie_Poisson} Consider a closed $G$-invariant
two-form $\mathcal{B}$ on $G$. The Poisson reduced space for the right
cotangent
lifted action of $G$ on $(T^*G,\Omega_{\rm can}-\pi_G^*\mathcal{B})$ is
$\mathfrak{g}^*$ with Poisson bracket given by
\[
\{f,g\}_\mathcal{B}(\mu)=\left\langle\mu,\left[\frac{\delta
f}{\delta\mu},\frac{\delta
g}{\delta\mu}\right]\right\rangle+\mathcal{B}(e)\left(\frac{\delta
f}{\delta\mu},\frac{\delta g}{\delta\mu}\right)
\]
for $f,g\in\mathcal{F}(\mathfrak{g}^*)$.
\end{theorem}
Note that for a Hamiltonian $h:\mathfrak{g}^*\to\mathbb{R}$, the corresponding
Hamiltonian vector field is given by
\[
X_h(\mu)=-\operatorname{ad}^*_{\frac{\delta
h}{\delta\mu}}\mu-\mathcal{B}(e)\left(\frac{\delta h}{\delta\mu},\cdot\right).
\]
When the magnetic term $\mathcal{B}$ is absent, we recover the standard
Lie-Poisson bracket on $\mathfrak{g}^*$. In this particular case, the
symplectic
reduced spaces $\mathbf{J}_R^{-1}(\mu)/G_\mu$ are symplectically diffeomorphic
to the coadjoint orbits $\mathcal{O}_\mu=\{\operatorname{Ad}^*_g(\mu)\mid g\in
G\}$, where $\mathbf{J}_R$ is the standard right momentum map for the
$G$-action
relative to the canonical symplectic form and $G_\mu$ is the coadjoint isotropy
group of $\mu\in\mathfrak{g}^*$. The symplectic diffeomorphism is induced by
the
map
\[
\varphi :
\mathbf{J}_R^{-1}(\mu)\rightarrow\mathcal{O}_\mu,\quad\varphi(\alpha_g)=T^*R_g(\alpha_g)=\operatorname{Ad}_{g^{-1}}^*\mu.
\]
When a magnetic term $\mathcal{B}$ is present, the existence of a momentum map
is not guaranteed, and we shall use the result of Theorem \ref{Magnetic_term}.
Recall that the infinitesimal generator of $\xi$ for the right action is the
left invariant extension of $\xi$; that is, $\xi_G=\xi^L$. Suppose that there
is
a smooth map $\phi : G\rightarrow \mathfrak{g}^*$ that satisfies
\[
\mathbf{i}_{\xi^L}\mathcal{B}=\mathbf{d}\langle\phi,\xi\rangle
\]
for all $\xi\in\mathfrak{g}$. Since $\phi$ is determined by this equation only
up to a constant, we can always impose the condition $\phi(e)=0$. From Theorem
\ref{Magnetic_term}, the map $\mathbf{J}=\mathbf{J}_R-\phi\circ\pi_G$ is a
momentum map for the cotangent lifted action of $G$ on $T^*G$ with symplectic
form $\Omega_{\rm can}-\pi^*_G\mathcal{B}$.  In this case, using the relation
\eqref{nonequivariance_cocycle}, the nonequivariance cocycle is simply given by
$\sigma=-\phi$. We denote by $G_\mu^{\,\sigma}$ the isotropy group of $\mu$
relative to the affine action
$\theta_g(\mu)=\operatorname{Ad}^*_g(\mu)+\sigma(g)$. The following result (see
\cite{MaMiOrPeRa2007}, Theorem 7.2.2) shows that the reduction of a magnetic
cotangent bundle of a Lie group at a given point $\mu\in\mathfrak{g}^*$ is
symplectically diffeomorphic with the affine coadjoint orbit
$\mathcal{O}^{\,\sigma}_\mu$ passing through $\mu$.

\begin{theorem}\label{Affine_coadjoint_orbit} Consider a closed $G$-invariant
two-form $\mathcal{B}$ on $G$, suppose that there is a smooth map $\phi :
G\rightarrow \mathfrak{g}^*$ that satisfies
\[
\mathbf{i}_{\xi^L}\mathcal{B}=\mathbf{d}\langle\phi,\xi\rangle
\]
for all $\xi\in\mathfrak{g}$, and consider the momentum map
$\mathbf{J}=\mathbf{J}_{\operatorname{can}}-\phi\circ\pi_Q$. Then for each
$\mu\in\mathfrak{g}^*$, the symplectic reduced space
$\mathbf{J}^{-1}(\mu)/G_\mu^{\,\sigma}$ is symplectically diffeomorphic to
\[
\mathcal{O}^{\,\sigma}_\mu=\left\{\theta_g(\mu)=\operatorname{Ad}^*_g\mu+\sigma(g)\mid
g\in G\right\},
\]
the affine orbit through $\mu$. The tangent space at
$\lambda=\theta_g^{\,\sigma}(\mu)\in\mathcal{O}^{\,\sigma}_\mu$ to
$\mathcal{O}^{\,\sigma}_\mu$ is given by
\[
T_\lambda\mathcal{O}^{\,\sigma}_\mu=\left\{\operatorname{ad}^*_\xi\lambda-\Sigma(\xi,\cdot)\mid\xi\in\mathfrak{g}\right\},
\]
where $\Sigma(\xi,\cdot):=-T_e\sigma(\xi)=-\mathcal{B}(e)(\xi,\cdot)$. The
symplectic structure on $\mathcal{O}^{\,\sigma}_\mu$ has the expression
\begin{align*}
\omega^+_\mathcal{B}(\lambda)\left(\operatorname{ad}^*_\xi\lambda\right.&-\Sigma(\xi,\cdot),\left.\operatorname{ad}^*_\eta\lambda-\Sigma(\eta,\cdot)
\right)\\
&=\langle\lambda,[\xi,\eta]\rangle-\Sigma(\xi,\eta),
\end{align*}
which we call the magnetic orbit symplectic form. 
\end{theorem}

The symplectic diffeomorphism between the symplectic reduced spaces and the
affine coadjoint orbits is constructed as follows. Consider the smooth map
$\varphi : \mathbf{J}^{-1}(\mu)\rightarrow\mathcal{O}^{\,\sigma}_\mu$ defined
for $\alpha_g\in\mathbf{J}^{-1}(\mu)$ by
\[
\varphi(\alpha_g):=\theta_{g^{-1}}(\mu)=\operatorname{Ad}^*_{g^{-1}}\mu+\sigma(g^{-1}).
\]
Then $\varphi$ is $G^{\,\sigma}_\mu$-invariant and induces a symplectic
diffeomorphism
\[
\overline{\varphi} :
(\mathbf{J}^{-1}(\mu)/G_\mu^{\,\sigma},\Omega_\mu)\rightarrow(\mathcal{O}^{\,\sigma}_\mu,\omega^+_\mathcal{B}).
\]
Note that we have $\varphi(\alpha_g)=T^*R_g(\alpha_g)$. Indeed, since
$\alpha_g\in\mathbf{J}^{-1}(\mu)$, we have $\mu=T^*L_g(\alpha_g)-\phi(g)$,
therefore we obtain
\begin{align*}
\varphi(\alpha_g)&=\operatorname{Ad}^*_{g^{-1}}\mu+\sigma(g^{-1})\\
&=T^*R_g(\alpha_g)+\operatorname{Ad}^*_{g^{-1}}(\sigma(g))+\sigma(g^{-1})\\
&=T^*R_g(\alpha_g).
\end{align*}
The general theory of symplectic reduction implies that the affine coadjoint
orbits $(\mathcal{O}^{\,\sigma}_\mu,\omega^+_\mathcal{B})$ are the symplectic
leaves of the Poisson manifold
$\left(\mathfrak{g}^*,\{\,,\}_\mathcal{B}\right)$, where $\{\,,\}_\mathcal{B}$
denotes the Poisson bracket in Theorem \ref{Magnetic_Lie_Poisson}.

The following proposition shows that when the magnetic term $\mathcal{B}$ is an
exact two-form, then the magnetic cotangent bundle is symplectomorphic to the
canonical symplectic cotangent bundle. Through this symplectomorphism, the
cotangent lift is transformed into an affine action. In the particular case of
a
Lie group $G$ acting on its cotangent bundle by right cotangent lift, this
affine action is of the form \eqref{affine_cot_lift}.

\begin{proposition}\label{magnetic_translation} Assume that all the hypotheses
of Theorem \ref{Magnetic_term} are satisfied and suppose that
$\mathcal{B}=\mathbf{d}\alpha$. Let $t_\alpha : (T^*Q,\Omega_{\rm
can})\rightarrow (T^*Q,\Omega_{\rm can}-\pi^*_Q\mathcal{B})$ be the fiber
translation defined by
\[
t_\alpha(\beta_q):=\beta_q-\alpha(q).
\]
Then the following hold:
\begin{enumerate}
\item[\bf{(i)}] $t_\alpha$ is a symplectic fiber translation.
\item[\bf{(ii)}] The symplectic action $\Psi$ on $(T^*Q,\Omega_{\rm can})$
induced
by $\Phi^{T^*}$ through the map $t_\alpha$, that is, $\Psi_g : = t _\alpha
^{-1}
\circ \Phi \circ t _\alpha$ for any $g \in G $, is the affine action given by
\[
\Psi_g(\beta_q)=\Phi^{T^*}_g(\beta_q)+C_g(q),\quad \text{where} \quad
C_g(q):=\alpha(\Phi_g(q))-\Phi^{T^*}(\alpha(q)).
\]
\item[\bf{(iii)}] A momentum map relative to the $G$-action $\Psi$ on
$(T^*Q,\Omega_{\rm can})$ is given by
\[
\mathbf{J}_\alpha=\mathbf{J}\circ t_\alpha=\mathbf{J}_{\operatorname{can}}\circ
t_\alpha-\phi\circ\pi_Q;
\]
its nonequivariance cocycle equals the nonequivariance cocycle of $\mathbf{J}$.
\end{enumerate}
\end{proposition}
\textbf{Proof.}  \textbf{(i)} See Proposition 6.6.2 in \cite{MaRa1999}.

\textbf{(ii)} We have 
\begin{align*}
\Psi_g(\beta_q)&=t_\alpha^{-1}(\Phi_g^{T^*}(t_\alpha(\beta_q)))=t_\alpha^{-1}(\Phi_g^{T^*}(\beta_q-\alpha(q)))\\
&=\Phi_g^{T^*}(\beta_q-\alpha(q))+\alpha(\Phi_g(q))\\
&=\Phi_g^{T^*}(\beta_q)+\alpha(\Phi_g(q))-\Phi_g^{T^*}(\alpha(q))\\
&=\Phi_g^{T^*}(\beta_q)+C_g(q).
\end{align*}
\textbf{(iii)} This is a consequence of the fact that
$\mathbf{J}_\alpha=\mathbf{J}\circ t_\alpha$. The nonequivariance one-cocycle
of
$\mathbf{J}_\alpha$ is
\begin{align*}
\sigma_\alpha(g)&=\mathbf{J}_\alpha(\Psi_g(\beta_q))-\operatorname{Ad}^*_g(\mathbf{J}_\alpha(\beta_q))\\
&=\mathbf{J}(\Phi_g^{T^*}(t_\alpha(\beta_q)))-\operatorname{Ad}^*_g(\mathbf{J}(t_\alpha(\beta_q)))\\
&=\sigma(g).\qquad\blacksquare
\end{align*}
Note that $\mathbf{d}\alpha$ is assumed to be $G$-invariant. When the one-form
$\alpha$ is also $G$-invariant, then the affine term $C$ vanishes. Therefore,
the most interesting case happens when $\alpha$ is not $G$-invariant but
$\mathbf{d}\alpha$ is.

\medskip  

\noindent \textbf{Affine Lie-Poisson reduction.}
We apply now the previous results concerning the reduction with magnetic terms
to our initial problem, that is, the Lie-Poisson reduction of the canonical
cotangent bundle $(T^*G,\Omega_{\rm can})$ with respect to an affine action of
the form \eqref{affine_cot_lift}.
We obtain below, as an easy consequence of the previous theorems, the main
result of this section.

\medskip

\begin{theorem}\label{affine_LiePoisson_reduction} 
Consider the symplectic manifold $(T^*G,\Omega_{\rm can})$, and the affine
action
\[
\Psi_g(\beta_f):=R^{T^*}_g(\beta_f)+C_g(f),
\]
where $C\in\mathcal{C}(G)$. Let $\alpha\in\Omega^1_0(G)$ be the one-form
associated to $\Psi_g$. Then the following hold:
\begin{enumerate}
\item[\bf{(i)}] The fiber translation $t_\alpha : (T^*G,\Omega_{\rm
can})\rightarrow (T^*G,\Omega_{\rm can}-\pi^*_G\mathbf{d}\alpha)$ is a
symplectic map. The action induced by $\Psi_g$ on $(T^*G,\Omega_{\rm
can}-\pi^*_G\mathbf{d}\alpha)$ through $t_\alpha$ is simply the cotangent lift
$R^{T^*}_g$.
\item[\bf{(ii)}] Suppose that $\mathbf{d}\alpha$ is $G$-invariant. Then the
action
$\Psi_g$ is symplectic relative to the canonical symplectic form $\Omega_{\rm
can}$.
\item[\bf{(iii)}] Suppose that there is a smooth map $\phi : G\rightarrow
\mathfrak{g}^*$ that satisfies
\[
\mathbf{i}_{\xi^L}\mathbf{d}\alpha=\mathbf{d}\langle\phi,\xi\rangle
\]
for all $\xi\in\mathfrak{g}$. Then the map $\mathbf{J}_\alpha=\mathbf{J}_R\circ
t_\alpha-\phi\circ\pi_G$ is a momentum map for the action $\Psi_g$ relative to
the canonical symplectic form. We can always choose $\phi$ such that
$\phi(e)=0$. In this case, the nonequivariance one-cocycle of
$\mathbf{J}_\alpha$ is $\sigma=-\phi$.
\item[\bf{(iv)}] The symplectic reduced space
$(\mathbf{J}_\alpha^{-1}(\mu)/G_\mu^{\,\sigma},\Omega_\mu)$ is symplectically
diffeomorphic to the affine coadjoint orbit
$(\mathcal{O}^{\,\sigma}_\mu,\omega_\mathcal{B}^+)$, the symplectic
diffeomorphism being induced by the $G_\mu^{\,\sigma}$-invariant smooth map
\[
\psi :
\mathbf{J}_\alpha^{-1}(\mu)\rightarrow\mathcal{O}^{\,\sigma}_\mu,\quad\psi(\alpha_g):=\Psi_{g^{-1}}(\alpha_g).
\]
\end{enumerate}
\end{theorem}
\textbf{Proof.} 
\textbf{(i)} That $t_\alpha$ is a symplectic map follows from item $\bf{(i)}$ in
Proposition \ref{magnetic_translation}. From item $\bf{ii}$ in Proposition
\ref{magnetic_translation}, we know that the action induced on
$(T^*G,\Omega_{\rm can})$ by the right cotangent lift on $(T^*G,\Omega_{\rm
can}-\pi^*_Q\mathbf{d}\alpha)$ through the map $t_\alpha$ is the affine action
whose affine term is given by $\alpha$. This is precisely the action $\Psi_g$.
Thus we conclude that $\Psi_g$ induces the right cotangent lifted action on
$(T^*G,\Omega_{can}-\pi^*_Q\mathbf{d}\alpha)$.

\textbf{(ii)} When $\mathbf{d}\alpha$ is $G$-invariant, we know that the right
cotangent lift is symplectic relative to $\Omega-\pi^*_Q\mathbf{d}\alpha$.
Since
$t_\alpha$ is symplectic, we conclude the result.

\textbf{(iii)} This follows from item $\bf{i}$ and $\bf{ii}$ in Theorem
\ref{Magnetic_term}. Since $\phi$ can be chosen modulo a constant term, we can
impose the condition $\phi(e)=0$. From the relation
\eqref{nonequivariance_cocycle} we obtain the equality $\phi=-\sigma$.

\textbf{(iv)} The symplectic diffeomorphism $t_\alpha$ induces a symplectic
diffeomorphism between the reduced spaces. Therefore, by Theorem
\ref{Affine_coadjoint_orbit}, we obtain that the point reduced space
$(\mathbf{J}_\alpha^{-1}(\mu)/G_\mu^{\,\sigma},\Omega_\mu)$ is symplectically
diffeomorphic to the affine coadjoint orbit
$(\mathcal{O}^{\,\sigma}_\mu,\omega_\mu^{\,\sigma})$, the symplectic
diffeomorphism being induced by the map $\psi:=\varphi\circ t_\alpha :
\mathbf{J}_\alpha^{-1}(\mu)\rightarrow\mathcal{O}^{\,\sigma}_\mu$. We have
\begin{align*}
\psi(\beta_g)&=\varphi(t_\alpha(\beta_g))
=T^*R_g(\beta_g-\alpha(g))
=R^{T^*}_{g^{-1}}(\beta_g)-R^{T^*}_{g^{-1}}(\alpha(g))\\
&=R^{T^*}_{g^{-1}}(\beta_g)+C_{g^{-1}}(g)
=\Psi_g(\beta_g),
\end{align*}
where in the fourth equality we used
$C_g(f)=\alpha(fg)-R^{T^*}_g(\alpha(f)).\qquad\blacksquare$
\medskip

The affine coadjoint orbits $ \left( \mathcal{O} ^\sigma _\mu , \omega _\sigma ^+ \right) $ are symplectic leaves in the affine Lie-Poisson space $ \left( \mathfrak{g}^\ast , \{\,,\} _\sigma ^+ \right) $, where 
\[
\{f,g\} _\sigma ^+( \mu ) = \left\langle \mu \left[ \frac{\delta f}{ \delta \mu}, 
\frac{\delta g}{ \delta \mu} \right]  \right\rangle  - \Sigma \left( \frac{\delta f}{ \delta \mu}, \frac{\delta g}{ \delta \mu} \right) .
\]
Note that this bracket is the Lie-Poisson bracket on the Poisson submanifold $ \mathfrak{g}^\ast \times  \{1\} \subset  \widehat{ \mathfrak{g} } ^\ast  : = \mathfrak{g}^\ast \times  \mathbb{R} $, where $ \widehat{ \mathfrak{g} } $ is the one-dimensional central extension of $ \mathfrak{g} $ defined by the cocycle $ - \Sigma $. 

Assume that $ - \Sigma $ integrates to a group two-cocycle $ B : G \times  G \rightarrow  \mathbb{R} $, that is, 
\[
- \Sigma ( \xi , \eta ) = \left.\frac{d^2}{dt ds}\right|_{t=s=0} \left( B( g (t) , h (s) ) - B ( h (s) , g (t) )  \right) ,
\] 
where $ t \mapsto g (t) $ and $ s \mapsto h (s) $ are smooth curves through $ e \in  G $ with tangent vectors $ \xi = \left.\frac{d g (t) }{dt}\right|_{t=0} $ and $ \eta = \left.\frac{dh (s) }{ds}\right|_{s=0} $. Let $ \widehat{G} $ be the central extension of $ G $ defined by the two-cocycle $ B $ and recall that the Lie algebra of $ \widehat{G} $ equals $ \widehat{\mathfrak{g} } $.  Then the affine coadjoint orbit $ \mathcal{O} _\mu ^\sigma $ is obtained by usual Lie-Poisson reduction of $ T ^\ast \widehat{G} $ relative to the lift of right translation at $ ( \mu , 1) $. See \cite{MaMiOrPeRa2007}, \S6.2 for more details.


\section{Affine Hamiltonian Semidirect Product Theory}\label{AHSPT}

This is the Hamiltonian version of Section \ref{Lagrangian_formulation}. More
precisely, we carry out the Poisson and symplectic reductions of a canonical
cotangent bundle $(T^*S,\Omega_{\rm can})$, where $S=G\,\circledS\,V$ is the
semidirect product of a Lie group $G$ and a vector space $V$ and where $S$ acts
on its cotangent bundle by cotangent lift \textit{plus an affine term}. We will
see that this process is a particular case of the theory developed in the
previous section.

Consider the semidirect product Lie group $S:=G\,\circledS\,V$ associated to a
right representation $\rho : G\rightarrow\operatorname{Aut}(V)$. The cotangent
lift of the right translation is given by
\begin{align*}
R^{T^*}_{(g,v)}(\alpha_f,(u,a)):&=T_{(f,u)(g,v)}^*R_{(g,v)^{-1}}(\alpha_f,(u,a))\\
&=(T_{fg}^*R_{g^{-1}}(\alpha_f),v+\rho_g(u),\rho_{g^{-1}}^*(a))\\
&=(R^{T^*}_g(\alpha_f),v+\rho_g(u),\rho_{g^{-1}}^*(a))\in
T^*_{(f,u)(g,v)}S.
\end{align*}
We modify this cotangent lifted action by an affine term of the form
\begin{equation}\label{C}
C_{(g,v)}(f,u):=(0_{fg},v+\rho_g(u),c(g)),
\end{equation}
for a group one-cocycle $c\in\mathcal{F}(G,V^*)$, that is, verifying the
property $c(fg)=\rho_{g^{-1}}^*(c(f))+c(g)$, as in Section
\ref{Lagrangian_formulation}. The resulting affine right action on $T^*S$ is
therefore given by
\begin{align}\label{Psi}
\Psi_{(g,v)}(\alpha_f,(u,a)):&=R^{T^*}_{(g,v)}(\alpha_f,(u,a))+C_{(g,v)}(f,u)\nonumber\\
&=(R^{T^*}_g(\alpha_f),v+\rho_g(u),\rho_{g^{-1}}^*(a)+c(g))
\end{align}
This action is clearly of the form \eqref{affine_cot_lift}. We now check that
property \eqref{cocycle_property} holds. This will prove that $\Psi_{(g,v)}$ is
a right action. Indeed, 
\begin{align*}
&C_{(h,w)}((f,u)(g,v))+R^{T^*}_{(h,w)}(C_{(g,v)}(f,u))\\
&=C_{(h,w)}(fg,v+\rho_g(u))+R^{T^*}_{(h,w)}(0_{fg},v+\rho_g(u),c(g))\\
&=(0_{fgh},w+\rho_h(v+\rho_g(u)),c(h))+(R^{T^*}_h(0_{fg}),w+\rho_h(v+\rho_g(u)),\rho_{h^{-1}}^*(c(g)))\\
&=(0_{fgh},w+\rho_h(v+\rho_g(u)),c(h)+\rho_{h^{-1}}^*(c(g)))\\
&=\left(0_{fgh},w+\rho_h(v)+\rho_{gh}(u),c(gh)\right)\\
&=C_{(gh,w+\rho_h(v))}(f,u)=C_{(g,v)(h,w)}(f,u).
\end{align*}

In the following lemmas, we compute the one-form $\alpha\in\Omega^1_0(S)$
associated to $C$ and we show that it verifies the hypotheses of Theorem
\ref{affine_LiePoisson_reduction}. Recall that $\alpha$ is defined by
$\alpha(g,v):=C_{(g,v)}(e,0)$.

\begin{lemma}\label{dalpha} The one-form $\alpha \in \Omega^1_0(S) $ associated
to the affine term \eqref{C} is given by
\begin{equation}
\label{group_alpha}
\alpha(g,v)(\xi_g,(v,u))=\langle c(g),u\rangle,
\end{equation}
for $(\xi_g,(v,u))\in T_{(g,v)}S$. Moreover $\mathcal{B}:=\mathbf{d}\alpha$ is
$S$-invariant and its value at the identity is given by
\begin{equation}
\label{group_b}
\mathcal{B}(e,0)((\xi,u),(\eta,w))=\langle\mathbf{d}c(\xi),w\rangle-\langle
\mathbf{d}c(\eta),u\rangle.
\end{equation}
\end{lemma}
\textbf{Proof.} For $(\xi_g,(v,u))\in T_{(g,v)}S$ we have
$\alpha(g,v):=C_{(g,v)}(e,0)=(0_g,v,c(g))$. Therefore we obtain the equality
$\alpha(g,v)(\xi_g,(v,u))=\langle c(g),u\rangle$. We now prove the
right-invariance of $\mathbf{d}\alpha$. For $(\xi,u)\in\mathfrak{s}$ define the
associated right-invariant vector field $(\xi,u)^R\in\mathfrak{X}(S)$ by
\[
(\xi,u)^R(g,v):=TR_{(g,v)}(\xi,u)=(\xi^R(g),v,\rho_g(u)),
\]
where $\xi^R(g):=T_eR_g(\xi)$. Note that we have
$\alpha\left((\xi,u)^R\right)(g,v)=\langle c(g),\rho_g(u)\rangle=-\langle
c(g^{-1}),u\rangle$. Therefore
\begin{align*}
R_{(g,v)}^*&\mathbf{d}\alpha(e,0)((\xi,u),(\eta,w))=\mathbf{d}\alpha(g,v)(TR_{(g,v)}(\xi,u),TR_{(g,v)}(\eta,w))\\
&=\mathbf{d}\alpha(g,v)\left((\xi,u)^R(g,v),(\eta,w)^R(g,v)\right)\\
&=\mathbf{d}\left(\alpha\left((\eta,w)^R\right)\right)(g,v)\left((\eta,w)^R(g,v)\right) \\
&\qquad -\mathbf{d}\left(\alpha\left((\eta,w)^R\right)\right)(g,v)\left((\eta,w)^R(g,v)\right)
-\alpha\left(\left[(\xi,u)^R,(\eta,w)^R\right]\right)(g,v)\\
&=\left.\frac{d}{dt}\right|_{t=0}\alpha\left((\eta,w)^R\right)(\operatorname{exp}(t\xi)g,v+\rho_g(ut))\\
& \qquad -\left.\frac{d}{dt}\right|_{t=0}\alpha\left((\xi,u)^R\right)(\operatorname{exp}(t\eta)g,v+\rho_g(wt))\\
&\qquad+\alpha\left(([\xi,\eta],u\eta-w\xi)^R\right)(g,v)\\
&=-\left.\frac{d}{dt}\right|_{t=0}\left\langle
c(g^{-1}\operatorname{exp}(t\xi)^{-1}),w\right\rangle+\left.\frac{d}{dt}\right|_{t=0}\left\langle
c(g^{-1}\operatorname{exp}(t\eta)^{-1}),u\right\rangle\\
&\qquad+\alpha(g,v)\left([\xi,\eta]^R(g),\rho_g(u\eta-w\xi)\right)\\
&=\langle c(g^{-1})\xi+\mathbf{d}c(\xi),w\rangle-\langle
c(g^{-1})\eta+\mathbf{d}c(\eta),u\rangle-\langle c(g^{-1}),u\eta-w\xi\rangle\\
&=\langle\mathbf{d}c(\xi),w\rangle-\langle
\mathbf{d}c(\eta),u\rangle=\mathbf{d}\alpha(e,0)((\xi,u),(\eta,w)),
\end{align*}
where in the fourth equality we used the identity
\[
\left[(\xi,u)^R,(\eta,w)^R\right]=-\left[(\xi,u),(\eta,w)\right]^R=-([\xi,\eta],u\eta-w\xi)^R,
\]
and for the sixth equality the identity
\begin{align*}
\left.\frac{d}{dt}\right|_{t=0}\left\langle
c(g^{-1}\operatorname{exp}(t\xi)^{-1}),w\right\rangle&=\left.\frac{d}{dt}\right|_{t=0}\left\langle
\rho^*_{\operatorname{exp}(t\xi)}(c(g^{-1}))+c(\operatorname{exp}(t\xi)^{-1}),w\right\rangle\\
&=-\langle c(g^{-1})\xi+\mathbf{d}c(\xi),w\rangle.\qquad\blacksquare
\end{align*}

This lemma shows that hypothesis \textbf{(ii)} of Theorem
\ref{affine_LiePoisson_reduction} is verified. This implies that the action
$\Psi_{(g,v)}$ is symplectic with respect to the canonical symplectic form
$\Omega_{\rm can}$ on $T^*S$. We now check hypothesis \textbf{(iii)} of Theorem
\ref{affine_LiePoisson_reduction}. This implies that the action $\Psi_{(g,v)}$
admits a momentum map relative to the canonical symplectic form.

\begin{lemma}\label{phi} The map $\phi : S\rightarrow\mathfrak{s}^*$ defined by
\[
\phi(g,v)=(\mathbf{d}c^T(v)-v\diamond c(g),-c(g)),
\]
verifies the property
\[
\mathbf{i}_{(\xi,u)^L}\mathbf{d}\alpha=\mathbf{d}\langle\phi,(\xi,u)\rangle,
\]
where $(\xi,u)^L\in\mathfrak{X}(S)$ is the left-invariant vector field induced
by $(\xi,u)\in\mathfrak{s}$.
\end{lemma}
\textbf{Proof.} We will use the following formulas:
\begin{align*}\langle\phi(g,v),(\xi,u)\rangle&=-\langle
c(g),u\rangle+\langle v,c(g)\xi+\mathbf{d}c(\xi)\rangle,\\ 
(\xi,u)^L(g,v)&=TL_{(g,v)}(\xi,u)=(TL_g(\xi),v,u+v\xi),\\
TR_{(g,v)^{-1}}\left((\xi,u)^L(g,v)\right)&=(\operatorname{Ad}_g\xi,\rho_{g^{-1}}(u+v\xi)).
\end{align*}
By the right-invariance of $\mathbf{d}\alpha$, we have
\begin{align*}
\mathbf{i}_{(\xi,u)^L}\mathbf{d}\alpha&(g,v)\left((\eta,w)^L(g,v)\right)=\mathbf{d}\alpha(g,v)\left((\xi,u)^L(g,v),(\eta,w)^L(g,v)\right)\\
&=\mathbf{d}\alpha(e,0)\left((\operatorname{Ad}_g\xi,\rho_{g^{-1}}(u+v\xi)),(\operatorname{Ad}_g\eta,\rho_{g^{-1}}(w+v\eta))\right)\\
&=\langle\mathbf{d}c(\operatorname{Ad}_g\xi),\rho_{g^{-1}}(w+v\eta)\rangle-\langle\mathbf{d}c(\operatorname{Ad}_g\eta),\rho_{g^{-1}}(u+v\xi)\rangle\\
&=\langle T_gc(TL_g(\xi)),w+v\eta\rangle-\langle
T_gc(TL_g(\eta)),u+v\xi\rangle.
\end{align*}
For the last equality, we use that
\begin{align*}
\rho_{g^{-1}}^*(\mathbf{d}c(\operatorname{Ad}_g\xi))&=\rho_{g^{-1}}^*\left.\frac{d}{dt}\right|_{t=0}c(g\operatorname{exp}(t\xi)g^{-1})
=\left.\frac{d}{dt}\right|_{t=0}(c(g\operatorname{exp}(t\xi))-c(g))\\
&=T_gc(TL_g(\xi)).
\end{align*}
On the other hand we have
\begin{align*}
&\mathbf{d}\langle\phi,(\xi,u)\rangle(g,v)\left((\eta,w)^L(g,v)\right)=\left.\frac{d}{dt}\right|_{t=0}\langle\phi(g\operatorname{exp}(t\eta),wt+\rho_{\operatorname{exp}(t\eta)}(v)),(\xi,u)\rangle\\
&\quad=\left.\frac{d}{dt}\right|_{t=0}-\langle
c(g\operatorname{exp}(t\eta)),u\rangle+\langle
wt+\rho_{\operatorname{exp}(t\eta)}(v),c(g\operatorname{exp}(t\eta))\xi+\mathbf{d}c(\xi)\rangle\\
&\quad=-\langle T_gc(TL_g(\eta)),u\rangle+\langle
w+v\eta,c(g)\xi+\mathbf{d}c(\xi)\rangle+\left.\frac{d}{dt}\right|_{t=0}\langle
v,c(g\operatorname{exp}(t\eta))\xi\rangle\\
&\quad=-\langle T_gc(TL_g(\eta)),u\rangle+\langle
w+v\eta,T_gc(TL_g(\xi))\rangle-\langle v\xi, T_gc(TL_g(\eta))\rangle\\
&\quad=\langle T_gc(TL_g(\xi)), w+v\eta\rangle-\langle
T_gc(TL_g(\eta)),u+v\xi\rangle.
\end{align*}
Thus we obtain that
\[
\mathbf{i}_{(\xi,u)^L}\mathbf{d}\alpha=\mathbf{d}\langle\phi,(\xi,u)\rangle.\qquad\blacksquare
\]

\medskip  

\noindent \textbf{The momentum map.} By item $\bf{(iii)}$ of Theorem
\ref{affine_LiePoisson_reduction} and using $\alpha \in \Omega^1_0(S) $ given
by
\eqref{group_alpha}, we obtain that a momentum map for the right-action
\[
\Psi_{(g,v)}(\alpha_f,(u,a))=(R^{T^*}_g(\alpha_f),v+\rho_g(u),\rho_{g^{-1}}^*(a)+c(g))
\]
is given by
\begin{align}\label{momentum_map}
\mathbf{J}_\alpha&(\beta_f,(u,a))=\mathbf{J}_R(t_\alpha(\beta_f,(u,a)))-\phi(f,u)\nonumber\\
&=\mathbf{J}_R(\beta_f,(u,a-c(f)))-\phi(f,u)\nonumber\\
&=T^*L_{(f,u)}(\beta_f,(u,a-c(f)))-\phi(f,u)\nonumber\\
&=(T^*L_f(\beta_f)+u\diamond (a-c(f)),a-c(f))-(\mathbf{d}c^T(u)-u\diamond
c(f),-c(f))\nonumber\\
&=(T^*L_f(\beta_f)+u\diamond a-\mathbf{d}c^T(u),a),
\end{align}
with nonequivariance one-cocycle
\begin{equation}\label{nonequivariance}
\sigma(f,u)=-\phi(f,u)=(u\diamond c(f)-\mathbf{d}c^T(u),c(f))\in\mathfrak{s}^*.
\end{equation}

\medskip  

\noindent \textbf{Poisson bracket and Hamiltonian vector fields.} Using Theorem
\ref{Magnetic_Lie_Poisson} and the expression of $\mathcal{B}$ given in Lemma
\ref{dalpha}, we obtain that the reduced Poisson bracket on $\mathfrak{s}^*$ is
given by
\begin{align*}
\{f,g\}_\mathcal{B}(\mu,a)&=\left\langle(\mu,a),\left[\left(\frac{\delta
f}{\delta\mu},\frac{\delta f}{\delta a}\right),\left(\frac{\delta
g}{\delta\mu},\frac{\delta g}{\delta
a}\right)\right]\right\rangle \\
& \qquad +\mathcal{B}(e,0)\left(\left(\frac{\delta
f}{\delta\mu},\frac{\delta f}{\delta a}\right),\left(\frac{\delta
g}{\delta\mu},\frac{\delta g}{\delta a}\right)\right)\\
&=\left\langle\mu,\left[\frac{\delta f}{\delta\mu},\frac{\delta
g}{\delta\mu}\right]\right\rangle+\left\langle a,\frac{\delta f}{\delta
a}\frac{\delta g}{\delta\mu}-\frac{\delta g}{\delta a}\frac{\delta
f}{\delta\mu}\right\rangle\\
&\qquad+\left\langle\mathbf{d}c\left(\frac{\delta
f}{\delta\mu}\right),\frac{\delta g}{\delta
a}\right\rangle-\left\langle\mathbf{d}c\left(\frac{\delta
g}{\delta\mu}\right),\frac{\delta f}{\delta a}\right\rangle.
\end{align*}
Given a Hamiltonian function $h:\mathfrak{s}^*\to\mathbb{R}$, the corresponding
Hamiltonian vector field with respect to the bracket $\{\,,\}_\mathcal{B}$, is
given by
\begin{align*}
X_h(\mu,a)&=-\operatorname{ad}^*_{\left(\frac{\delta
h}{\delta\mu},\frac{\delta h}{\delta
a}\right)}(\mu,a)-\mathcal{B}(e,0)\left(\left(\frac{\delta
h}{\delta\mu},\frac{\delta h}{\delta a}\right),\cdot\right)\\
&=\left(-\operatorname{ad}^*_{\frac{\delta h}{\delta \mu}}\mu-\frac{\delta
h}{\delta a}\diamond a+\mathbf{d}c^T\left(\frac{\delta h}{\delta
a}\right),-a\frac{\delta h}{\delta \mu}-\mathbf{d}c\left(\frac{\delta h}{\delta
\mu}\right)\right).
\end{align*}
\medskip 

\noindent\textbf{The symplectic reduced spaces.} By item $\bf{iv}$ of Theorem
\ref{affine_LiePoisson_reduction}, the reduced space 
\[
\left(\mathbf{J}_\alpha^{-1}(\mu,a)/S^{\,\sigma}_{(\mu,a)},\Omega_{(\mu,a)}\right)
\]
is symplectically diffeomorphic to the affine coadjoint orbit
$\left(\mathcal{O}^{\,\sigma}_{(\mu,a)},\omega^+_\mathcal{B}\right)$. More
precisely, we have
\begin{align}\label{affine_coadj}
\mathcal{O}^{\,\sigma}_{(\mu,a)}&=\left\{\left.\operatorname{Ad}^*_{(g,u)}(\mu,a)+\sigma(g,u)\right|
(g,u)\in S\right\}\nonumber\\
&=\left\{\left.\left(\operatorname{Ad}^*_g\mu+u\diamond\theta_g(a) -\mathbf{d}c^T(u),\theta_g(a) \right)\right|
(g,u)\in S \right\},
\end{align}
where $\theta_g(a): = \rho_{g^{-1}}^*(a)+c(g)$ is the affine action of $G$ 
on $V ^\ast$. 
Note also that the bilinear form $\Sigma$ appearing in the formula of the
affine
orbit symplectic form (see Theorem \ref{Affine_coadjoint_orbit}), is given by
\begin{align*}
\Sigma((\xi,u),\cdot)&=-T_{(e,0)}\sigma(\xi,u)
=-\left.\frac{d}{dt}\right|_{t=0}\left(tu\diamond
c(\operatorname{exp}(t\xi))-\mathbf{d}c^T(tu),c(\operatorname{exp}(t\xi))\right)\\
&=(\mathbf{d}c^T(u),-\mathbf{d}c(\xi)),
\end{align*}
where $(\xi,u)\in\mathfrak{s}$. The tangent space to the affine coadjoint orbit
$\mathcal{O}^{\,\sigma}_{(\mu,a)}$ at $(\lambda,b)$ is equal to
\begin{align*}
T_{(\lambda,b)}\mathcal{O}^{\,\sigma}_{(\mu,a)}&=\left\{\left.\operatorname{ad}^*_{(\xi,u)}(\lambda,b)-\Sigma((\xi,u),\cdot)\right|
(\xi,u)\in\mathfrak{s}\right\}\\
&=\left\{\left.\left(\operatorname{ad}^*_\xi\lambda+u\diamond
b-\mathbf{d}c^T(u),b\xi+\mathbf{d}c(\xi)\right)\right|
(\xi,u)\in\mathfrak{s}\right\}.
\end{align*}
The symplectic form on $\mathcal{O}^{\,\sigma}_{(\mu,a)}$ is given by
\begin{align}\label{affineKKS}
\omega^+_\mathcal{B}(\lambda,b)&\left(\left(\operatorname{ad}^*_\xi\lambda+u\diamond
b-\mathbf{d}c^T(u),b\xi+\mathbf{d}c(\xi)\right),  
\phantom{\frac{A}{B}} \right. \nonumber \\
& \qquad  \left. \phantom{\frac{A}{B}}
\left(\operatorname{ad}^*_\eta\lambda+w\diamond
b-\mathbf{d}c^T(w),b\eta+\mathbf{d}c(\eta)\right)\right)\nonumber\\
&=\left\langle
(\lambda,b),\left[(\xi,u),(\eta,w)\right]\right\rangle-\Sigma((\xi,u),(\eta,w))\nonumber\\
&=\left\langle \lambda,[\xi,\eta]\right\rangle+\left\langle
b,u\eta-w\xi\right\rangle+\langle\mathbf{d}c(\eta),u\rangle-\langle\mathbf{d}c(\xi),w\rangle.
\end{align}
Recall that the affine coadjoint orbits are the symplectic leaves of the
Poisson
manifold $\left(\mathfrak{s}^*,\{\,,\}_\mathcal{B}\right)$.

We now apply these results to the main result of this section, that is, the
Hamiltonian counterpart of Theorem \ref{AEPSD}

Consider a Hamiltonian function $H: T ^\ast G \times V ^\ast \rightarrow
\mathbb{R}$ right-invariant under the $G$-action
\begin{equation}\label{affine_action}
(\alpha_h,a)\mapsto(R^{T^*}_g(\alpha_h),\theta_g(a)):=(R^{T^*}_g(\alpha_h),\rho_{g^{-1}}^*(a)+c(g)).
\end{equation}
This $G $-action on $T ^\ast G \times V ^\ast$ is induced by the $S $-action
\eqref{Psi}  on $T ^* S $ given by
\begin{align*}
\Psi_{(g,v)}(\alpha_h,(u,a))&=R^{T^*}_{(g,v)}(\alpha_h,(u,a))+C_{(g,v)}(h,u)\\
&=\left(R^{T^*}_g(\alpha_h),v+\rho_g(u),\rho_{g^{-1}}^*(a)+c(g)\right).
\end{align*}
Note also that we can think of this Hamiltonian $H:T^*G\times V^*\to\mathbb{R}$
as being the Poisson reduction of a $S$-invariant Hamiltonian
$\overline{H}:T^*S\to\mathbb{R}$ by the normal subgroup $\{e\} \times V $ since
$(T ^\ast S)/(\{e\} \times V)  \cong T ^\ast G \times V^\ast$.

In particular, the
function $H_{a_0}: = H|_{T ^\ast G\times \{a_0\}}: T ^\ast G
\rightarrow \mathbb{R}$ is invariant under the induced action
of the isotropy subgroup $G_{a_0}^c$ of $a_0$ relative to the affine action
$\theta$, for any $a_0 \in V ^\ast$. Recall that $\theta_g(a): = 
\rho_{g^{-1}}^*(a)+c(g)$ for any $g \in G$ and $a \in V^\ast$. The following
theorem is a generalization of Theorem \ref{LPSD} and is also a consequence of 
the reduction
by stages method for nonequivariant momentum maps, together with the results
obtained in section \ref{AMLPR} and at the beginning of the present section.

\begin{theorem}\label{ALPSD}
For $\alpha(t)\in T^*_{g(t)}G$ and
$\mu(t):=T_e^*R_{g(t)}(\alpha(t))\in\mathfrak{g}^*$, the following are
equivalent:
\begin{enumerate}
\item[\bf{(i)}] $\alpha(t)$ satisfies Hamilton's equations for
$H_{a_0}$ on $T^*G$.
\item[\bf{(ii)}] The following affine Lie-Poisson equation holds on
$\mathfrak{s}^*$:
\[
\frac{\partial}{\partial t}(\mu,a)=\left(-\operatorname{ad}^*_{\frac{\delta
h}{\delta \mu}}\mu-\frac{\delta h}{\delta a}\diamond
a+\mathbf{d}c^T\left(\frac{\delta h}{\delta a}\right),-a\frac{\delta h}{\delta
\mu}-\mathbf{d}c\left(\frac{\delta h}{\delta \mu}\right)\right),
\]
$a(0)=a_0 $, where $\mathfrak{s}$ is the semidirect product Lie algebra
$\mathfrak{s}=\mathfrak{g}\,\circledS\, V$. The associated Poisson bracket is
the following affine Lie-Poisson bracket on the dual $\mathfrak{s}^*$
\begin{align*}
\{f,g\}_\mathcal{B}(\mu,a)&=\left\langle\mu,\left[\frac{\delta
f}{\delta\mu},\frac{\delta g}{\delta\mu}\right]\right\rangle+\left\langle
a,\frac{\delta f}{\delta a}\frac{\delta g}{\delta\mu}-\frac{\delta g}{\delta
a}\frac{\delta f}{\delta\mu}\right\rangle\\
&\qquad+\left\langle\mathbf{d}c\left(\frac{\delta
f}{\delta\mu}\right),\frac{\delta g}{\delta
a}\right\rangle-\left\langle\mathbf{d}c\left(\frac{\delta
g}{\delta\mu}\right),\frac{\delta f}{\delta a}\right\rangle.
\end{align*}
\end{enumerate}
As on the Lagrangian side, the evolution of the advected quantities is given by
$a(t)=\theta_{g(t)^{-1}}(a_0)$.
\end{theorem}
\textbf{Proof.} Recall that the momentum map relative to the canonical
symplectic form on $T^*S$ and to the action 
\[
\Psi_{(g,v)}(\beta_f,(u,a))=\left(R^{T^*}_g(\beta_f),v+\rho_g(u),\rho_{g^{-1}}^*(a)+c(g)\right)
\]
is given by
\[
\mathbf{J}_\alpha(\beta_f,(u,a))=(T^*L_f(\beta_f)+u\diamond
a-\mathbf{d}c^T(u),a).
\]
The action $\Psi_{(g,v)}$ of $S$ induces an action of $V$ given by
\[
(\beta_f,(u,a))\mapsto(\beta_f,(v+u,a)).
\]
Since $V$ is a closed normal subgroup of $S$, this action admits a momentum map
given by
\[
\mathbf{J}_V(\beta_f,(u,a))=a.
\]
Since $V$ is an Abelian group, the coadjoint isotropy group of $a\in V^*$ is
$V_{a}=V$ and the first reduced space $(T^*S)_{a_a}=\mathbf{J}^{-1}(a)/V$ is
symplectically diffeomorphic to the canonical symplectic manifold
$(T^*G,\Omega_{\rm can})$. The action $\Psi$ of $S$ on $T^*S$ restricts to an
action $\Psi^a$ of $G_a^c\,\circledS\,V$ on $\mathbf{J}_V^{-1}(a)$, where
\[
G^c_a:=\{g\in G\mid\rho^*_{g^{-1}}(a)+c(g)=a\}.
\]
 Passing to the quotient
spaces, this action induces an action of $G^c_a$ on $(T^*S)_a$, which is
readily
seen to be the cotangent lifted action of $G^c_a$ on $T^*G$. We denote by
$\mathbf{J}_a : (T^*S)_a\to(\mathfrak{g}^c_a)^*$ the associated equivariant
momentum map, where $\mathfrak{g}_a^c$ is the Lie algebra of $G_a^c$. Reducing
$(T^*S)_a$ at the point $\mu_a:=\mu|\mathfrak{g}^c_a$, we get the second
reduced
space $\left((T^*S)_a\right)_{\mu_a}=\mathbf{J}_a^{-1}(\mu_a)/(G_a^c)_{\mu_a}$,
with symplectic form denoted by $(\Omega_a)_{\mu_a}$.

By the Reduction by Stages Theorem for nonequivariant momentum maps, the second
reduced space is symplectically diffeomorphic to the reduced space
\[
\left(\mathbf{J}_\alpha^{-1}(\mu,a)/S^\sigma_{(\mu,a)},\Omega_{(\mu,a)}\right)
\]
obtained by reducing $T^*S$ by the whole group $S$ at the point
$(\mu,a)\in\mathfrak{s}^*$. Here $S^\sigma_{(\mu,a)}$ denotes the isotropy
group
of the affine coadjoint action with cocycle $\sigma$ given in
\eqref{nonequivariance}.

As shown at the beginning of this section, this reduced space is symplectically
diffeomorphic to the affine coadjoint orbit $\left(\mathcal{O}^\sigma_{(\mu,a)},\omega_\mathcal{B}^+\right)$
endowed with the affine orbit symplectic symplectic form described in
\eqref{affineKKS}. These affine coadjoint orbits are the symplectic leaves of
the Poisson manifold $(\mathfrak{s}^*,\{\,,\}_\mathcal{B})$. 

Note finally that we can consider the right-invariant Hamiltonian $H:T^*G\times
V^*\to\mathbb{R}$, as coming from a $S$-invariant Hamiltonian
$\overline{H}:T^*S\to\mathbb{R}$.

The theorem is then a consequence of all these
observations.$\qquad\blacksquare$

\medskip  

\noindent \textbf{Reconstruction of dynamics.} We give now some details concerning the
passage from the reduced formulation $({\bf ii})$ to the 
canonical formulation $({\bf i})$. Let $(\mu(t) ,a(t))\in\mathfrak{g}^*\times
V^*$ be the solution of the affine Lie-Poisson equations, with initial
condition
$(\mu_0,a_0)$. Then the curve $\alpha(t):=T^*R_{g(t)^{-1}}(\mu(t))$, where
$g(t)\in G$ satisfies the linear ordinary differential equation with
time-dependent coefficients
\[
\dot g(t)=TR_{g(t)}\left(\frac{\delta h}{\delta \mu}\right),\quad g(0)=e,
\]
is the solution of Hamilton's equations associated to $H_{a_0}$ on $T^*G$ and
with initial condition $\alpha(0)=\mu_0$. Note that for the curve $g(t)$ define
above, we have $a(t)=\theta_{g(t)^{-1}}(a_0)$.

\medskip

The preceding theorem is compatible with Theorem \ref{AEP}. Indeed, we can
start
with a Lagrangian $L_{a_0}:TG\rightarrow\mathbb{R}$ as in 
Section \ref{Lagrangian_formulation}, that is, we have a function $L:TG\times
V^*\rightarrow\mathbb{R}$ which is right $G$-invariant under the affine action
$(v_h,a)\mapsto
(T_hR_g(v_h),\theta_g(a))=(T_hR_g(v_h),\rho^*_{g^{-1}}(a)+c(g))$, such that
$L_{a_0}(v_g)=L(v_g,a_0)$. Then $L_{a_0}$ is right invariant under the lift to
$TG$ of the right action of $G_{a_0}^c$ on $G$. Suppose that the Legendre
transformation $\mathbb{F}L_{a_0}$ is
invertible and form the corresponding Hamiltonian
$H_{a_0}=E_{a_0}\circ\mathbb{F}L_{a_0}^{-1}$, where $E_{a_0}$ is the energy of
$L _{a_0}$. Then the function $H: T^\ast G \times V ^\ast
\rightarrow \mathbb{R}$ so defined is $S $-invariant and one can apply this
theorem. At the level of the reduced space, to a reduced Lagrangian
$l:\mathfrak{g}\times V^*\rightarrow\mathbb{R}$ we associate the reduced
Hamiltonian $h:\mathfrak{g}^*\,\circledS\,V^*\rightarrow\mathbb{R}$ given by
\[
h(\mu,a):=\langle\mu,\xi\rangle-l(\xi,a),\quad\mu=\frac{\delta l}{\delta\xi}.
\]
Since
\[
\frac{\delta h}{\delta\mu}=\xi\quad\text{and}\quad\frac{\delta h}{\delta
a}=-\frac{\delta l}{\delta a},
\]
we see that the affine Lie-Poisson equations for $h$ on $\mathfrak{s}^*$ are
equivalent to the affine Euler-Poincar\'e equation \eqref{AEP} for $l$ together
with the affine advection equation 
\[
\dot{a}+a\xi+\mathbf{d}c(\xi)=0.
\]


\section{Hamiltonian Approach to Continuum Theories of Perfect Complex
Fluids}\label{Hamiltonian_PCF}

This section is the Hamiltonian version of Section \ref{Lagrangian_PCF}. Recall
that in Section \ref{Lagrangian_PCF} we have applied Theorem \ref{AEPSD} to the
case of complex fluids. Here we apply the Hamiltonian analogue of this theorem,
namely Theorem \ref{ALPSD}.

Recall that for complex fluids we apply the abstract theory to the Lie group 
$G=\operatorname{Diff}(\mathcal{D})\,\circledS\,\mathcal{F}(\mathcal{D},\mathcal{O})$
and $V^*=V_1^*\oplus V_2^*$. The representations of $G$ on $V$ and $V^*$ are of
the form
\[
\rho_{(\eta,\chi)}(v,w)=(v\eta,w(\eta,\chi)),\quad\text{and}\quad\rho^*_{(\eta,\chi)^{-1}}(a,\gamma)=(a\eta,\gamma(\eta,\chi)).
\]
This implies that the infinitesimal actions of $\mathfrak{g}$ on $V$ and $V^*$
are of the form 
\[
(v,w)(\mathbf{u},\nu)=(v\mathbf{u},w\mathbf{u}+w\nu) \quad \text{and} \quad
(a,\gamma)(\mathbf{u},\nu)=(a\mathbf{u},\gamma\mathbf{u}+\gamma\nu).
\]
 We therefore obtain the diamond operation 
 \[
 (v,w)\diamond (a,\gamma)=(v\diamond a+w\diamond_1\gamma,w\diamond_2\gamma).
 \]
  Since the affine term has the particular form $c(\eta,\chi)=(0,C(\chi))$, we
obtain the equalities 
  \[
  \mathbf{d}c(\mathbf{u},\nu)=(0,\mathbf{d}C(\nu)) \quad \text{and} \quad
\mathbf{d}c^T(v,w)=(0,\mathbf{d}C^T(w)).
  \]

We now compute some useful expressions in the particular case of complex fluids
by using the general formulas of section \ref{AHSPT}.

By Theorem \ref{ALPSD}, we obtain that the affine Lie-Poisson bracket for
complex fluids is 
\begin{align*}
\{f,g\}(\mathbf{m},\kappa,a,\gamma)&=\int_\mathcal{D}\mathbf{m}\cdot\left[\frac{\delta
f}{\delta\mathbf{m}},\frac{\delta g}{\delta\mathbf{m}}\right]\mu\\
&\quad+\int_\mathcal{D}\kappa\cdot\left(\operatorname{ad}_{\frac{\delta
f}{\delta\kappa}}\frac{\delta g}{\delta\kappa}+\mathbf{d}\frac{\delta
f}{\delta\kappa}\cdot\frac{\delta g}{\delta\mathbf{m}}-\mathbf{d}\frac{\delta
g}{\delta\kappa}\cdot\frac{\delta f}{\delta\mathbf{m}}\right)\mu\\
&\quad+\int_\mathcal{D}a\cdot\left(\frac{\delta f}{\delta a}\frac{\delta
g}{\delta\mathbf{m}}-\frac{\delta g}{\delta a}\frac{\delta
f}{\delta\mathbf{m}}\right)\mu\\
&\quad+\int_\mathcal{D}\gamma\cdot\left(\frac{\delta
f}{\delta\gamma}\frac{\delta g}{\delta\mathbf{m}}+\frac{\delta
f}{\delta\gamma}\frac{\delta g}{\delta\kappa}-\frac{\delta
g}{\delta\gamma}\frac{\delta f}{\delta\mathbf{m}}-\frac{\delta
g}{\delta\gamma}\frac{\delta f}{\delta\kappa}\right)\mu\\
&\quad+\int_\mathcal{D}\left(\mathbf{d}C\left(\frac{\delta
f}{\delta\kappa}\right)\cdot\frac{\delta
g}{\delta\gamma}-\mathbf{d}C\left(\frac{\delta
g}{\delta\kappa}\right)\cdot\frac{\delta f}{\delta\gamma}\right)\mu.
\end{align*}
The first four terms give the Lie-Poisson bracket on the dual Lie algebra
\[
\left( \left[\mathfrak{X}( \mathcal{D}) \,\circledS\, \mathcal{F}( \mathcal{D},
\mathfrak{o}) \right] \,\circledS\, \left[V _1\oplus V _2 \right] \right)^\ast
\cong 
\Omega^1(\mathcal{D}) \times \mathcal{F}(\mathcal{D},\mathfrak{o}^*) \times
V^*_1\ \times V_2^*.
\]
The last term is due to the presence of the affine term $C$ in the
representation. Since $C$ depends only on the group
$\mathcal{F}(\mathcal{D},\mathcal{O})$, this term does not involve the
functional derivatives with respect to $\mathbf{m}$.

The symplectic leaves of this bracket are the affine coadjoint orbits in the
dual Lie algebra
$\left( \left[\mathfrak{X}( \mathcal{D}) \,\circledS\, \mathcal{F}(
\mathcal{D},
\mathfrak{o}) \right] \,\circledS\, \left[V _1\oplus V _2 \right] \right)^\ast
$. 
The expression of the tangent spaces and of the affine orbit symplectic forms
involves the bilinear form $\Sigma$ which is defined in this case on
$\left[\mathfrak{X}(\mathcal{D})\,\circledS\,\mathcal{F}(\mathcal{D},\mathfrak{o})\right]\,\circledS\,\left[V_1\,\oplus\,V_2\right]$
by
\[
\Sigma((\mathbf{u}_1,\nu_1,v_1,w_1),(\mathbf{u}_2,\nu_2,v_2,w_2))=\mathbf{d}C(\nu_2)\cdot
w_1-\mathbf{d}C(\nu_1)\cdot w_2
\] 

For a Hamiltonian $h=h(\mathbf{m},\kappa,a,\gamma):\Omega^1(\mathcal{D}) \times
\mathcal{F}(\mathcal{D},\mathfrak{o}^*) \times V^*_1\ \times V_2^*
\to\mathbb{R}$, the affine Lie-Poisson equations of Theorem \ref{ALPSD} become
\begin{equation}\label{ALP_PCF}
\left\{
\begin{array}{ll}
\vspace{0.1cm}\displaystyle\frac{\partial}{\partial
t}\mathbf{m}=-{\boldsymbol{\pounds}}_{\frac{\delta h}{\delta
\mathbf{m}}}\mathbf{m}-\operatorname{div}\left(\frac{\delta h}{\delta
\mathbf{m}}\right)\mathbf{m}-\kappa\cdot\mathbf{d}\frac{\delta h}{\delta
\kappa}-\frac{\delta h}{\delta a}\diamond a-\frac{\delta h}{\delta
\gamma}\diamond_1\gamma\\
\vspace{0.1cm}\displaystyle\frac{\partial}{\partial
t}\kappa=-\operatorname{ad}^*_{\frac{\delta h}{\delta
\kappa}}\kappa-\operatorname{div}\left(\frac{\delta h}{\delta
\mathbf{m}}\kappa\right)-\frac{\delta
h}{\delta\gamma}\diamond_2\gamma+\mathbf{d}C^T\left(\frac{\delta
h}{\delta\gamma}\right)\\
\vspace{0.1cm}\displaystyle\frac{\partial}{\partial t}a=-a\frac{\delta
h}{\delta
\mathbf{m}}\\
\vspace{0.1cm}\displaystyle\frac{\partial}{\partial
t}\gamma=-\gamma\frac{\delta
h}{\delta \mathbf{m}}-\gamma\frac{\delta h}{\delta
\kappa}-\mathbf{d}C\left(\frac{\delta h}{\delta \kappa}\right).
\end{array}
\right.
\end{equation}

As explained in the previous section, when the reduced Hamiltonian $h$ is
defined by a Lagrangian $l$ through the Legendre transformation 
\[
h(\mathbf{m},\kappa,a,\gamma)=\langle(\mathbf{m},\kappa),(\mathbf{u},\nu)\rangle-l(\mathbf{u},\nu,a,\gamma),\quad(\mathbf{m},\kappa)=\left(\frac{\delta
l}{\delta \mathbf{u}},\frac{\delta l}{\delta \nu}\right),
\]
we have
\[
\frac{\delta h}{\delta \mathbf{m}}=\mathbf{u},\quad \frac{\delta h}{\delta
\kappa}=\nu,\quad \frac{\delta h}{\delta a}=-\frac{\delta l}{\delta a},\quad
\frac{\delta h}{\delta \gamma}=-\frac{\delta l}{\delta \gamma}.
\]
Using these equalities, we see directly that the affine Lie-Poisson equations
\eqref{ALP_PCF} are equivalent to the affine Euler-Poincar\'e equations
\eqref{AEP_PCF}, together with the advection equations \eqref{advection_PCF}.

Using formula \eqref{momentum_map}, the momentum map of the affine right action of the semidirect product
$[\operatorname{Diff}(\mathcal{D})\,\circledS\,\mathcal{F}(\mathcal{D},\mathcal{O})]\,\circledS\,[V_1\oplus
V_2]$ on its cotangent bundle is computed to be
\begin{align*}
&\mathbf{J}_\alpha(\mathbf{m}_\eta,\kappa_\chi,(v,w),(a,\gamma))
=\left(T^*\eta\circ\mathbf{m}_\eta+T^*\chi\circ\kappa_\chi+v\diamond
a+w\diamond_1\gamma, \phantom{C^T} \right.\\
& \qquad  \qquad \qquad  \qquad \qquad  \qquad  \qquad
\left. T^*L_\chi\circ\kappa_\chi+w\diamond_2\gamma-\mathbf{d}C^T(w),(a,\gamma)\right).
\end{align*}
In this formula we need to elaborate on the meaning of the expression
$T^*\chi\circ\kappa_\chi \in \Omega^1( \mathcal{D})$. Thus, by definition, for
any $u _x\in T _x\mathcal{D}$ we set $\left\langle (T^*\chi\circ\kappa_\chi
)(x), u_x \right\rangle : = \left\langle T_x^*\chi \left(\kappa_\chi(x)\right)
,
u _x \right\rangle = \left\langle \kappa_\chi(x), T_x\chi (u _x) \right\rangle
$. In this last expression, recall that $\kappa_\chi: \mathcal{D} \rightarrow T
^\ast \mathcal{O}$ covers $\chi: \mathcal{D} \rightarrow \mathcal{O}$. 

By the general theory, the nonequivariance cocycle of $\mathbf{J}_\alpha$ is
given by $\sigma=-\phi$, where $\phi$ is computed (using Lemma \ref{phi}) to be
\[
\phi :
[\operatorname{Diff}(\mathcal{D})\,\circledS\,\mathcal{F}(\mathcal{D},\mathcal{O})]\,\circledS\,[V_1\oplus
V_2]\to\Omega^1(\mathcal{D}) \times \mathcal{F}(\mathcal{D},\mathfrak{o}^*)
\times V^*_1\ \times V_2^*,
\]
\begin{align*}
\phi(\eta,\chi,v,w)&=(\mathbf{d}c^T(v,w)-(v,w)\diamond
c(\eta,\chi),-c(\eta,\chi))\\
&=(-w\diamond_1 C(\chi),\mathbf{d}C^T(w)-w\diamond_2 C(\chi),0,-C(\chi)).
\end{align*}
One can also compute the momentum map $\mathbf{J}_{(a_0,\gamma_0)}$ appearing
in
the proof of Theorem \ref{ALPSD} at the second stage of reduction. It is
associated to the cotangent lifted action of the isotropy group
\[
\left(\operatorname{Diff}(\mathcal{D})\,\circledS\,\mathcal{F}(\mathcal{D},\mathcal{O})\right)_{(a_0,\gamma_0)}=\{(\eta,\chi)\mid
(a_0\eta,\gamma_0(\eta,\chi)+C(\chi))=(a_0,\gamma_0)\}
\]
on the canonical cotangent bundle
$T^*\left(\operatorname{Diff}(\mathcal{D})\,\circledS\,\mathcal{F}(\mathcal{D},\mathcal{O})\right)$.
It is given by
\begin{align}\label{second_stage_momentum_map}
&\mathbf{J}_{(a_0,\gamma_0)}(\mathbf{m}_\eta,\kappa_\chi)=\left(T^*\eta\circ\mathbf{m}_\eta+T^*\chi\circ\kappa_\chi,T^*L_\chi\circ\kappa_\chi\right).
\end{align}

\medskip  

\noindent \textbf{Basic example.} As in the example given in section
\ref{Lagrangian_PCF}, we consider the particular case when
$V_2^*=\Omega^1(\mathcal{D},\mathfrak{o})$ and the affine representation is
\begin{equation}\label{example}
(a,\gamma)\mapsto(a\eta,\operatorname{Ad}_{\chi^{-1}}\eta^*\gamma+\chi^{-1}T\chi).
\end{equation}
Recall that in this particular case we have
\[
\gamma\mathbf{u}={\boldsymbol{\pounds}}_\mathbf{u}\gamma=\mathbf{d}^\gamma(\gamma(\mathbf{u}))+\mathbf{i}_\mathbf{u}\mathbf{d}^\gamma\gamma,
\]
\[
\quad\gamma\nu=-\operatorname{ad}_\nu\gamma,\quad\mathbf{d}C(\nu)=\mathbf{d}\nu,\quad\text{and}\quad\mathbf{d}C^T(w)=-\operatorname{div}(w).
\]
The diamond operations are given by
\[
w\diamond_1\gamma=(\operatorname{div}^\gamma
w)\cdot\gamma-w\cdot\mathbf{i}_{\_\,}\mathbf{d}^\gamma\gamma\quad\text{and}\quad
w\diamond_2\gamma=-\operatorname{Tr}\left(\operatorname{ad}^*_\gamma w\right).
\]
The affine Lie-Poisson equations \eqref{ALP_PCF} become
\begin{equation}\label{ALP_PerfectComplexFluid}
\left\{
\begin{array}{ll}
\vspace{0.1cm}\displaystyle\frac{\partial}{\partial
t}\mathbf{m}=-{\boldsymbol{\pounds}}_{\frac{\delta h}{\delta
\mathbf{m}}}\mathbf{m}-\operatorname{div}\left(\frac{\delta h}{\delta
\mathbf{m}}\right)\mathbf{m}-\kappa\cdot\mathbf{d}\frac{\delta h}{\delta
\kappa}-\frac{\delta h}{\delta a}\diamond a \\
\vspace{0.1cm} \qquad \quad \; \displaystyle 
-\left(\operatorname{div}^\gamma \frac{\delta h}{\delta
\gamma}\right)\gamma+\frac{\delta h}{\delta
\gamma}\cdot\mathbf{i}_{\_\,}\mathbf{d}^\gamma\gamma\\
\vspace{0.1cm}\displaystyle\frac{\partial}{\partial
t}\kappa=-\operatorname{ad}^*_{\frac{\delta h}{\delta
\kappa}}\kappa-\operatorname{div}\left(\frac{\delta h}{\delta
\mathbf{m}}\kappa\right)-\operatorname{div}^\gamma\frac{\delta
h}{\delta\gamma}\\
\vspace{0.1cm}\displaystyle\frac{\partial}{\partial t}a=-a\frac{\delta
h}{\delta
\mathbf{m}}\\
\vspace{0.1cm}\displaystyle\frac{\partial}{\partial
t}\gamma=-\mathbf{d}^\gamma\left(\gamma\left(\frac{\delta h}{\delta
\mathbf{m}}\right)\right)-\mathbf{i}_\frac{\delta h}{\delta
\mathbf{m}}\mathbf{d}^\gamma\gamma-\mathbf{d}^\gamma\frac{\delta h}{\delta
\kappa}\\
\end{array}
\right.
\end{equation}
So we recover, by a reduction from a canonical situation, the equations (3.44)
of \cite{Ho2002}, up to sign conventions, as well as their Hamiltonian
structure. In matrix notation and with respect to local coordinates we have
\begin{equation}\label{hamiltonian_matrix}
\left[\begin{array}{c}
\dot m_i\\
\dot \kappa_a\\
\dot a\\
\dot \gamma^a_i\end{array}\right]\!=\!-\!\left[\begin{array}{cccc}
\!m_k\partial_i+\partial_km_i\!     &\kappa_b\partial_i&(\square\diamond
a)_i&\partial_j\gamma^b_i-\gamma^b_{j,i}\\
\partial_k\kappa_a&\kappa_cC^c_{ba}&0&\!\delta^b_a\partial_j-C^b_{ca}\gamma^c_j\!\\
a\square\partial_k&0&0&0\\
\gamma_k^a\partial_i+\gamma_{i,k}^a&\delta_b^a\partial_i+C^a_{cb}\gamma_i^c&0&0\end{array}\right]\!\displaystyle\!\left[\begin{array}{c}
\!(\delta h/\delta  m)^k\!\\
(\delta h/\delta\kappa)^b\\
\delta h/\delta a\\
(\delta h/\delta \gamma)_b^j\end{array}\right]
\end{equation}
This matrix appears in \cite{HoKu1988} (formula (2.26a)) and in \cite{Ho2002}
(formula (3.46)) as the common Hamiltonian structure for various hydrodynamical
systems. For another derivation of this Hamiltonian structure based on reduction see See \cite{CeMaRa2003}.

The associated affine Lie-Poisson bracket is
\begin{align}\label{ALPB}
\{f,g\}(\mathbf{m},\kappa,a,\gamma)&=\int_\mathcal{D}\mathbf{m}\cdot\left[\frac{\delta
f}{\delta\mathbf{m}},\frac{\delta g}{\delta\mathbf{m}}\right]\mu\\
&\quad+\int_\mathcal{D}\kappa\cdot\left(\operatorname{ad}_{\frac{\delta
f}{\delta\kappa}}\frac{\delta g}{\delta\kappa}+\mathbf{d}\frac{\delta
f}{\delta\kappa}\cdot\frac{\delta g}{\delta\mathbf{m}}-\mathbf{d}\frac{\delta
g}{\delta\kappa}\cdot\frac{\delta f}{\delta\mathbf{m}}\right)\mu\nonumber\\
&\quad+\int_\mathcal{D}a\cdot\left(\frac{\delta f}{\delta a}\frac{\delta
g}{\delta\mathbf{m}}-\frac{\delta g}{\delta a}\frac{\delta
f}{\delta\mathbf{m}}\right)\mu\nonumber\\
&\quad+\int_\mathcal{D}\left[\left(\mathbf{d}^\gamma\frac{\delta
f}{\delta\kappa}+{\boldsymbol{\pounds}}_{\frac{\delta
f}{\delta\mathbf{m}}}\gamma\right)\cdot\frac{\delta
g}{\delta\gamma}-\left(\mathbf{d}^\gamma\frac{\delta
g}{\delta\kappa}+{\boldsymbol{\pounds}}_{\frac{\delta
g}{\delta\mathbf{m}}}\gamma\right)\cdot\frac{\delta
f}{\delta\gamma}\right]\mu\nonumber.
\end{align}
The momentum map is computed to be
\begin{align*}
&\mathbf{J}_\alpha(\mathbf{m}_\eta,\kappa_\chi,(v,w),(a,\gamma))\\
&\qquad=\left(T^*\eta\circ\mathbf{m}_\eta+T^*\chi\circ\kappa_\chi+v\diamond
a+(\operatorname{div}^\gamma
w)\cdot\gamma-w\cdot\mathbf{i}_{\_\,}\mathbf{d}^\gamma\gamma,\right.\\
&\qquad\qquad\qquad\qquad\qquad \left.T^*L_\chi\circ\kappa_\chi+\operatorname{div}^\gamma
w,(a,\gamma)\right).
\end{align*}

\medskip  

\noindent \textbf{The $B$-representation.} As on the Lagrangian side, we consider the
case where the Hamiltonian is given in terms of the curvature
$B=\mathbf{d}^\gamma\gamma$. Recall that the \textit{affine\/} representation
of
$\operatorname{Diff}(\mathcal{D})\,\circledS\,\mathcal{F}(\mathcal{D},\mathcal{O})$
on a connection one-form on the trivial principal $\mathcal{O}$-bundle
$\mathcal{O}\times \mathcal{D}\rightarrow \mathcal{D}$ induced by $\gamma\in
\Omega^1(\mathcal{D}, \mathfrak{o})$ is
\[
\gamma( \eta, \chi) : =
\operatorname{Ad}_{\chi^{-1}}\eta^*\gamma+\chi^{-1}T\chi.
\]
It induces the \textit{linear\/} representation 
\[
B(\eta, \chi) : =
\operatorname{Ad}^*_{\chi^{-1}}\eta^*B
\]
on the curvature $B=\mathbf{d}^\gamma\gamma$. The Lie-Poisson reduction for
semidirect products (see Theorem \ref{LPSD}) gives the equations
\begin{equation}\label{ALP_PerfectComplexFluid_B_representation}
\left\{
\begin{array}{ll}
\vspace{0.1cm}\displaystyle\frac{\partial}{\partial
t}\mathbf{m}=-{\boldsymbol{\pounds}}_{\frac{\delta h}{\delta
\mathbf{m}}}\mathbf{m}-\operatorname{div}\left(\frac{\delta h}{\delta
\mathbf{m}}\right)\mathbf{m}-\kappa\cdot\mathbf{d}\frac{\delta h}{\delta
\kappa}-\frac{\delta h}{\delta a}\diamond a \\
\vspace{0.1cm} \qquad \qquad \displaystyle
-\operatorname{div}\frac{\delta h}{\delta
B}\cdot\mathbf{i}_{\_\,}B+\frac{\delta
h}{\delta B}\cdot\mathbf{i}_{\_\,}\mathbf{d}B\\
\vspace{0.1cm}\displaystyle\frac{\partial}{\partial
t}\kappa=-\operatorname{ad}^*_{\frac{\delta h}{\delta
\kappa}}\kappa-\operatorname{div}\left(\frac{\delta h}{\delta
\mathbf{m}}\kappa\right)+\operatorname{Tr}\left(\operatorname{ad}^*_B\frac{\delta
h}{\delta B}\right)\\
\vspace{0.1cm}\displaystyle\frac{\partial}{\partial t}a=-a\frac{\delta
h}{\delta
\mathbf{m}}\\
\vspace{0.1cm}\displaystyle\frac{\partial}{\partial
t}B=-{\boldsymbol{\pounds}}_{\frac{\delta h}{\delta
\mathbf{m}}}B+\operatorname{ad}_{\frac{\delta h}{\delta \kappa}}B\\
\end{array}
\right.
\end{equation}
Using formulas \eqref{compatible_diamonds}, we can prove, as on the Lagrangian
side, that the Lie-Poisson equations
\eqref{ALP_PerfectComplexFluid_B_representation} are compatible with the affine
Lie-Poisson equations \eqref{ALP_PerfectComplexFluid}. In matrix notation and
with respect to local coordinates, the Lie-Poisson equations read
\begin{equation*}
\left[\begin{array}{c}
\dot m_i\\
\dot \kappa_a\\
\dot a\\
\dot B^a_{ij}\end{array}\right] =
-\left[\begin{array}{cccc}
m_k\partial_i+\partial_km_i     &\kappa_b\partial_i&(\square\diamond
a)_i&M_{ikl}^b\\
\partial_k\kappa_a&\kappa_cC^c_{ba}&0&-C^b_{ca}B^c_{kl}\\
a\square\partial_k&0&0&0\\
N^a_{ijk}&C^a_{cb}B_{ij}^c&0&0\end{array}\right]\displaystyle\left[\begin{array}{c}
(\delta h/\delta  m)^k\\
(\delta h/\delta\kappa)^b\\
\delta h/\delta a\\
(\delta h/\delta B)_b^{kl}\end{array}\right],
\end{equation*}
where
\[
M_{ikl}^b=-B^b_{kl,i}+\partial_kB^b_{il}-\partial_lB^b_{ik}\qquad\text{and}\qquad
N_{ijk}^a=B^a_{ij,k}+B^a_{kj}\partial_i-B^a_{ki}\partial_j.
\]
As before, we recover by a reduction from a canonical cotangent bundle, the
Hamiltonian structures appearing in \cite{HoKu1988} (formula (2.28)) and
\cite{Ho2002} (p.152), and we have explained in which sense this matrix is
Lie-Poisson, as already noted in these papers. More precisely, we have found
the
Lie group which corresponds to the Lie algebra underlying this Hamiltonian
structure. This group is given by
\[
\left[\operatorname{Diff}(\mathcal{D})\,\circledS\,\mathcal{F}(\mathcal{D},\mathcal{O})\right]\,\circledS\,\left[V_1^*\,\oplus\,\Omega^2(\mathcal{D},\mathfrak{o})\right],
\]
where
$\operatorname{Diff}(\mathcal{D})\,\circledS\,\mathcal{F}(\mathcal{D},\mathcal{O})$
acts on $\Omega^2(\mathcal{D},\mathfrak{o})$ by the representation
\[
B\mapsto \operatorname{Ad}_{\chi^{-1}}\eta^*B,
\]
and where the space $V_1^*$ is only acted upon by the subgroup
$\operatorname{Diff}(\mathcal{D})$.

The Lie-Poisson bracket is computed to be
\begin{align*}
\{f,g\}(\mathbf{m},&\kappa,a,B)=\int_\mathcal{D}\mathbf{m}\cdot\left[\frac{\delta
f}{\delta\mathbf{m}},\frac{\delta g}{\delta\mathbf{m}}\right]\mu\\
&\quad+\int_\mathcal{D}\kappa\cdot\left(\operatorname{ad}_{\frac{\delta
f}{\delta\kappa}}\frac{\delta g}{\delta\kappa}+\mathbf{d}\frac{\delta
f}{\delta\kappa}\cdot\frac{\delta g}{\delta\mathbf{m}}-\mathbf{d}\frac{\delta
g}{\delta\kappa}\cdot\frac{\delta f}{\delta\mathbf{m}}\right)\mu\\
&\quad+\int_\mathcal{D}a\cdot\left(\frac{\delta f}{\delta a}\frac{\delta
g}{\delta\mathbf{m}}-\frac{\delta g}{\delta a}\frac{\delta
f}{\delta\mathbf{m}}\right)\mu\\
&\quad+\int_\mathcal{D}\left[\left({\boldsymbol{\pounds}}_{\frac{\delta
f}{\delta\mathbf{m}}}B-\operatorname{ad}_{\frac{\delta
f}{\delta\kappa}}B\right)\cdot\frac{\delta g}{\delta
B}-\left({\boldsymbol{\pounds}}_{\frac{\delta
g}{\delta\mathbf{m}}}B-\operatorname{ad}_{\frac{\delta
g}{\delta\kappa}}B\right)\cdot\frac{\delta f}{\delta B}\right]\mu.
\end{align*}
Using formulas \eqref{link_infinitesimal_actions} and
\[
\frac{\delta f}{\delta\gamma}=-\operatorname{div}^\gamma\frac{\delta f}{\delta
B},
\]
we obtain that the map
\[
(\mathbf{m},\nu,a,\gamma)\mapsto (\mathbf{m},\nu,a,\mathbf{d}^\gamma\gamma)
\]
is a Poisson map relative to the affine Lie-Poisson bracket $\{\,,\}$ given in
\eqref{ALPB} and the Lie-Poisson bracket associated to the $B$-representation.


\section{The Circulation Theorems}\label{Kelvin_Noether}

The Kelvin-Noether theorem is a version of the Noether theorem that holds for
solutions of the Euler-Poincar\'e equations. For example, an application of
this
theorem to
the compressible adiabatic fluid gives the Kelvin
circulation theorem
\[
\frac{d}{dt}\oint_{\gamma_t}\mathbf{u}^\flat=\oint_{\gamma_t}T\textbf{d}s,
\]
where $\gamma_t\subset \mathcal{D}$ is a closed curve which moves with the
fluid
velocity
$\mathbf{u}$, $T=\partial e/\partial s$  is the temperature, and $e,s$ denote
respectively the specific internal energy and the specific entropy. The
Kelvin-Noether theorem associated to Euler-Poincar\'e reduction for semidirect
product is presented in \cite{HoMaRa1998}. We now adapt this result to the case
of affine Euler-Poincar\'e reduction.

\medskip  

\noindent \textbf{Kelvin-Noether Theorem.}  We work under the hypotheses and the
notations of \S\ref{Lagrangian_formulation}. Let
$\mathcal{C}$ be a manifold
on which $G$ acts on the left and suppose we have an equivariant map
$\mathcal{K}:\mathcal{C}\times V^*\rightarrow\mathfrak{g}^{**}$, that is, for
all $g\in G, a\in V^*, c\in\mathcal{C}$, we have
\[
\langle\mathcal{K}(gc,\theta_g(a)),\mu\rangle=\langle\mathcal{K}(c,a),\operatorname{Ad}_g^*\mu\rangle,
\]
where $gc$ denotes the action of $G$ on $\mathcal{C}$, and $\theta_g$ is the
affine action of $G$ on $V^*$.

Define the Kelvin-Noether quantity $I:\mathcal{C}\times\mathfrak{g}\times
V^*\rightarrow\mathbb{R}$ by
\[
I(c,\xi,a):=\left\langle\mathcal{K}(c,a),\frac{\delta
l}{\delta\xi}(\xi,a)\right\rangle.
\]
The same proof as in \cite{HoMaRa1998} yields the following result.

\begin{theorem}\label{KN}
Fixing $c_0\in\mathcal{C}$, let $\xi(t),a(t)$ satisfy the
affine Euler-Poincar\'e equations \eqref{AEP} and define $g(t)$ to be the
solution of
$\dot{g}(t)=TR_{g(t)}\xi(t)$ and, say, $g(0)=e$. Let $c(t)=g(t)c_0$ and
$I(t):=I(c(t),\xi(t),a(t))$. Then
\[
\frac{d}{dt}I(t)=\left\langle\mathcal{K}(c(t),a(t)),\frac{\delta l}{\delta
a}\diamond a-\mathbf{d}c^T\left(\frac{\delta l}{\delta
a}\right)\right\rangle.
\]
\end{theorem}

As we will see in the applications, some examples do not admit a Lagrangian
formulation. Nevertheless, a Kelvin-Noether theorem is still valid for the
Hamiltonian formulation. Keeping the same notations as before, the
Kelvin-Noether quantity is now the mapping
$J:\mathcal{C}\times\mathfrak{g}^*\times V^*\rightarrow\mathbb{R}$ defined by
\[
J(c,\mu,a):=\langle\mathcal{K}(c,a),\mu\rangle,
\]
and we have the following result.

\begin{theorem}\label{KN_Hamiltonian}
Fixing $c_0\in\mathcal{C}$, let $\mu(t),a(t)$ satisfy the
affine Lie-Poisson equations of Theorem \ref{ALPSD} and define $g(t)$ to be the
solution of
\[
\dot{g}(t)=TR_{g(t)}\left(\frac{\delta h}{\delta\mu}\right),\quad g(0)=e.
\]
Let $c(t)=g(t)c_0$ and
$J(t):=J(c(t),\mu(t),a(t))$. Then
\[
\frac{d}{dt}J(t)=\left\langle\mathcal{K}(c(t),a(t)),-\frac{\delta h}{\delta
a}\diamond a+\mathbf{d}c^T\left(\frac{\delta h}{\delta
a}\right)\right\rangle.
\]
\end{theorem}

This result follows from the reconstruction of dynamics as described in 
\S\ref{AHSPT}. Of course, when the Hamiltonian $h$ comes from a Lagrangian by
Legendre transformation, then this theorem is a corollary of Theorem \ref{KN}.

In the case of dynamics on the group $G=\operatorname{Diff}(\mathcal{D})$, the
standard choice for the equivariant map $\mathcal{K}$ is
\begin{equation}\label{K}
\langle\mathcal{K}(c,a),\mathbf{m}\rangle:=\oint_c\frac{1}{\rho}\mathbf{m},
\end{equation}
where $c\in\mathcal{C}=\operatorname{Emb}(S^1,\mathcal{D})$, the manifold of
all
embeddings of the circle $S^1$ in $\mathcal{D}$,
$\mathbf{m}\in\Omega^1(\mathcal{D})$, and $\rho$ is advected as
$(J\eta)(\rho\circ\eta)$. There is a generalization of this map in the case of
the group
$G=\operatorname{Diff}(\mathcal{D})\,\circledS\,\mathcal{F}(\mathcal{D},\mathcal{O})$,
see \S7 in \cite{GBRa2008}. Therefore, Theorems \ref{KN} and
\ref{KN_Hamiltonian} can be applied in the case of the affine Euler-Poincar\'e
and Lie-Poisson equations \eqref{AEP_PCF} and \eqref{ALP_PCF}. Nevertheless we
shall not use this point of view here and we apply the Kelvin-Noether theorem
only to the first component of the group $G$, namely the group
$\operatorname{Diff}(\mathcal{D})$, and we obtain the following result.

\begin{theorem} Consider the affine Euler-Poincar\'e equations for complex
fluids \eqref{AEP_PCF}. Suppose that one of the linear advected variables, say
$\rho$, is advected as $(J\eta)(\rho\circ\eta)$. Then, using the map \eqref{K},
we have
\[
\frac{d}{dt}\oint_{c_t}\frac{1}{\rho}\frac{\delta
l}{\delta\mathbf{u}}=\oint_{c_t}\frac{1}{\rho}\left(-\frac{\delta l}{\delta
\nu}\cdot\mathbf{d}\nu+\frac{\delta l}{\delta a}\diamond a+\frac{\delta
l}{\delta \gamma}\diamond_1 \gamma\right),
\]
where $c_t$ is a loop in $\mathcal{D}$ which moves with the fluid velocity
$\mathbf{u}$.

Similarly, consider the affine Lie-Poisson equations for complex fluids
\eqref{ALP_PCF}. Suppose that one of the linear advected variables
is advected as $(J\eta)(\rho\circ\eta)$. Then, using the map 
\eqref{K}, we have
\[
\frac{d}{dt}\oint_{c_t}\frac{1}{\rho}\mathbf{m}=\oint_{c_t}\frac{1}{\rho}\left(-\kappa\cdot\mathbf{d}\frac{\delta
h}{\delta \kappa}-\frac{\delta h}{\delta a}\diamond a-\frac{\delta h}{\delta
\gamma}\diamond_1 \gamma\right),
\]
where $c_t$ is a loop in $\mathcal{D}$ which moves with the fluid velocity
$\mathbf{u}$, defined be the equality
\[
\mathbf{u}:=\frac{\delta h}{\delta\mathbf{m}}.
\]
\end{theorem}

\medskip  

\noindent \textbf{$\gamma$-circulation.} The $\gamma$-circulation is associated to the
equation
\[
\frac{\partial}{\partial
t}\gamma+\boldsymbol{\pounds}_\mathbf{u}\gamma=-\mathbf{d}\nu+\operatorname{ad}_\nu\gamma.
\]
Let $\eta_t$ be the flow of the vector field $\mathbf{u}$, let $c_0$ be a loop
in $\mathcal{D}$ and let $c_t:=\eta_t \circ c_0$. Then, by change of variables,
we have
\[
\frac{d}{dt}\oint_{c_t}\gamma=\frac{d}{dt}\oint_{c_0}\eta_t^*\gamma=\oint_{c_0}\eta_t^*(\dot\gamma+\boldsymbol{\pounds}_\mathbf{u}\gamma)=\oint_{c_0}\eta_t^*(-\mathbf{d}\nu+\operatorname{ad}_\nu\gamma)=\oint_{c_t}\operatorname{ad}_\nu\gamma\in\mathfrak{o}.
\]


\section{Applications}\label{applications}

\subsection{Spin Systems}
\label{subsec:spin_chains}

We thank  D. Holm for challenging us with this example as a simple model for
many applications. It illustrates the applicability of Theorems \ref{AEPSD} and
\ref{ALPSD} in a very simple situation that exhibits, nevertheless, some of the
key difficulties of more complicated fluid models. Let $\mathcal{D}$ be a
manifold and $\mathcal{O}$ a Lie group thought of as the ``order
parameters" of some fluid model. Take
$G=\mathcal{F}(\mathcal{D},\mathcal{O})\ni\chi$ and
$V^*=\Omega^1(\mathcal{D},\mathfrak{o})\ni\gamma$. Consider the affine $G
$-representation on $V ^\ast$ given by  
\[
\theta_\chi(\gamma):=\operatorname{Ad}_{\chi^{-1}}\gamma+\chi^{-1}T\chi.
\]
This is simply the gauge transformation of the connection on the trivial
principal $\mathcal{O}$-bundle $\mathcal{O} \times \mathcal{D} \rightarrow
\mathcal{D}$ induced by $\gamma$ (see \eqref{connection_gamma}). The associated
diamond operation is given by
\[
w\diamond\gamma=-\operatorname{Tr}(\operatorname{ad}^*_\gamma w).
\]
Since $c(\chi)=\chi^{-1}T\chi$ we obtain, as in the case of complex fluids,
\[
\mathbf{d}c(\nu)=\mathbf{d}\nu\quad\text{and}\quad\mathbf{d}c^T(w)=-\operatorname{div}w.
\]
As before we use the notations $\nu\in\mathcal{F}(\mathcal{D},\mathfrak{o})$
and
$w\in\mathfrak{X}(\mathcal{D},\mathfrak{o}^*)$. The affine Euler-Poincar\'e
equations \eqref{AEP} become
\begin{equation}\label{AEP_spin_chain}
\frac{\partial}{\partial t}\frac{\delta
l}{\delta\nu}=-\operatorname{ad}^*_\nu\frac{\delta
l}{\delta\nu}+\operatorname{div}^\gamma\frac{\delta l}{\delta\gamma},
\end{equation}
where
$\left(\operatorname{ad}^*_\nu\kappa\right)(x)=\operatorname{ad}^*_{\nu(x)}(\kappa
(x))$. The evolution equation for $\gamma$ is
\[
\frac{\partial}{\partial t}\gamma+\mathbf{d}^\gamma\nu=0.
\]
Similarly, the affine Lie-Poisson equations of Theorem \ref{ALPSD} become
\begin{equation}\label{ALP_spin_chain}
\left\lbrace
\begin{array}{ll}
\vspace{0.2cm}\displaystyle
\frac{\partial}{\partial t}\kappa=-\operatorname{ad}^*_{\frac{\delta
h}{\delta\kappa}}\kappa-\operatorname{div}^\gamma\frac{\delta
h}{\delta\gamma}\\
\displaystyle\frac{\partial}{\partial t}\gamma+\mathbf{d}^\gamma\frac{\delta
h}{\delta\kappa}=0.
\end{array}\right.
\end{equation}

In the particular case $\mathcal{D}=\mathbb{R}^3$ and $\mathcal{O}=SO(3)$, this system of equations appears in the context of the \textit{macroscopic description of spin glasses}, see equations (28) and (29) in \cite{Dz1980} and references therein. See also equations (3.9), (3.10) in \cite{IsKoPe1994}, system (1) in \cite{Iv2000} and references therein for an application to \textit{magnetic media}. In this context, the variable $\kappa$ is interpreted as the \textit{spin density}, $\nu$ is the \textit{infinitesimal spin rotation} and the curvature $\mathbf{d}^\gamma\gamma$ is the \textit{disclination density}.

Interesting choices for the Lagrangian are
\[
l_\perp(\nu,\gamma)=\frac{1}{2}\int_\mathcal{D}\|[\nu,\gamma]\|^2\mu,\qquad l_{\|}(\nu,\gamma)=\frac{1}{2}\int_\mathcal{D}\|k(\nu,\gamma)\|^2\mu,
\]
and
\begin{equation}\label{spin_glass_lagrangian}
l_{SG}(\nu,\gamma)=\frac{\epsilon}{2}\int_\mathcal{D}\|\nu\|^2\mu-\frac{\rho}{2}\int_\mathcal{D}\|\gamma\|^2\mu.
\end{equation}
In the expression of the Lagrangian $l_\perp$, $[\nu,\gamma]$ denotes the
$\mathfrak{o}$-valued one-form on $\mathcal{D}$ given by $v_x\mapsto
[\nu(x),\gamma(x)(v_x)]\in\mathfrak{o}$. The norm is associated to the metric
$(gk)$, on the vector bundle of $\mathfrak{o}$-valued one-forms on
$\mathcal{D}$, induced by a Riemannian metric $g$ on $\mathcal{D}$ and by an
$\operatorname{Ad}$-invariant inner product $k$ on $\mathfrak{o}$.
More precisely, the metric $(gk)$ on the vector bundle of $\mathfrak{o}$-valued
$k$-forms on $\mathcal{D}$ is given in the following way. The Riemannian metric
$g$ induces a Riemannian metric $\overline{g}$ on the
vector bundles $\Lambda^k\mathcal{D}\rightarrow \mathcal{D}$ of exterior $k
$-forms on $\mathcal{D}$. For
$\alpha_x, \beta_x\in\Lambda^k(\mathcal{D},\mathfrak{o})_x$, we can write
$\alpha_x=\alpha^af_a$ and $\beta_x=\beta^af_a$, where $\{f_a\}$ is a
basis of $\mathfrak{o}$ and
$\alpha^a,\beta^a\in(\Lambda^k M)_x$. So we define
\[
(gk)_x(\alpha_x,\beta_x):=k_{ab}\,\overline{g}(\alpha^a,\beta^b),
\]
where $k_{ab}:=k(f_a,f_b)$. This construction is
independent of the basis. If the manifold is
taken to be $\mathbb{R}$ and if $\mathcal{O}=SO(3)$, then the Lagrangian $l_\perp$ reads
\[
l_\perp(\nu,\gamma)=\frac{1}{2}\int_{-\infty}^\infty\|\nu\times\gamma\|^2\mu.
\]
This choice is reminiscent of the \textit{Skyrme model}, a nonlinear
topological model of pions in nuclear physics, see \cite{Sk1961}.

In the expression of the Lagrangian $l_{\|}$, $k(\nu,\gamma)$ denotes the
one-form on $\mathcal{D}$ given by $v_x\mapsto k(\nu(x),\gamma(x)(v_x))$. The
norm is taken relative to the Riemannian metric $g$ on $\mathcal{D}$.

Interestingly, when the Lagrangians $l_\perp$ or $l_{\|}$ are used, the affine
Euler-Poincar\'e equations \eqref{AEP_spin_chain} simplify to
\begin{equation}
\frac{\partial}{\partial t}\frac{\delta
l}{\delta\nu}=\operatorname{div}\frac{\delta l}{\delta\gamma},
\end{equation}
where on the right hand side we have the usual divergence operator in $\mathbb{R}^3$.

In the expression of the Lagrangian $l_{SG}$, we used the symbols $\epsilon$ and $\rho$ for the \textit{constants of susceptibility} and the \textit{rigidity}. The norms are respectively associated to the inner product $k$ and to the metric $(gk)$. The associated Hamiltonian reads
\[
h_{SG}(\kappa,\gamma)=\frac{1}{2\epsilon}\int_\mathcal{D}\|\kappa\|^2\mu+\frac{\rho}{2}\int_\mathcal{D}\|\gamma\|^2\mu,
\]
and is used, with $\mathcal{O}=SO(3)$, in the context of the \textit{macroscopic description of spin glasses}, see expression (26) in \cite{Dz1980}. In this case the affine Lie-Poisson equation \eqref{ALP_spin_chain} reads
\begin{equation}\label{spin_glass_dzyaloshinskii}
\left\lbrace
\begin{array}{ll}
\vspace{0.2cm}\displaystyle
\frac{\partial}{\partial t}\kappa=-\rho\operatorname{div}\gamma^\sharp\\
\displaystyle\frac{\partial}{\partial t}\gamma+\frac{1}{\epsilon}\mathbf{d}^\gamma\kappa^\sharp=0,
\end{array}\right.
\end{equation}
where $\gamma^\sharp\in\mathfrak{X}(\mathcal{D},\mathfrak{o}^*)$ and $\kappa^\sharp\in\mathcal{F}(\mathcal{D},\mathfrak{o})$ are associated to $\gamma$ and $\kappa$ via the metrics.

More general expressions, such as
\[
h(\kappa,\gamma)=\frac{1}{2\epsilon}\int_\mathcal{D}\|\kappa\|^2\mu+\frac{1}{2\epsilon_1}\int_\mathcal{D}\|\kappa\|^4\mu+\frac{\rho}{2}\int_\mathcal{D}\|\gamma\|^2\mu+\frac{\rho_1}{2}\int_\mathcal{D}\|\gamma\|^4\mu+\frac{q}{2}\int_\mathcal{D}\|\kappa\!\cdot\!\gamma\|^2\mu,
\]
are used in the theory of magnetic media, see e.g. equation (3) in \cite{Iv2000}.

\medskip  

\noindent \textbf{Lagrangian reduction.} Consider a Lagrangian
\[
L:
T\mathcal{F}(\mathcal{D},\mathcal{O})\times\Omega^1(\mathcal{D},\mathfrak{o})\rightarrow\mathbb{R},\;L=L(\nu_\chi,\gamma)
\]
such that, for all $\psi\in\mathcal{F}(\mathcal{D},\mathcal{O})$ we have
\[
L(TR_\psi\circ\nu_\chi,\operatorname{Ad}_{\psi^{-1}}\gamma+\psi^{-1}T\psi)=L(\nu_\chi,\gamma).
\]
Let $\chi$ be a curve in $\mathcal{F}(\mathcal{D},\mathcal{O})$ and consider
the
curve $\nu:=TR_{\chi^{-1}}\circ\dot\chi$. For
$\gamma_0\in\Omega^1(\mathcal{D},\mathfrak{o})$ consider the solution $\gamma$
of the equation
\[
\frac{\partial}{\partial t}\gamma+\mathbf{d}^\gamma\nu=0
\]
with initial condition $\gamma_0$. This solution is given by
$\gamma=\operatorname{Ad}_\chi\gamma_0+\chi T\chi^{-1}$. Then, by Theorem
\ref{AEPSD},  $\chi$ is a solution of the Euler-Lagrange equations associated
to
$L_{\gamma_0}$ if and only if $\nu$ is solution of \eqref{AEP_spin_chain}. Note that in the special case $\gamma_0=0$, the evolution of $\gamma$ is given by the important relation
\[
\gamma=\chi T\chi^{-1}
\]
and the disclination density vanishes, that is, $\mathbf{d}\gamma+[\gamma,\gamma]=0$. This hypothesis is usually assumed in the examples treated in \cite{Dz1980} and \cite{Iv2000}, and is referred to as the \textit{Maurer-Cartan constraint}. Recall that the vanishing of the curvature is preserved by the flow of \eqref{ALP_spin_chain}. In our approach the variable $\chi$ can be interpreted as the \textit{Lagrangian evolution of the spin}.

\medskip  

\noindent \textbf{Hamiltonian reduction.} Consider a Hamiltonian
\[
H:
T^*\mathcal{F}(\mathcal{D},\mathcal{O})\times\Omega^1(\mathcal{D},\mathfrak{o})\rightarrow\mathbb{R},\;H=H(\kappa_\chi,\gamma)
\]
such that, for all $\psi\in\mathcal{F}(\mathcal{D},\mathcal{O})$ we have
\[
H(T^*R_{\psi^{-1}}\circ\kappa_\chi,\operatorname{Ad}_{\psi^{-1}}\gamma+\psi^{-1}T\psi)=H(\kappa_\chi,\gamma).
\]
Then, by Theorem \ref{ALPSD}, a curve $\kappa_\chi\in
T^*\mathcal{F}(\mathcal{D},\mathcal{O})$ is a solution of the Hamilton
equations
associated to $H_{\gamma_0}$ if and only if the curve
\[
\kappa:=T^*R_\chi\circ\kappa_{\chi}
\]
is a solution of the affine Lie-Poisson equation \eqref{ALP_spin_chain}. The associated affine Lie-Poisson bracket is given by

\begin{align}\label{ALPB_spin_chain}
\{f,g\}(\kappa,\gamma)=\int_\mathcal{D}\kappa\cdot\left(\operatorname{ad}_{\frac{\delta
f}{\delta\kappa}}\frac{\delta
g}{\delta\kappa}\right)\mu+\int_\mathcal{D}\left(\mathbf{d}^\gamma\frac{\delta
f}{\delta\kappa}\cdot\frac{\delta
g}{\delta\gamma}-\mathbf{d}^\gamma\frac{\delta
g}{\delta\kappa}\cdot\frac{\delta f}{\delta\gamma}\right)\mu\nonumber.
\end{align}
Thus, we have recovered the Poisson bracket of \cite{Dz1980}, by reduction of the canonical structure. The momentum map is
\[
\mathbf{J}_\alpha(\kappa_\chi,(w,\gamma))=(T^*L_\chi\circ\kappa_\chi+\operatorname{div}^\gamma
w,\gamma)
\]
and the reduced symplectic spaces are affine coadjoint orbits in the dual Lie algebra\\ $(\mathcal{F}(\mathcal{D},\mathfrak{o})\,\circledS\,\Omega^1(\mathcal{D},\mathfrak{o}))^*$.
Using formula \eqref{affine_coadj}, we obtain that they are given by
\[
\mathcal{O}^\sigma_{(\kappa,\gamma)}=\left\{(\operatorname{Ad}^*_\chi\kappa+\operatorname{div}^{\theta_\chi(\gamma)}w,\theta_\chi(\gamma))\mid
(\chi,w)\in\mathcal{F}(\mathcal{D},\mathcal{O})\,\circledS\,\mathfrak{X}(\mathcal{D},\mathfrak{o}^*)\right\}.
\]


\subsection{Yang-Mills Magnetohydrodynamics}\label{YM_MHD}

Magnetohydrodynamics models the motion of an electrically charged and perfectly
conducting fluid. In the balance of momentum law, one must add the Lorentz
force
of the magnetic field created by the fluid in motion. In addition, the
hypothesis of infinite conductivity leads one to the conclusion that magnetic
lines are frozen in the fluid, i.e. that they are transported along the
particle
paths. This hypothesis leads to the equation
\[
\frac{\partial}{\partial t}B+\boldsymbol{\pounds}_\mathbf{u}B=0.
\]
This model can be extend to incorporate nonabelian Yang-Mills interactions and
is known under the name of Yang-Mills magnetohydrodynamics; see \cite{HoKu1984}
for a derivation of this model. Recall that for Yang-Mills theory, the field
$B$
is seen as the curvature of a connection $A$ on a principal bundle. Clearly the
connection $A$ represents the variable $\gamma$ in the general theory developed
previously, on which the automorphism group acts by affine transformations.
This
shows that the abstract formalism developed previously is very natural in the
context of Yang-Mills theory. Note that there is a more general model of fluid
motion with Yang-Mills charged particles, namely the Euler-Yang-Mills
equations.
The Hamiltonian structure of these equations is given in \cite{GiHoKu1983}, see
also \cite{GBRa2008} for the associated Lagrangian and Hamiltonian reductions.

As remarked in \cite{HoKu1988}, at the reduced level, the Hamiltonian structure
of Yang-Mills magnetohydrodynamics is given by the matrix
\eqref{hamiltonian_matrix}. In this paragraph we carry out the corresponding
affine Lie-Poisson reduction.

The group $G$ is chosen to be the semidirect product of the diffeomorphism
group
with the group of $\mathcal{O}$-valued function on $\mathcal{D}$, that is,
$G$ is the group $\operatorname{Diff}(\mathcal{D})\,\circledS\,\mathcal{F}(\mathcal{D},\mathcal{O})$.
The order parameter Lie group $\mathcal{O} $ represents here the
\textit{symmetry group of the particles interaction}. For example,
$\mathcal{O}=S^1$ corresponds to electromagnetism, $
\mathcal{O}=SU(2)$ and $\mathcal{O}=SU(3)$ correspond to weak and strong
interactions, respectively. The advected quantities are the \textit{mass
density} $\rho$, the \textit{specific entropy} $s$, and the \textit{potential
of
the Yang-Mills field} $A$. Therefore, we set
\[
a=(\rho,s)\in V_1^*=\mathcal{F}(\mathcal{D})\times\mathcal{F}(\mathcal{D})
\quad\text{and}\quad A=\gamma\in\Omega^1(\mathcal{D},\mathfrak{o}).
\]
The action of $(\eta,\chi)\in G$ on $(\rho,s)\in V_1^*$ is the usual right
representation of the fluid relabeling group on the mass density and entropy.
It
is given by
\[
(\rho,s)(\eta,\chi)=(J\eta(\rho\circ\eta),s\circ\eta).
\]
The right affine action of $(\eta,\chi)\in G$ on $A\in
\Omega^1(M,\mathfrak{o})$
is given, as in the example \eqref{example}, by
\[
A \mapsto  
\operatorname{Ad}_{\chi^{-1}}\eta^*A+\chi^{-1}T\chi.
\]
Since the variable $\kappa\in\mathcal{F}(\mathcal{D},\mathfrak{o}^*)$ is
interpreted as the \textit{gauge-charge density}, we use the notations
$Q\in\mathcal{F}(\mathcal{D},\mathfrak{o}^*)=T_0^*\mathcal{F}(\mathcal{D},\mathcal{O})$
and $Q_\chi\in T_\chi^*\mathcal{F}(\mathcal{D},\mathcal{O})$.

The Hamiltonian 
$H_{(\rho,s,A)}:T^*(\operatorname{Diff}(\mathcal{D})\,\circledS\,\mathcal{F}(\mathcal{D},\mathcal{O}))\rightarrow\mathbb{R}$,
is given by
\begin{align}\label{YM-MHD_Hamiltonian}
H_{(\rho,s,A)}(\mathbf{m}_\eta,Q_\chi)=&\frac{1}{2}\int_\mathcal{D}\frac{1}{\rho}\|\mathbf{m}_\eta\|^2\mu+\int_\mathcal{D}\rho
e(\rho (J\eta^{-1}),s)\mu\nonumber\\
&\qquad+\frac{1}{2}\int_\mathcal{D}\|\mathbf{d}^AA\!\cdot\!T\eta^{-1}\|^2(J\eta)\mu,
\end{align}
where $e$ denotes the \textit{specific internal energy}, the norm in the first
term is associated to a Riemannian metric $g$ on $\mathcal{D}$, and the norm in
the third term is associated to the metric $(gk)$, on the vector bundle of
$\mathfrak{o}$-valued $k$-forms on $\mathcal{D}$, induced by the metric $g$ and
by an $\operatorname{Ad}$-invariant inner product $k$ on $\mathfrak{o}$. For
details see, \S\ref{subsec:spin_chains}.

The metric $(gk)$ can be used to identify
$\Omega_k(\mathcal{D},\mathfrak{o}^*)$
and its dual $\Omega^k(\mathcal{D},\mathfrak{o})$, by raising and lowering
indices. Through this identification, the operators $\operatorname{div}$ and
$\operatorname{div}^A$ act also on $\Omega^k(\mathcal{D},\mathfrak{o})$.
The Hamiltonian $H(\mathbf{m}_\eta,Q_\chi,\rho,s,A)$ is invariant under the
right action of $(\varphi,\psi)$ given by
\begin{align*}  
&(\mathbf{m}_\eta,Q_\chi,\rho,s,A) \\
&\mapsto\!\left(J\varphi(\mathbf{m}_\eta\circ\varphi),J\varphi(T^*R_{\psi^{-1}}\!\circ
Q_\chi\circ\varphi),J\varphi(\rho\circ\varphi),s\circ\varphi,\operatorname{Ad}_{\psi^{-1}}\varphi^*A+\psi^{-1}T\psi\right).
\end{align*}
Therefore, the hypotheses of Theorem \ref{ALPSD} are satisfied and the reduced
Hamiltonian
$h:\Omega^1(\mathcal{D})\times\mathcal{F}(\mathcal{D},\mathfrak{o}^*)\times
V_1^*\times V_2^*\rightarrow\mathbb{R}$ is given by
\[
h(\mathbf{m},Q,\rho,s,A)=\frac{1}{2}\int_\mathcal{D}\frac{1}{\rho}
\|\mathbf{m}\|^2\mu+\int_\mathcal{D}\rho
e(\rho,s)\mu+\frac{1}{2}\int_\mathcal{D}\|\mathbf{d}^AA\|^2\mu,
\]
where the norms are respectively associated to the metrics $g$ and $(gk)$. We
now compute the affine Lie-Poisson equations \eqref{ALP_PerfectComplexFluid}
associated to this Hamiltonian. The functional derivatives are
\[
\mathbf{u}:=\frac{\delta h}{\delta
\mathbf{m}}=\frac{1}{\rho}\mathbf{m}^\sharp,\quad\nu:=\frac{\delta h}{\delta
Q}=0
\]
and
\[
\quad\frac{\delta h}{\delta
\rho}=-\frac{1}{2}\|\mathbf{u}\|^2+e+\rho\frac{\partial
e}{\partial\rho},\quad\frac{\delta h}{\delta s}=\rho\frac{\partial e}{\partial
s},\quad\frac{\delta h}{\delta A}=-\operatorname{div}^A\mathbf{d}^AA,
\]
where $\operatorname{div}^A $ is defined by
\eqref{divergence_connection_formula}.
The advection equation are
\[
\displaystyle\frac{\partial}{\partial t}\rho+\operatorname{div}(\rho
\mathbf{u})=0,\quad\displaystyle\frac{\partial}{\partial
t}s+\mathbf{d}s(\mathbf{u})=0,\quad\displaystyle\frac{\partial}{\partial
t}A+\mathbf{d}^A(A(\mathbf{u}))+\mathbf{i}_\mathbf{u}B=0.
\]
The equation for the gauge charge is
\[
\frac{\partial}{\partial
t}Q=-\operatorname{div}(Q\mathbf{u})+\operatorname{div}^A\operatorname{div}^A\mathbf{d}^AA=-\operatorname{div}(Q\mathbf{u}).
\]
Indeed, for all $f\in\mathcal{F}(\mathcal{D},\mathfrak{o})$ we have
\[
\int_\mathcal{D}k\left(\operatorname{div}^A\operatorname{div}^A\mathbf{d}^AA,f\right)\mu\!=\!\int_\mathcal{D}(gk)\left(\mathbf{d}^AA,\mathbf{d}^A\mathbf{d}^Af\right)\mu\!=\!\int_\mathcal{D}(gk)\left(B,[B,f]\right)\mu\!=0,
\]
where we used the equality we have $\mathbf{d}^A\mathbf{d}^Af=[B,f]$ for
$B=\mathbf{d}^AA$ and the fact that, in an orthonormal frame with respect to
$g$, we have
\begin{align*}
(gk)\left(B,[B,f]\right)&=B^a_{ij}[B,f]^b_{ij}k_{ab}=B^a_{ij}[B_{ij},f]^bk_{ab}\\
&=k(B_{ij},[B_{ij},f])=-k([B_{ij},B_{ij}],f)=0.
\end{align*}
This proves that $\operatorname{div}^A\operatorname{div}^A\mathbf{d}^AA=0$.

\medskip
Using the advection equation for $\rho$ we obtain the equality
\[
\left(\frac{\partial}{\partial
t}\mathbf{m}+{\boldsymbol{\pounds}}_\mathbf{u}\mathbf{m}+(\operatorname{div}\mathbf{u})\mathbf{m}\right)^\sharp=\rho\left(\frac{\partial}{\partial
t}\mathbf{u}+\nabla_\mathbf{u}\mathbf{u}+\nabla\mathbf{u}^T\cdot\mathbf{u}\right).
\]
We also have
\begin{align*}
-\left(\frac{\delta h}{\delta\rho}\diamond\rho+\frac{\delta h}{\delta
s}\diamond
s\right)^\sharp&=\frac{\rho}{2}\operatorname{grad}\|\mathbf{u}\|^2-\rho\operatorname{grad}e-\rho\operatorname{grad}\left(\rho\frac{\partial
e}{\partial\rho}\right)+\rho\frac{\partial e}{\partial s}\operatorname{grad}s\\
&=\rho(\nabla\mathbf{u}^T\cdot\mathbf{u})-\rho\operatorname{grad}\left(\rho\frac{\partial
e}{\partial\rho}\right)-\rho\frac{\partial
e}{\partial\rho}\operatorname{grad}\rho\\
&=\rho(\nabla\mathbf{u}^T\cdot\mathbf{u})-\operatorname{grad}\left(\rho^2\frac{\partial
e}{\partial\rho}\right)
\end{align*}
and
\begin{align*}
\frac{\delta h}{\delta
A}\diamond_1A=&\left(\operatorname{div}^A\frac{\delta
h}{\delta A}\right)A-\frac{\delta h}{\delta
A}\cdot\mathbf{i}_{\_\,}\mathbf{d}^AA\\
&=0+(gk)(\operatorname{div}^A\mathbf{d}^AA,\mathbf{i}_{\_\,}\mathbf{d}^AA)
\end{align*}
Therefore we obtain that the first line of \eqref{ALP_PerfectComplexFluid}
becomes
\[
\frac{\partial}{\partial
t}\mathbf{u}+\nabla_\mathbf{u}\mathbf{u}=-\frac{1}{\rho}\left(\operatorname{grad}p+(gk)\left(\mathbf{i}_{\_\,}B,\operatorname{div}^AB\right)^\sharp\right),
\]
where $B=\mathbf{d}^AA$ and $p=\rho^2\frac{\partial e}{\partial\rho}$. In
summary, we have obtained the equations of Yang-Mills magnetohydrodynamics
\begin{equation}\label{ALP_YMMHD_A}
\left\lbrace
\begin{array}{ll}
\vspace{0.2cm}\displaystyle\frac{\partial}{\partial t}
\mathbf{u}+\nabla_\mathbf{u}\mathbf{u}=-\frac{1}{\rho}\left(\operatorname{grad}p+(gk)\left(\mathbf{i}_{\_\,}B,\operatorname{div}^AB\right)^\sharp\right),\quad
B=\mathbf{d}^AA,\\
\vspace{0.2cm}\displaystyle\frac{\partial}{\partial
t}Q+\operatorname{div}(Q\mathbf{u})=0,\qquad\displaystyle\frac{\partial}{\partial
t}A+\mathbf{d}^A(A(\mathbf{u}))+\mathbf{i}_\mathbf{u}B=0,\\
\vspace{0.2cm}\displaystyle\frac{\partial}{\partial
t}\rho+\operatorname{div}(\rho
\mathbf{u})=0,\qquad\,\,\,\displaystyle\frac{\partial}{\partial
t}s+\mathbf{d}s(\mathbf{u})=0.
\end{array} \right.
\end{equation}

We now treat the particular case of magnetohydrodynamics, that is, the case
$\mathcal{O}=S^1$. In order to recover the standard equations we suppose that
$\mathcal{D}$ is three dimensional. In this case we can define the
\textit{magnetic potential} $\mathbf{A}:=A^\sharp \in \mathfrak{X}(
\mathcal{D})$ and the \textit{magnetic field} $\mathbf{B}:=(\star B)^\sharp \in
\mathfrak{X}( \mathcal{D}) $. Since the group is Abelian, covariant
differentiation coincides with usual differentiation and the equality
$\mathbf{d}^AA=B$ reads $\operatorname{curl}\mathbf{A}=\mathbf{B}$. Using the
identities $(\operatorname{div}B)^\sharp=-\operatorname{curl}\mathbf{B}$ and
$(\mathbf{i}_\mathbf{u}B)^\sharp=\mathbf{B}\times\mathbf{u}$ we obtain
\[
g\left(\mathbf{i}_{\_\,}B,\operatorname{div}B\right)^\sharp=-\left(\mathbf{i}_{(\operatorname{div}B)^\sharp}B\right)^\sharp=\mathbf{B}\times\operatorname{curl}\mathbf{B}.
\]
Suppose that all particles have mass $m$. The electric charge $q$ is such that
$Q=\rho\frac{q}{m}$, therefore the equation for $Q$ in \eqref{ALP_YMMHD_A}
becomes
\[
\frac{\partial}{\partial t}q+\mathbf{d}q(\mathbf{u})=0.
\]
If we suppose that at time $t=0$ all the particle have the same charge, then
this charge remains constant for all time. By making use af these remarks and
hypotheses, equations \eqref{ALP_YMMHD_A} become
\begin{equation}\label{MHD_A}
\left\lbrace
\begin{array}{ll}
\vspace{0.2cm}\displaystyle\frac{\partial}{\partial t}
\mathbf{u}+\nabla_\mathbf{u}\mathbf{u}=-\frac{1}{\rho}\left(\operatorname{grad}p+\mathbf{B}\times\operatorname{curl}\mathbf{B}\right),\quad
\mathbf{B}=\operatorname{curl}\mathbf{A},\\
\vspace{0.2cm}\displaystyle\frac{\partial}{\partial
t}\mathbf{A}+\operatorname{grad}[g(\mathbf{A},\mathbf{u})]+\mathbf{B}\times\mathbf{u}=0,\\
\vspace{0.2cm}\displaystyle\frac{\partial}{\partial
t}\rho+\operatorname{div}(\rho
\mathbf{u})=0,\qquad\,\,\,\displaystyle\frac{\partial}{\partial
t}s+\mathbf{d}s(\mathbf{u})=0.
\end{array} \right.
\end{equation}
Thus, we have recovered the equations for magnetohydrodynamics. Turning back to the general case and using Theorem \ref{ALPSD}, we obtain the
following result.

\medskip  

\noindent \textbf{Hamiltonian reduction for Yang-Mills magnetohydrodynamics.} 
A smooth path $(\mathbf{m}_\eta,Q_\chi)\in
T^*[\operatorname{Diff}(\mathcal{D})\,\circledS\,\mathcal{F}(\mathcal{D},\mathcal{O})]$
is a solution of Hamilton's equations associated to the Hamiltonian
$H_{(\rho_0,s_0,A_0)}$ given in \eqref{YM-MHD_Hamiltonian} if and only if the
curve
\[
(\rho\mathbf{u}^\flat,Q)=:(\mathbf{m},Q):=J(\eta^{-1})\left(\mathbf{m}_\eta\circ\eta^{-1},T^*R_{\chi\circ\eta^{-1}}(Q_\chi\circ\eta^{-1})\right)
\]
is a solution of the system \eqref{ALP_YMMHD_A}
with initial conditions $(\rho_0,s_0,A_0)$.

The evolution of the advected quantities is given by
\[
\rho=J(\eta^{-1})(\rho_0\circ\eta^{-1}),\qquad s=s_0\circ\eta^{-1},
\]
\[
A=\operatorname{Ad}_{\chi\circ\eta^{-1}}\eta_*A_0+(\chi\circ\eta^{-1})T(\chi^{-1}\circ\eta^{-1})=\eta_*\left(\operatorname{Ad}_\chi A_0+\chi T\chi^{-1}\right).
\]
This theorem is interesting from two points of view. Firstly, it allows us to
recover the non-canonical Hamiltonian structure given in \cite{HoKu1988} by a
reduction from a canonical cotangent bundle. Secondly, it generalizes to the
nonabelian case the Hamiltonian reduction for magnetohydrodynamics given in
\cite{MaRaWe1984}.

The associated affine Lie-Poisson bracket is that given in \eqref{ALPB}, where
the third term takes the explicit form
\[
\int_\mathcal{D}\rho\left(\textbf{d}\left(\frac{\delta f}{\delta
\rho}\right)\frac{\delta g}{\delta 
\mathbf{m}}-\textbf{d}\left(\frac{\delta g}{\delta
\rho}\right)\frac{\delta f}{\delta  \mathbf{m}}\right)\mu+\int_\mathcal{D}
s\left(\operatorname{div}\left(\frac{\delta f}{\delta
s}\frac{\delta g}{\delta 
\mathbf{m}}\right)-\operatorname{div}\left(\frac{\delta g}{\delta
s}\frac{\delta f}{\delta  \mathbf{m}}\right)\right)\mu.
\]
Since the Hamiltonian depends on $A$ only through its curvature, the equations
can be formulated using the $B$-representation. One simply replaces the
equation
for $A$ by those for its curvature $B$, namely,
\[
\frac{\partial}{\partial t}B+\boldsymbol{\pounds}_\mathbf{u}B=0.
\]
In particular, we obtain that the force term
$(gk)\left(\mathbf{i}_{\_\,}B,\operatorname{div}^AB\right)^\sharp$ depends only
on the curvature $B$ and not on the connection one-form $A $.

In the particular case of magnetohydrodynamics, the evolution equation of
$\mathbf{B}$ reads
\[
\frac{\partial}{\partial
t}\mathbf{B}+\operatorname{curl}(\mathbf{B}\times\mathbf{u})=0.
\]

In the general case of Yang-Mills magnetohydrodynamics, the Kelvin-Noether
theorem gives
\[
\frac{d}{dt}\oint_{c_t}\mathbf{u}^\flat=\oint_{c_t}T\mathbf{d}s-\oint_{c_t}\frac{1}{\rho}(gk)\left(\mathbf{i}_{\_\,}B,\operatorname{div}^AB\right),
\]
and the $\gamma$-circulation gives
\[
\frac{d}{dt}\oint_{c_t}A=0,
\]
where $c_t$ is a loop which move with the fluid velocity $\mathbf{u}$, that is,
$c_t=\eta_t \circ  c_0$.


\subsection{Hall Magnetohydrodynamics}

As we will see, Hall magnetohydrodynamics does not require the use of the
affine
Lie-Poisson reduction developed in this paper. However, in view of the next
paragraph about superfluids, we quickly recall here from \cite{Ho1987} the
Hamiltonian formulation of these equations. We will obtain this Hamiltonian
structure by a Lie-Poisson reduction for semidirect products, associated to the
direct product group
$G:=\operatorname{Diff}(\mathcal{D})\times\operatorname{Diff}(\mathcal{D})$.
The
advected quantities are
\[
(\rho,s;n)\in\mathcal{F}(\mathcal{D})\times\mathcal{F}(\mathcal{D})\times\mathcal{F}(\mathcal{D}).
\]
The variables $\rho$ and $s$ are, as before, the \textit{mass density} and the
\textit{specific entropy}, on which only the first diffeomorphism group acts as
\[
(\rho,s)\mapsto((J\eta)(\rho\circ\eta),s\circ\eta).
\]
The variable $n$ is the \textit{electron charge density}, on which only the
second diffeomorphism group acts as
\[
n\mapsto (J\xi)(n\circ\xi).
\]
By Lie-Poisson reduction, for a Hamiltonian
$h=h(\mathbf{m},\rho,s;\mathbf{n},n)$ defined on the dual Lie-algebra
\begin{align*}
&\big([\mathfrak{X}(\mathcal{D})\,\circledS\,(\mathcal{F}(\mathcal{D})\times\mathcal{F}(\mathcal{D}))]\times[\mathfrak{X}(\mathcal{D})\,\circledS\,\mathcal{F}(\mathcal{D})]\big)^* \\
& \qquad \cong[\Omega^1(\mathcal{D})\times\mathcal{F}(\mathcal{D})\times\mathcal{F}(\mathcal{D})]\times[\Omega^1(\mathcal{D})\times\mathcal{F}(\mathcal{D})],
\end{align*}
we obtain the coupled Lie-Poisson equations
\begin{equation}\label{LP_H-MHD_1}
\left\{
\begin{array}{ll}
\vspace{0.1cm}\displaystyle\frac{\partial}{\partial
t}\mathbf{m}=-{\boldsymbol{\pounds}}_{\frac{\delta h}{\delta
\mathbf{m}}}\mathbf{m}-\operatorname{div}\left(\frac{\delta h}{\delta
\mathbf{m}}\right)\mathbf{m}-\frac{\delta h}{\delta \rho}\diamond
\rho-\frac{\delta h}{\delta s}\diamond s\\
\vspace{0.1cm}\displaystyle\frac{\partial}{\partial
t}\rho=-\operatorname{div}\left(\frac{\delta h}{\delta \mathbf{m}}\rho\right)\\
\vspace{0.1cm}\displaystyle\frac{\partial}{\partial
t}s=-\mathbf{d}s\left(\frac{\delta h}{\delta \mathbf{m}}\right)
\end{array}
\right.
\end{equation}
and
\begin{equation}\label{LP_H-MHD_2}
\left\{
\begin{array}{ll}
\vspace{0.1cm}\displaystyle\frac{\partial}{\partial
t}\mathbf{n}=-{\boldsymbol{\pounds}}_{\frac{\delta h}{\delta
\mathbf{n}}}\mathbf{n}-\operatorname{div}\left(\frac{\delta h}{\delta
\mathbf{n}}\right)\mathbf{n}-\frac{\delta h}{\delta n}\diamond n\\
\vspace{0.1cm}\displaystyle\frac{\partial}{\partial
t}n=-\operatorname{div}\left(\frac{\delta h}{\delta \mathbf{n}}n\right).
\end{array}
\right.
\end{equation}

From the second Lie-Poisson system we obtain that the evolution of
$\mathbf{n}/n$ is given by
\begin{equation}\label{evol_N/n}
\frac{\partial}{\partial
t}\left(\mathbf{n}/n\right)=-\boldsymbol{\pounds}_{\frac{\delta h}{\delta
\mathbf{n}}}(\mathbf{n}/n)-\operatorname{grad}\frac{\delta h}{\delta n}.
\end{equation}
The Hamiltonian for Hall magnetohydrodynamics is
\begin{equation}\label{Hamiltonian_H-MHD}
h(\mathbf{m},\rho,s;\mathbf{n},n):=\frac{1}{2}\int_\mathcal{D}\frac{1}{\rho}\left\|\mathbf{m}-\frac{a\rho}{R}A\right\|^2\mu+\int_\mathcal{D}
\rho e(\rho,s)\mu+\frac{1}{2}\int_\mathcal{D}\|\mathbf{d}A\|^2\mu,
\end{equation}
where the one-from $A$, defined by
\[
A:=R\frac{\mathbf{n}}{n} \in\Omega^1(\mathcal{D}),
\]
is the \textit{magnetic vector potential}, the constants $a, R$ are
respectively
the \textit{ion charge-to-mass ratio} and the \textit{Hall scaling parameter},
and the norms are taken with respect to a fixed Riemannian metric $g$ on
$\mathcal{D}$. The functional derivatives are computed to be
\[
\mathbf{u}:=\frac{\delta h}{\delta
\mathbf{m}}=\frac{1}{\rho}\mathbf{m}^\sharp-\frac{a}{R}A^\sharp,\quad\frac{\delta
h}{\delta
\rho}=-\frac{1}{2}\|\mathbf{u}\|^2-\frac{a}{R}A\cdot\mathbf{u}+e+\rho\frac{\partial
e}{\partial \rho},\quad\frac{\delta h}{\delta s}=\rho\frac{\partial e}{\partial
s},
\]
\[
\mathbf{v}:=\frac{\delta h}{\delta
\mathbf{n}}=-\frac{a\rho}{n}\mathbf{u}-\frac{R}{n}(\operatorname{div}B)^\sharp,\qquad \frac{\delta
h}{\delta
n}=\frac{a\rho}{n^2}\mathbf{n}\cdot\mathbf{u}+\frac{R}{n^2}\mathbf{n}\cdot(\operatorname{div}B)^\sharp=-\frac{1}{R}A\cdot\mathbf{v},
\]
where $B:=\mathbf{d}A$. Recall from \eqref{divergence_connection_formula} that $\operatorname{div}$ is
defined on $\Omega_k(\mathcal{D}) $ as the negative of the adjoint of the
exterior differential $\mathbf{d}$ on $\Omega^k( \mathcal{D}) $. However, the
metric $g $ on $\mathcal{D}$ gives an identification of 
$\Omega_k(\mathcal{D}) $ with $\Omega^k(\mathcal{D})$ and hence we can regard
$\operatorname{div} $ as defined also on $\Omega^k(\mathcal{D})$. It follows
that on $\Omega^k( \mathcal{D}) $ we have  $\operatorname{div}  =
-\delta$, where $\delta
$ is the usual codifferential induced by  $g $ and $\mathbf{d}$.

The variable $\mathbf{v}$ is interpreted as the \textit{electron fluid
velocity}. The advection equations for $\rho,s$, and $n$ are given by
\[
\frac{\partial}{\partial t}\rho+\operatorname{div}(\rho\mathbf{u})=0,\quad
\frac{\partial}{\partial
t}s+\mathbf{d}s(\mathbf{u})=0,\quad\text{and}\quad\frac{\partial}{\partial
t}n+\operatorname{div}(n\mathbf{v})=0.
\]
Using the expression of $\mathbf{v}$ in terms of $\mathbf{u}$ we obtain that
$\operatorname{div}(n
\mathbf{v})=-a\operatorname{div}(\rho\mathbf{u})$ (since $B = \mathbf{d}A $
implies that $
\operatorname{div}\left((\operatorname{div} B)^\sharp \right) = 0 $) which
proves that
\[
\frac{\partial}{\partial t}(a\rho+n)=0.
\]
Thus, if we assume that $a\rho_0+n_0=0$ for the initial conditions, we have
$a\rho+n=0$ for all time.

Using the definition of $A$ and \eqref{evol_N/n}, we obtain that the equation
for $A$ is given by
\[
\frac{\partial}{\partial t}A=-\mathbf{i}_{\mathbf{v}}B.
\] 
Using the equations for $A$ and $\rho$, we obtain the following equation for
$\mathbf{u}$:
\[
\frac{\partial}{\partial
t}\mathbf{u}+\nabla_\mathbf{u}\mathbf{u}=-\frac{1}{\rho}\left(\operatorname{grad}p-\frac{a\rho}{R}\left(\mathbf{i}_{\mathbf{v}-\mathbf{u}}B\right)^\sharp\right),\quad
p=\rho^2\frac{\partial e}{\partial\rho}.
\]
Suppose that the initial conditions $\rho_0$ and $n_0$ verify the equality
$a\rho_0+n_0=0$. As we have seen above, the equality remains valid for all
times, and we obtain
$\mathbf{v}=\mathbf{u}+\frac{R}{a\rho}(\operatorname{div}B)^\sharp$. Using
this,
the equations above simplify and we obtain the system
\begin{equation}\label{H-MHD}
\left\{
\begin{array}{ll}
\vspace{0.1cm}\displaystyle\frac{\partial}{\partial
t}\mathbf{u}+\nabla_\mathbf{u}\mathbf{u}=-\frac{1}{\rho}\left(\operatorname{grad}p-\left(\mathbf{i}_{(\operatorname{div}B)^\sharp}B\right)^\sharp\right)\\
\vspace{0.1cm}\displaystyle\frac{\partial}{\partial
t}\rho=-\operatorname{div}\left(\rho\mathbf{u}\right),\quad\displaystyle\frac{\partial}{\partial
t}s=-\mathbf{d}s\left(\mathbf{u}\right)\\
\vspace{0.1cm}\displaystyle\frac{\partial}{\partial
t}A=-\mathbf{i}_\mathbf{u}B-\frac{R}{a\rho}\mathbf{i}_{(\operatorname{div}B)^\sharp}B
\end{array}
\right.
\end{equation}

When $\mathcal{D}$ is three dimensional, we can define the \textit{magnetic
potential} $\mathbf{A}:=A^\sharp$ and the \textit{magnetic field}
$\mathbf{B}:=(\star B)^\sharp$. In this case the previous equations read
\begin{equation}\label{H-MHD_3D}
\left\{
\begin{array}{ll}
\vspace{0.1cm}\displaystyle\frac{\partial}{\partial
t}\mathbf{u}+\nabla_\mathbf{u}\mathbf{u}=-\frac{1}{\rho}\left(\operatorname{grad}p+\mathbf{B}\times\operatorname{curl}\mathbf{B}\right)\\
\vspace{0.1cm}\displaystyle\frac{\partial}{\partial
t}\rho=-\operatorname{div}\left(\rho\mathbf{u}\right),\quad\displaystyle\frac{\partial}{\partial
t}s=-\mathbf{d}s\left(\mathbf{u}\right)\\
\vspace{0.1cm}\displaystyle\frac{\partial}{\partial
t}\mathbf{A}=\mathbf{u}\times
\mathbf{B}+\frac{R}{a\rho}\mathbf{B}\times\operatorname{curl}\mathbf{B}.
\end{array}
\right.
\end{equation}
These are the classical equations of Hall magnetohydrodynamics. Note that we
can
pass from the equations for magnetohydrodynamics to those for Hall
magnetohydrodynamics by simply replacing the advection law for $A$ by the
\textit{Ohm's law}. In terms of the magnetic field $B$, one simply replace the
advection law
\[
\frac{\partial}{\partial t}B+\boldsymbol{\pounds}_\mathbf{u}B=0,
\]
where $\mathbf{u}$ is the fluid velocity, by the equation
\[
\frac{\partial}{\partial t}B+\boldsymbol{\pounds}_\mathbf{v}B=0,
\]
where $\mathbf{v}$ is the electron fluid velocity.

In fact, the Hamiltonian $h$ given in \eqref{Hamiltonian_H-MHD}, is the value
at
the identity of the right invariant Hamiltonian
$H(\mathbf{m}_\eta,\rho,s;\mathbf{n}_\xi,n)=H_{(\rho,s;n)}(\mathbf{m}_\eta;\mathbf{n}_\xi)$,
where
\[
H_{(\rho,s;n)} :
T^*(\operatorname{Diff}(\mathcal{D})\times\operatorname{Diff}(\mathcal{D}))\rightarrow\mathbb{R}
\]
is given by
\begin{align*}
H_{(\rho,s;n)}(\mathbf{m}_\eta;\mathbf{n}_\xi)&=\frac{1}{2}\int_\mathcal{D}
\left\|\mathbf{m}_\eta-a\rho\frac{\mathbf{n}_\xi\circ\xi^{-1}\circ\eta}{n\circ\xi^{-1}\circ\eta}\right\|^2\mu+\int_\mathcal{D}
\rho e(\rho J\eta^{-1},s)\mu \nonumber\\
 & \quad +\frac{R^2}{2}\int_\mathcal{D}
\left\|\mathbf{d}\left(\frac{\mathbf{n}_\xi\circ\xi^{-1}}{n\circ\xi^{-1}}\right)\right\|^2\mu.
\end{align*}

\medskip  

\noindent \textbf{Hamiltonian reduction for Hall magnetohydrodynamics.} Suppose that
$a\rho_0+n_0=0$. A curve $(\mathbf{m}_\eta,\mathbf{n}_\xi)\in
T^*(\operatorname{Diff}(\mathcal{D})\times\operatorname{Diff}(\mathcal{D}))$ is
a solution of Hamilton's equations associated to $H_{(\rho_0,s_0;n_0)}$ if and
only if the curve
\[
(\mathbf{m},\mathbf{n}):=(J(\eta^{-1})(\mathbf{m}_\eta\circ\eta^{-1}),J(\xi^{-1})(\mathbf{n}_\xi\circ\xi^{-1}))
\]
is a solution of the equations \eqref{H-MHD} where
$A:=\frac{R}{n}\mathbf{n}=-\frac{R}{a\rho}\mathbf{n}$, since $a\rho+n=0$.
Moreover the evolution of the advected quantities is given by
\[
\rho=J(\eta^{-1})(\rho_0\circ\eta^{-1}),\quad s=s_0\circ\eta^{-1},\quad
n=J(\xi^{-1})(n_0\circ\xi^{-1}).
\]
Let us assume from now on that the initial conditions $\rho_0$ and  $n_0$ are
related by $a \rho_0 + n _0 = 0 $. We have seen that this implies that 
$a\rho+n=0$. From the relations above we conclude the interesting result that
the action of $\xi^{-1}\circ\eta$ fixes $n_0$, that is, $J(\xi^{-1}\circ
\eta)(n_0 \circ \xi^{-1}\circ \eta) = n_0$. Conversely, given this relation and
the condition $a \rho_0 + n _0 = 0 $, it is
easily seen that $a\rho
+n=0$. 

The Lie-Poisson bracket associated to these equations is clearly the sum of two
Lie-Poisson brackets associated to the semidirect products
$\operatorname{Diff}(\mathcal{D})\,\circledS\,[\mathcal{F}(\mathcal{D})\times\mathcal{F}(\mathcal{D})]$
and $\operatorname{Diff}(\mathcal{D})\,\circledS\,\mathcal{F}(\mathcal{D})$.

The Kelvin-Noether theorem associated to the variable $\mathbf{m}$ gives
\[
\frac{d}{dt}\oint_{c_t}\left(\mathbf{u}^\flat+\frac{a}{R}A\right)=\oint_{c_t}T\mathbf{d}s,
\]
which can be rewritten as
\[
\frac{d}{dt}\oint_{c_t}\mathbf{u}^\flat=\oint_{c_t}T\mathbf{d}s+\oint_{c_t}\frac{1}{\rho}\mathbf{i}_{(\operatorname{div}B)^\sharp}B,
\]
where $c_t$ is a loop which moves with the \textit{fluid velocity}
$\mathbf{u}$,
that is, $c_t=\eta_t \circ c_0$. The Kelvin-Noether theorem associated to the
variable $\mathbf{n}$ gives
\[
\frac{d}{dt}\oint_{d_t}A=0,
\]
where $d_t$ is a loop which moves with the \textit{electron fluid velocity}
$\mathbf{v}$, that is, $d_t=\xi_t \circ d_0$.


\subsection{Multivelocity Superfluids}
\label{sec:multivelocity}

Superfluidity is a rare state of matter encountered in few fluids at extremely
low temperatures. Such materials exhibit strange behavior such as the lack of
viscosity, the ability to flow through very small channels that are impermeable
to ordinary fluids, and the fact that it can form a layer whose thickness is
that of one atom on the walls of the container in which it is placed. In
addition, the rotational speed of a superfluid is quantized, that is, the fluid
can rotate only at certain values of the speed. Superfluidity is considered to
be a manifestation of quantum mechanical effects at macroscopic level. Typical
examples of superfluids are ${}^3$He, whose atoms are fermions and the
superfluid transition occurs by Cooper pairing between atoms rather than
electrons, and ${}^4$He, whose atoms are bosons and the superfluidity is a
consequence of Bose-Einstein condensation in an interacting system.

For example at temperatures close to absolute zero a solution of ${}^3$He and
${}^4$He has its hydrodynamics described by three velocities: two superfluid
velocities $\mathbf{v}_s ^1, \mathbf{v}_s^2 $ and one normal fluid velocity
$\mathbf{v}_n $. If other kinds of superfluid are present, one needs to
introduce additional superfluid velocities. For a history of the equations
considered below and the Hamiltonian structure for multivelocity superfluids
see
\cite{HoKu1987}.

\medskip  

\noindent \textbf{The two-fluid model.}   We first treat the two-fluid model, that is,
the case of one superfluid velocity $\mathbf{v}_s$ and one normal-fluid
velocity
$\mathbf{v}_n$. Remarkably, this Hamiltonian structure can be obtained by
affine
Lie-Poisson reduction, with order parameter Lie group $\mathcal{O}=S^1$. In
this
paragraph we also carry out the corresponding Lagrangian formulation, by
applying the general theory of affine Euler-Poincar\'e reduction to the
semidirect product group 
$\operatorname{Diff}(\mathcal{D})\,\circledS\,\mathcal{F}(\mathcal{D},S^1)$.

The linear advected quantity is the \textit{entropy density} $S$ on which a
diffeomorphism $\eta$ acts as
\[
S\mapsto (J\eta)(S\circ\eta).
\]
The affine advected quantity is the \textit{superfluid velocity}
$\mathbf{v}_s$,
on which the element
$(\eta,\chi)\in\operatorname{Diff}(\mathcal{D})\,\circledS\,\mathcal{F}(\mathcal{D},S^1)$
acts as
\[
\mathbf{v}_s\mapsto
\left(\eta^*\mathbf{v}_s^\flat+\mathbf{d}\chi\right)^\sharp.
\]
This action is simply the affine representation
\eqref{affine_representation_gamma} for the Lie group $\mathcal{O}=S^1$. Here
the advected quantity $\mathbf{v}_s $ is a vector field and not a one-form,
since it represents a velocity and hence the formula 
\eqref{affine_representation_gamma} was changed accordingly. As will be seen, 
in this formalism, the \textit{mass density} does not appear as an advected
quantity in the representation space $V^*$; it is a momentum, that is, one of
the variables in the dual Lie algebra $\mathfrak{g}^\ast =
\Omega^1(\mathcal{D})
\times \mathcal{F}( \mathcal{D})$.

The reduced Lagrangian for superfluids is
\[
l:[\mathfrak{X}(\mathcal{D})\,\circledS\,\mathcal{F}(\mathcal{D})]\,\circledS\,[\mathcal{F}(\mathcal{D})\oplus\mathfrak{X}(\mathcal{D})] \longrightarrow\mathbb{R}
\]
given by
\begin{equation}\label{lagrangian_superfluids}
l(\mathbf{v}_n,\nu,S,\mathbf{v}_s):=\frac{1}{2}\int_\mathcal{D}\rho\|\mathbf{v}_n\|^2\mu+\int_\mathcal{D}(\rho\nu)\mu-\int_\mathcal{D}\varepsilon(\rho,S,\mathbf{v}_s-\mathbf{v}_n)\mu,
\end{equation}
where $\mathbf{v}_n$ is the \textit{velocity of the normal flow}. The internal
energy density $\varepsilon$ is seen here as a function of three variables
$\varepsilon=\varepsilon(\rho,S,\mathbf{r}):\mathbb{R}\times\mathbb{R}\times
T\mathcal{D}\rightarrow\mathbb{R}$. The norm in the first term is taken
relative
to a fixed Riemannian metric $g$ on $\mathcal{D}$. Note that for superfluids it
is more convenient to work with the internal energy and entropy \textit{per
unit
volume} and not \textit{per unit mass} as in the preceding examples. We make
the
following definitions:
\begin{align*}
\mu_{\rm
chem}&:=\frac{\partial\varepsilon}{\partial\rho}(\rho,S,\mathbf{v}_s-\mathbf{v}_n)\in\mathcal{F}(\mathcal{D}),\quad
T:=\frac{\partial\varepsilon}{\partial
S}(\rho,S,\mathbf{v}_s-\mathbf{v}_n)\in\mathcal{F}(\mathcal{D}),\\
\mathbf{p}&:=\frac{\partial\varepsilon}{\partial\mathbf{r}}(\rho,S,\mathbf{v}_s-\mathbf{v}_n)\in\Omega^1(\mathcal{D}).
\end{align*}
The interpretation of the quantities $\mu_{\rm chem}, T$, and $\mathbf{p}$ is
obtained from the following thermodynamic derivative identity for the internal
energy (superfluid first law):
\[
\mathbf{d}(\varepsilon(\rho,S,\mathbf{v}_s-\mathbf{v}_n))=\mu_{\rm
chem}\mathbf{d}\rho+T\mathbf{d}S+\mathbf{p}\cdot\nabla_{\_\,}(\mathbf{v}_s-\mathbf{v}_n)\in\Omega^1(\mathcal{D}),
\]
where $\nabla$ denotes the Levi-Civita covariant derivative associated to the
metric $g$. The function $\mu_{\rm chem}$ is the \textit{chemical potential},
$T$ is the \textit{temperature}, and $\mathbf{p}$ is the \textit{relative
momentum density}. The mass density $\rho$ is the function
$\rho=\rho(\mathbf{v}_n,\nu,S,\mathbf{v}_s)$ defined implicitly by the
condition
\begin{equation}\label{definition_rho}
\mu_{\rm
chem}:=\frac{\partial\varepsilon}{\partial\rho}(\rho,S,\mathbf{v}_s-\mathbf{v}_n)=\frac{1}{2}\|\mathbf{v}_n\|^2+\nu.
\end{equation}
Therefore, $\rho$ is not a variable in this approach. By the implicit function
theorem, the relation above defines a unique function $\rho$, provided the
function $\varepsilon$ verifies the condition
\begin{equation}\label{condition_rho}
\frac{\partial^2\varepsilon}{\partial\rho^2}(r,s,v_x)\neq 0,\;\text{for
all}\;(r,s,v_x)\in\mathbb{R}\times\mathbb{R}\times T\mathcal{D}.
\end{equation}

\medskip  

\noindent \textbf{The affine Euler-Poincar\'e equations for the two-fluid model.}
Using
the definition \eqref{definition_rho} of the function $\rho$, we compute below
the functional derivatives of $l$. We have
\begin{align*}
\frac{\delta
l}{\delta\mathbf{v}_n}&=\rho\mathbf{v}_n^\flat+\frac{1}{2}\|\mathbf{v_n}\|^2\frac{\partial\rho}{\partial\mathbf{v}_n}+\nu\frac{\partial\rho}{\partial\mathbf{v}_n}-\mu_{\rm
chem}\frac{\partial\rho}{\partial\mathbf{v}_n}+\mathbf{p}=\rho\mathbf{v}_n^\flat+\mathbf{p}=:\mathbf{m},\\
\frac{\delta
l}{\delta\nu}&=\frac{1}{2}\|\mathbf{v_n}\|^2\frac{\partial\rho}{\partial\nu}+\rho+\nu\frac{\partial\rho}{\partial\nu}-\mu_{\rm
chem}\frac{\partial\rho}{\partial\nu}=\rho,\\
\frac{\delta l}{\delta
S}&=\frac{1}{2}\|\mathbf{v_n}\|^2\frac{\partial\rho}{\partial
S}+\nu\frac{\partial\rho}{\partial S}-\mu_{\rm
chem}\frac{\partial\rho}{\partial
S}-T=-T,\\
\frac{\delta
l}{\delta\mathbf{v}_s}&=\frac{1}{2}\|\mathbf{v_n}\|^2\frac{\partial\rho}{\partial
\mathbf{v}_s}+\nu\frac{\partial\rho}{\partial \mathbf{v}_s}-\mu_{\rm
chem}\frac{\partial\rho}{\partial \mathbf{v}_s}-\mathbf{p}=-\mathbf{p}.
\end{align*}
Using the affine Euler-Poincar\'e equations
\eqref{EPComplexFluid_covariant_form}, we obtain the following equations for
$\rho, S$, and $\mathbf{v}_s$:
\begin{align*}
&\frac{\partial}{\partial
t}\rho+\operatorname{div}(\rho\mathbf{v}_n+\mathbf{p}^\sharp)=0,\quad\frac{\partial}{\partial
t}S+\operatorname{div}(S\mathbf{v}_n)=0,\\
&\frac{\partial}{\partial
t}\mathbf{v}_s+\operatorname{grad}\left(g(\mathbf{v}_s,\mathbf{v}_n)+\mu_{\rm
chem}-\frac{1}{2}\|\mathbf{v}_n\|^2\right)+\left(\mathbf{i}_{\mathbf{v}_n}\mathbf{d}\mathbf{v}_s^\flat\right)^\sharp=0.
\end{align*}
The last equation can be rewritten as
\begin{equation}\label{equ_v_s}
\frac{\partial}{\partial
t}\mathbf{v}_s+\nabla_{\mathbf{v}_s}\mathbf{v}_s=-\operatorname{grad}\left(\mu_{\rm
chem}-\frac{1}{2}\|\mathbf{v}_s-\mathbf{v}_n\|^2\right)+\left(\mathbf{i}_{\mathbf{v}_s-\mathbf{v}_n}\mathbf{d}\mathbf{v}_s^\flat\right)^\sharp
\end{equation}
When $\mathcal{D}$ is three dimensional, the last term reads
\[
\left(\mathbf{i}_{\mathbf{v}_s-\mathbf{v}_n}\mathbf{d}\mathbf{v}_s^\flat\right)^\sharp=\left(\star\mathbf{d}\mathbf{v}_s^\flat\right)^\sharp\times
(\mathbf{v}_s-\mathbf{v}_n)=\operatorname{curl}\mathbf{v}_s\times(\mathbf{v}_s-\mathbf{v}_n)=(\mathbf{v}_n-\mathbf{v}_s)\times\operatorname{curl}\mathbf{v}_s.
\]
Using the equality
\[
-\rho\mathbf{d}\left(\mu_{\rm
chem}-\frac{1}{2}\|\mathbf{v}_n\|^2\right)-S\mathbf{d}T=-\mathbf{d}p-\mathbf{p}\cdot\nabla_{\_\,}\mathbf{v}_s+\mathbf{m}\cdot\nabla_{\_\,}\mathbf{v}_n,
\]
where $p:=-\varepsilon(\rho,S,\mathbf{v}_s-\mathbf{v}_n)+\mu_{\rm chem}\rho+ST$
is the \textit{Euler pressure law}, the equation for $\mathbf{m}$ is computed
as follows:
\begin{align*}
\frac{\partial}{\partial
t}\mathbf{m}&=-\boldsymbol{\pounds}_{\mathbf{v}_n}\mathbf{m}-\operatorname{div}(\mathbf{v}_n)\mathbf{m}-\rho\mathbf{d}\left(\mu_{\rm
chem}-\frac{1}{2}\|\mathbf{v}_n\|^2\right)-S\mathbf{d}T\\
& \qquad -\operatorname{div}(\mathbf{p}^\sharp)\mathbf{v}_s^\flat+\mathbf{d}\mathbf{v}_s^\flat(\_,\mathbf{p}^\sharp)\\
&=-\boldsymbol{\pounds}_{\mathbf{v}_n}\mathbf{m}-\operatorname{div}(\mathbf{v}_n)\mathbf{m}+\mathbf{m}\cdot\nabla_{\_\,}\mathbf{v}_n-\mathbf{d}p-\mathbf{p}\cdot\nabla_{\_\,}\mathbf{v}_s\\
& \qquad -\operatorname{div}(\mathbf{p}^\sharp)\mathbf{v}_s^\flat+\mathbf{d}\mathbf{v}_s^\flat(\_,\mathbf{p}^\sharp)\\
&=-\nabla_{\mathbf{v}_n}\mathbf{m}-\operatorname{div}(\mathbf{v}_n)\mathbf{m}-\mathbf{d}p-\nabla_{\mathbf{p}^\sharp}\mathbf{v}_n^\flat-\operatorname{div}(\mathbf{p}^\sharp)\mathbf{v}_n^\flat\\
&=-\operatorname{Div}\mathbf{T},
\end{align*}
where $\mathbf{T}$ is the $(1,1)$ \textit{superfluid stress tensor} defined by
\[
\mathbf{T}:=\mathbf{v}_n\otimes\mathbf{m}+\mathbf{p}^\sharp\otimes\mathbf{v}_s^\flat+p\,\delta,
\]
$\delta$ is the Kronecker $(1,1)$ tensor, and $\operatorname{Div}$ is the
divergence of a $(1,1)$ tensor, defined as the trace of the bilinear map
\[
(\alpha,v)\mapsto\nabla_v\mathbf{T}(\alpha,\_)\in\Omega^1(\mathcal{D}).
\]
In coordinates, we have
\[
\mathbf{T}^i_j=\mathbf{v}_n^i\mathbf{m}_j+\mathbf{p}^i\mathbf{v}_{s
j}+p\,\delta^i_j
\]
and
\[
(\operatorname{Div}\mathbf{T})_j=(\nabla\mathbf{T})^i_{ji}=\partial_i\mathbf{T}^i_j+\mathbf{T}^l_j\Gamma^i_{il}-\mathbf{T}^i_l\Gamma^l_{ij}.
\]
The equations for superfluid dynamics are therefore given by
\begin{equation}\label{superfluids}
\left\{
\begin{array}{ll}
\vspace{0.1cm}\displaystyle\frac{\partial}{\partial
t}\mathbf{m}=-\operatorname{Div}\mathbf{T},\\
\vspace{0.1cm}\displaystyle\frac{\partial}{\partial
t}\rho+\operatorname{div}(\rho\mathbf{v}_n+\mathbf{p}^\sharp)=0,\qquad\frac{\partial}{\partial
t}S+\operatorname{div}(S\mathbf{v}_n)=0,\\
\vspace{0.1cm}\displaystyle\frac{\partial}{\partial
t}\mathbf{v}_s+\operatorname{grad}\left(g(\mathbf{v}_s,\mathbf{v}_n)+\mu_{\rm
chem}-\frac{1}{2}\|\mathbf{v}_n\|^2\right)+\left(\mathbf{i}_{\mathbf{v}_n}\mathbf{d}\mathbf{v}_s^\flat\right)^\sharp=0,
\end{array}
\right.
\end{equation}
where
\[
\mathbf{T}:=\mathbf{v}_n\otimes\mathbf{m}+\mathbf{p}^\sharp\otimes\mathbf{v}_s^\flat+p\,\delta,\quad\text{and}\quad
p:=-\varepsilon(\rho,S,\mathbf{v}_s-\mathbf{v}_n)+\mu_{\rm chem}\rho+ST.
\]
Thus we have recovered the equations (1a) - (1d) in \cite{HoKu1987}, in the
particular case of the two-fluids model. By Legendre transformation of the
reduced Lagrangian \eqref{lagrangian_superfluids}, we obtain the reduced
Hamiltonian
\begin{equation}\label{hamiltonian_superfluids} 
h(\mathbf{m},\rho,S,\mathbf{v}_s)=-\frac{1}{2}\int_\mathcal{D}\rho\|\mathbf{v}_n\|^2\mu+\int_\mathcal{D}(\mathbf{m}\cdot\mathbf{v}_n)\mu+\int_\mathcal{D}\varepsilon(\rho,S,\mathbf{v}_s-\mathbf{v}_n)\mu,
\end{equation}
where $\mathbf{v}_n=\mathbf{v}_n(\mathbf{m},\rho,S,\mathbf{v}_s)$ is the vector
field defined by the implicit condition
\begin{equation}\label{definition_v_n}
\mathbf{m}-\rho\mathbf{v}_n^\flat=\frac{\partial\varepsilon}{\partial\mathbf{r}}(\rho,S,\mathbf{v}_s-\mathbf{v}_n)=:\mathbf{p}.
\end{equation}
By the implicit function theorem, the above relation defines a unique vector
field $\mathbf{v}_n$, provided the function $\varepsilon$ verifies the
condition
that
\[
u_x\mapsto\frac{\partial^2\varepsilon}{\partial\mathbf{r}^2}(r,s,v_x)\cdot
u_x-ru_x
\]
is a bijective linear map. If \eqref{condition_rho} holds, then this condition
is equivalent to saying that the Legendre transformation is invertible. Of
course, the functions $l$ and $h$ are the values at the identity of the
corresponding unreduced right invariant Lagrangian and Hamiltonian
$L(\mathbf{v}_\eta,\nu_\chi,S,\mathbf{v}_s)=L_{(S,\mathbf{v}_s)}(\mathbf{v}_\eta,\nu_\chi)$ and
$H(\mathbf{m}_\eta,\rho_\chi,S,\mathbf{v}_s)=H_{(S,\mathbf{v}_s)}(\mathbf{m}_\eta,\rho_\chi)$,
where $
L_{(S,\mathbf{v}_s)}:T(\operatorname{Diff}(\mathcal{D})\,\circledS\,\mathcal{F}(\mathcal{D},S^1))\rightarrow\mathbb{R}$ and $
H_{(S,\mathbf{v}_s)}:T^*(\operatorname{Diff}(\mathcal{D})\,\circledS\,\mathcal{F}(\mathcal{D},S^1))\rightarrow\mathbb{R}$.

\noindent \textbf{Lagrangian reduction for superfluids.}  A curve
$(\eta,\chi)\in\operatorname{Diff}(\mathcal{D})\,\circledS\,\mathcal{F}(\mathcal{D},S^1)$
is a solution of the Euler-Lagrange equations associated to the Lagrangian
$L_{(S_0,\mathbf{v}_{s0})}$ if and only if the curve
\[
(\mathbf{v}_n,\nu):=\left(\dot\eta\circ\eta^{-1},(TR_{\chi^{-1}}\dot\chi)\circ\eta^{-1}\right)
\]
is a solution of the superfluids equations \eqref{superfluids}
with initial conditions $(S_0,\mathbf{v}_{s 0})$.

The evolution of the advected quantities is given by
\[
S=J(\eta^{-1})(S_0\circ\eta^{-1})\quad\text{and}\quad\mathbf{v}_s=\left(\eta_*\left(\mathbf{v}_{s
0}^\flat+\mathbf{d}\chi^{-1}\right)\right)^\sharp.
\]
Note that the evolution of the \textit{superfluid vorticity} is given by
\[
\mathbf{d}\mathbf{v}_s^\flat=\eta_*\mathbf{d}\mathbf{v}_{s 0}^\flat,
\]
therefore, the irrotationality condition $\mathbf{d}\mathbf{v}_s=0$
($\operatorname{curl}\mathbf{v}_s=0$ for the three dimensional case) is
preserved.

\medskip 

 \noindent \textbf{Hamiltonian reduction for superfluids.} A curve
$(\mathbf{m}_\eta,\rho_\chi)$ in the cotangent bundle $
T^*[\operatorname{Diff}(\mathcal{D})\,\circledS\,\mathcal{F}(\mathcal{D},S^1)]$
is a solution of Hamilton's equations associated to the superfluid Hamiltonian
$H_{(S_0,\mathbf{v}_{s 0})}$ if and only if the curve
\[
(\mathbf{m},\rho):=J(\eta^{-1})\left(\mathbf{m}_\eta\circ\eta^{-1},\rho_\chi\circ\eta^{-1}\right)
\]
is a solution of the system \eqref{superfluids}
with initial conditions $(S_0,\mathbf{v}_{s 0})$.

The associated Poisson bracket for superfluids is
\begin{align}\label{superfluids_bracket}
\{f,g\}(\mathbf{m},&\rho,S,\mathbf{v}_s)=\int_\mathcal{D}\mathbf{m}\cdot\left[\frac{\delta
f}{\delta\mathbf{m}},\frac{\delta
g}{\delta\mathbf{m}}\right]\mu+\int_\mathcal{D}\rho\cdot\left(\mathbf{d}\frac{\delta
f}{\delta\rho}\cdot\frac{\delta g}{\delta\mathbf{m}}-\mathbf{d}\frac{\delta
g}{\delta\rho}\cdot\frac{\delta f}{\delta\mathbf{m}}\right)\mu\nonumber\\
&+\int_\mathcal{D}S\cdot\left(\mathbf{d}\frac{\delta f}{\delta
S}\cdot\frac{\delta g}{\delta\mathbf{m}}-\mathbf{d}\frac{\delta g}{\delta
S}\cdot\frac{\delta f}{\delta\mathbf{m}}\right)\mu \nonumber\\
&+\int_\mathcal{D}\left[\left(\mathbf{d}\frac{\delta
f}{\delta\rho}+{\boldsymbol{\pounds}}_{\frac{\delta
f}{\delta\mathbf{m}}}\mathbf{v}_s^\flat\right)\cdot\frac{\delta
g}{\delta\mathbf{v}_s}^\sharp-\left(\mathbf{d}\frac{\delta
g}{\delta\rho}+{\boldsymbol{\pounds}}_{\frac{\delta
g}{\delta\mathbf{m}}}\mathbf{v}_s^\flat\right)\cdot\frac{\delta
f}{\delta\mathbf{v}_s}^\sharp\right]\mu.
\end{align}

\medskip  

\noindent \textbf{Multivelocity superfluids.} We now quickly explain how to generalize
the preceding approach to the case of superfluids with $m$ velocities. Consider
the semidirect product
\[
G:=\operatorname{Diff}(\mathcal{D})\,\circledS\,\underbrace{\left[\mathcal{F}(\mathcal{D},S^1)\times
... \times \mathcal{F}(\mathcal{D},S^1)\right]}_{m\;\; \text{times}},
\]
where the group on the right is the direct product of the groups
$\mathcal{F}(\mathcal{D},S^1)$. The semidirect product is associated to the
right action of $\operatorname{Diff}(\mathcal{D})$ given by
\[
(\chi^1,...,\chi^m)\mapsto (\chi^1\circ\eta,...,\chi^m\circ\eta).
\]
The affine advected quantities are the $m$ superfluid velocities
$(\mathbf{v}_s^1,...,\mathbf{v}_s^m)\in\mathfrak{X}(\mathcal{D})^m$ on which an
element $(\eta,\chi^1,...,\chi^m)$ acts as
\[
(\mathbf{v}_s^1,...,\mathbf{v}_s^m)\mapsto
(\eta^*(\mathbf{v}_s^1)^\flat+\mathbf{d}\chi^1,...,\eta^*(\mathbf{v}_s^m)^
\flat+\mathbf{d}\chi^m)^\sharp.
\]
The reduced Lagrangian is defined on
$[\mathfrak{X}(\mathcal{D})\,\circledS\,\mathcal{F}(\mathcal{D})^m]\,\circledS\,[\mathcal{F}(\mathcal{D})\oplus\mathfrak{X}(\mathcal{D})^m]$
and is given by
\[
l(\mathbf{v}_n,(\nu^\alpha),S,(\mathbf{v}_s^\alpha)):=\frac{1}{2}\int_\mathcal{D}\rho\|\mathbf{v}_n\|\mu+\sum_{\alpha=1}^m\int_\mathcal{D}\rho^\alpha\nu^\alpha-\int_\mathcal{D}\varepsilon((\rho^\alpha),S,(\mathbf{v}_s^\alpha-\mathbf{v}_n))\mu,
\]
where $\rho^\alpha$, the \textit{mass density} of the condensate particles with
flow $\mathbf{v}_s^\alpha$, is the function
$\rho^\alpha=\rho^\alpha(\mathbf{v}_n,(\nu^\alpha),S,(\mathbf{v}_s^\alpha))$
defined by the condition
\[
\mu_{\rm
chem}^\alpha:=\frac{\partial\varepsilon}{\partial\rho^\alpha}((\rho^\alpha),S,(\mathbf{v}^\alpha_s-\mathbf{v}_n))=\frac{1}{2}\|\mathbf{v}_n\|^2+\nu^\alpha.
\]
By the implicit function theorem, these conditions uniquely determine
$\rho^\alpha$, provided the matrix
\[
\left(\frac{\partial^2\varepsilon}{\partial\rho^\alpha\partial\rho^\beta}((r^i),s,(v_x^j))\right)_{\alpha\beta}
\]
is invertible for all $r^i,s,v^j_x\in\mathbb{R}\times\mathbb{R}\times
T\mathcal{D}$. The variable $\rho$ is defined by $\rho:=\rho^1+...+\rho^m$ and
denotes the \textit{total mass density}. Using the notations
\begin{align*}
\mu_{\rm
chem}^\alpha&:=\frac{\partial\varepsilon}{\partial\rho^\alpha}((\rho^\alpha),S,(\mathbf{v}_s^\alpha-\mathbf{v}_n))\in\mathcal{F}(\mathcal{D}),\\
T&:=\frac{\partial\varepsilon}{\partial
S}((\rho^\alpha),S,(\mathbf{v}_s^\alpha-\mathbf{v}_n))\in\mathcal{F}(\mathcal{D}),\\
\mathbf{p}^\alpha&:=\frac{\partial\varepsilon}{\partial\mathbf{r}^\alpha}((\rho^\alpha),S,(\mathbf{v}_s^\alpha-\mathbf{v}_n))\in\Omega^1(\mathcal{D}),
\end{align*}
we obtain the functional derivatives
\[
\frac{\delta l}{\delta
\mathbf{v}_n}=\rho\mathbf{v}_n^\flat+\sum_{\alpha=1}^m\mathbf{p}^\alpha,\;\;
\frac{\delta l}{\delta\nu^\alpha }=\rho^\alpha,\;\; \frac{\delta l}{\delta
S}=-T,\;\;\frac{\delta l}{\delta\mathbf{v}^\alpha_s}=-\mathbf{p}^\alpha
\]
and the thermodynamic derivative identity for the internal energy
\[
\mathbf{d}(\varepsilon((\rho^\alpha),S,(\mathbf{v}^\alpha_s-\mathbf{v}_n)))=\sum_{\alpha=1}^m\mu_{\rm
chem}^\alpha\mathbf{d}\rho^\alpha+T\mathbf{d}S+\sum_{\alpha=1}^m\mathbf{p}^\alpha\cdot\nabla_{\_\,}(\mathbf{v}^\alpha_s-\mathbf{v}_n)\in\Omega^1(\mathcal{D}).
\]
Using the affine Euler-Poincar\'e equations associated to 
$\operatorname{Diff}(\mathcal{D})\,\circledS\,[\mathcal{F}(\mathcal{D},S^1)\times...\times\mathcal{F}(\mathcal{D},S^1)]$,
we obtain the equations for multivelocity superfluids
\begin{equation}\label{multivelocity_superfluids}
\left\{
\begin{array}{ll}
\vspace{0.1cm}\displaystyle\frac{\partial}{\partial
t}\mathbf{m}=-\operatorname{Div}\mathbf{T},\\
\vspace{0.1cm}\displaystyle\frac{\partial}{\partial
t}\rho^\alpha+\operatorname{div}\left(\rho^\alpha\mathbf{v}_n+(\mathbf{p}^\alpha)^\sharp\right)=0,\qquad\frac{\partial}{\partial
t}S+\operatorname{div}(S\mathbf{v}_n)=0,\\
\vspace{0.1cm}\displaystyle\frac{\partial}{\partial
t}\mathbf{v}^\alpha_s+\operatorname{grad}\left(g(\mathbf{v}^\alpha_s,\mathbf{v}_n)+\mu_{\rm
chem}^ \alpha
-\frac{1}{2}\|\mathbf{v}_n\|^2\right)+\left(\mathbf{i}_{\mathbf{v}_n}\mathbf{d}(\mathbf{v}_s^\alpha)^\flat\right)^\sharp=0,
\end{array}
\right.
\end{equation}
where $\alpha=1,...,m$. The stress tensor $\mathbf{T}$ and the pressure $p$ are
given by
\begin{align*}
\mathbf{T}:&=\mathbf{v}_n\otimes\mathbf{m}+\sum_{\alpha=1}^m(\mathbf{p}^\alpha)^\sharp\otimes(\mathbf{v}^\alpha_s)^\flat+p\,\delta,\\
p:&=-\varepsilon((\rho^\alpha),S,(\mathbf{v}^\alpha_s-\mathbf{v}_n))+\sum_{\alpha=1}^m\mu^\alpha_{\rm
chem}\rho^\alpha+ST.
\end{align*}
By Legendre transformation, we obtain the Hamiltonian
\[
h\left(\mathbf{m},(\rho^\alpha),S,(\mathbf{v}_s^\alpha)\right)=-\frac{1}{2}\int_\mathcal{D}\rho\|\mathbf{v}_n\|^2\mu+\int_\mathcal{D}(\mathbf{m}\cdot\mathbf{v}_n)\mu+\int_\mathcal{D}\varepsilon((\rho^\alpha),S,(\mathbf{v}_s^\alpha-\mathbf{v}_n))\mu,
\]
where
$\mathbf{v}_n=\mathbf{v}_n(\mathbf{m},(\rho^\alpha),S,(\mathbf{v}_s^\alpha))$
is
the vector field defined by the implicit condition
\[
\mathbf{m}-\rho\mathbf{v}_n^\flat=\sum_{\alpha=1}^m\frac{\partial\varepsilon}{\partial\mathbf{r}^\alpha}((\rho^\alpha),S,(\mathbf{v}_s^\alpha-\mathbf{v}_n)).
\]
By the implicit function theorem, the above relation defines a unique function
$\mathbf{v}_n$, provided the function $\varepsilon$ verifies the condition that
the linear map
\[
u_x\mapsto\sum_{\alpha,\beta=1}^m\frac{\partial^2\varepsilon}{\partial\mathbf{r}^\alpha\partial\mathbf{r}^\beta}(r,s,v_x)\cdot
u_x-ru_x
\]
is bijective.

Lagrangian and Hamiltonian reductions hold as in the two-fluids model. The
evolutions of $S$ and $\mathbf{v}^\alpha_s$ are given by
\[
S=J(\eta^{-1})(S_0\circ\eta^{-1})\;\;\text{and}\;\;\mathbf{v}_s^\alpha=\eta_*\left((\mathbf{v}_{s
0}^\alpha)^\flat+\mathbf{d}(\chi^\alpha)^{-1}\right)^\sharp,
\]
and the irrotationality condition $\mathbf{d}(\mathbf{v}_s^\alpha)^\flat=0$ is
preserved.

The associated Poisson bracket is given by
\begin{align}\label{multivelocity_superfluids_bracket}
&
\!\!\!\!
\{f,g\}(\mathbf{m},\rho,S,\mathbf{v}_s)=\int_\mathcal{D}\mathbf{m}\cdot\left[\frac{\delta
f}{\delta\mathbf{m}},\frac{\delta
g}{\delta\mathbf{m}}\right]\mu \nonumber \\
&\!\!\!\! +\sum_{\alpha=1}^m\int_\mathcal{D}\rho^\alpha\cdot\left(\mathbf{d}\frac{\delta
f}{\delta\rho^\alpha}\cdot\frac{\delta
g}{\delta\mathbf{m}}-\mathbf{d}\frac{\delta
g}{\delta\rho^\alpha}\cdot\frac{\delta
f}{\delta\mathbf{m}}\right)\mu\nonumber\\
& \!\!\!\!  +\int_\mathcal{D}S\cdot\left(\mathbf{d}\frac{\delta f}{\delta
S}\cdot\frac{\delta g}{\delta\mathbf{m}}-\mathbf{d}\frac{\delta g}{\delta
S}\cdot\frac{\delta f}{\delta\mathbf{m}}\right)\mu\nonumber\\
& \!\!\!\!  +\sum_{\alpha=1}^m\int_\mathcal{D}\left[\left(\mathbf{d}\frac{\delta
f}{\delta\rho^\alpha}+{\boldsymbol{\pounds}}_{\frac{\delta
f}{\delta\mathbf{m}}}(\mathbf{v}_s^\alpha)^\flat\right)\cdot\frac{\delta
g}{\delta\mathbf{v}_s^\alpha}^\sharp-\left(\mathbf{d}\frac{\delta
g}{\delta\rho^\alpha}+{\boldsymbol{\pounds}}_{\frac{\delta
g}{\delta\mathbf{m}}}(\mathbf{v}_s^\alpha)^\flat\right)\cdot\frac{\delta
f}{\delta\mathbf{v}_s^\alpha}^\sharp\right]\mu.
\end{align}
The $\gamma$-circulation gives
\[
\frac{d}{dt}\oint_{c_t}(\mathbf{v}_s^\alpha)^\flat=0,\quad \text{for all} \quad
\alpha=1,...,m,
\]
where $c_t$ is a loop which moves with the \textit{normal fluid velocity}
$\mathbf{v}_n$.


\subsection{Superfluid Yang-Mills Magnetohydrodynamics}

In this paragraph we combine the Hamiltonian structures of Yang-Mills
magnetohydrodynamics and superfluid dynamics, to obtain a new physical model
for
the theory of superfluids Yang-Mills magnetohydrodynamics as well as the
corresponding Hamiltonian structure. In the Abelian case we recover the theory
and the Hamiltonian structure derived in \cite{HoKu1987}. We need a slight
generalization of the geometric framework developed in \S\ref{Lagrangian_PCF}
and \S\ref{Hamiltonian_PCF}, namely we consider the group semidirect product
\[
\operatorname{Diff}(\mathcal{D})\,\circledS\,(\mathcal{F}(\mathcal{D},\mathcal{O})\times\mathcal{F}(\mathcal{D},S^1)),
\]
where $\mathcal{F}(\mathcal{D},\mathcal{O})\times\mathcal{F}(\mathcal{D},S^1)$
is a direct product of groups on which $\operatorname{Diff}(\mathcal{D})$ acts
as
\[
(\chi_1,\chi_2)\mapsto (\chi_1\circ\eta,\chi_2\circ\eta).
\]
The affine advected quantities are the \textit{potential of the Yang-Mills
fluid} $A$ and the \textit{superfluid velocity} $\mathbf{v}_s$, on which
$(\eta,\chi_1,\chi_2)$ acts as
\[
A\mapsto
\operatorname{Ad}_{\chi_1^{-1}}\eta^*A+\chi_1^{-1}T\chi_1\quad\text{and}\quad
\mathbf{v}_s\mapsto (\eta^*\mathbf{v}_s^\flat+\mathbf{d}\chi_2)^\sharp.
\]
The reduced Hamiltonian is defined on the dual of the Lie algebra
\[
[\mathfrak{X}(\mathcal{D})\,\circledS\,(\mathcal{F}(\mathcal{D},\mathfrak{o})\times\mathcal{F}(\mathcal{D}))]\,\circledS\,(\mathcal{F}(\mathcal{D})\times\Omega^1(\mathcal{D},\mathfrak{o})\times\mathfrak{X}(\mathcal{D}))
\]
and is given by
\begin{align*}
h(\mathbf{m},Q,\rho,S,A,\mathbf{v}_s)&=-\frac{1}{2}\int_\mathcal{D}\rho\|\mathbf{v}_n\|^2\mu+\int_\mathcal{D}(\mathbf{m}\cdot\mathbf{v}_n)\mu \\
& \qquad \qquad  +\int_\mathcal{D}\varepsilon(\rho,S,\mathbf{v}_s-\mathbf{v}_n)\mu+\frac{1}{2}\int_\mathcal{D}\|\mathbf{d}^AA\|^2\mu,
\end{align*}
where $\mathbf{v}_n$ is the \textit{normal fluid velocity} defined as in
\eqref{definition_v_n}. This is simply the Hamiltonian
\eqref{hamiltonian_superfluids} plus the energy of the Yang-Mills field. The
norms are respectively associated to the metrics $g$ and $(gk)$, where $g$ is a
Riemannian metric on $\mathcal{D}$ and $k$ is an $\operatorname{Ad}$-invariant
inner product on $\mathfrak{o}$. The affine Lie-Poisson equations associated to
this Hamiltonian are computed to be
\begin{equation}\label{YM_superfluids}
\left\{
\begin{array}{ll}
\vspace{0.1cm}\displaystyle\frac{\partial}{\partial
t}\mathbf{m}=-\operatorname{Div}\mathbf{T},\qquad\frac{\partial}{\partial
t}\rho+\operatorname{div}(\rho\mathbf{v}_n+\mathbf{p}^\sharp)=0,\\
\vspace{0.1cm}\displaystyle\frac{\partial}{\partial
t}Q+\operatorname{div}(Q\mathbf{v}_n)=0,\qquad\frac{\partial}{\partial
t}S+\operatorname{div}(S\mathbf{v}_n)=0,\\
\vspace{0.1cm}\displaystyle\frac{\partial}{\partial
t}\mathbf{v}_s+\operatorname{grad}\left(g(\mathbf{v}_s,\mathbf{v}_n)+\mu_{\rm
chem}-\frac{1}{2}\|\mathbf{v}_n\|^2\right)+\left(\mathbf{i}_{\mathbf{v}_n}\mathbf{d}\mathbf{v}_s^\flat\right)^\sharp=0,\\
\vspace{0.1cm}\displaystyle\frac{\partial}{\partial
t}A+\mathbf{d}^A(A(\mathbf{v}_n))+\mathbf{i}_{\mathbf{v}_n}B=0,\quad
B:=\mathbf{d}^AA.
\end{array}
\right.
\end{equation}
For superfluid Yang-Mills magnetohydrodynamics the stress tensor is given by
\[
\mathbf{T}:=\mathbf{v}_n\otimes\mathbf{m}+\mathbf{p}^\sharp\otimes\mathbf{v}_s^\flat+B\!\cdot\!
B+ p\,\delta,
\]
where $B\!\cdot\! B$ is the $(1,1)$ tensor field defined by 
\[
(B\!\cdot\! B)^i_j:=B^b_{lj}B_b^{li},
\]
and where the \textit{pressure} is given by
$p:=-\varepsilon(\rho,S,\mathbf{v}_s-\mathbf{v}_n)+\mu_{\rm
chem}\rho+ST-\frac{1}{2}\|B\|^2$.

The corresponding Hamiltonian reduction and affine Lie-Poisson bracket can be
found as before and the evolutions of the advected quantities are given by
\[
S=J(\eta^{-1})(S_0\circ\eta^{-1}),\;\; A=\eta_*\left(\operatorname{Ad}_\chi
A_0+\chi_1
T\chi_1^{-1}\right),\;\;\mathbf{v}_s=\left(\eta_*\left(\mathbf{v}_{s
0}^\flat+\mathbf{d}\chi_2^{-1}\right)\right)^\sharp.
\]
As in the preceding example, it is possible to generalize this approach to
multivelocity superfluids. In this case, the $\gamma$-circulation gives
\[
\frac{d}{dt}\oint_{c_t}(\mathbf{v}_s^\alpha)^\flat=0,\quad \text{for all} \quad
\alpha=1,...,m, \quad \text{and}\quad \frac{d}{dt}\oint_{c_t}A=0,
\]
where $c_t$ is a loop which moves with the \textit{normal fluid velocity}
$\mathbf{v}_n$.


\subsection{Superfluid Hall Magnetohydrodynamics}
\label{sec:HallMHD}

The Hamiltonian formulation of superfluid Hall magnetohydrodynamics is given in
\cite{HoKu1987}. As one can guess, the Hamiltonian structure of these equations
combines the Hamiltonian structures of Hall magnetohydrodynamics and of
superfluids. This is still true at the group level and we will obtain the
equations by affine Lie-Poisson reduction associated to the group
\[
G:=\left[\operatorname{Diff}(\mathcal{D})\,\circledS\,\mathcal{F}(\mathcal{D},S^1)\right]\times\operatorname{Diff}(\mathcal{D}).
\]
In this expression, the symbol $\times$ denotes the \textit{direct product} of
the two groups. The advected quantities are
\[
(S,\mathbf{u};n)\in\mathcal{F}(\mathcal{D})\times\mathfrak{X}(\mathcal{D})\times\mathcal{F}(\mathcal{D}).
\]
The variable $S$ is the \textit{entropy density} of the normal flow, the other
variables will be interpreted later. The action of $(\eta,\chi;\xi)\in G$ is
given by
\[
(S,\mathbf{u};n)\mapsto(J\eta(S\circ\eta),(\eta^*\mathbf{u}^\flat+\mathbf{d}\chi)^\sharp;J\xi(n\circ\xi)).
\]
The resulting affine Lie-Poisson equations consist of two systems, the affine
Lie-Poisson equations associated to the variables
$(\mathbf{m},\rho,S,\mathbf{u})$ and the Lie-Poisson equations associated to
the
variables $(\mathbf{n},n)$.

The Hamiltonian of superfluid Hall magnetohydrodynamics is defined on the dual
Lie algebra
\begin{align*}
&\big(\left[(\mathfrak{X}(\mathcal{D})\,\circledS\,\mathcal{F}(\mathcal{D}))\,\circledS\,(\mathcal{F}(\mathcal{D})\oplus\mathfrak{X}(\mathcal{D}))\right]\times[\mathfrak{X}(\mathcal{D})\,\circledS\,\mathcal{F}(\mathcal{D})]
\big)^\ast \\
&\quad\cong
\Omega^1(\mathcal{D})\times\mathcal{F}(\mathcal{D})\times\mathcal{F}(\mathcal{D})\times\Omega^1(\mathcal{D})\times\Omega^1(\mathcal{D})\times\mathcal{F}(\mathcal{D})
\end{align*}
and is given by
\begin{align*}
h(\mathbf{m},\rho,S,\mathbf{u};\mathbf{n},n):=&-\frac{1}{2}\int_\mathcal{D}\rho\|\mathbf{v}_n\|^2\mu+\int_\mathcal{D}\left(\left(\mathbf{m}-\frac{a\rho}{R}
A\right)\cdot\mathbf{v}_n\right)\mu\\
&+\int_\mathcal{D}\varepsilon(\rho,S,\mathbf{v}_s-\mathbf{v}_n)\mu+\frac{1}{2}\int_\mathcal{D}\|\mathbf{d}A\|^2\mu,
\end{align*}
where $\mathbf{v}_n$ is the \textit{velocity of the normal flow},
$\mathbf{v}_s:=\mathbf{u}-\frac{a}{R}A^\sharp$ is the \textit{superfluid
velocity}, and $\varepsilon$ is the \textit{internal energy density}. The
one-form $A$ is defined by
\[
A:=R\frac{\mathbf{n}}{n}.
\]
The norm in the first term is taken with respect to a fixed Riemannian metric
$g$ on $\mathcal{D}$. The velocity $\mathbf{v}_n$ is the function
$\mathbf{v}_n=\mathbf{v}_n(\mathbf{m},\rho,S,\mathbf{u};\mathbf{n},n)$ defined
by the implicit condition
\[
\mathbf{m}-\rho\mathbf{v}_n^\flat-\frac{a\rho}{R}A=\frac{\partial\varepsilon}{\partial\mathbf{r}}(\rho,S,\mathbf{v}_s-\mathbf{v}_n)=:\mathbf{p}.
\]
By the implicit function theorem, the above relation defines a unique function
$\mathbf{v}_n$, provided the function $\varepsilon$ verifies the condition that
the linear map
\[
u_x\mapsto\frac{\partial^2\varepsilon}{\partial\mathbf{r}^2}(r,s,v_x,w_x)\cdot
u_x-ru_x
\]
is bijective for all $(r,s,v_x,w_x)\in\mathbb{R}\times\mathbb{R}\times
T\mathcal{D}\times T\mathcal{D}$.

Using the notations
\begin{align*}
\mu_{\rm
chem}&:=\frac{\partial\varepsilon}{\partial\rho}(\rho,S,\mathbf{v}_s-\mathbf{v}_n)\in\mathcal{F}(\mathcal{D}),\quad
T:=\frac{\partial\varepsilon}{\partial
S}(\rho,S,\mathbf{v}_s-\mathbf{v}_n)\in\mathcal{F}(\mathcal{D}),\\
\mathbf{p}&:=\frac{\partial\varepsilon}{\partial\mathbf{r}}(\rho,S,\mathbf{v}_s-\mathbf{v}_n)\in\Omega^1(\mathcal{D}),
\end{align*}
the functional derivatives of $h$ are computed to be
\[
\frac{\delta h}{\delta\mathbf{m}}=\mathbf{v}_n,\;\;\frac{\delta
h}{\delta\rho}=-\frac{1}{2}\|\mathbf{v}_n\|^2-\frac{a}{R}A\cdot\mathbf{v}_n+\mu_{\rm
chem} ,\;\;\frac{\delta h}{\delta S}=T,\;\;\frac{\delta
h}{\delta\mathbf{u}}=\mathbf{p},
\]
\[
\mathbf{v}:=\frac{\delta
h}{\delta\mathbf{n}}=-\frac{1}{n}\left(a\rho\mathbf{v}_n+a\mathbf{p}^\sharp+R(\operatorname{div}B)^\sharp\right),\;\;\frac{\delta
h}{\delta n}=-\frac{1}{R}A\cdot\mathbf{v}.
\]
The vector field $\mathbf{v}$ is interpreted as the \textit{electron fluid
velocity}. The equations for $\rho, S$, and $n$ are given by
\[
\frac{\partial}{\partial
t}\rho+\operatorname{div}(\rho\mathbf{v}_n+\mathbf{p}^\sharp)=0,\;\;\frac{\partial}{\partial
t}S+\operatorname{div}(S\mathbf{v}_n)=0,\;\;\text{and}\;\;\frac{\partial}{\partial
t}n+\operatorname{div}(n\mathbf{v})=0.
\]
Using the expression of $\mathbf{v}$ in terms of $\mathbf{v}_n$ we obtain that
$\operatorname{div}(n\mathbf{v})=-a\operatorname{div}(\rho\mathbf{v}_n+\mathbf{p}^\sharp)$
which proves that
\[
\frac{\partial}{\partial t}(a\rho+n)=0.
\]
Thus, if we assume that the initial conditions verify $a\rho_0+n_0=0$, then we
have $a\rho+n=0$ for all time. The equations for $A$ and $\mathbf{u}$ are
computed to be
\[
\frac{\partial}{\partial
t}A=-\mathbf{i}_{\mathbf{v}_n}B-\frac{1}{\rho}\mathbf{i}_{\mathbf{p}^\sharp}B-\frac{R}{a\rho}\mathbf{i}_{(\operatorname{div}B)^\sharp}B
\]
\[
\frac{\partial}{\partial
t}\mathbf{u}+\operatorname{grad}\left(g(\mathbf{v}_s,\mathbf{v}_n)+\mu_{\rm
chem}-\frac{1}{2}\|\mathbf{v}_n\|^2\right)+\left(\mathbf{i}_{\mathbf{v}_n}\mathbf{d}\mathbf{u}^\flat\right)^\sharp=0.
\]
From these two equations we obtain the evolution of the superfluid velocity
$\mathbf{v}_s=\mathbf{u}-\frac{a}{R}A^\sharp$ as
\begin{align*}
\frac{\partial}{\partial
t}\mathbf{v}_s = &-\operatorname{grad}\left(g(\mathbf{v}_s,\mathbf{v}_n)+\mu_{\rm chem}-\frac{1}{2}\|\mathbf{v}_n\|^2\right) \\
& +\left(\frac{a}{R\rho}\mathbf{i}_{\mathbf{p}^\sharp}B+\frac{1}{\rho}\mathbf{i}_{(\operatorname{div}B)^\sharp}B-\mathbf{i}_{\mathbf{v}_n}\mathbf{d}\mathbf{v}_s^\flat\right)^\sharp.
\end{align*}
Doing computations similar to those for superfluids we obtain that the equation
for $\mathbf{m}+\mathbf{n}$ is given by
\[
\frac{\partial}{\partial
t}(\mathbf{m}+\mathbf{n})=-\operatorname{Div}\mathbf{T},
\]
where the stress tensor $\mathbf{T}$ is given by
\[
\mathbf{T}=\mathbf{v}_n\otimes(\rho\mathbf{v}_n^\flat+\mathbf{p})+\mathbf{p}^\sharp\otimes\mathbf{v}_s^\flat+B\!\cdot\!B+p\delta,\quad
p=\rho\mu_{\rm chem}+ST-\varepsilon-\frac{1}{2}\|B\|^2.
\]
Thus, we have obtained the following equations
\begin{equation}\label{superfluid_H_MHD}
\left\{
\begin{array}{ll}
\vspace{0.1cm}\displaystyle\frac{\partial}{\partial
t}(\mathbf{m}+\mathbf{n})=-\operatorname{Div}\mathbf{T},\\
\vspace{0.1cm}\displaystyle\frac{\partial}{\partial
t}\rho+\operatorname{div}(\rho\mathbf{v}_n+\mathbf{p}^\sharp)=0,\qquad\frac{\partial}{\partial
t}S+\operatorname{div}(S\mathbf{v}_n)=0,\\
\vspace{0.1cm}\displaystyle\frac{\partial}{\partial
t}A=-\mathbf{i}_{\mathbf{v}_n}B-\frac{1}{\rho}\mathbf{i}_{\mathbf{p}^\sharp}B-\frac{R}{a\rho}\mathbf{i}_{(\operatorname{div}B)^\sharp}B,\\
\vspace{0.1cm}\displaystyle\frac{\partial}{\partial
t}\mathbf{v}_s=-\operatorname{grad}\left(g(\mathbf{v}_s,\mathbf{v}_n)+\mu_{\rm
chem}-\frac{1}{2}\|\mathbf{v}_n\|^2\right)\\
\qquad \quad  \;
\;+\left(\frac{a}{R\rho}\mathbf{i}_{\mathbf{p}^\sharp}B+\frac{1}{\rho}\mathbf{i}_{(\operatorname{div}B)^\sharp}B-\mathbf{i}_{\mathbf{v}_n}\mathbf{d}\mathbf{v}_s^\flat\right)^\sharp.
\end{array}
\right.
\end{equation}
These are the equations for superfluid Hall magnetohydrodynamics as given in
\cite{HoKu1987} equations (35a)--(35e). When $\mathcal{D}$ is three
dimensional,
the two last equations read
\[
\frac{\partial}{\partial
t}A^\sharp=\left(\mathbf{v}_n+\frac{1}{\rho}\mathbf{p}^\sharp-\frac{R}{a\rho}\operatorname{curl}\mathbf{B}\right)\times\mathbf{B}\qquad \text{and}
\]
\begin{align*}
\frac{\partial}{\partial
t}\mathbf{v}_s=&-\operatorname{grad}\left(g(\mathbf{v}_s,\mathbf{v}_n)+\mu_{\rm
chem}-\frac{1}{2}\|\mathbf{v}_n\|^2\right)+\mathbf{v}_n\times\operatorname{curl}\mathbf{v}_s \\
& +\frac{1}{\rho}\left(\operatorname{curl}\mathbf{B}-\frac{a}{R}\mathbf{p}^\sharp\right)\times\mathbf{B}.
\end{align*}
\medskip  

\noindent \textbf{Hamiltonian reduction for superfluid Hall magnetohydrodynamics.}
Consider the right-invariant Hamiltonian function
$H(\mathbf{m}_\eta,\rho_\chi,S,\mathbf{u};\mathbf{n}_\xi,n)=H_{(S,\mathbf{u};n)}(\mathbf{m}_\eta,\rho_\chi;\mathbf{n}_\xi)$
induced by $h$ and suppose that we have $a\rho_0+n_0=0$. A smooth curve
\[
(\mathbf{m}_\eta,\rho_\chi;\mathbf{n}_\xi)\in
T^*\left[(\operatorname{Diff}(\mathcal{D})\,\circledS\,\mathcal{F}(\mathcal{D},S^1))\times\operatorname{Diff}(\mathcal{D})\right]
\]
is a solution of Hamilton's equations associated to
$H_{(S_0,\mathbf{u}_0;n_0)}$
and with the initial condition $\rho_0$ if and only if the curve
\[
(\mathbf{m},\rho;\mathbf{n}):=\left(J(\eta^{-1})(\mathbf{m}_\eta\circ\eta^{-1}),J(\eta^{-1})(\rho_\chi\circ\eta^{-1});J(\xi^{-1})(\mathbf{n}\circ\xi^{-1})\right)
\]
is a solution of the equations \eqref{superfluid_H_MHD}, where
$\mathbf{v}_s=\mathbf{u}- a A^\sharp/R=\mathbf{u}-a\mathbf{n}/n$.

The Poisson bracket for superfluid Hall magnetohydrodynamics is the sum of the
affine Lie-Poisson bracket associated to the variables
$(\mathbf{m},\rho,S,\mathbf{u})$ and the Lie-Poisson bracket associated to the
variables $(\mathbf{n},n)$.

\medskip

The $\gamma$-circulation gives
\[
\frac{d}{dt}\oint_{c_t}\mathbf{u}^\flat=0.
\]
Using the definition $\mathbf{v}_s:=\mathbf{u}-\frac{a}{R}A^\sharp$, we obtain
\[
\frac{d}{dt}\oint_{c_t}\left(\mathbf{v}_s^\flat+\frac{a}{R}A^\sharp\right)=0,
\]
where $c_t$ is a loop which moves with the \textit{normal fluid velocity}
$\mathbf{v}_n$. The Kelvin-Noether theorem associated to the variable
$\mathbf{n}$ gives
\[
\frac{d}{dt}\oint_{d_t}A=0,
\]
where $d_t$ is a loop which moves with the \textit{electron fluid velocity}
$\mathbf{v}$.


\subsection{HVBK Dynamics for Superfluid ${}^4$He with Vortices}\label{HVBK}

The Hall-Vinen-Bekarevich-Khalatnikov (HVBK) equations describe superfluid
Helium turbulence. We consider the version of HVBK equations, together with its
Hamiltonian structure, as given in \cite{Ho2001}. It turns out that this
Hamiltonian structure is the same as that of superfluid Hall
magnetohydrodynamics, that is, it is obtained by affine Lie-Poisson reduction
associated to the group
\[
G:=\left[\operatorname{Diff}(\mathcal{D})\,\circledS\,\mathcal{F}(\mathcal{D},S^1)\right]\times\operatorname{Diff}(\mathcal{D}).
\]
As before, the advected quantities are
\[
(S,\mathbf{u};n)\in\mathcal{F}(\mathcal{D})\times\mathfrak{X}(\mathcal{D})\times\mathcal{F}(\mathcal{D}),
\]
where $S$ is the \textit{entropy density of the normal flow}. The action of
$(\eta,\chi;\xi)\in G$ is given by
\[
(S,\mathbf{u};n)\mapsto(J\eta(S\circ\eta),(\eta^*\mathbf{u}^\flat+\mathbf{d}\chi)^\sharp;J\xi(n\circ\xi)).
\]
The resulting affine Lie-Poisson equations consist of two systems, the affine
Lie-Poisson equations associated to the variables
$(\mathbf{m},\rho,S,\mathbf{u})$ and the Lie-Poisson equations associated to
the
variables $(\mathbf{n},n)$.

For simplicity we assume that the manifold $\mathcal{D}$ is three dimensional.
The Hamiltonian of HVBK dynamics is defined on the dual Lie algebra
\begin{align*}
&\big(\left[(\mathfrak{X}(\mathcal{D})\,\circledS\,\mathcal{F}(\mathcal{D}))\,\circledS\,(\mathcal{F}(\mathcal{D})\oplus\mathfrak{X}(\mathcal{D}))\right]\times[\mathfrak{X}(\mathcal{D})\,\circledS\,\mathcal{F}(\mathcal{D})]\big)^*\\
&\quad\cong
\Omega^1(\mathcal{D})\times\mathcal{F}(\mathcal{D})\times\mathcal{F}(\mathcal{D})\times\Omega^1(\mathcal{D})\times\Omega^1(\mathcal{D})\times\mathcal{F}(\mathcal{D})
\end{align*}
and is given by
\[
h(\mathbf{m},\rho,S,\mathbf{u};\mathbf{n},n):=-\frac{1}{2}\int_\mathcal{D}\rho\|\mathbf{v}_n\|^2\mu+\int_\mathcal{D}(\mathbf{m}-\rho
A)\cdot\mathbf{v}_n\mu+\int_\mathcal{D}\varepsilon(\rho,S,\mathbf{v}_s-\mathbf{v}_n,\boldsymbol{\omega})\mu,
\]
where $\mathbf{v}_n$ is the \textit{velocity of the normal flow},
$\mathbf{v}_s:=\mathbf{u}-A^\sharp$ is the \textit{superfluid velocity},
$\boldsymbol{\omega}:=\operatorname{curl}\mathbf{v}_s$ is the
\textit{superfluid
vorticity}, and $\varepsilon$ is the \textit{internal energy density}. The
one-form $A$ is defined by
\[
A:=-\frac{\mathbf{n}}{n}.
\]
Note that the Hamiltonian of HVBK dynamics is similar to that of Hall
magnetohydrodynamics, for $R=a=-1$. The norm in the first term and the operator
$\operatorname{curl}$ are taken with respect to a fixed Riemannian metric $g$
on
$\mathcal{D}$. The velocity $\mathbf{v}_n$ is the function
$\mathbf{v}_n=\mathbf{v}_n(\mathbf{m},\rho,S,\mathbf{u};\mathbf{n},n)$ defined
by the implicit condition
\[
\mathbf{m}-\rho\mathbf{v}_n^\flat-\rho
A=\frac{\partial\varepsilon}{\partial\mathbf{r}}(\rho,S,\mathbf{v}_s-\mathbf{v}_n,\boldsymbol{\omega})=:\mathbf{p}.
\]
By the implicit function theorem, the above relation defines a unique function
$\mathbf{v}_n$, provided the function $\varepsilon$ verifies the condition that
the linear map
\[
u_x\mapsto\frac{\partial^2\varepsilon}{\partial\mathbf{r}^2}(r,s,v_x,w_x)\cdot
u_x-ru_x
\]
is bijective, for all $(r,s,v_x,w_x)\in\mathbb{R}\times\mathbb{R}\times
T\mathcal{D}\times T\mathcal{D}$.

We make the following definitions
\begin{align*}
\mu_{\rm
chem}&:=\frac{\partial\varepsilon}{\partial\rho}(\rho,S,\mathbf{v}_s-\mathbf{v}_n,\boldsymbol{\omega})\in\mathcal{F}(\mathcal{D}),\quad
T:=\frac{\partial\varepsilon}{\partial
S}(\rho,S,\mathbf{v}_s-\mathbf{v}_n,\boldsymbol{\omega})\in\mathcal{F}(\mathcal{D}),\\
\mathbf{p}&:=\frac{\partial\varepsilon}{\partial\mathbf{r}}(\rho,S,\mathbf{v}_s-\mathbf{v}_n,\boldsymbol{\omega})\in\Omega^1(\mathcal{D}),\quad\lambda:=\frac{\partial\varepsilon}{\partial\boldsymbol{\omega}}(\rho,S,\mathbf{v}_s-\mathbf{v}_n,\boldsymbol{\omega})\in\Omega^1(\mathcal{D}).
\end{align*}
The interpretation of the quantities $\mu_{\rm chem}, T,\mathbf{p}$, and
$\lambda$ is obtained from the following thermodynamic derivative identity for
the internal energy:
\[
\mathbf{d}(\varepsilon(\rho,S,\mathbf{v}_s-\mathbf{v}_n,\boldsymbol{\omega}))=\mu_{\rm
chem}\mathbf{d}\rho+T\mathbf{d}S+\mathbf{p}\cdot\nabla_{\_\,}(\mathbf{v}_s-\mathbf{v}_n)+\lambda\cdot\nabla_{\_\,}\boldsymbol{\omega}\in\Omega^1(\mathcal{D}),
\]
where $\nabla$ denotes the Levi-Civita covariant derivative associated to the
metric $g$. The functional derivatives of $h$ are computed to be
\[
\frac{\delta h}{\delta\mathbf{m}}=\mathbf{v}_n,\;\;\frac{\delta
h}{\delta\rho}=-\frac{1}{2}\|\mathbf{v}_n\|^2-A\cdot\mathbf{v}_n+\mu_{\rm
chem},\;\;\frac{\delta h}{\delta S}=T,\;\;\frac{\delta
h}{\delta\mathbf{u}}=\mathbf{p}+\operatorname{curl}\lambda,
\]
\[
\mathbf{v}_l:=\frac{\delta
h}{\delta\mathbf{n}}=\frac{1}{n}\left(\rho\mathbf{v}_n+\mathbf{p}^\sharp+\operatorname{curl}\lambda\right),\;\;\frac{\delta
h}{\delta n}=A\cdot\mathbf{v}_l.
\]
The vector field $\mathbf{v}_l$ will be interpreted as the \textit{vortex line
velocity}. The equations for $\rho, S$, and $n$ are given by
\[
\frac{\partial}{\partial
t}\rho+\operatorname{div}(\rho\mathbf{v}_n+\mathbf{p}^\sharp+\operatorname{curl}\lambda)=0,\;\;\frac{\partial}{\partial
t}S+\operatorname{div}(S\mathbf{v}_n)=0,\;\;\text{and}\;\;\frac{\partial}{\partial
t}n+\operatorname{div}(n\mathbf{v}_l)=0.
\]
Using the expression of $\mathbf{v}_l$ in terms of $\mathbf{v}_n$ we obtain
that
$\operatorname{div}(n\mathbf{v}_l)=\operatorname{div}(\rho\mathbf{v}_n+\mathbf{p}^\sharp+\operatorname{curl}\lambda)$
which proves that
\[
\frac{\partial}{\partial t}(\rho-n)=0.
\]
Thus, if we assume that the initial conditions verify $\rho_0=n_0$, then we
have
$\rho=n$ for all time. The equations for $A$ and $\mathbf{u}$ are computed to
be
\[
\frac{\partial}{\partial
t}A^\sharp=\frac{1}{\rho}(\rho\mathbf{v}_n+\mathbf{p}^\sharp+\operatorname{curl}\lambda)\times\operatorname{curl}A^\sharp=\mathbf{v}_l\times\operatorname{curl}A^\sharp
\]
\[
\frac{\partial}{\partial
t}\mathbf{u}+\operatorname{grad}\left(g(\mathbf{v}_s,\mathbf{v}_n)+\mu_{\rm
chem}-\frac{1}{2}\|\mathbf{v}_n\|^2\right)+\operatorname{curl}\mathbf{u}\times\mathbf{v}_n=0.
\]
As in superfluid dynamics, this equation preserves the condition
$\operatorname{curl}\mathbf{u}=0$, and we will suppose that it holds initially:
$\operatorname{curl}\mathbf{u}_0=0$. In this case we have
$\boldsymbol{\omega}=-\operatorname{curl}A^\sharp$ and the equations above read
\begin{equation}\label{A_and_u}
\frac{\partial}{\partial
t}A^\sharp+\mathbf{v}_l\times\boldsymbol{\omega}=0\;\;\text{and}\;\;\frac{\partial}{\partial
t}\mathbf{u}+\operatorname{grad}\left(g(\mathbf{v}_s,\mathbf{v}_n)+\mu_{\rm
chem}-\frac{1}{2}\|\mathbf{v}_n\|^2\right)=0.
\end{equation}
From these two equations we obtain that the evolution of the superfluid
velocity
$\mathbf{v}_s=\mathbf{u}-A^\sharp$ is given by
\begin{equation}\label{v_s}
\frac{\partial}{\partial
t}\mathbf{v}_s+\boldsymbol{\omega}\times\mathbf{v}_l=-\operatorname{grad}\left(g(\mathbf{v}_s,\mathbf{v}_n)+\mu_{\rm
chem}-\frac{1}{2}\|\mathbf{v}_n\|^2\right),
\end{equation}
which can be rewritten as
\[
\frac{\partial}{\partial
t}\mathbf{v}_s+\nabla_{\mathbf{v}_s}\mathbf{v}_s=-\operatorname{grad}\left(\mu_{\rm
chem}-\frac{1}{2}\|\mathbf{v}_s-\mathbf{v}_n\|^2\right)+\mathbf{f},\quad\mathbf{f}=(\mathbf{v}_l-\mathbf{v}_s)\times\boldsymbol{\omega},
\]
(compare to \eqref{equ_v_s}). Doing computations similar to those for
superfluids we obtain that the equation for $\mathbf{m}+\mathbf{n}$ is given by
\[
\frac{\partial}{\partial
t}(\mathbf{m}+\mathbf{n})=-\operatorname{Div}\mathbf{T},
\]
where $\mathbf{T}$ is the \textit{HVBK stress tensor} given by
\[
\mathbf{T}=\mathbf{v}_n\otimes(\rho\mathbf{v}_n^\flat+\mathbf{p})+\mathbf{p}^\sharp\otimes\mathbf{v}_s^\flat-\boldsymbol{\omega}\otimes\lambda^\flat+p\delta,\quad
p=\rho\mu_{\rm chem}+ST-\varepsilon+\boldsymbol{\omega}^\flat\cdot\lambda.
\]
Thus, we have obtained the following equations
\begin{equation}\label{HVBK_superfluids}
\left\{
\begin{array}{ll}
\vspace{0.1cm}\displaystyle\frac{\partial}{\partial
t}\mathbf{J}=-\operatorname{Div}\mathbf{T},\\
\vspace{0.1cm}\displaystyle\frac{\partial}{\partial
t}\rho+\operatorname{div}\mathbf{J}=0,\qquad\frac{\partial}{\partial
t}S+\operatorname{div}(S\mathbf{v}_n)=0,\\
\vspace{0.1cm}\displaystyle\frac{\partial}{\partial
t}\mathbf{v}_s+\nabla_{\mathbf{v}_s}\mathbf{v}_s=-\operatorname{grad}\left(\mu_{\rm
chem}-\frac{1}{2}\|\mathbf{v}_s-\mathbf{v}_n\|^2\right)+\mathbf{f},
\end{array}
\right.
\end{equation}
where $\mathbf{f}=(\mathbf{v}_l-\mathbf{v}_s)\times\boldsymbol{\omega} $ and we have used the notation
\[
\mathbf{J}:=\rho\mathbf{v}_n^\flat+\mathbf{p}=\mathbf{m}+\mathbf{n}
\]
for the \textit{total momentum density}. These are the equations of HBVK
dynamics as given in  equation (1) of \cite{Ho2001} with $R =0 $.

\medskip  

\noindent \textbf{Hamiltonian reduction for HVBK dynamics.} Consider the
right-invariant Hamiltonian
$H(\mathbf{m}_\eta,\rho_\chi,S,\mathbf{u};\mathbf{n}_\xi,n)=H_{(S,\mathbf{u};n)}(\mathbf{m}_\eta,\rho_\chi;\mathbf{n}_\xi)$
induced by $h$. Suppose that $\rho_0=n_0$ and
$\operatorname{curl}\mathbf{u}_0=0$ and let
$(\mathbf{m}_\eta,\rho_\chi;\mathbf{n}_\xi)$ be a smooth curve in
$T^*\left[(\operatorname{Diff}(\mathcal{D})\,\circledS\,\mathcal{F}(\mathcal{D},S^1))\times\operatorname{Diff}(\mathcal{D})\right]$
solution of the Hamilton equations associated to $H_{(S_0,\mathbf{u}_0;n_0)}$
and with the initial condition $\rho_0$. Then the curve
\[
(\mathbf{m},\rho;\mathbf{n}):=\left(J(\eta^{-1})(\mathbf{m}_\eta\circ\eta^{-1}),J(\eta^{-1})(\rho_\chi\circ\eta^{-1});J(\xi^{-1})(\mathbf{n}\circ\xi^{-1})\right)
\]
is a solution of the HVBK equations \eqref{HVBK_superfluids}, where
$\mathbf{v}_s=\mathbf{u}-A^\sharp=\mathbf{u}+\mathbf{n}/n$. To obtain the
converse of this assertion, it is not enough to assume that
$(\mathbf{m},\rho,S,\mathbf{v}_s,\mathbf{n})$ verifies
\eqref{HVBK_superfluids}.
It is also required that $A$ or $\mathbf{u}$ verify the corresponding equation
in \eqref{A_and_u}.

Note that from the equalities $\mathbf{v}_s=\mathbf{u}-A^\sharp$ and
$\operatorname{curl}\mathbf{u}=0$ we obtain that the variables $\mathbf{u}$ and
$A^\sharp$ are interpreted respectively as the \textit{potential} and the
\textit{rotational}  components of the superfluid velocity $\mathbf{v}_s$.

\medskip
As in \S\ref{sec:HallMHD}, we have
\[
\frac{d}{dt}\oint_{c_t}(\mathbf{v}_s^\flat+A)=0,\;\;\text{and}\;\;\frac{d}{dt}\oint_{d_t}A=0,
\]
where $c_t$ and $d_t$ are loops which move with the \textit{normal fluid
velocity} $\mathbf{v}_n$ and the \textit{vortex line velocity} $\mathbf{v}_l$,
respectively. Using the equation \eqref{v_s} for the evolution of the
superfluid
velocity $\mathbf{v}_s$, we obtain the \textit{vortex Kelvin theorem}
\[
\frac{d}{dt}\oint_{d_t}\mathbf{v}_s^\flat=0.
\]
By the Stokes theorem, this can be rewritten as
\[
\frac{d}{dt}\iint_{S_t}(\boldsymbol{\omega}^\flat\cdot\mathbf{n})dS=0,
\]
where $\mathbf{n}$ is the unit vector normal to the surface $S_t$ whose
boundary
$\partial S_t$ is a loop which moves with the vortex line velocity
$\mathbf{v}_l$. This is the conservation of the flux of superfluid vorticity
through any surface whose boundary moves with the velocity $\mathbf{v}_l $.

\medskip

In \cite{Ho2001} it is also supposed that
$\frac{\partial\varepsilon}{\partial\mathbf{r}}$ is collinear to $\mathbf{r}$,
more precisely, that there exists a positive function $\rho_s$ such that
\[
\mathbf{p}=\rho_s(\mathbf{v}_s-\mathbf{v}_n).
\]
The function $\rho_s$ is interpreted as the \textit{superfluid mass density},
and the \textit{density of the normal fluid} is given by $\rho_n:=\rho-\rho_s$.
Using these notations, the total momentum $\mathbf{J}$ and the stress tensor
$\mathbf{T}$ can be rewritten as
\[
\mathbf{J}=\rho_n\mathbf{v}_n+\rho_s\mathbf{v}_s
\]
and
\[
\mathbf{T}=\mathbf{v}_n\otimes\rho_n\mathbf{v}_n^\flat+\mathbf{v}_s\otimes\rho_s\mathbf{v}_s^\flat-\boldsymbol{\omega}\otimes\lambda^\flat+p\,\delta.
\]


\subsection{Classical Fluids Versus Superfluids}\label{fluids_vs_superlfuids}

We summarize below the examples that have been studied in the previous
sections.

\medskip  

\noindent \textbf{Classical fluids.} In order to compare the two theories, we use the
notation $\mathbf{v}_n$ for the fluid velocity. We also express the dynamics in
terms of the entropy density $S$ and internal energy density $\varepsilon$ (and
not in terms of the specific entropy $s$ and internal energy $e$). We have
\[
\varepsilon(\rho,S)=\rho e(\rho,S/\rho),
\]
and the first law of thermodynamics
\[
\mathbf{d}e=\frac{p}{\rho^2}\mathbf{d}\rho+T\mathbf{d}s,\quad
p=\rho^2\frac{\partial e}{\partial \rho}
\]
reads
\[
\mathbf{d}\varepsilon=\mu_{\rm chem}\mathbf{d}\rho+T\mathbf{d}S,\quad
p=\rho\mu_{\rm chem}+ST-\varepsilon.
\]

\begin{enumerate}
\item[(i)] \textbf{Basic hydrodynamics}\\
Symmetry group $\operatorname{Diff}(\mathcal{D})$, momentum
$\mathbf{m}\in\Omega^1(\mathcal{D})$\\
Advected quantities
$(\rho,S)\in\mathcal{F}(\mathcal{D})\times\mathcal{F}(\mathcal{D})$\\
Hamiltonian
\[
h(\mathbf{m},\rho,S)=\frac{1}{2}\int_\mathcal{D}\frac{1}{\rho}\|\mathbf{m}\|^2\mu+\int_\mathcal{D}\varepsilon(\rho,S)\mu,
\]
\[
\mathbf{m}=\rho\mathbf{v}_n^\flat.
\]
Stress tensor formulation $\dot{\mathbf{m}}=-\operatorname{Div}\mathbf{T}$,
where
\[
\mathbf{T}=\mathbf{v_n}\otimes\rho\mathbf{v}_n^\flat+p\,\delta,\quad
p=\rho\mu_{\rm chem}+ST-\varepsilon.
\]
\item[(ii)] \textbf{Yang-Mills magnetohydrodynamics}\\
Symmetry group
$\operatorname{Diff}(\mathcal{D})\,\circledS\,\mathcal{F}(\mathcal{D},\mathcal{O})$\\
Momenta
$(\mathbf{m},Q)\in\Omega^1(\mathcal{D})\times\mathcal{F}(\mathcal{D},\mathfrak{o}^*)$\\
Advected quantities
$(\rho,S,A)\in\mathcal{F}(\mathcal{D})\times\mathcal{F}(\mathcal{D})\times\Omega^1(\mathcal{D},\mathfrak{o})$\\
Hamiltonian
\[
h(\mathbf{m},Q,\rho,S,A)=\frac{1}{2}\int_\mathcal{D}\frac{1}{\rho}\|\mathbf{m}\|^2\mu+\int_\mathcal{D}\varepsilon(\rho,S)\mu+\frac{1}{2}\int_\mathcal{D}\|\mathbf{d}^AA\|^2\mu,
\]
\[
\mathbf{m}=\rho\mathbf{v}_n^\flat.
\]
Stress tensor formulation $\dot{\mathbf{m}}=-\operatorname{Div}\mathbf{T}$,
where
\[
\mathbf{T}=\mathbf{v_n}\otimes\rho\mathbf{v}_n^\flat+B\!\cdot\!
B+p\,\delta,\quad p=\rho\mu_{\rm chem}+ST-\varepsilon-\frac{1}{2}\|B\|^2.
\]
\item[(iii)] \textbf{Hall magnetohydrodynamics}\\
Symmetry group
$\operatorname{Diff}(\mathcal{D})\times\operatorname{Diff}(\mathcal{D})$,
momenta
$(\mathbf{m},\mathbf{n})\in\Omega^1(\mathcal{D})\times\Omega^1(\mathcal{D})$\\
Advected quantities
$(\rho,S;n)\in\mathcal{F}(\mathcal{D})\times\mathcal{F}(\mathcal{D})\times\mathcal{F}(\mathcal{D})$\\
Hamiltonian
\[
h(\mathbf{m},\rho,S;\mathbf{n},n)=\frac{1}{2}\int_\mathcal{D}\frac{1}{\rho}\left\|\mathbf{m}-\frac{a\rho}{R}A\right\|^2\mu+\int_\mathcal{D}\varepsilon(\rho,S)\mu+\frac{1}{2}\int_\mathcal{D}\|\mathbf{d}A\|^2\mu,
\]
where $A:=R\frac{\mathbf{n}}{n}$,
\[
\mathbf{m}=\rho\mathbf{v}_n^\flat+\frac{a\rho}{R}A.
\]
Initial conditions
\[
a\rho_0+n_0=0\Longrightarrow \left\{a\rho+n=0\;\;\text{and}\;\;
\mathbf{m}+\mathbf{n}=\rho\mathbf{v}_n^\flat\right\}.
\]
Stress tensor formulation
$\dot{\mathbf{m}}+\dot{\mathbf{n}}=-\operatorname{Div}\mathbf{T}$, where
\[
\mathbf{T}=\mathbf{v_n}\otimes\rho\mathbf{v}_n^\flat+B\!\cdot\!
B+p\,\delta,\quad p=\rho\mu_{\rm chem}+ST-\varepsilon-\frac{1}{2}\|B\|^2.
\]
\end{enumerate}

\medskip 

 \noindent \textbf{Superfluids.} As before, we denote by $\mathbf{v}_n$ the velocity of
the normal flow and by $\mathbf{v}_s$ the superfluid velocity. For simplicity
we
treat the two fluid model in this summary. Generalization to multi fluids
models
follows as above in the examples. The superfluid first law reads
\[
\mathbf{d}\varepsilon=\mu_{\rm
chem}\mathbf{d}\rho+T\mathbf{d}S+\mathbf{p}\cdot\nabla_{\_\,}(\mathbf{v}_s-\mathbf{v}_n),
\]
where $\varepsilon$ is the internal energy density. For HBVK dynamics the term
$\lambda\cdot\nabla_{\_\,}\boldsymbol{\omega}$ has to be added.
\begin{enumerate}
\item [(i)]\textbf{Basic superfluid hydrodynamics}\\
Symmetry group
$\operatorname{Diff}(\mathcal{D})\,\circledS\,\mathcal{F}(\mathcal{D},S^1)$,
momenta
$(\mathbf{m},\rho)\in\Omega^1(\mathcal{D})\times\mathcal{F}(\mathcal{D})$\\
Advected quantities
$(S,\mathbf{v}_s)\in\mathcal{F}(\mathcal{D})\times\mathfrak{X}(\mathcal{D})$.\\
Hamiltonian
\[
h(\mathbf{m},\rho,S,\mathbf{v}_s)=-\frac{1}{2}\int_\mathcal{D}\rho\|\mathbf{v}_n\|^2\mu+\int_\mathcal{D}(\mathbf{m}\cdot\mathbf{v}_n)\mu+\int_\mathcal{D}\varepsilon(\rho,S,\mathbf{v}_s-\mathbf{v}_n)\mu,
\]
\[
\mathbf{m}=\rho\mathbf{v}_n^\flat+\mathbf{p}.
\]
Stress tensor formulation $\dot{\mathbf{m}}=-\operatorname{Div}\mathbf{T}$,
where
\[
\mathbf{T}=\mathbf{v}_n\otimes(\rho\mathbf{v}_n^\flat+\mathbf{p})+\mathbf{p}^\sharp\otimes\mathbf{v}_s^\flat+p\,\delta,\quad
p=\rho\mu_{\rm chem}+ST-\varepsilon.
\]
\item[(ii)] \textbf{Superfluid Yang-Mills magnetohydrodynamics}\\
Symmetry group
$\operatorname{Diff}(\mathcal{D})\,\circledS\,(\mathcal{F}(\mathcal{D},\mathcal{O})\times\mathcal{F}(\mathcal{D},S^1))$\\
Momenta $(\mathbf{m},Q,\rho)\in\Omega^1(\mathcal{D})\times\mathcal{F}(\mathcal{D},\mathfrak{o}^*)\times\mathcal{F}(\mathcal{D})$\\
Advected quantities
$(S,A,\mathbf{v}_s)\in\mathcal{F}(\mathcal{D})\times\Omega^1(\mathcal{D},\mathfrak{o})\times\mathfrak{X}(\mathcal{D})$\\
Hamiltonian
\begin{align*}
&h(\mathbf{m},Q,\rho,S,A,\mathbf{v}_s)\\
&=-\frac{1}{2}\int_\mathcal{D}\rho\|\mathbf{v}_n\|^2\mu+\int_\mathcal{D}(\mathbf{m}\cdot\mathbf{v}_n)\mu+\int_\mathcal{D}\varepsilon(\rho,S,\mathbf{v}_s-\mathbf{v}_n)\mu+\frac{1}{2}\int_\mathcal{D}\|\mathbf{d}^AA\|^2\mu,
\end{align*}
\[
\mathbf{m}=\rho\mathbf{v}_n^\flat+\mathbf{p}.
\]
Stress tensor formulation $\dot{\mathbf{m}}=-\operatorname{Div}\mathbf{T}$,
where
\[
\mathbf{T}=\mathbf{v_n}\otimes(\rho\mathbf{v}_n^\flat+\mathbf{p})
+\mathbf{p}^\sharp\otimes\mathbf{v}_s^\flat+B\!\cdot\!
B+p\,\delta,\quad p=\rho\mu_{\rm chem}+ST-\varepsilon-\frac{1}{2}\|B\|^2.
\]
\item[(iii)] \textbf{Superfluid Hall magnetohydrodynamics}\\
Symmetry group
$\left[\operatorname{Diff}(\mathcal{D})\,\circledS\,\mathcal{F}(\mathcal{D},S^1)\right]\times\operatorname{Diff}(\mathcal{D})$\\
Momenta
$(\mathbf{m},\rho,\mathbf{n})\in\Omega^1(\mathcal{D})\times\Omega^1(\mathcal{D})\times\mathcal{F}(\mathcal{D})$\\
Advected quantities
$(S,\mathbf{u};n)\in\mathcal{F}(\mathcal{D})\times\mathfrak{X}(\mathcal{D})\times\mathcal{F}(\mathcal{D})$\\
Hamiltonian
\begin{align*}
h(\mathbf{m},\rho,S,\mathbf{u};\mathbf{n},n)
=&-\frac{1}{2}\int_\mathcal{D}\rho\|\mathbf{v}_n\|^2\mu+\int_\mathcal{D}\left(\left(\mathbf{m}-\frac{a\rho}{R}
A\right)\cdot\mathbf{v}_n\right)\mu\\
& +\int_\mathcal{D}\varepsilon(\rho,S,\mathbf{v}_s-\mathbf{v}_n)\mu
+\frac{1}{2}\int_\mathcal{D}\|\mathbf{d}A\|^2\mu, \quad \text{where}\\
&\qquad
A:=R\frac{\mathbf{n}}{n},\quad\mathbf{v}_s:=\mathbf{u}-\frac{a}{R}A^\sharp,
\end{align*}
\[
\mathbf{m}=\rho\mathbf{v}_n^\flat+\frac{a\rho}{R} A+\mathbf{p}.
\]
Initial conditions
\begin{align*}
&a\rho_0+n_0=0\Longrightarrow\left\{
a\rho+n=0\;\;\text{and}\;\;\mathbf{m}+\mathbf{n}=\rho\mathbf{v}_n^\flat+\mathbf{p}\right\},
\\
&\operatorname{curl}\mathbf{u}_0=0\Longrightarrow\operatorname{curl}\mathbf{u}=0.
\end{align*}
Stress tensor formulation
$\dot{\mathbf{m}}+\dot{\mathbf{n}}=-\operatorname{Div}\mathbf{T}$, where
\[
\mathbf{T}=\mathbf{v}_n\otimes(\rho\mathbf{v}_n^\flat+\mathbf{p})+\mathbf{p}^\sharp\otimes\mathbf{v}_s^\flat+B\!\cdot\!B+p\,\delta,\quad
p=\rho\mu_{\rm chem}+ST-\varepsilon-\frac{1}{2}\|B\|^2.
\]

\item[(iv)] \textbf{HBVK hydrodynamics}\\
Symmetry group
$\left[\operatorname{Diff}(\mathcal{D})\,\circledS\,\mathcal{F}(\mathcal{D},S^1)\right]\times\operatorname{Diff}(\mathcal{D})$\\
Momenta
$(\mathbf{m},\rho,\mathbf{n})\in\Omega^1(\mathcal{D})\times\Omega^1(\mathcal{D})\times\mathcal{F}(\mathcal{D})$\\
Advected quantities
$(S,\mathbf{u};n)\in\mathcal{F}(\mathcal{D})\times\mathfrak{X}(\mathcal{D})\times\mathcal{F}(\mathcal{D})$\\
Hamiltonian
\begin{align*}
h(\mathbf{m},\rho,S,\mathbf{u};\mathbf{n},n)=&-\frac{1}{2}\int_\mathcal{D}\rho\|\mathbf{v}_n\|^2\mu+\int_\mathcal{D}((\mathbf{m}-\rho
A)\cdot\mathbf{v}_n)\mu \\
&
+\int_\mathcal{D}\varepsilon(\rho,S,\mathbf{v}_s-\mathbf{v}_n,\boldsymbol{\omega})\mu,
\end{align*}
\[
A:=-\frac{\mathbf{n}}{n},\quad\mathbf{v}_s:=\mathbf{u}-A^\sharp,\quad\boldsymbol{\omega}:=\operatorname{curl}\mathbf{v}_s,
\]
\[
\mathbf{m}=\rho\mathbf{v}_n^\flat+\rho A+\mathbf{p}.
\]
Initial conditions
\begin{align*}
&\rho_0=n_0\Longrightarrow\left\{
\rho=n\;\;\text{and}\;\;\mathbf{m}+\mathbf{n}=\rho\mathbf{v}_n^\flat+\mathbf{p}=:\mathbf{J}\right\},\\
&\operatorname{curl}\mathbf{u}_0=0\Longrightarrow\operatorname{curl}\mathbf{u}=0.
\end{align*}
Stress tensor formulation
$\dot{\mathbf{m}}+\dot{\mathbf{n}}=-\operatorname{Div}\mathbf{T}$, where
\[
\mathbf{T}=\mathbf{v}_n\otimes(\rho\mathbf{v}_n^\flat+\mathbf{p})+\mathbf{p}^\sharp\otimes\mathbf{v}_s^\flat-\boldsymbol{\omega}\otimes\lambda^\flat+p\,\delta,\quad
p=\rho\mu_{\rm chem}+ST-\varepsilon+\boldsymbol{\omega}^\flat\cdot\lambda.
\]
The hypothesis $\mathbf{p}=\rho_s(\mathbf{v}_s-\mathbf{v}_n)$ implies the
equalities $\mathbf{J}=\rho_n\mathbf{v}_n+\rho_s\mathbf{v}_s$ and 
\[
\mathbf{T}=\mathbf{v}_n\otimes\rho_n\mathbf{v}_n^\flat+\mathbf{v}_s\otimes\rho_s\mathbf{v}_s^\flat+\boldsymbol{\omega}\otimes\lambda^\flat+p\,\delta.
\]
\end{enumerate}


\subsection{Volovik-Dotsenko Theory of Spin Glasses}\label{VD_spin_glasses}

In \cite{VoDo1980}, the authors use a Poisson bracket approach to derive the equations of nonplanar magnet with disclinations and of spin glass. These models are referred to as the \textit{Volovik-Dotsenko spin glasses}. In \cite{HoKu1988}, the Hamiltonian structure of the Volovik-Dotsenko spin glasses is shown to be isomorphic to that of Yang-Mills magnetohydrodynamics. Thus, it can be obtained using the affine Lie-Poisson reduction developed in the present paper. In this section we also carry out the Lagrangian version of the approach given in \cite{HoKu1988}, to which we refer for additional comments on the physics of spin glasses.

The advected variables are 
\[
\rho\in V^*_1=\mathcal{F}(\mathcal{D})\quad\text{and}\quad \gamma\in
V^*_2=\Omega^1(\mathcal{D},\mathfrak{o}).
\]
The variable $\rho$ is the \textit{defect inertial-mass density} and
the curvature $B=\mathbf{d}^\gamma \gamma$ is interpreted as the \textit{disclination
density}.

The reduced Lagrangian
$l:[\mathfrak{X}(D)\,\circledS\,\mathcal{F}(\mathcal{D},\mathfrak{o})]\,\circledS\,[V_1^*\,\oplus\,V_2^*]\rightarrow\mathbb{R}$
is given by
\begin{equation}\label{Volovik-Dotsenko_Lagr}
l(\mathbf{u},\nu,\rho,\gamma)=\frac{1}{2}\int_\mathcal{D}\rho
\|\mathbf{u}\|^2\mu+\frac{\epsilon}{2}\int_\mathcal{D}\|\gamma(\mathbf{u})+\nu\|^2\mu-\frac{1}{2}\int_\mathcal{D}\rho\|\gamma\|^2\mu,
\end{equation}
where the norms are associated to the metrics $g, k$, and $(gk)$, respectively,
and $\epsilon$ is the \textit{constant of susceptibility}. The unreduced
Lagrangian $L$ is the right-invariant function induced by $l$ on the cotangent
bundle. Since it has a complicated expression, we do not give the formula for
$L$. We will justify the choice of this Lagrangian by showing that its Legendre transformation yields the Hamiltonian of the Volovik-Dotsenko spin glasses.

This Lagrangian appears as a generalization of the Lagrangian
\begin{equation}\label{Dz_spin-glass_Lagr}
l_{SG}(\nu,\gamma)=\frac{\epsilon}{2}\int_\mathcal{D}\|\nu\|^2\mu-\frac{1}{2}\int_\mathcal{D}\rho\|\gamma\|^2\mu,
\end{equation}
associated to the macroscopic description of spin glasses in \cite{Dz1980}, where $\rho$ is the \textit{constant of rigidity}, see \S\ref{subsec:spin_chains}. In order to understand mathematically the passage form the Lagrangian \eqref{Dz_spin-glass_Lagr} to the Lagrangian \eqref{Volovik-Dotsenko_Lagr}, we consider the following general situation.

\medskip  

\noindent \textbf{General Case.} Consider a Lagrangian $l_\rho:\mathcal{F}(\mathcal{D},\mathfrak{o})\oplus\Omega^1(\mathcal{D},\mathfrak{o})\rightarrow \mathbb{R},\;l_\rho=l_\rho(\nu,\gamma)$ associated to a spin system and depending on a parameter $\rho$ interpreted as the \textit{spin rigidity}. Recall that the affine Euler-Poincar\'e and advection equations are (see \S\ref{subsec:spin_chains})
\begin{equation}\label{AEP_spin_chain_2}
\left\lbrace
\begin{array}{l}
\vspace{0.2cm}\displaystyle\frac{\partial}{\partial t}\frac{\delta
l_\rho}{\delta\nu}=-\operatorname{ad}^*_\nu\frac{\delta
l_\rho}{\delta\nu}+\operatorname{div}^\gamma\frac{\delta l_\rho}{\delta\gamma},\\
\displaystyle\frac{\partial }{\partial t}\gamma+\mathbf{d}^\gamma\nu=0.
\end{array}
\right.
\end{equation}
To $l_\rho$ we associate the Lagrangian $l:\left[\mathfrak{X}(\mathcal{D})\,\circledS\,\mathcal{F}(\mathcal{D},\mathfrak{o})\right]\,\circledS\,\left[\mathcal{F}(\mathcal{D})\oplus\Omega^1(\mathcal{D},\mathfrak{o})\right] \rightarrow \mathbb{R}$ given by
\[
l(\mathbf{u},\nu,\rho,\gamma):=\frac{1}{2}\int_\mathcal{D}\rho\|\mathbf{u}\|^2\mu+l_\rho(\bar\nu,\gamma),
\]
where $\bar\nu:=\gamma(\mathbf{u})+\nu$ and $\rho$ is now a variable. Remark the analogy with the process of minimal coupling. In the case where $l_\rho$ is given by \eqref{Dz_spin-glass_Lagr}, we recover the Volovik-Dotsenko Lagrangian \eqref{Volovik-Dotsenko_Lagr}.

The functional derivatives of $l$ are computed to be
\[
\kappa:=\frac{\delta
l}{\delta\nu}=\frac{\delta
l_\rho}{\delta\bar\nu}\in\mathcal{F}(\mathcal{D},\mathfrak{o}^*),\;\;\mathbf{m}:=\frac{\delta l}{\delta\mathbf{u}}=\rho\mathbf{u}^\flat+\frac{\delta l_\rho}{\delta\bar\nu}\!\cdot \! \gamma\in\Omega^1(\mathcal{D}),
\]
and
\[
\frac{\delta
l}{\delta\rho}=\frac{1}{2}\|\mathbf{u}\|^2+\frac{\delta l_\rho}{\delta\rho}\in\mathcal{F}(\mathcal{D}),\quad\frac{\delta
l}{\delta \gamma}=\frac{\delta l_\rho}{\delta\bar\nu}\mathbf{u}+\frac{\delta l_\rho}{\delta\gamma}\in\mathfrak{X}(\mathcal{D},\mathfrak{o}^*).
\]
The advection equations are
\[
\frac{\partial}{\partial
t}\rho+\operatorname{div}(\rho\mathbf{u})=0\;\;\text{and}\;\;\frac{\partial}{\partial
t}\gamma+\mathbf{i}_\mathbf{u}B+\mathbf{d}^\gamma\bar\nu=0.
\]
We now compute the affine Euler-Poincar\'e equations. We have
\begin{align*}
\frac{\partial}{\partial
t}\frac{\delta
l_\rho}{\delta\bar\nu}&=-\operatorname{ad}^*_{\nu}\frac{\delta
l_\rho}{\delta\bar\nu}-\operatorname{div}\left(\frac{\delta
l_\rho}{\delta\bar\nu}\mathbf{u}\right)+\operatorname{div}^\gamma\left(\frac{\delta l_\rho}{\delta\bar\nu}\mathbf{u}+\frac{\delta
l_\rho}{\delta\gamma}\right)\\
&=-\operatorname{ad}^*_{\bar\nu-\gamma(\mathbf{u})}\frac{\delta
l_\rho}{\delta\bar\nu}-\operatorname{div}\left(\frac{\delta
l_\rho}{\delta\bar\nu}\mathbf{u}\right)+\operatorname{div}\left(\frac{\delta l_\rho}{\delta\bar\nu}\mathbf{u}\right)\\
&\qquad-\operatorname{Tr}\left(\operatorname{ad}^*_\gamma\frac{\delta l_\rho}{\delta\bar\nu}\mathbf{u}\right)+\operatorname{div}^\gamma\left(\frac{\delta
l_\rho}{\delta\gamma}\right)\\
&=-\operatorname{ad}^*_{\bar\nu}\frac{\delta
l_s}{\delta\bar\nu}+\operatorname{div}^\gamma\left(\frac{\delta
l_s}{\delta\gamma}\right).
\end{align*} 
Using the equations for $\rho, \gamma, \frac{\delta
l_\rho}{\delta\bar\nu}$, the relation
$\mathbf{m}=\rho\mathbf{u}^\flat+ \frac{\delta
l_\rho}{\delta\bar\nu}\cdot \gamma$, and the identity
\[
\pounds_\mathbf{u}\left( \frac{\delta
l_\rho}{\delta\bar\nu}\!\cdot\! \gamma\right)= \mathbf{d}\frac{\delta
l_\rho}{\delta\bar\nu}(\mathbf{u})\!\cdot\!
\gamma+\frac{\delta
l_\rho}{\delta\bar\nu}\!\cdot\! \pounds_\mathbf{u}\gamma
\]
we obtain 
\begin{align*}
&\left(\frac{\partial}{\partial
t}\mathbf{m}+\pounds_\mathbf{u}\mathbf{m}+(\operatorname{div}\mathbf{u})\mathbf{m}\right)^\sharp\\
&\quad=\rho\left(\frac{\partial}{\partial
t}\mathbf{u}+\nabla_\mathbf{u}\mathbf{u}+\nabla\mathbf{u}^T\!\cdot\!\mathbf{u}\right)+\frac{\partial}{\partial
t}\left(\frac{\delta
l_\rho}{\delta\bar\nu}\!\cdot\! \gamma\right)+\pounds_\mathbf{u}\left( \frac{\delta
l_\rho}{\delta\bar\nu}\!\cdot\!
\gamma\right)+(\operatorname{div}\mathbf{u}) \frac{\delta
l_\rho}{\delta\bar\nu}\!\cdot\! \gamma\\
&\quad=\rho\left(\frac{\partial}{\partial
t}\mathbf{u}+\nabla_\mathbf{u}\mathbf{u}+\nabla\mathbf{u}^T\!\cdot\!\mathbf{u}\right)-\operatorname{ad}^*_{\bar\nu}\frac{\delta
l_\rho}{\delta\bar\nu}\!\cdot\! \gamma+\operatorname{div}^\gamma\left(\frac{\delta
l_\rho}{\delta\gamma}\right)\!\cdot\! \gamma-\frac{\delta
l_\rho}{\delta\bar\nu}\!\cdot\!\mathbf{i}_\mathbf{u}B\\
&\qquad\qquad\qquad\qquad-\frac{\delta
l_\rho}{\delta\bar\nu}\!\cdot\!\mathbf{d}^\gamma\bar\nu+\mathbf{d}\frac{\delta
l_\rho}{\delta\bar\nu}(\mathbf{u})\!\cdot\!
\gamma+\frac{\delta
l_\rho}{\delta\bar\nu}\!\cdot\! \pounds_\mathbf{u}\gamma+(\operatorname{div}\mathbf{u}) \frac{\delta
l_\rho}{\delta\bar\nu}\!\cdot\! \gamma\\
&\quad=\rho\left(\frac{\partial}{\partial
t}\mathbf{u}+\nabla_\mathbf{u}\mathbf{u}+\nabla\mathbf{u}^T\!\cdot\!\mathbf{u}\right)-\operatorname{ad}^*_{\bar\nu}\frac{\delta
l_\rho}{\delta\bar\nu}\!\cdot\! \gamma+\operatorname{div}^\gamma\left(\frac{\delta
l_\rho}{\delta\gamma}\right)\!\cdot\! \gamma\\
&\qquad\qquad\qquad\qquad+\frac{\delta
l_\rho}{\delta\bar\nu}\!\cdot\!\mathbf{d}^\gamma(\gamma(\mathbf{u})-\bar\nu)+\operatorname{div}\left(\frac{\delta
l_\rho}{\delta\bar\nu}\mathbf{u}\right)\!\cdot\!\gamma,
\end{align*}
where in the last equality we used the identity \eqref{Covariant_Cartan} and the definition of the curvature $B $. We also have
\[
\left(\frac{\delta l}{\delta\rho}\diamond
\rho\right)^\sharp=\rho\left(\nabla\mathbf{u}^T\cdot\mathbf{u}+\operatorname{grad}\frac{\delta
l_\rho}{\delta\rho}\right)
\]
and
\[
\frac{\delta l}{\delta \gamma}\diamond_1 \gamma=\operatorname{div}^\gamma\left(\frac{\delta
l_\rho}{\delta\bar\nu}\mathbf{u}+\frac{\delta
l_\rho}{\delta\gamma}\right)\!\cdot\!\gamma-\left(\frac{\delta
l_\rho}{\delta\bar\nu}\mathbf{u}+\frac{\delta
l_\rho}{\delta\gamma}\right)\!\cdot\!\mathbf{i}_{\_\,}B.
\]
Thus the affine Euler-Poincar\'e equation for the variable $\mathbf{u}$ is 
\begin{align*}
&\rho\left(\frac{\partial}{\partial
t}\mathbf{u}+\nabla_\mathbf{u}\mathbf{u}+\nabla\mathbf{u}^T\!\cdot\!\mathbf{u}\right)-\operatorname{ad}^*_{\bar\nu}\frac{\delta
l_\rho}{\delta\bar\nu}\!\cdot\! \gamma+\operatorname{div}^\gamma\left(\frac{\delta
l_\rho}{\delta\gamma}\right)\!\cdot\! \gamma\\
&\qquad\qquad\qquad\qquad\qquad+\frac{\delta
l_\rho}{\delta\bar\nu}\!\cdot\!\mathbf{d}^\gamma(\gamma(\mathbf{u})-\bar\nu)+\operatorname{div}\left(\frac{\delta
l_\rho}{\delta\bar\nu}\mathbf{u}\right)\!\cdot\!\gamma\\
&=-\frac{\delta l_\rho}{\delta\bar\nu}\!\cdot\!\mathbf{d}\nu+\rho\left(\nabla\mathbf{u}^T\cdot\mathbf{u}+\operatorname{grad}\frac{\delta
l_\rho}{\delta\rho}\right)+\operatorname{div}^\gamma\left(\frac{\delta
l_\rho}{\delta\bar\nu}\mathbf{u}+\frac{\delta
l_\rho}{\delta\gamma}\right)\!\cdot\!\gamma\\
&\qquad\qquad\qquad\qquad\qquad-\left(\frac{\delta
l_\rho}{\delta\bar\nu}\mathbf{u}+\frac{\delta
l_\rho}{\delta\gamma}\right)\!\cdot\!\mathbf{i}_{\_\,}B,
\end{align*}
which, after remarkable cancellations, reads:
\[
\frac{\partial}{\partial
t}\mathbf{u}+\nabla_\mathbf{u}\mathbf{u}=\operatorname{grad}\frac{\delta
l_\rho}{\delta\rho} - \frac{1}{ \rho} \left(\frac{\delta
l_\rho}{\delta\bar\nu}\mathbf{u} + \frac{\delta
l_\rho}{\delta\gamma}\right)\!\cdot\!\mathbf{i}_{\_\,}B.
\]
Thus the affine Euler-Poincar\'e equations associated to $l$ are given by
\begin{equation}\label{general_case}
\left\lbrace
\begin{array}{ll}
\vspace{0.2cm}\displaystyle\frac{\partial}{\partial
t}\mathbf{u}+\nabla_\mathbf{u}\mathbf{u}=\operatorname{grad}\frac{\delta
l_\rho}{\delta\rho}-\frac{1}{\rho}\left[\left(\frac{\delta
l_\rho}{\delta\bar\nu}\mathbf{u}+\frac{\delta
l_\rho}{\delta\gamma}\right)\!\cdot\!\mathbf{i}_{\_\,}B\right]^\sharp,\\
\vspace{0.2cm}\displaystyle\frac{\partial}{\partial
t}\frac{\delta
l_\rho}{\delta\bar\nu}=-\operatorname{ad}^*_{\bar\nu}\frac{\delta
l_\rho}{\delta\bar\nu}+\operatorname{div}^\gamma\left(\frac{\delta
l_\rho}{\delta\gamma}\right),\qquad\displaystyle\frac{\partial}{\partial
t}\gamma+\mathbf{i}_\mathbf{u}B+\mathbf{d}^\gamma\bar\nu=0,\\
\vspace{0.2cm}\displaystyle\frac{\partial}{\partial
t}\rho+\operatorname{div}(\rho \mathbf{u})=0,\qquad\,\,\,\displaystyle
B=\mathbf{d}^\gamma\gamma,\qquad\,\,\,\bar\nu=\gamma(\mathbf{u})+\nu.
\end{array} \right.
\end{equation}

This system of equations can be seen as a generalization of equations \eqref{AEP_spin_chain_2}. If $l_\rho$ is hyperregular, then $l$ is also hyperregular, and the associated Hamiltonian is given by
\begin{equation}\label{assoc_Hamiltonian}
\int_\mathcal{D}\frac{1}{2\rho}\|\mathbf{m}-\kappa\!\cdot\!\gamma\|^2\mu+h_\rho(\kappa,\gamma),
\end{equation}
where $h_\rho$ is the Legendre transformation of $l_\rho$.

\medskip  

\noindent \textbf{The case of Volovik-Dotsenko Spin Glasses.} We now particularize the previous discussion to the case of the Volovik-Dotsenko spin glasses, that is, we consider the Lagrangian \eqref{Volovik-Dotsenko_Lagr}. In this case, the momenta are given by
\[
\kappa=\epsilon(\gamma(\mathbf{u})+\nu)\quad\text{and}\quad\mathbf{m}=\rho\mathbf{u}^\flat+\kappa\!\cdot\!\gamma,
\]
and equations \eqref{general_case} read

\begin{equation}\label{spin_glass}
\left\lbrace
\begin{array}{ll}
\vspace{0.2cm}\displaystyle\frac{\partial}{\partial t}
\mathbf{u}+\nabla_\mathbf{u}\mathbf{u}=-\operatorname{grad}\frac{1}{2}\|\gamma\|^2+\left[\left(\gamma^\sharp-\frac{1}{\rho}\kappa\mathbf{u}\right)\mathbf{i}_{\_\,}B\right]^\sharp,\\
\vspace{0.2cm}\displaystyle\frac{\partial}{\partial
t}\kappa+\operatorname{div}^\gamma(\rho
\gamma^\sharp)=0,\qquad\displaystyle\frac{\partial}{\partial
t}\gamma+\mathbf{i}_\mathbf{u}B+\frac{1}{\epsilon}\mathbf{d}^\gamma\kappa^\sharp=0,\\
\vspace{0.2cm}\displaystyle\frac{\partial}{\partial
t}\rho+\operatorname{div}(\rho \mathbf{u})=0,\qquad\,\,\,\displaystyle
B=\mathbf{d}^\gamma\gamma.
\end{array} \right.
\end{equation}
In the case of the spin glass dynamics described by the equations above, the
term $-\operatorname{grad}\frac{1}{2}\|\gamma\|^2$ is an analogue of the
electrostatic force and the term
$\left(\gamma^\sharp-\frac{1}{\rho}\kappa\mathbf{u}\right)\mathbf{i}_{\_\,}B$ is an
analogue of the Lorentz force.

\medskip  

\noindent \textbf{Lagrangian reduction for the Volovik-Dotsenko spin glasses.} A curve
$(\eta,\chi)\in\operatorname{Diff}(\mathcal{D})\,\circledS\,\mathcal{F}(\mathcal{D},\mathcal{O})$
is a solution of the Euler-Lagrange equations associated to the Lagrangian
$L_{(\rho_0,\gamma_0)}$ if and only if the curve
\[
(\mathbf{u},\nu):=\left(\dot\eta\circ\eta^{-1},(TR_{\chi^{-1}}\dot\chi)\circ\eta^{-1}\right)
\]
is a solution of the spin glasses equations \eqref{spin_glass}
with initial conditions $(\rho_0,\gamma_0)$, and where
$\kappa=\epsilon(\gamma(\mathbf{u})+\nu)$.

The evolution of the advected quantities is given by
\[
\rho=J(\eta^{-1})(\rho_0\circ\eta^{-1})\;\;\text{and}\;\;
\gamma=\eta_*\left(\operatorname{Ad}_\chi \gamma_0+\chi T\chi^{-1}\right),
\]
and the evolution of the disclination density is
\[
B=\eta_*\left(\operatorname{Ad}_\chi \mathbf{d}^{\gamma_0}\gamma_0\right).
\]
The variable $\chi$ is the \textit{orientation of Lagrangian particles in their
reference configuration}.

By Legendre transforming $L_{(\rho_0,\gamma_0)}$ and $l$, we obtain the
right-invariant Hamiltonian $H_{(\rho_0,\gamma_0)}$ and the reduced Hamiltonian $h$,
given by
\[
h(\mathbf{m},\kappa,\rho,\gamma)
=
\frac{1}{2}\int\frac{1}{\rho}\|\mathbf{m}
-\kappa\!\cdot\! \gamma\|^2\mu
+
\frac{1}{2\epsilon}\int\|\kappa\|^2\mu
+\frac{1}{2}\int\rho\|\gamma\|^2\mu,
\]
see also \eqref{assoc_Hamiltonian}. Note that this Hamiltonian differs by a sign in the first term from the
Hamiltonian (2.26b) in \cite{HoKu1988}. This is due to our convention in the
Hamiltonian structure which also differs from theirs. This justifies the choice we made for the Volovik-Dotsenko Lagrangian \eqref{Volovik-Dotsenko_Lagr}.

\medskip  

\noindent \textbf{Hamiltonian reduction for the Volovik-Dotsenko spin glasses.} A curve
$(\mathbf{m}_\eta,\kappa_\chi)\in
T^*[\operatorname{Diff}(\mathcal{D})\,\circledS\,\mathcal{F}(\mathcal{D},\mathcal{O})]$
is a solution of Hamilton's equations associated to the spin glasses
Hamiltonian
$H_{(\rho_0,\gamma_0)}$ if and only if the curve
\[
(\rho\mathbf{u}^\flat+\kappa\!\cdot\!
\gamma,\kappa)=:(\mathbf{m},\kappa):=J(\eta^{-1})\left(\mathbf{m}_\eta\circ\eta^{-1},T^*R_{\chi\circ\eta^{-1}}(\kappa_\chi\circ\eta^{-1})\right)
\]
is a solution of the system \eqref{spin_glass}
with initial conditions $(\rho_0,\gamma_0)$.

The associated Poisson bracket is identical to that of Yang-Mills
magnetohydrodynamics, except the fact that the variable $s$ is not present in
this case. Notice that the variables have not the same physical meaning in the two theories; see \cite{HoKu1988}.

\medskip

The Kelvin-Noether theorem gives the complicated expression
\[
\frac{d}{dt}\oint_{c_t}\left(\mathbf{u}^\flat+\frac{\kappa}{\rho}\gamma\right)=\oint_{c_t}\frac{1}{\rho}\left(-\kappa\mathbf{d}\nu+\operatorname{div}^\gamma(w)\gamma-w\cdot\mathbf{i}_{\_\,}B\right),
\]
where
\[
\nu:=\frac{\kappa}{\epsilon}-\gamma(\mathbf{u}),\;\;w:=\kappa\mathbf{u}-\rho \gamma^\sharp,
\]
which can be rewritten in the simpler form
\[
\frac{d}{dt}\oint_{c_t}\mathbf{u}^\flat=\oint_{c_t}\left(\gamma^\sharp-\frac{1}{\rho}\kappa\mathbf{u}\right)\mathbf{i}_{\_\,}B.
\]
The $\gamma$-circulation gives
\[
\frac{d}{dt}\oint_{c_t}\gamma=\oint_{c_t}\operatorname{ad}_{(\kappa/\epsilon)
-\gamma(\mathbf{u})}\gamma.
\]


\subsection{Microfluids}\label{MF}

Microfluids are fluids whose material points are \textit{small deformable
particles}. Examples of microfluids include \textit{liquid crystals, blood,
polymer melts, bubbly fluids, suspensions with deformable particles, biological
fluids}. In this section we find the Hamiltonian structure of the equations
governing the motion of non-dissipative microfluids in Eringen's formulation by
showing that they appear by Euler-Poincar\'e and Lie-Poisson reduction.

We quickly recall from \cite{Er2001} some needed facts about microfluids. A
material particle $P$ in the fluid is characterized by its position $X$ and by
a
vector $\Xi$ attached to $P$ that denotes the orientation and intrinsic
deformation of $P$. Both $X$ and $\Xi$ have their own motions, $X\mapsto
x=\eta(X,t)$ and $\Xi\mapsto \xi=\chi(X,\Xi,t)$, called respectively the
\textit{macromotion} and \textit{micromotion}. Since the material particles are
considered to be of very small size, a linear approximation in $\Xi$ is
permissible for the micromotion. Therefore, we can write
\[
\xi=\chi(X,t)\Xi,
\]
where $\chi(X,t)\in GL(3)^+:=\{A\in GL(3)\mid\operatorname{det}(A)>0\}$. The
classical Eringen theory considers only three possible groups in the
description
of the micromotion of the particles:
$GL(3)^+ \supset CSO(3) \supset SO(3) $. These cases correspond to
\textit{micromorphic}, \textit{microstretch}, and \textit{micropolar} fluids,
respectively, all of them discussed in detail below. The Lie group $CSO(3) $ is a
certain closed subgroup of $GL(3)^+$ that is associated to rotations and
stretch. Of course, the general theory developed in this paper admits other
groups describing the micromotion.

\medskip 

 \noindent \textbf{Micromorphic fluids.} A fluid in a domain $\mathcal{D}$ is called
\textit{micromorphic} if its macromotion and micromotion are described
respectively by a diffeomorphism $\eta\in\operatorname{Diff}(\mathcal{D})$ and
a
function $\chi\in\mathcal{F}(\mathcal{D},GL(3)^+)$. As a consequence of this
description, the configuration manifold of micromorphic fluids is isomorphic to
the product of the two groups $\operatorname{Diff}(\mathcal{D})$ and
$\mathcal{F}(\mathcal{D},GL(3)^+)$. We will show the remarkable facts that, as
in the previous examples, the relevant group structure on the configuration
manifold is given by the semidirect product
$\operatorname{Diff}(\mathcal{D})\,\circledS\,\mathcal{F}(\mathcal{D},GL(3)^+)$.

The general equations for \textit{micromorphic continua} are given by 
\begin{equation}\label{Eringen_micromorphic_continua}
\left\lbrace
\begin{array}{ll}
\vspace{0.2cm}\displaystyle
t_{kl,k}+\rho\left(f_l-\frac{D}{dt}\mathbf{u}_l\right)=0,\\
\vspace{0.2cm}\displaystyle
m_{klm,k}+t_{ml}-s_{ml}+\rho(l_{lm}-\sigma_{lm})=0,\\
\vspace{0.2cm}\displaystyle\frac{D}{dt}\rho+\rho\operatorname{div}\mathbf{u}=0,\qquad\displaystyle
\frac{D}{dt}i_{kl}-i_{kr}\nu_{lr}-i_{lr}\nu_{kr}=0,
\end{array} \right.
\end{equation}
where $D/dt$ denotes the material derivative.

Since this is the first time this symbol $D/dt$ appears, it is useful to make
some comments and to place this discussion in the larger context of complex fluid motion on general Riemannian manifolds.
The domain $ \mathcal{D} \subset  \mathbb{R} ^3 $ is endowed with the usual
Riemannian metric given by the inner product in Euclidean space. This Riemannian
metric has an associated Levi-Civita connection $ \nabla $ whose covariant
derivative along an arbitrary vector field $ \mathbf{u} $ is defined by $ \nabla_ \mathbf{u} : = \mathbf{u} \cdot  \nabla $. Of course, if 
$ \mathcal{D} $ is replaced by a general Riemannian manifold, then $ \nabla _ \mathbf{u} $  is the covariant derivative of the Levi-Civita connection associated with the given metric and does not have this simple expression. The material derivative of a \textit{vector field\/} $ \mathbf{v} $ along $\mathbf{u}$ is given, in general, by
\[
\frac{D}{dt}\mathbf{v}=\frac{\partial}{\partial
t}\mathbf{v}+\nabla_\mathbf{u}\mathbf{v}.
\]
On \textit{functions\/} $f\in\mathcal{F}(\mathcal{D},V)$, where $V$ is a vector
space, the material derivative is given by
\[
\frac{D}{dt}f=\frac{\partial}{\partial t}f+\mathbf{d}f(\mathbf{u})
\]
which is metric independent.

In \eqref{Eringen_micromorphic_continua} the material derivative operator is applied to the \textit{vector field\/} $\mathbf{u}$ and to the \textit{functions\/}  $\rho$  and $ i_{kl} $,
even though $ i_{kl} $ are the components of the symmetric tensor $ i $.
Thus, one should not interpret the last equation in
\eqref{Eringen_micromorphic_continua} as the covariant derivative of the tensor
$ i $. This convention is in force throughout the rest of the paper and we shall
comment on this in the relevant places.

Equations \eqref{Eringen_micromorphic_continua} are given in \cite{Er2001},
equations (2.2.10), (2.2.11),
(2.2.31), and (2.2.32). The variable $\rho$ is the \textit{mass density},
$\nu_{kl}$ is the \textit{microgyration tensor}, and the symmetric tensor
$i_{kl}$ is the \textit{microinertia}. The vector field $f_k$ is the
\textit{body force density} and $l_{lm}$ is the \textit{body couple density}.
The \textit{spin inertia} $\sigma_{kl}$ is given by
\[
\sigma_{kl}=i_{ml}\left(\frac{D}{dt}\nu_{km}+\nu_{kn}\nu_{nm}\right).
\]
The tensors $t_{kl}, m_{klm}$, and $s_{kl}$ are respectively the \textit{stress
tensor}, the \textit{couple stress tensor}, and the \textit{microstress
tensor};
see chapters 1, 2, and 17 in \cite{Er2001} for details. The four equations in
\eqref{Eringen_micromorphic_continua} correspond respectively to the balance of
momentum, the balance of momentum moments, the conservation of mass, and the
conservation of microinertia.

The \textit{constitutive equations} for \textit{non-dissipative} micromorphic
fluids are given by
\begin{equation}\label{constitutive_equations_micromorphic}
t_{kl}=\frac{\partial\Psi}{\partial\rho^{-1}}\delta_{kl},\quad
m_{klm}=0,\quad\text{and}\quad
s_{kl}=\frac{\partial\Psi}{\partial\rho^{-1}}\delta_{kl}+2\rho\frac{\partial\Psi}{\partial
i_{rk}}i_{rl},
\end{equation}
see (3.4.6) and (3.4.7) in \cite{Er2001}. Here the function
$\Psi=\Psi(\rho^{-1},i):\mathbb{R}\times Sym(3)\rightarrow\mathbb{R}$ is the
\textit{free energy} and $Sym(3)$ denotes the space of symmetric $3 \times 3$
matrices. It is usually assumed that
\[
\Psi(\rho^{-1},A^{-1}iA)=\Psi(\rho^{-1},i),\quad\text{for all $A\in O(3)$.}
\]
This condition is imposed by the \textit{axiom of objectivity}. Since $i $ is
symmetric, this implies that the free energy depends on $i$ only through the
quantities $\operatorname{Tr}(i), \operatorname{Tr}(i^2)$ and
$\operatorname{Tr}(i^3)$.

Using the constitutive equations \eqref{constitutive_equations_micromorphic}
and
assuming that $f_k=0, l_{kl}=0$, equations
\eqref{Eringen_micromorphic_continua}
read
\begin{equation}\label{Eringen_micromorphic_fluids}
\left\lbrace
\begin{array}{ll}
\vspace{0.2cm}\displaystyle\rho\frac{D}{dt}\mathbf{u}_l=\partial_l\frac{\partial\Psi}{\partial\rho^{-1}},\qquad
\sigma_{lm}=-2\frac{\partial \Psi}{\partial i_{rm}}i_{rl},\\
\vspace{0.2cm}\displaystyle\frac{D}{dt}\rho+\rho\operatorname{div}\mathbf{u}=0,\qquad\displaystyle
\frac{D}{dt}i_{kl}-i_{kr}\nu_{lr}-i_{lr}\nu_{kr}=0.
\end{array} \right.
\end{equation}
These are the equations for \textit{non-dissipative micromorphic fluids}.

\medskip  

\noindent \textbf{Microstretch fluids.}  A \textit{microstretch fluid} is a
micromorphic fluid whose micromotion $\chi$ takes values in the four
dimensional
real Lie group $CSO(3)$ defined by
\[
CSO(3)=\left\{A\in GL(3)^+ \mid \text{there exists $\lambda\in\mathbb{R}$ such
that $A A^T=\lambda I_3$}\right\}
\]
and called the \textit{conformal special orthogonal group}. The Lie algebra of $CSO(3)$ is
\[
\mathfrak{cso}(3)=\left\{\nu\in \mathfrak{gl}(3)\mid \text{there exists
$\mu\in\mathbb{R}$ such that $\nu+\nu^T=\mu I_3$}\right\}.
\]
Note that in the relations above we necessarily have
$\lambda^3=\operatorname{det}(A)^2$ and $3\mu=2\operatorname{Tr}(\nu)$.
Note also that each $\nu\in\mathfrak{cso}(3)$ decomposes uniquely as
\begin{equation}\label{decomposition}
\nu_{ij}=\nu_0\delta_{ij}-\varepsilon_{ijk}\boldsymbol{\nu}_k,
\end{equation}
where $\nu_0\in\mathbb{R}$, and
$(\boldsymbol{\nu}_1,\boldsymbol{\nu}_2,\boldsymbol{\nu}_3)=:\boldsymbol{\nu}$
is a vector in $\mathbb{R}^3$. The equality \eqref{decomposition} can be
rewritten as
\[
\nu=\nu_0I_3+\widehat{\boldsymbol{\nu}},
\]
where $\widehat{\boldsymbol{\nu}} \in \mathfrak{so}(3) $ is the matrix whose
entries are given by $\widehat{\boldsymbol{\nu}}_{ij} = -
\varepsilon_{ijk}\boldsymbol{\nu}_k$.

The material particles of microstretch fluids have seven degrees of freedom:
three for \textit{translations}, three for \textit{rotations}, and one for
\textit{stretch}.

The general equations for \textit{microstretch continua} are given by 
\begin{equation}\label{Eringen_microstretch_continua}
\left\lbrace
\begin{array}{ll}
\vspace{0.2cm}\displaystyle
t_{kl,k}+\rho\left(f_l-\frac{D}{dt}\mathbf{u}_l\right)=0,\\
\vspace{0.2cm}\displaystyle m_{k,k}+t-s+\rho(l-\sigma)=0,\quad
m_{kl,k}+\varepsilon_{lmn}t_{mn}+\rho(l_l-\sigma_l)=0,\\
\vspace{0.2cm}\displaystyle\frac{D}{dt}\rho+\rho\operatorname{div}\mathbf{u}=0,\\
\vspace{0.2cm}\displaystyle\frac{D}{dt}j_0-2j_0\nu_0=0,\qquad
\frac{D}{dt}j_{kl}-2\nu_0j_{kl}+(\varepsilon_{kpr}j_{lp}+\varepsilon_{lpr}j_{kp})\boldsymbol{\nu}_r=0.
\end{array} \right.
\end{equation}
These equations are given \cite{Er2001}, equations (2.2.38) to (2.2.41).
Following the conventions introduced in the discussion following 
\eqref{Eringen_micromorphic_continua}, the material derivative operator 
$D /dt$ acts on the \textit{vector field} $\mathbf{u} $ and  on the \textit{functions\/} $\nu_0$, $\boldsymbol{\nu}_l$,
 $ \rho $, $ j_0$, and $ j_{kl}$. As before, the variable
$\rho$ is the \textit{mass density}. The variables $\nu_0$ and
$\boldsymbol{\nu}$ are respectively the \textit{microstretch rate} and the
\textit{microrotation rate}, constructed from $\nu_{kl}$ as in
\eqref{decomposition}. The \textit{microstretch microinertia} $j_0$ and the
\textit{microinertia} $j_{kl}$ are constructed from $i_{kl}$ as follows:
\[
j_0:=2i_{kk}\quad\text{and}\quad j_{kl}:=\frac{1}{2}j_0\delta_{kl}-i_{kl}.
\]
The \textit{microstretch spin inertia} $\sigma$ and the \textit{spin inertia}
$\sigma_k$ are given by
\[
\sigma_k:=j_{kl}\frac{D}{dt}\boldsymbol{\nu}_l+2\nu_0j_{kl}\boldsymbol{\nu}_l+\varepsilon_{klm}j_{mn}\boldsymbol{\nu}_l\boldsymbol{\nu}_n\quad\text{and}\quad\sigma:=\frac{1}{2}\left(\frac{D}{dt}\nu_0+\nu_0^2\right)-j_{kl}\boldsymbol{\nu}_k\boldsymbol{\nu}_l.
\]
The \textit{microstretch vector} $m_k$, the \textit{couple stress tensor}
$m_{kl}$, the \textit{microstretch force density} $l$, and the \textit{couple
density} $l$ are defined from $m_{klm}$ and $l_{kl}$ by the decompositions
\[
m_{klm}=\frac{1}{3}m_k\delta_{lm}-\frac{1}{2}\varepsilon_{lmr}m_{kr}\quad\text{and}\quad
l_{kl}=\frac{1}{3}l\delta_{kl}-\frac{1}{2}\varepsilon_{klr}l_r.
\]
We also used the notations $t:=t_{kk}$ and $s:=s_{kk}$. Using these definitions
and the fact that the fluid has the microstretch property, one can obtain the
equations \eqref{Eringen_microstretch_continua} from the equations
\eqref{Eringen_micromorphic_continua}.

The \textit{constitutive equations} for \textit{non-dissipative} microstretch
fluids are given by
\[
t_{kl}=\frac{\partial\Psi}{\partial\rho^{-1}}\delta_{kl},\quad m_{kl}=0,\quad
m_k=0,\quad\text{and}\quad s-t=2\rho\left(\frac{\partial\Psi}{\partial
j_{kl}}j_{kl}+\frac{\partial\Psi}{\partial j_0}j_0\right),
\]
where $\Psi=\Psi(\rho^{-1},j,j_0)$ is the \textit{free energy}, see (3.4.16) in
\cite{Er2001}.

\medskip  

\noindent \textbf{Micropolar fluids.} A \textit{micropolar fluid} is a micromorphic
fluid whose micromotion $\chi$ takes values in the Lie group $SO(3)$. The
material particles of micropolar fluids have six degrees of freedom: three for
\textit{translations} and three for \textit{rotations}. Micropolar fluids are,
therefore, a particular case of microstretch fluids.

The general equations for \textit{micropolar continua} are given by 
\begin{equation}\label{Eringen_micropolar_continua}
\left\lbrace
\begin{array}{ll}
\vspace{0.2cm}\displaystyle
t_{kl,k}+\rho\left(f_l-\frac{D}{dt}\mathbf{u}_l\right)=0,\\
\vspace{0.2cm}\displaystyle
m_{kl,k}+\varepsilon_{lmn}t_{mn}+\rho(l_l-\sigma_l)=0,\\
\vspace{0.2cm}\displaystyle\frac{D}{dt}\rho+\rho\operatorname{div}\mathbf{u}=0,\\
\vspace{0.2cm}\displaystyle
\frac{D}{dt}j_{kl}+(\varepsilon_{kpr}j_{lp}+\varepsilon_{lpr}j_{kp})\boldsymbol{\nu}_r=0,
\end{array} \right.
\end{equation}
where the variables and tensors have the same meaning as in the previous
examples.
These equations are given \cite{Er2001}, equations (2.2.43) to (2.2.46). They 
are derived from those of microstretch continua, using that the microgyration
tensor $\nu$ takes values in the Lie subalgebra $\mathfrak{so}(3)$ of
$\mathfrak{cso}(3)$, that is, $\nu_0=0$. In particular, the \textit{spin inertia}
$\sigma_k$ is given by
\[
\sigma_k:=j_{kl}\frac{D}{dt}\boldsymbol{\nu}_l+\varepsilon_{klm}j_{mn}\boldsymbol{\nu}_l\boldsymbol{\nu}_n=\frac{D}{dt}(j_{kl}\boldsymbol{\nu}_l).
\]
The \textit{constitutive equations} for \textit{non-dissipative} micropolar
fluids are given by
\[
t_{kl}=\frac{\partial\Psi}{\partial\rho^{-1}}\delta_{kl}\quad\text{and}\quad
m_{kl}=0,
\]
where $\Psi=\Psi(\rho^{-1},j)$ is the \textit{free energy}, see (3.4.27) in
\cite{Er2001}.

\medskip  

\noindent \textbf{Lagrangian and Hamiltonian formulation for micromorphic fluids.} We
now show that the equations for \textit{non-dissipative micromorphic fluids}
can
be obtained by Euler-Poincar\'e reduction associated to the semidirect product
$\operatorname{Diff}(\mathcal{D})\,\circledS\,\mathcal{F}(\mathcal{D},GL(3)^+)$.
The advected quantities are the \textit{mass density} $\rho\in
\mathcal{F}(\mathcal{D})$ and the \textit{microinertia tensor} $i\in
\mathcal{F}(\mathcal{D},Sym(3))$. The symmetry group
$\operatorname{Diff}(\mathcal{D})\,\circledS\,\mathcal{F}(\mathcal{D},GL(3)^+)$
acts \textit{linearly} on the advected variables $(\rho,i)$ by
\[
(\rho,i)\mapsto \left(J\eta(\rho\circ\eta),\chi^T (i\circ\eta)\chi\right).
\]
The choice of this group representation is dictated by the form of the
advection
equations for the mass density and the microinertia in equations
\eqref{Eringen_micromorphic_fluids}.
The infinitesimal actions $\rho(\mathbf{u}, \nu) = \rho\mathbf{u}$ and $i (
\mathbf{u}, \nu) $  of $(\mathbf{u},\nu)\in
\mathfrak{X}(\mathcal{D})\,\circledS\,\mathcal{F}(\mathcal{D},\mathfrak{gl}(3))$
on $\rho $ and $i $ are given, respectively,  by
\[
\rho\mathbf{u}=\operatorname{div}(\rho\mathbf{u})\;\;\text{and}\;\;
i(\mathbf{u},\nu)=\mathbf{d}i(u)+\nu^Ti+i\nu.
\]
Given two matrices $a,b \in \mathfrak{gl}(3)$ we denote by $ab$ their product
and by $a\!\cdot\!b:=\operatorname{Tr}(a^T b)$ the contraction. By identifying
the dual of $\mathcal{F}(\mathcal{D},Sym(3))$ with itself through the pairing
\[
\langle m,i\rangle =\int_\mathcal{D}m(x)\!\cdot\!i(x)\mu,
\]
we obtain the diamond operations
\[
m\diamond_1 i=-m\!\cdot\!\mathbf{d}i\quad\text{and}\quad m\diamond_2 i=-2im.
\]
In order to obtain the equations for micromorphic fluids we consider the
Lagrangian $l:\left[
\mathfrak{X}(\mathcal{D})\,\circledS\,\mathcal{F}(\mathcal{D},\mathfrak{gl}(3))\right]\,\circledS\,\left[\mathcal{F}(\mathcal{D})\oplus\mathcal{F}(\mathcal{D},Sym(3))\right]\rightarrow\mathbb{R}$
given by
\begin{equation}\label{micromorphic_lagrangian}
l(\mathbf{u},\nu,\rho,i):=\frac{1}{2}\int_\mathcal{D}\rho\|\mathbf{u}\|^2\mu+\frac{1}{2}\int_\mathcal{D}\rho(i\nu\!\cdot\!\nu)\mu-\int_\mathcal{D}\rho\Psi(\rho^{-1},i)\mu,
\end{equation}
where the function $\Psi$ represents the \textit{free energy}, and where the
norm
in the first term is taken relative to a fixed Riemannian metric $g$ on
$\mathcal{D}$. The functional derivatives are 
\[
\frac{\delta l}{\delta \mathbf{u}}=\rho\mathbf{u}^\flat,\quad\frac{\delta
l}{\delta \nu}=\rho i\nu,\quad\frac{\delta l}{\delta
\rho}=\frac{1}{2}\|\mathbf{u}\|^2+\frac{1}{2}i\nu\!\cdot\!\nu-\Psi(\rho^{-1},i)+\frac{1}{\rho}\frac{\partial\Psi}{\partial\rho^{-1}}(\rho^{-1},i)
\]
\[
\text{and}\quad\frac{\delta l}{\delta
i}=\frac{1}{2}\rho\nu\nu^T-\rho\frac{\partial \Psi}{\partial i}(\rho^{-1},i).
\]
A computation, involving remarkable cancellations, shows that the
Euler-Poincar\'e equations \eqref{EP} associated to the Lagrangian
\eqref{micromorphic_lagrangian} and to the group
$G=\operatorname{Diff}(\mathcal{D})\,\circledS\,\mathcal{F}(\mathcal{D},GL(3)^+)$
acting linearly on
$V^*=\mathcal{F}(\mathcal{D})\oplus\mathcal{F}(\mathcal{D},Sym(3))$ are given
by
\begin{equation}\label{micromorphic_fluids}
\left\lbrace
\begin{array}{ll}
\vspace{0.2cm}\displaystyle\frac{\partial}{\partial t}
\mathbf{u}+\nabla_\mathbf{u}\mathbf{u}=\frac{1}{\rho}\operatorname{grad}\frac{\partial
\Psi}{\partial\rho^{-1}},\\
\vspace{0.2cm}\displaystyle i\left(\frac{\partial}{\partial
t}\nu+\mathbf{d}\nu(\mathbf{u})-\nu\nu\right)=2 i \frac{\partial\Psi}{\partial
i},\\
\vspace{0.2cm}\displaystyle\frac{\partial}{\partial
t}\rho+\operatorname{div}(\rho \mathbf{u})=0,\qquad\,\,\,\displaystyle
\frac{\partial}{\partial t}i+\mathbf{d}i(\mathbf{u})+\nu^Ti+i\nu=0.
\end{array} \right.
\end{equation}
Thus we have recovered the equations \eqref{Eringen_micromorphic_continua} by
Euler-Poincar\'e reduction, up to a change of variables, replacing $\nu$ by
$-\nu^T$.

\medskip

Consider the right-invariant Lagrangian
$L_{(\rho_0,i_0)}$ on $T\left[\operatorname{Diff}(\mathcal{D})\,\circledS\,\mathcal{F}(\mathcal{D},GL(3)^+)\right]$
induced by the Lagrangian \eqref{micromorphic_lagrangian}. A curve
$(\eta,\chi)\in
\operatorname{Diff}(\mathcal{D})\,\circledS\,\mathcal{F}(\mathcal{D},GL(3)^+)$
is a solution of the Euler-Lagrange equations associated to $L_{(\rho_0,i_0)}$
if and only if the curve
\[
(\mathbf{u},\nu):=(\dot\eta\circ\eta^{-1},\dot\chi\chi^{-1}\circ\eta^{-1})\in\mathfrak{X}(\mathcal{D})\,\circledS\,\mathcal{F}(\mathcal{D},GL(3)^+)
\]
is a solution of the micromorphic fluid equations \eqref{micromorphic_fluids}
with initial conditions $(\rho_0,i_0)$. Note that, in our approach, the
relation
$\nu=(\dot\chi\chi^{-1})\circ\eta^{-1}$ between the microgyration tensor and
the
micro and macromotions is due to the passage from the Lagrangian to the spatial
representation. This relation coincides with that given in \cite{Er2001}, up to
conventions. It is considered there as a definition of $\nu$.

The evolution of the mass density $\rho$ and the microinertia $i$ is given by
\[
\rho=J(\eta^{-1})(\rho_0 \circ\eta^{-1})\quad\text{and}\quad
i=\left(\left(\chi^T\right)^{-1}i_0\chi^{-1}\right)\circ\eta^{-1}.
\]
Remark that the evolution of the determinant of $i$ is 
\[
\operatorname{det}(i)=\frac{\operatorname{det}(i_0)}{\operatorname{det}(\chi)^2}\circ\eta^{-1}.
\]
Therefore, if the initial microinertia $i_0$ is invertible, then $i$ is
invertible for all time. Under this hypothesis we can take the Legendre
transformation of the Lagrangian $l$ and we obtain the Hamiltonian
\[
h(\mathbf{m},\kappa,\rho,i)=\frac{1}{2}\int_\mathcal{D}\frac{1}{\rho}\|\mathbf{m}\|^2\mu+\frac{1}{2}\int_\mathcal{D}\frac{1}{\rho}(i^{-1}\kappa\!\cdot\!\kappa)\mu+\int_\mathcal{D}\rho\Psi(\rho^{-1},i)\mu,
\]
where $(\mathbf{m},\kappa,\rho,i) \in \Omega^1( \mathcal{D}) \times
\mathcal{F}(\mathcal{D}, \mathfrak{gl}(3)) \times \mathcal{F}( \mathcal{D})
\times \mathcal{F}( \mathcal{D}, Sym(3))$, 
which consists of the sum of the kinetic energy due to macromotion and
micromotion and the free energy. By Lie-Poisson reduction, we obtain that the
equations \eqref{micromorphic_fluids} are Hamiltonian with respect to the
Lie-Poisson bracket
\begin{align}\label{Poisson_bracket_micromorphic}
\{f,g\}(\mathbf{m},\kappa,\rho,i)&=\int_\mathcal{D}\mathbf{m}\cdot\left[\frac{\delta
f}{\delta\mathbf{m}},\frac{\delta g}{\delta\mathbf{m}}\right]\mu\nonumber\\
&\quad+\int_\mathcal{D}\kappa\cdot\left(\operatorname{ad}_{\frac{\delta
f}{\delta\kappa}}\frac{\delta g}{\delta\kappa}+\mathbf{d}\frac{\delta
f}{\delta\kappa}\cdot\frac{\delta g}{\delta\mathbf{m}}-\mathbf{d}\frac{\delta
g}{\delta\kappa}\cdot\frac{\delta f}{\delta\mathbf{m}}\right)\mu\nonumber\\
&\quad+\int_\mathcal{D}\rho\left(\textbf{d}\left(\frac{\delta f}{\delta
\rho}\right)\frac{\delta g}{\delta 
\mathbf{m}}-\textbf{d}\left(\frac{\delta g}{\delta
\rho}\right)\frac{\delta f}{\delta  \mathbf{m}}\right)\mu\\
&\quad+\int_\mathcal{D}
i\!\cdot\!\left(\operatorname{div}\left(\frac{\delta
f}{\delta i}\frac{\delta g}{\delta \mathbf{m}}\right)-\frac{\delta g}{\delta
\kappa}\frac{\delta f}{\delta i}-\frac{\delta f}{\delta i}\left(\frac{\delta
g}{\delta \kappa}\right)^T\right.\nonumber\\
&\qquad\qquad\qquad\left.-\operatorname{div}\left(\frac{\delta g}{\delta
i}\frac{\delta f}{\delta \mathbf{m}}\right)+\frac{\delta f}{\delta
\kappa}\frac{\delta g}{\delta i}+\frac{\delta g}{\delta i}\left(\frac{\delta
f}{\delta \kappa}\right)^T\right)\mu\nonumber.
\end{align}

The Kelvin-Noether circulation theorem applied to micromorphic fluids yields
the
simple relation
\[
\frac{d}{dt}\oint_{c_t}\mathbf{u}^\flat=\oint_{c_t}\frac{\partial\Psi}{\partial
i}\!\cdot\!\mathbf{d}i.
\]
 
\medskip  

\noindent \textbf{Lagrangian and Hamiltonian formulation for microstretch fluids.} The
symmetry group of microstretch fluid dynamics is
$\operatorname{Diff}(\mathcal{D})\,\circledS\,\mathcal{F}(\mathcal{D},CSO(3))$.
As
before, the advected quantities are the \textit{mass density} $\rho\in
\mathcal{F}(\mathcal{D})$ and the \textit{microinertia tensor} $i\in
\mathcal{F}(\mathcal{D},Sym(3))$, on which the  symmetry group acts
\textit{linearly} as in the micromorphic case.

The Lagrangian of the microstretch fluid has the same expression as that of the
micromorphic fluid, namely,
\[
l:\left[
\mathfrak{X}(\mathcal{D})\,\circledS\,\mathcal{F}(\mathcal{D},\mathfrak{cso}(3))\right]\,\circledS\,\left[\mathcal{F}(\mathcal{D})\oplus\mathcal{F}(\mathcal{D},Sym(3))\right]\rightarrow\mathbb{R}
\]
is given by
\[
l(\mathbf{u},\nu,\rho,i):=\frac{1}{2}\int_\mathcal{D}\rho\|\mathbf{u}\|^2\mu+\frac{1}{2}\int_\mathcal{D}\rho
(i\nu\!\cdot\!\nu)\mu-\int_\mathcal{D}\rho\Psi(\rho^{-1},i)\mu.
\]
Using the change of variables $i\mapsto  j:=\operatorname{Tr}(i)I_3-i$, this
Lagrangian reads
\begin{equation}\label{microstretch_lagrangian}
l(\mathbf{u},\nu,\rho,j)=\frac{1}{2}\int_\mathcal{D}\rho\|\mathbf{u}\|^2\mu+\frac{1}{2}\int_\mathcal{D}\rho\left(\frac{1}{2}j_0\nu_0^2+j\boldsymbol{\nu}\!\cdot\!\boldsymbol{\nu}\right)\mu-\int_\mathcal{D}\rho\Psi(\rho^{-1},j)\mu,
\end{equation}
where $j_0:=\operatorname{Tr}(j)$. The functions
$\nu_0\in\mathcal{F}(\mathcal{D})$ and
$\boldsymbol{\nu}=(\boldsymbol{\nu}_1,\boldsymbol{\nu}_2,\boldsymbol{\nu}_3)\in\mathcal{F}(\mathcal{D},\mathbb{R}^3)$
are defined as in the equation \eqref{decomposition}. The expression
\eqref{microstretch_lagrangian} for the Lagrangian has the advantage of giving
the expressions of the energy due to microstretch and microrotation separately.

The associated Euler-Poincar\'e equations are the same as in
\eqref{micromorphic_fluids}. We now rewrite these equations, using the fact
that, for microstretch fluids, the microgyration tensor $\nu$ has values in the
Lie subalgebra $\mathfrak{cso}(3)$. We will make use of the following two lemmas.

\begin{lemma} Suppose that the free energy $\Psi$ verifies the axiom of
objectivity, that is,
\[
\Psi(\rho^{-1},A^{-1}iA)=\Psi(\rho^{-1},i),\;\text{for all $A\in O(3)$.}
\]
Then the matrix $i\frac{\partial \Psi}{\partial i}$ is symmetric.
\end{lemma}
\textbf{Proof.} Consider a curve $A(t)\in SO(3)$ such that $A(0) = I_3$ and 
$\dot A(0)=\xi\in\mathfrak{so}(3)$. Differentiating the equality
$\Psi(A(t)^{-1}iA(t))=\Psi(i)$ at $t=0 $, we obtain the condition
\[
\mathbf{D}\Psi(i)(i\xi-\xi i)=0,\;\text{for all $\xi\in \mathfrak{so}(3)$.}
\]
Using the equalities $\mathbf{D}\Psi(i)(i\xi-\xi
i)=\operatorname{Tr}\left(\frac{\partial \Psi}{\partial i}(i\xi-\xi
i)\right)=\operatorname{Tr}\left(\left(\frac{\partial \Psi}{\partial
i}i-i\frac{\partial \Psi}{\partial i}\right)\xi\right)$, we obtain that the
matrix $\frac{\partial \Psi}{\partial i}i-i\frac{\partial \Psi}{\partial i}$ is
symmetric.
Since this matrix is clearly also antisymmetric, we obtain $\frac{\partial
\Psi}{\partial i}i-i\frac{\partial \Psi}{\partial i}=0.\qquad\blacksquare$

\begin{lemma} For $i\in\mathcal{F}(\mathcal{D},Sym(3))$ define
\[
j:=\operatorname{Tr}(i)I_3-i\in\mathcal{F}(\mathcal{D},Sym(3)).
\]
For $\nu\in\mathcal{F}(\mathcal{D},\mathfrak{cso}(3))$ define
$\nu_0\in\mathcal{F}(\mathcal{D})$ and
$\boldsymbol{\nu}\in\mathcal{F}(\mathcal{D},\mathbb{R}^3)$ by the condition
\[
\nu=\nu_0I_3+\widehat{\boldsymbol{\nu}}.
\]
Then,
\begin{itemize}
\item the equation
\begin{equation}\label{advection_nu}
i\left(\frac{D}{d t}\nu-\nu\nu\right)=2 i \frac{\partial\Psi}{\partial i}
\end{equation}
is equivalent to the system
\begin{equation}\label{system}
\left\lbrace
\begin{array}{l}
\displaystyle\frac{j_0}{2}\left(\frac{D}{d
t}\nu_0-\nu_0^2\right)+(j\boldsymbol{\nu})\!\cdot\!\boldsymbol{\nu}=2\left(j_0\frac{\partial
\Psi}{\partial j_0}+j\!\cdot\!\frac{\partial \Psi}{\partial j}\right),\\
\displaystyle j\frac{D}{d
t}\boldsymbol{\nu}-2\nu_0j\boldsymbol{\nu}-(j\boldsymbol{\nu})\times\boldsymbol{\nu}=0,
\end{array}\right.
\end{equation}
\item the equation 
\[
\frac{D}{d t}i+\nu^Ti+i\nu=0
\]
is equivalent to the equation
\[
\frac{D}{d t}j+2\nu_0 j+[j,\widehat{\boldsymbol{\nu}}]=0.
\]
\end{itemize}
\end{lemma}
\textbf{Proof.} The results follow by direct computations. The two equations in
\eqref{system} are obtained by taking respectively the trace and the
antisymmetric part of the equation \eqref{advection_nu}. For the computation of
the trace, we use the equalities
\[
\operatorname{Tr}(i\nu)=\nu_0\operatorname{Tr}(i)=\frac{1}{2}\nu_0j_0,\quad
\operatorname{Tr}(i\nu\nu)=\frac{1}{2}\nu_0^2-(j\boldsymbol{\nu})\!\cdot\!\boldsymbol{\nu},\quad\text{and}
\]
\[
\operatorname{Tr}\left(i \frac{\partial\Psi}{\partial
i}\right)=j_0\frac{\partial \Psi}{\partial j_0}+j\!\cdot\!\frac{\partial
\Psi}{\partial j}.
\]
For the computation of the antisymmetric part, we use the equalities
\begin{align*}
i\nu-(i\nu)^T&=\widehat{j\boldsymbol{\nu}},\\
i\nu\nu-(i\nu\nu)^T&=2\nu_0(i\widehat{\boldsymbol{\nu}}-(i\widehat{\boldsymbol{\nu}})^T)+(i\widehat{\boldsymbol{\nu}}\widehat{\boldsymbol{\nu}}-(i\widehat{\boldsymbol{\nu}}\widehat{\boldsymbol{\nu}})^T)=2\nu_0\widehat{j\boldsymbol{\nu}}+\widehat{(j\boldsymbol{\nu})\times\boldsymbol{\nu}},
\end{align*}
and the fact that the matrix $i\frac{\partial\Psi}{\partial i}$ is symmetric,
by
the preceding lemma.$\qquad\blacksquare$

\bigskip

Using this lemma, we obtain that the Euler-Poincar\'e equations
\eqref{micromorphic_fluids} become
\begin{equation}\label{microstretch_fluids}
\left\lbrace
\begin{array}{ll}
\vspace{0.2cm}\displaystyle\frac{\partial}{\partial t}
\mathbf{u}+\nabla_\mathbf{u}\mathbf{u}=\frac{1}{\rho}\operatorname{grad}\frac{\partial
\Psi}{\partial\rho^{-1}},\\
\vspace{0.2cm}\displaystyle\frac{j_0}{2}\left(\frac{D}{d
t}\nu_0-\nu_0^2\right)+(j\boldsymbol{\nu})\!\cdot\!\boldsymbol{\nu}=2\left(j_0\frac{\partial
\Psi}{\partial j_0}+j\!\cdot\!\frac{\partial \Psi}{\partial j}\right),\\
\vspace{0.2cm}\displaystyle j\frac{D}{d
t}\boldsymbol{\nu}-2\nu_0j\boldsymbol{\nu}-(j\boldsymbol{\nu})\times\boldsymbol{\nu}=0,\\
\vspace{0.2cm}\displaystyle\frac{\partial}{\partial
t}\rho+\operatorname{div}(\rho \mathbf{u})=0,\qquad\,\,\,\displaystyle
\frac{D}{d t}j+2\nu_0 j+[j,\widehat{\boldsymbol{\nu}}]=0.
\end{array} \right.
\end{equation}
The third equation can be rewritten as
\[
\frac{D}{dt}(j\boldsymbol{\nu})=0.
\]
From the last equation, we deduce that the conservation law for $j_0$ reads
\[
\frac{D}{dt}j_0+2\nu_0j_0=0.
\]
Thus we have recovered the equations \eqref{Eringen_microstretch_continua} by
Euler-Poincar\'e reduction, up to a change of variables, replacing $\nu$ by
$-\nu^T$.

The Lagrangian reduction for microstretch fluids follows from that of
micromorphic fluids and so we do not repeat it. Making use of the fact that the
micromotion $\chi$ takes values in $CSO(3)$, we obtain that the evolution of the
microinertia $i$ with initial value $i_0$ is
\[
i=\left(\left(\chi^T\right)^{-1} i_0
\chi^{-1}\right)\circ\eta^{-1}=\left(\frac{1}{\operatorname{det}(\chi)^{2/3}}\chi
i_0\chi^{-1}\right)
\circ\eta^{-1}.
\]
It follows that the evolution of the variable $j$ has the same expression and
that the evolution of the microstretch inertia $j_0$, with initial value
$(j_0)_0$, is given by
\[
j_0=\frac{1}{\operatorname{det}(\chi)^{2/3}}(j_0)_0\circ\eta^{-1}.
\]
As in the micromorphic case, if the initial microinertia tensor $i_0$ is
invertible, we can obtain the equations and its associated Poisson bracket by
Lie-Poisson reduction.

\medskip  

\noindent \textbf{Lagrangian and Hamiltonian formulation for micropolar fluids.} The
symmetry group of micropolar fluid dynamics is
$\operatorname{Diff}(\mathcal{D})\,\circledS\,\mathcal{F}(\mathcal{D},SO(3))$.
In terms of the variable $j$, the Lagrangian reads
\[
l(\mathbf{u},\boldsymbol{\nu},\rho,j)=\frac{1}{2}\int_\mathcal{D}\rho\|\mathbf{u}\|^2\mu+\frac{1}{2}\int_\mathcal{D}\rho\left(j\boldsymbol{\nu}\!\cdot\!\boldsymbol{\nu}\right)\mu-\int_\mathcal{D}\rho\Psi(\rho^{-1},j)\mu.
\]
Using that $\nu_0=0$ for micropolar fluids, we obtain that the Euler-Poincar\'e
equations are given by
\begin{equation}\label{micropolar_fluids}
\left\lbrace
\begin{array}{ll}
\vspace{0.2cm}\displaystyle\frac{\partial}{\partial t}
\mathbf{u}+\nabla_\mathbf{u}\mathbf{u}=\frac{1}{\rho}\operatorname{grad}\frac{\partial
\Psi}{\partial\rho^{-1}},\qquad\,\,\, j\frac{D}{d
t}\boldsymbol{\nu}-(j\boldsymbol{\nu})\times\boldsymbol{\nu}=0,\\
\vspace{0.2cm}\displaystyle\frac{\partial}{\partial
t}\rho+\operatorname{div}(\rho \mathbf{u})=0,\qquad\,\,\,\displaystyle
\frac{D}{d t}j+[j,\widehat{\boldsymbol{\nu}}]=0.
\end{array} \right.
\end{equation}
Note the analogy between the second equation and the equation for the rigid
body. 
 
The Lagrangian reduction for micropolar fluids follows from that of
micromorphic
fluids so we shall not repeat it. Making use of the fact that the micromotion
$\chi$ takes values in $SO(3)$, we obtain that the evolution of the
microinertia
$j$ with initial value $j_0$ is
\[
j=\left(\chi j_0\chi^{-1}\right)
\circ\eta^{-1}.
\]
Thus, the evolution of its determinant is given by
\[
\operatorname{det}(j)=\operatorname{det}(j_0)\circ\eta^{-1}
\]
which shows that $j $ is invertible if and only if $j_0$ is invertible. Thus,
if
the initial microinertia tensor $j_0$ is invertible, the equations of motion
\eqref{micropolar_fluids} can be obtained by Lie-Poisson reduction. Using the
equalities
\[
\mathbf{m}:=\frac{\delta
l}{\delta\mathbf{u}}=\rho\mathbf{u}^\flat\quad\text{and}\quad
\boldsymbol{\kappa}:=\frac{\delta l}{\delta\boldsymbol{\nu}}=\rho
j\boldsymbol{\nu},
\]
the Legendre transformation yields the Hamiltonian
\[
h(\mathbf{m},\boldsymbol{\kappa},\rho,j)=\frac{1}{2}\int_\mathcal{D}\frac{1}{\rho}\|\mathbf{m}\|^2\mu+\frac{1}{2}\int_\mathcal{D}\frac{1}{\rho}(j^{-1}\boldsymbol{\kappa}\!\cdot\!\boldsymbol{\kappa})\mu+\int_\mathcal{D}\rho\Psi(\rho^{-1},j)\mu
\]
representing the total energy of the system. In the particular case of
micropolar fluids, and in terms of the variable $j$, the Poisson bracket
\eqref{Poisson_bracket_micromorphic} becomes
\begin{align}\label{Poisson_bracket_micropolar}
\{f,g\}(\mathbf{m},&\boldsymbol{\kappa},\rho,j)=\int_\mathcal{D}\mathbf{m}\cdot\left[\frac{\delta
f}{\delta\mathbf{m}},\frac{\delta g}{\delta\mathbf{m}}\right]\mu\nonumber\\
&+\int_\mathcal{D}\kappa\cdot\left(\operatorname{ad}_{\frac{\delta
f}{\delta\kappa}}\frac{\delta g}{\delta\kappa}+\mathbf{d}\frac{\delta
f}{\delta\kappa}\cdot\frac{\delta g}{\delta\mathbf{m}}-\mathbf{d}\frac{\delta
g}{\delta\kappa}\cdot\frac{\delta f}{\delta\mathbf{m}}\right)\mu\\
&+\int_\mathcal{D}\rho\left(\textbf{d}\left(\frac{\delta f}{\delta
\rho}\right)\frac{\delta g}{\delta 
\mathbf{m}}-\textbf{d}\left(\frac{\delta g}{\delta
\rho}\right)\frac{\delta f}{\delta  \mathbf{m}}\right)\mu\nonumber\\
&+\int_\mathcal{D}
j\!\cdot\!\left(\operatorname{div}\left(\frac{\delta
f}{\delta j}\frac{\delta g}{\delta \mathbf{m}}\right)+\left[\frac{\delta
f}{\delta j},\frac{\delta g}{\delta
\kappa}\right]-\operatorname{div}\left(\frac{\delta g}{\delta j}\frac{\delta
f}{\delta \mathbf{m}}\right)-\left[\frac{\delta g}{\delta j},\frac{\delta
f}{\delta \kappa}\right]\right)\mu\nonumber,
\end{align}
where the brackets in the last term denote the usual commutator bracket of
matrices.

\medskip  

\noindent \textbf{A quaternionic point of view on microstretch and micropolar fluids.} We now show that, in the case of microstretch fluids, we can use the Lie group $\mathbb{H}^\times$ of invertible quaternions to describe the micromotion of the particles. To see this, recall that there is a $2$ to $1$ surjective group homomorphism
\[
\pi : S^3\cong SU(2)\longrightarrow SO(3)
\]
given by
\[
\alpha+\textbf{j}\,\beta \;\cong \;\left[\begin{array}{cc}\alpha&-\bar\beta\\ \beta &\bar\alpha\end{array}\right]\longmapsto\left[\begin{array}{ccc}\operatorname{Re}(\alpha^2-\beta^2)&\operatorname{Im}(\alpha^2-\beta^2)& 2\operatorname{Re}(\alpha\bar\beta)\\
-\operatorname{Im}(\alpha^2+\beta^2)&\operatorname{Re}(\alpha^2+\beta^2)&-2\operatorname{Im}(\alpha\bar\beta)\\
-2\operatorname{Re}(\alpha\beta)&-2\operatorname{Im}(\alpha\beta)&|\alpha|^2-|\beta|^2\end{array}\right],
\]
where the universal covering group $S^3=\{\alpha+\textbf{j}\,\beta \in\mathbb{H}\mid |\alpha|^2+|\beta|^2=1\}\subset \mathbb{H}^\times$ denotes the Lie group of unit quaternions. The tangent map of $\pi$ at the identity is the Lie algebra isomorphism
\[
\mathbf{p}=\mathbf{k}(\boldsymbol{\nu}_1+\mathbf{i}\boldsymbol{\nu}_2+\mathbf{j}\boldsymbol{\nu}_3)\in T_1S^3 \longmapsto 2\hat{\boldsymbol{\nu}}\in\mathfrak{so}(3),
\]
where $\boldsymbol{ \nu} = (\boldsymbol{ \nu}_1, \boldsymbol{ \nu}_2, \boldsymbol{ \nu}_3) $ and the Lie algebra $T_1S^3$ consists of pure quaternions.

Remarkably, the map $\pi$ extends to a $2$ to $1$ surjective group homomorphism
\[
\pi : \mathbb{H}^\times\longrightarrow CSO(3)
\]
given by the same expression. Thus the group $\mathbb{H}^\times$ of invertible quaternions can be seen as the universal covering group of the conformal special orthogonal rotations. The Lie algebra isomorphism reads
\[
\mathbf{p}=\nu_0+\mathbf{k}(\boldsymbol{\nu}_1+\mathbf{i}\boldsymbol{\nu}_2+\mathbf{j}\boldsymbol{\nu}_3)\in \mathbb{H} \longmapsto 2\left(\nu_0 I_3+\hat{\boldsymbol{\nu}}\right)\in\mathfrak{cso}(3).
\]
These observations show that the micromotion of a microstretch fluid can be described by a map $\chi:\mathcal{D}\rightarrow \mathbb{H}^\times$ with values in the group of invertible quaternions. 

The same remark applies to micropolar fluids where one can use the group of unit quaternions $S ^3 $ instead of $SO(3)$. In both cases, the only difference in the equations of motion is that the variable $\nu$ takes values in $\mathbb{H}$ or in the Lie algebra of pure quaternions, respectively.

\medskip  

\noindent \textbf{Remarks on the use of other groups and the anisotropic cases.} As is clear from the general theory in \S\ref{Lagrangian_PCF} and 
\S\ref{Hamiltonian_PCF}, the Lagrangian and Hamiltonian theory of microfluids can be formulated for any order parameter subgroup 
$\mathcal{O} \subset GL(3)^+$ to model different kinds of micromotion. We have chosen two specific subgroups relevant to anisotropy. The first one, $SO(K)$, was suggested to us by D. Holm (see \cite{Holm2008}, Vol I, \S1.10.2) and the second, $CSO(K)$, is its conformal version. 

For $K$ an invertible and symmetric $3\times 3$ matrix, we define the group
\[
SO(K)=\left\{A\in GL(3)^+\mid A^TKA=K\right\}.
\]
This is a three dimensional Lie group, called the \textit{special $K$-orthogonal group} with Lie algebra given by
\[
\mathfrak{so}(K)=\left\{\nu\in\mathfrak{gl}(3)\mid \nu^T K+K\nu=0\right\}.
\]
Note that $SO(K)$ is the group of orthogonal linear maps relative to the (possibly indefinite) inner product $\boldsymbol{u}^TK\boldsymbol{v}$. 

Recall than when $K = I_3$ the linear isomorphism given by the hat map $\widehat{\;} : \mathbb{R}^3\rightarrow\mathfrak{so}(3)$ pulls back the commutator bracket on $\mathfrak{so}(3)$ to the cross product on $\mathbb{R}^3$. We shall search for the analogue of this map for $\mathfrak{so}(K) $. For an arbitrary invertible symmetric matrix $L$ we define
$\widehat{\;\;}^K : \mathbb{R}^3\rightarrow\mathfrak{so}(K)$ by
\[
\widehat{\boldsymbol{u}}^K:=K^{-1}\widehat{L\boldsymbol{u}}\in\mathfrak{so}(K).
\]
Working in this generality allows us later to make specific choices of $L $ that simplify certain formulas. The adjoint and coadjoint actions of $SO(K)$ on $\mathbb{R}^3$ are computed to be
\[
\operatorname{Ad}_A\boldsymbol{u}=L^{-1}AL\boldsymbol{u}\quad\text{and}\quad \operatorname{Ad}^*_{A^{-1}}\boldsymbol{v}=LA^{-T}L^{-1}\boldsymbol{v}.
\]
By differentiating the adjoint action we find the expression of the Lie bracket on $\mathbb{R}^3$ associated to the group $SO(K)$. This Lie bracket generalizes the cross product $\boldsymbol{u}\times\boldsymbol{v}$ associated to $SO(3)$, thus we shall use the notation $\boldsymbol{u}\times_K\boldsymbol{v}$ and we call it the \textit{cross-product associated to $K$}. We have
\begin{align*}
\boldsymbol{u}\times_K\boldsymbol{v}:&=\left.\frac{d}{dt}\right|_{t=0}\operatorname{Ad}_{\operatorname{exp}\left(t\widehat{\boldsymbol{u}}^K\right)}\boldsymbol{v}=L^{-1}\widehat{\boldsymbol{u}}^KL\boldsymbol{v}=L^{-1}K^{-1}\widehat{L\boldsymbol{u}}L\boldsymbol{v}\\
&=L^{-1}K^{-1}(L\boldsymbol{u}\times L\boldsymbol{v})=\operatorname{det}(L)L^{-1}K^{-1}L^{-1}(\boldsymbol{u}\times \boldsymbol{v}),
\end{align*}
where we used the relation
\[
L\boldsymbol{u}\times L\boldsymbol{v}=\operatorname{det}(L)L^{-1}(\boldsymbol{u}\times \boldsymbol{v}),
\]
valid for any invertible and symmetric matrix $L$. To show this formula, we first note that it holds when $L$ is a diagonal matrix. Then it suffices to write $L=B^TDB$, where $B\in SO(3)$ and $D$ is the diagonal matrix of eigenvalues. 
Note that, by construction, we have the formula
\[
\widehat{\boldsymbol{u}\times _K\boldsymbol{v}}^K=\left[\widehat{\boldsymbol{u}}^K,\widehat{\boldsymbol{v}}^K\right].
\]
The infinitesimal coadjoint action is computed to be
\[
\operatorname{ad}^*_{\boldsymbol{u}}\boldsymbol{w}=\operatorname{det}(L)(L^{-1}K^{-1}L^{-1}\boldsymbol{w})\times\boldsymbol{u}.
\]

Note that $\mathfrak{so}(K) = [\mathfrak{so}(K) ,\mathfrak{so}(K)]$,  as is easily seen using the formula for $\times_K$. Thus, since $\dim \mathfrak{so}(K) = 3$,  the Lie algebra $\mathfrak{so}(K) $ is simple (see \cite{Knapp1996}, example at the end of \S2, Chapter I). According to the classification of all three dimensional real Lie algebras (see \cite{Knapp1996}, problems 28--35 in  \S15, Chapter I), it follows that if $K $ is definite, then $\mathfrak{so}(K) \cong \mathfrak{so}(3)$ and if $K $ is indefinite, then $\mathfrak{so}(K) \cong \mathfrak{sl}(2, \mathbb{R}) $ (see also the discussion at the end of \S14.6 in \cite{MaRa1999}).

A Casimir function for the Lie-Poisson bracket on $(\mathbb{R}^3,\times_K)$ is given by 
\[
C(\boldsymbol{w})=\frac{\operatorname{det}(L)}{2}\boldsymbol{w}^TL^{-1}K^{-1}L^{-1}\boldsymbol{w},
\]
since we have
\[
\vspace{0.2cm}\operatorname{ad}^*_{\delta C/\delta \boldsymbol{w}}\boldsymbol{w}=0.
\]
This shows that the coadjoint orbits of $SO(K)$ are ellipsoids (when $K$ is definite) or hyperboloids (when $K$ is indefinite).

Note that if we choose $L=\operatorname{det}(K)K^{-1}$, the cross-product and the Casimir function take the simple expressions
\[
\boldsymbol{u}\times_K\boldsymbol{v}=K(\boldsymbol{u}\times\boldsymbol{v})
\]
and
\[
C(\boldsymbol{w})=\frac{1}{2}\boldsymbol{w}^TK\boldsymbol{w}.
\]
Other relevant choices for $L$ are $L=I_3$ and $L=K$.

When $K$ is positive definite, the group $\mathcal{O}=SO(K)$ can be used for the description of the \textit{anisotropic version} of the theory of micropolar fluids. In such micropolar fluids, the particles have preferred axes of deformations. Note that if $\boldsymbol{u} \in \mathbb{R}^3$ is given, then the $SO(K)$-orbit $\{A \boldsymbol{u} \mid A \in SO(K) \}$ through $\boldsymbol{u} $ in $\mathbb{R}^3$ coincides with the ellipsoid $\mathcal{E}_c: = \{\boldsymbol{v} \in \mathbb{R}^3 \mid \boldsymbol{v}^T K \boldsymbol{v} = c \}$ containing $\boldsymbol{u} $. Typical cases are given by
\[
K=\left[\begin{array}{ccc}\; 1\;&0\;&0\;\\ \;0\;&1\;&0\;\\ \;0\;&0\;&\varepsilon\;\end{array}\right]\quad\text{or}\quad K=\left[\begin{array}{ccc} \;\varepsilon\;&0\;&0\;\\ \;0\;&\varepsilon\;&0\;\\ \;0\;&0\;&1\;\end{array}\right], \quad 0<\varepsilon\ll 1,
\]
corresponding to extremely oblate and prolate ellipsoids.
\medskip 

As in the micropolar case, the Lagrangian of the \textit{anisotropic micropolar fluid} is deduced from that of the micromorphic case. For the microrotation energy, we have identity
\[
\operatorname{Tr}((i\nu)^T\nu)=j_K\boldsymbol{\nu}\!\cdot\boldsymbol{\nu},
\]
where $\widehat{\boldsymbol{\nu}}^K=\nu$ and where
\[
j_K:=\operatorname{Tr}\left(iK^{-2}\right)L^{2}-LK^{-1}iK^{-1}L\in Sym(3).
\]
If we choose $L=K$, then we obtain
\[
j_K:=\operatorname{Tr}\left(iK^{-2}\right)K^{2}-i\in Sym(3),
\]
which is the anisotropic generalization of the relation
\[
j=\operatorname{Tr}(i)I_3-i
\]
that was used in the micropolar case. Thus the Lagrangian of the anisotropic micropolar fluid can be written as
\[
l(\mathbf{u},\boldsymbol{\nu},\rho,j_K)=\frac{1}{2}\int_\mathcal{D}\rho\|\mathbf{u}\|\mu+\frac{1}{2}\int_\mathcal{D}\rho(j_K\boldsymbol{\nu}\!\cdot\boldsymbol{\nu})\mu-\int_\mathcal{D}\rho\Psi(\rho^{-1},j_K)\mu.
\]
In terms of the variables $(\mathbf{u},\nu,\rho,i)$ the associated motion equations are given by \eqref{micromorphic_fluids}, except that now $\nu$ takes values in $\mathfrak{so}(K)$. One can write these equations in terms of the variables $(\mathbf{u},\boldsymbol{\nu},\rho,j_K)$ by simply replacing $\nu$ by $\widehat{\boldsymbol{\nu}}^K$ and $i$ by
\[
\frac{1}{2}\operatorname{Tr}(j_KL^{-2})K^2-KL^{-1}j_KL^{-1}K.
\]
in the micromorphic fluid's equations \eqref{micromorphic_fluids}.

\medskip

We can also consider an anisotropic version of the microstretch theory, by choosing as internal symmetry group the Lie group
\[
CSO(K)=\{ A\in GL^+(3)\mid \text{there exists $\lambda\in\mathbb{R}$ such that $A^TKA=\lambda K$}\} 
\]
called the \textit{conformal special $K$-orthogonal group}. The Lie algebra of $CSO(K)$ is
\[
\mathfrak{cso}(K)=\{\nu\in\mathfrak{gl}(3)\mid \text{there exists $\mu\in\mathbb{R}$ such that $\nu^TK+K\nu=\mu K$}\}.
\]
In such \textit{anisotropic microstetch fluids}, in addition to have preferred axes of deformations, the fluid particles can also stretch.

As in the microstretch case, we have, for all $\nu\in\mathfrak{cso}(K)$, the decomposition
\[
\nu=\widehat{\boldsymbol{\nu}}^K+\nu_0I_3,
\]
for a unique $\boldsymbol{\nu}\in\mathbb{R}^3$ and $\nu_0\in\mathbb{R}$. The adjoint action and Lie bracket read
\[
\operatorname{Ad}_A(\widehat{\boldsymbol{\nu}}^K+\nu_0I_3)=\frac{1}{\operatorname{det}(A)^{1/3}}\left(L^{-1}AL \boldsymbol{\nu}\right){\!\!\widehat{\phantom{Q}}}^K+\nu_0I_3\quad\text{and}\quad [\nu,\mu]=\widehat{\boldsymbol{\nu}\times_K \boldsymbol{\mu}}^K
\]
As in the microstretch case, the Lagrangian of the anisotropic microstretch fluid is deduced from that of the micromorphic case. In the anisotropic case the microrotation energy takes the complicate expression
\begin{align*}
\operatorname{Tr}((i\nu)^T\nu)&=j_K\boldsymbol{\nu}\!\cdot\!\boldsymbol{\nu}-2\nu_0\operatorname{Tr}\left((KL^{-1}j_KL^{-1}K)\widehat{\nu}^K\right)\\
&\qquad+\nu_0^2\frac{1}{2}\operatorname{Tr}(j_KL^{-2})\operatorname{Tr}(K^2)-\nu_0^2\operatorname{Tr}(KL^{-1}j_KL^{-1}K).
\end{align*}
If we choose $L=K$, then this equality reads
\[
\operatorname{Tr}((i\nu)^T\nu)=j_K\boldsymbol{\nu}\!\cdot\!\boldsymbol{\nu}-2\nu_0\operatorname{Tr}(j_K\widehat{\boldsymbol{\nu}}^K)+\nu_0^2\frac{1}{2}\operatorname{Tr}(j_KK^{-2})\operatorname{Tr}(K^2)-\nu_0^2\operatorname{Tr}(j_K),
\]
which generalizes the formula
\[
\operatorname{Tr}((i\nu)^T\nu)=j\boldsymbol{\nu}\!\cdot\!\boldsymbol{\nu}+\frac{1}{2}j_0\nu_0^2
\]
valid when $K=I_3$.


\subsection{Liquid Crystals}\label{LC}

The liquid crystal state is a distinct phase of matter observed between the
crystalline (solid) and isotropic (liquid) states. There are three main types
of
liquid crystal states, depending upon the amount of order in the material. The
\textit{nematic} liquid crystal phase is characterized by rod-like molecules
that have no positional order but tend to point in the same direction. In
\textit{cholesteric} (or \textit{chiral nematic}), molecules resemble helical
springs, which may have opposite chiralities. As for nematics, the molecules
exhibit a privileged direction, which is the axis of the helices.
\textit{Smectic} liquid crystals are essentially different form both nematics
and cholesterics, in that they have one more degree of orientational order.
Smectics generally form layers within which there is a loss of positional
order,
while orientational order is still preserved. See for example \cite{Ch1992},
\cite{GePr1993}, and \cite{Virga1994} for more information.

There are various approaches to the dynamics of liquid crystals. In this
section
we carry out the Lagrangian an Hamiltonian formulation for three of them:\\
- the \textit{director theory} due to Oseen, Frank, Z\"ocher, Ericksen and
Leslie,\\
- the \textit{micropolar} and \textit{microstretch theories\/}, due to Eringen,
which take into account the microinertia of the particles and which is
applicable, for example, to \textit{liquid crystal polymers},\\
- the \textit{ordered micropolar} approach, due to Lhuillier and Rey, which
combines the director theory with the micropolar
models.

For simplicity we suppose that the fluid container $\mathcal{D}$ is a domain in
$\mathbb{R}^3$ and all boundary conditions are ignored. This means that in all
integration by parts we assume that the boundary terms vanish.


\subsubsection{Director theory}

In this theory it is assumed that only the direction and not the sense of the
molecules matter in the description of the physical phenomena. Thus, 
the preferred orientation of the molecules around a point is described by a
unit
vector $\mathbf{n}:\mathcal{D}\rightarrow S^2$, called the {\it director}, and
$\mathbf{n}$ and $-\mathbf{n}$ are assumed to be equivalent.

This description is convenient for nematics and cholesterics. We will consider
the director as a map $\mathbf{n}:\mathcal{D}\rightarrow\mathbb{R}^3$, and we
will show that the condition $\|\mathbf{n}\|=1$ is preserved by the
Euler-Poincar\'e dynamics.

We shall obtain the Ericksen-Leslie equation by Euler-Poincar\'e and
Lie-Poisson
reduction. For nematic and cholesteric liquid crystals, the order parameter Lie
group is $\mathcal{O}=SO(3)$. In this paragraph we always identify the Lie
algebra $\mathfrak{so}(3)$ with $\mathbb{R}^3$, that is, we have
$\operatorname{ad}_uv=u\times v$. Identifying the dual through the canonical
inner product on $\mathbb{R}^3$, we have $\operatorname{ad}^*_uw=w\times u$.

The symmetry group is the semidirect product $\operatorname{Diff}(\mathcal{D})\,\circledS\,\mathcal{F}(\mathcal{D},SO(3))$.
In the original Erick\-sen-Leslie approach, the liquid crystal flow is supposed
to
be incompressible. In this case the subgroup
$\operatorname{Diff}(\mathcal{D})_{\rm{vol}}\,\circledS\,\mathcal{F}(\mathcal{D},SO(3))$
should be used. Here we treat general compressible flows. An element
$(\eta,\chi)$ in the group $\operatorname{Diff}(\mathcal{D})\,\circledS\,\mathcal{F}(\mathcal{D},SO(3))$
acts \textit{linearly} on the advected quantities
$(\rho,\mathbf{n})\in\mathcal{F}(\mathcal{D})\times\mathcal{F}(\mathcal{D},\mathbb{R}^3)$,
by
\[
(\rho,\mathbf{n})\mapsto
\left(J\eta(\rho\circ\eta),\chi^{-1}(\mathbf{n}\circ\eta)\right).
\]
The associated infinitesimal action and diamond operations are given by
\[
\mathbf{n}\mathbf{u}=\nabla\mathbf{n}\!\cdot\!\mathbf{u},\quad\mathbf{n}\boldsymbol{\nu}=\mathbf{n}\times\boldsymbol{\nu},\quad
\mathbf{m}\diamond_1\mathbf{n}=-\nabla\mathbf{n}^T\!\cdot\!\mathbf{m}\quad\text{and}\quad\mathbf{m}\diamond_2\mathbf{n}=\mathbf{n}\times\mathbf{m},
\]
where $\boldsymbol{\nu}, \mathbf{m},
\mathbf{n}\in\mathcal{F}(\mathcal{D},\mathbb{R}^3)$. We use here and in the rest
of the paper $\nabla \mathbf{n}$ instead of the usual derivative
$\mathbf{d}\mathbf{n}$ of the director
$\mathbf{n}\in\mathcal{F}(\mathcal{D},\mathbb{R}^3)$, since this notation is
standard in the liquid crystals literature. Thus $ \nabla \mathbf{n} $ is a $ 3
\times  3$ matrix whose rows are the vectors $ \nabla \mathbf{n} _1$, $\nabla
\mathbf{n}_2$, $\nabla \mathbf{n} _3$ and hence its columns are $ \partial_1
\mathbf{n} $, $\partial_2\mathbf{n}$, $\partial_3\mathbf{n}$;  $\nabla
\mathbf{n} ^T $ denotes the transpose of $ \nabla \mathbf{n} $. The
Euler-Poincar\'e equations \eqref{EP} associated to the group
$\left(
\operatorname{Diff}(\mathcal{D})\,\circledS\,\mathcal{F}(\mathcal{D},SO(3))\right)
\,\circledS\,\left(\mathcal{F}(\mathcal{D})\times\mathcal{F}(\mathcal{D},\mathbb{R}^3)
 \right)$ are
\begin{equation}
\label{EP_n}
\left\{
\begin{array}{ll}
\vspace{0.1cm}\displaystyle \frac{\partial}{\partial t}\frac{\delta
l}{\delta\mathbf{u}}=-{\boldsymbol{\pounds}}_\mathbf{u}\frac{\delta
l}{\delta\mathbf{u}}-\operatorname{div}\mathbf{u}\frac{\delta
l}{\delta\mathbf{u}}-\frac{\delta
l}{\delta\boldsymbol{\nu}}\!\cdot\!\mathbf{d}\boldsymbol{\nu}+\rho\,\mathbf{d}\frac{\delta
l}{\delta\rho}-\left(\nabla\mathbf{n}^T\!\cdot\!\frac{\delta
l}{\delta\mathbf{n}}\right)^\flat,\\
\vspace{0.1cm}\displaystyle \frac{\partial}{\partial t}\frac{\delta
l}{\delta\boldsymbol{\nu}}=\boldsymbol{\nu}\times\frac{\delta
l}{\delta\boldsymbol{\nu}}-\operatorname{div}\left(\frac{\delta
l}{\delta\boldsymbol{\nu}}\mathbf{u}\right)+\mathbf{n}\times\frac{\delta
l}{\delta\mathbf{n}},
\end{array}
\right.
\end{equation}
Note that the first equation is in $\Omega^1(\mathcal{D})$, the dual of 
the Lie algebra $\mathfrak{X}(\mathcal{D})$.
The advection equations are
\begin{equation}\label{advection_n}
\left\{
\begin{array}{ll}
\vspace{0.2cm}\displaystyle\frac{\partial}{\partial
t}\rho+\operatorname{div}(\rho\mathbf{u})=0,\\
\vspace{0.2cm}\displaystyle\frac{\partial}{\partial
t}\mathbf{n}+(\nabla\mathbf{n})\mathbf{u}+\mathbf{n}\times\boldsymbol{\nu}=0.
\end{array}
\right.
\end{equation}
The reduced Lagrangian for nematic and cholesteric liquid crystals is of the
form
\begin{equation}\label{Lagrangian_liquid_crystals}
l(\mathbf{u},\boldsymbol{\nu},\rho,\mathbf{n}):=\frac{1}{2}\int_\mathcal{D}\rho\|\mathbf{u}\|^2\mu+\frac{1}{2}\int_\mathcal{D}\rho
J\|\boldsymbol{\nu}\|^2\mu-\int_\mathcal{D}\rho
F(\rho^{-1},\mathbf{n},\nabla\mathbf{n})\mu,
\end{equation}
where the constant $J$ is the \textit{microinertia constant} and $F$ is the
\textit{free energy}. The axiom of objectivity requires that
\[
F(\rho^{-1},A^{-1}\mathbf{n},A^{-1}\nabla\mathbf{n}A)=F(\rho^{-1},\mathbf{n},\nabla\mathbf{n}),
\]
for all $A\in O(3)$ for nematics, or for all $A\in SO(3)$ for cholesterics.

A standard choice for $F$ is the \textit{Oseen-Z\"ocher-Frank free energy}
given by
\begin{align}\label{standard_energy}
\rho
F(\rho^{-1},\mathbf{n},\nabla\mathbf{n})&=K_2\underbrace{(\mathbf{n}\cdot\operatorname{curl}\mathbf{n})}_{\textbf{chirality}}+\frac{1}{2}K_{11}\underbrace{(\operatorname{div}\mathbf{n})^2}_{\textbf{splay}}+\frac{1}{2}K_{22}\underbrace{(\mathbf{n}\cdot\operatorname{curl}\mathbf{n})^2}_{\textbf{twist}}
\nonumber \\
& \qquad
+\frac{1}{2}K_{33}\underbrace{\|\mathbf{n}\times\operatorname{curl}\mathbf{n}\|^2}_{\textbf{bend}},
\end{align}
where $K_2\neq 0$ for cholesterics and $K_2=0$ for nematics. 
The free energy can also contain additional terms due to external
electromagnetic fields.
The constants $K_{11}, K_{22}, K_{33}$ are respectively associated to the three
principal distinct director axis deformations in nematic liquid crystals,
namely, splay, twist, and bend. In general, these constants are different, but
the expression of the resulting equations of motion can be greatly simplified
if
we make the \textit{one-constant approximation} $K_{11}=K_{22}=K_{33}=K$. In
this case the free energy becomes, up to the addition of a divergence,
\[
\rho
F(\rho^{-1},\mathbf{n},\nabla\mathbf{n})=\frac{1}{2}K\|\nabla\mathbf{n}\|^2.
\]
The functional derivatives of the Lagrangian \eqref{Lagrangian_liquid_crystals}
are computed to be
\[
\mathbf{m}:=\frac{\delta
l}{\delta\mathbf{u}}=\rho\mathbf{u}^\flat,\quad\boldsymbol{\kappa}:
=\frac{\delta l}{\delta\boldsymbol{\nu}}=\rho J\boldsymbol{\nu},
\]
and
\[
\frac{\delta
l}{\delta\rho}=\frac{1}{2}\|\mathbf{u}\|^2+\frac{1}{2}J\|\boldsymbol{\nu}\|^2-F+\frac{1}{\rho}\frac{\partial
F}{\partial\rho^{-1}},\quad-\mathbf{h}:=\frac{\delta
l}{\delta\mathbf{n}}=-\rho\frac{\partial
F}{\partial\mathbf{n}}+\partial_i\left(\rho\frac{\partial
F}{\partial\mathbf{n}_{,i}}\right).
\]
The vector field $\mathbf{h}$ is referred to as the \textit{molecular field}.
In
the case of the free energy \eqref{standard_energy} for nematics $(K_2=0)$, the
vector $\mathbf{h}$ is given by
\[
\mathbf{h}=K_{11}\operatorname{grad}\operatorname{div}\mathbf{n}-K_{22}(A\operatorname{curl}\mathbf{n}+\operatorname{curl}(A\mathbf{n}))+K_{33}(\mathbf{B}\times\operatorname{curl}\mathbf{n}+\operatorname{curl}(\mathbf{n}\times\mathbf{B})),
\]
where $A:=\mathbf{n}\!\cdot\!\operatorname{curl}\mathbf{n}$ and
$\mathbf{B}:=\mathbf{n}\times\operatorname{curl}\mathbf{n}$. In the case of the
one-constant approximation we have $\mathbf{h}=-K\Delta\mathbf{n}$.

Using the Lagrangian \eqref{Lagrangian_liquid_crystals}, the 
Euler-Poincar\'e
equations \eqref{EP_n} become
\begin{equation}
\label{EP_n_liquid_crystals}
\left\{
\begin{array}{ll}
\vspace{0.2cm}\displaystyle\rho\left(\frac{\partial}{\partial
t}\mathbf{u}+\nabla_\mathbf{u}\mathbf{u}\right)=\operatorname{grad}\frac{\partial
F}{\partial \rho^{-1}}-\partial_i\left(\rho\frac{\partial
F}{\partial\mathbf{n}_{,i}}\!\cdot\!\nabla\mathbf{n}\right),\\
\vspace{0.2cm}\displaystyle \rho
J\frac{D}{dt}\boldsymbol{\nu}=\mathbf{h}\times\mathbf{n},
\end{array}
\right.
\end{equation}
where we have used the standard notation $\frac{D}{dt}: =
\frac{\partial}{\partial t} + \mathbf{u}\!\cdot\!\nabla $ for the material
derivative acting on every component of $\boldsymbol{\nu} $. The advection
equations are
\begin{equation}\label{advection_liquid_crystals}
\left\{
\begin{array}{ll}
\vspace{0.1cm}\displaystyle\frac{\partial}{\partial
t}\rho+\operatorname{div}(\rho\mathbf{u})=0,\\
\vspace{0.1cm}\displaystyle\frac{D}{dt}\mathbf{n}=\boldsymbol{\nu}\times\mathbf{n}.
\end{array}
\right.
\end{equation}
The evolution of the advected quantities is given by
\[
\rho=J(\eta^{-1})(\rho_0\circ\eta^{-1})\;\;\text{and}\;\;\mathbf{n}=\left(\chi\mathbf{n}_0\right)\circ\eta^{-1}.
\]

We now show that under some conditions,  equations \eqref{EP_n_liquid_crystals}
and \eqref{advection_liquid_crystals} are equivalent to the Ericksen-Leslie
equations for liquid crystals. This will use the following lemma.

\begin{lemma} 
\label{liquid_crystals_n}
Let $\boldsymbol{\nu}$ and $\mathbf{n}$ be solutions of the Euler-Poincar\'e
equations \eqref{EP_n_liquid_crystals} and \eqref{advection_liquid_crystals}.
Then
\begin{enumerate}
\item[\bf{(i)}] $\|\mathbf{n}_0\|=1$ implies $\|\mathbf{n}\|=1$ for all time.
\vspace{0.2cm}
\item[\bf{(ii)}] $\displaystyle\frac{D}{dt}(\mathbf{n}\!\cdot\!\boldsymbol{\nu})=0$.
Therefore, $\mathbf{n}_0\!\cdot\!\boldsymbol{\nu}_0=0$ implies
$\mathbf{n}\!\cdot\!\boldsymbol{\nu}=0$ for all time.
\vspace{0.2cm}
\item[\bf{(iii)}] Suppose that $\mathbf{n}_0\!\cdot\!\boldsymbol{\nu}_0=0$ and
$\|\mathbf{n}_0\|=1$. Then
the second equation of \eqref{advection_liquid_crystals} reads
\[
\boldsymbol{\nu}=\mathbf{n}\times\frac{D}{dt}\mathbf{n}
\]
and the second equation of \eqref{EP_n_liquid_crystals} reads
\[
\rho J\frac{D^2}{dt^2}\mathbf{n}-2q\mathbf{n}+\mathbf{h}=0,
\]
where $2q: =  \mathbf{n}\cdot \mathbf{h} - \rho J\left\| \frac{D
\mathbf{n}}{dt}
\right\|^2 = \mathbf{n}\cdot \mathbf{h} - \rho J\left\|
\boldsymbol{\nu}\right\|^2 $.
\end{enumerate}
\end{lemma}
\textbf{Proof.} \textbf{(i)} This is clear from the evolution
$\mathbf{n}=\left(\chi\mathbf{n}_0\right)\circ\eta^{-1}$, since $\chi\in
SO(3)$.

\textbf{(ii)} Using the second equations in \eqref{EP_n_liquid_crystals} and
\eqref{advection_liquid_crystals}, we have
\[
\frac{D}{dt}(\mathbf{n}\!\cdot\!\boldsymbol{\nu})=\left(\frac{D}{dt}\mathbf{n}\right)\!\cdot\!\boldsymbol{\nu}+\mathbf{n}\!\cdot\!\left(\frac{D}{dt}\boldsymbol{\nu}\right)=(\boldsymbol{\nu}\times\mathbf{n})\!\cdot\!\boldsymbol{\nu}+\mathbf{n}\!\cdot\!\frac{1}{\rho
J}(\mathbf{h}\times\mathbf{n})=0.
\]
\textbf{(iii)} Using the first two  results, we obtain
\[
\mathbf{n}\times\frac{D}{dt}\mathbf{n}=\mathbf{n}\times(\boldsymbol{\nu}\times\mathbf{n})=(\mathbf{n}\!\cdot\!\mathbf{n})\boldsymbol{\nu}-(\mathbf{n}\!\cdot\!\boldsymbol{\nu})\mathbf{n}=\boldsymbol{\nu}
\]
which proves the first relation. 

To prove the second, we take the material time-derivative of the relation above
to rewrite the second equation of \eqref{EP_n_liquid_crystals} as $\rho
J\left(\mathbf{n}\times\frac{D^2}{dt^2}\mathbf{n}\right)=\mathbf{h}\times\mathbf{n}$.
Taking the cross product with $\mathbf{n}$ on the left we obtain the equation
\[
\rho J\left(\mathbf{n}\!\cdot\!\frac{D^2}{dt^2}\mathbf{n}\right)\mathbf{n}-\rho
J\frac{D^2}{dt^2}\mathbf{n}=\mathbf{h}-(\mathbf{n}\!\cdot\!\mathbf{h})\mathbf{n}.
\]
Defining $2q:=\mathbf{n}\!\cdot\!\mathbf{h}+\rho
J\left(\mathbf{n}\!\cdot\!\frac{D^2}{dt^2}\mathbf{n}\right)$, this equation
reads
\[
\rho J\frac{D^2}{dt^2}\mathbf{n}-2q\mathbf{n}+\mathbf{h}=0
\]
which is the second relation since $\mathbf{n}\!\cdot\!\frac{D^2}
{dt^2}\mathbf{n} = -\left\| \frac{D \mathbf{n}}{dt} \right\|^2= -
\|\boldsymbol{\nu}\|^2$. The last two equalities are obtained by taking two
consecutive material time-derivatives of $\|\mathbf{n}\|^2 = 1$ and using the
identity $\boldsymbol{\nu} = \mathbf{n} \times \frac{D}{dt} \mathbf{n}$,
respectively.
$\qquad\blacksquare$

\medskip

Thus, we obtain the following theorem.

\begin{theorem}\label{equivalence} Let
$(\mathbf{u},\boldsymbol{\nu},\rho,\mathbf{n})$ be a solution of the equations
\eqref{EP_n_liquid_crystals} and \eqref{advection_liquid_crystals}, with the
initial conditions $\mathbf{n}_0$ and $\boldsymbol{\nu}_0$ verifying
\[
\|\mathbf{n}_0\|=1\quad\text{and}\quad
\mathbf{n}_0\!\cdot\!\boldsymbol{\nu}_0=0.
\]
Then $(\mathbf{u},\rho,\mathbf{n})$ is a solution of the Ericksen-Leslie
equations
\begin{equation}\label{E_L}
\left\{
\begin{array}{ll}
\vspace{0.1cm}\displaystyle\rho\left(\frac{\partial}{\partial
t}\mathbf{u}+\nabla_\mathbf{u}\mathbf{u}\right)=\operatorname{grad}\frac{\partial
F}{\partial\rho^{-1}}-\partial_j\left(\rho\frac{\partial
F}{\partial\mathbf{n}_{,j}}\!\cdot\!\nabla\mathbf{n}\right),\\
\vspace{0.1cm}\displaystyle \rho
J\frac{D^2}{dt^2}\mathbf{n}-2q\mathbf{n}+\mathbf{h}=0,\\
\vspace{0.1cm}\displaystyle\frac{\partial}{\partial
t}\rho+\operatorname{div}(\rho\mathbf{u})=0,
\end{array}
\right.
\end{equation}
where $2q: =  \mathbf{n}\cdot \mathbf{h} - \rho J\left\| \frac{D
\mathbf{n}}{dt}
\right\|^2$.

Conversely, let $(\mathbf{u},\rho,\mathbf{n})$ be a solution of the
Ericksen-Leslie equations \eqref{E_L} such that $\|\mathbf{n}\|=1$, and define
\[
\boldsymbol{\nu}:=\mathbf{n}\times\frac{D}{dt}\mathbf{n}.
\]
Then $(\mathbf{u},\boldsymbol{\nu},\rho,\mathbf{n})$ is a solution of the
equations  \eqref{EP_n_liquid_crystals} and \eqref{advection_liquid_crystals}.
\end{theorem}

\medskip  

\noindent \textbf{Remark.} One should think of the function $q$ in the Ericksen-Leslie
equation the way one regards the pressure in ideal incompressible homogeneous
fluid dynamics, namely, the $q$ is an unknown function determined by the
imposed
constraint $\|\mathbf{n}\| = 1$ in the following way. The dot product of the
second equation of \eqref{E_L} with $\mathbf{n}$ yields the formula of $q$
given
in the statement of the theorem by imposing  $\|\mathbf{n}\| = 1$.

This $q$ does not appear in the Euler-Poincar\'e formulation relative to the
variables $(\mathbf{u}, \boldsymbol{\nu}, \rho, \mathbf{n})$,
since in this case, the constraint $\|\mathbf{n}\|=1$ is automatically
satisfied by Lemma \ref{liquid_crystals_n}(i). 
\medskip

As a consequence of Theorem \ref{equivalence}, we obtain the Ericksen-Leslie
equations for liquid crystals by Lagrangian reduction. Consider the
right-invariant Lagrangian 
\[
L_{(\rho_0,\mathbf{n}_0)}:T\left[\operatorname{Diff}(\mathcal{D})\,\circledS\,\mathcal{F}(\mathcal{D},SO(3))\right]\rightarrow\mathbb{R}
\]
induced by the Lagrangian \eqref{Lagrangian_liquid_crystals}, and assume that
$\|\mathbf{n}_0\|=1$ and $\boldsymbol{\nu}_0\!\cdot\!\mathbf{n}_0=0$. A curve 
$(\eta,\chi)\in\operatorname{Diff}(\mathcal{D})\,\circledS\,\mathcal{F}(\mathcal{D},SO(3))$
is a solution of the Euler-Lagrange equations associated to
$L_{(\rho_0,\mathbf{n}_0)}$, with initial condition $\boldsymbol{\nu}_0$ if and
only if the curve
\[
(\mathbf{u},\nu):=(\dot\eta\circ\eta^{-1},\dot\chi\chi^{-1}\circ\eta^{-1})
\]
is a solution of the Ericksen-Leslie equations \eqref{E_L}, where
\[
\rho=J(\eta^{-1})(\rho_0\circ\eta^{-1})\quad \text{and}\quad
\mathbf{n}=(\chi\mathbf{n}_0)\circ\eta^{-1}.
\]
As in the case of microfluids, the curve
$\eta\in\operatorname{Diff}(\mathcal{D})$ describes the \textit{Lagrangian
motion of the fluid} or \textit{macromotion}, and the curve
$\chi\in\mathcal{F}(\mathcal{D},SO(3))$ describes the \textit{local molecular
orientation relative to a fixed reference frame} or \textit{micromotion}. A
standard choice for the initial value of the director is
\[
\mathbf{n}_0(x):=(0,0,1),\;\;\text{for all $x\in\mathcal{D}$}.
\]
In this case we obtain that
\[
\mathbf{n}=\left[\begin{array}{l}\chi_{13}\\ \chi_{23}\\
\chi_{33}\end{array}\right]\circ\eta^{-1}.
\]
This relation is usually taken as a definition of the director, when the
$3$-axis is chosen as the reference axis of symmetry.
\medskip

By the Legendre transformation, the Hamiltonian for liquid crystals is given by
\[
h(\mathbf{m},\boldsymbol{\kappa},\rho,\mathbf{n}):=\frac{1}{2}\int_\mathcal{D}\frac{1}{\rho}\|\mathbf{m}\|^2\mu+\frac{1}{2J}\int_\mathcal{D}\frac{1}{\rho}\|\boldsymbol{\kappa}\|^2\mu+\int_\mathcal{D}\rho
F(\rho^{-1},\mathbf{n},\nabla\mathbf{n})\mu.
\]
The Poisson bracket for liquid crystals is given by
\begin{align}\label{liquid_crystals_bracket}
\{f,g\}(&\mathbf{m},\rho,\boldsymbol{\kappa},\mathbf{n})=\int_\mathcal{D}\mathbf{m}\cdot\left[\frac{\delta
f}{\delta\mathbf{m}},\frac{\delta g}{\delta\mathbf{m}}\right]\mu\\
&+\int_\mathcal{D}\boldsymbol{\kappa}\cdot\left(\frac{\delta
f}{\delta\boldsymbol{\kappa}}\times\frac{\delta
g}{\delta\boldsymbol{\kappa}}+\mathbf{d}\frac{\delta
f}{\delta\boldsymbol{\kappa}}\cdot\frac{\delta
g}{\delta\mathbf{m}}-\mathbf{d}\frac{\delta
g}{\delta\boldsymbol{\kappa}}\cdot\frac{\delta
f}{\delta\mathbf{m}}\right)\mu\nonumber\\
&+\int_\mathcal{D}\rho\left(\mathbf{d}\frac{\delta
f}{\delta\rho}\cdot\frac{\delta g}{\delta\mathbf{m}}-\mathbf{d}\frac{\delta
g}{\delta\rho}\cdot\frac{\delta f}{\delta\mathbf{m}}\right)\mu\nonumber\\
&+\int_\mathcal{D}\left[\left(\mathbf{n}\times\frac{\delta
f}{\delta\boldsymbol{\kappa}}+\nabla\mathbf{n}\cdot\frac{\delta
f}{\delta\mathbf{m}}\right)\frac{\delta
g}{\delta\mathbf{n}}-\left(\mathbf{n}\times\frac{\delta
g}{\delta\boldsymbol{\kappa}}+\nabla\mathbf{n}\cdot\frac{\delta
g}{\delta\mathbf{m}}\right)\frac{\delta
f}{\delta\mathbf{n}}\right]\mu.\nonumber
\end{align}

The Kelvin circulation theorem for liquid crystals reads
\[
\frac{d}{dt}\oint_{c_t}\mathbf{u}^\flat=\oint_{c_t}\frac{1}{\rho}\nabla\mathbf{n}^T\!\cdot\!\mathbf{h},
\]
where $\mathbf{h}$ is the molecular field defined by
\[
\mathbf{h}=\rho\frac{\partial
F}{\partial\mathbf{n}}-\partial_i\left(\rho\frac{\partial
F}{\partial\mathbf{n}_{,i}}\right).
\]


\subsubsection{Micropolar theory of liquid crystals}
\label{subsubsec:micropolar}

This approach is based on the equations for micropolar continua given in
\eqref{Eringen_micropolar_continua}. The difference from the micropolar fluid
treated before is that for liquid crystals the free energy $\Psi$ depends also
on a new variable
$\gamma=(\gamma^{ab}_i)\in\Omega^1(\mathcal{D},\mathfrak{so}(3))$ called the
\textit{wryness tensor}. This variable is denoted by
$\boldsymbol{\gamma}=(\boldsymbol{\gamma}^a_i)$ when it is seen as a form with
values in $\mathbb{R}^3$.

The \textit{constitutive equations} in the
\textit{non-dissipative} case are given by
\[
t_{kl}=\frac{\partial\Psi}{\partial\rho^{-1}}\delta_{kl}-\rho\frac{\partial\Psi}{\partial\boldsymbol{\gamma}^a_k}\boldsymbol{\gamma}^a_l\quad\text{and}\quad
m_{kl}=\rho\frac{\partial\Psi}{\partial\boldsymbol{\gamma}^l_k},
\]
according to (12.5.18) in \cite{Er2001}, where the function
$\Psi=\Psi(\rho^{-1},j,\gamma): \mathbb{R}\times
Sym(3)\times\mathfrak{gl}(3)\rightarrow\mathbb{R}$ denotes the \textit{free
energy}. The axiom of objectivity requires that
\[
\Psi(\rho^{-1},A^{-1}jA,A^{-1}\boldsymbol{\gamma}
A)=\Psi(\rho^{-1},j,\boldsymbol{\gamma}),
\]
for all $A\in O(3)$ (for nematics and nonchiral smectics), or for all $A\in
SO(3)$ (for cholesterics and chiral smectics). See paragraphs 12.6, 12.8 and
12.9 in \cite{Er2001} for the choice of the free energy for nematics,
cholesterics and smectics respectively.

Assuming that $f_l=0$ and $l_k=0$, the equation
\eqref{Eringen_micropolar_continua} for micropolar continua become
\begin{equation}\label{Micropolar_L_C_Eringen}
\left\lbrace
\begin{array}{l}\vspace{0.2cm}\displaystyle\rho\frac{D}{dt}\mathbf{u}_l=\partial_l\frac{\partial\Psi}{\partial\rho^{-1}}-\partial_k\left(\rho\frac{\partial\Psi}{\partial\boldsymbol{\gamma}^a_k}\boldsymbol{\gamma}^a_l\right),\qquad
\rho\sigma_l=\partial_k\left(\rho\frac{\partial\Psi}{\partial\boldsymbol{\gamma}^l_k}\right)-\varepsilon_{lmn}\rho\frac{\partial\Psi}{\partial\boldsymbol{\gamma}^a_m}\boldsymbol{\gamma}^a_n,\\
\vspace{0.2cm}\displaystyle\frac{D}{dt}\rho+\rho\operatorname{div}\mathbf{u}=0,\qquad\displaystyle
\frac{D}{dt}j_{kl}+(\varepsilon_{kpr}j_{lp}+\varepsilon_{lpr}j_{kp})\boldsymbol{\nu}_r=0.\end{array}\right.
\end{equation}
These are the equation for \textit{non-dissipative micropolar liquid crystals}
as given in section 12 of \cite{Er2001}. 

To these equations one needs to add the evolution of $\boldsymbol{\gamma}$
given
by 
\begin{equation}\label{gamma_evolution}
\frac{D}{dt}\boldsymbol{\gamma}^a_l =
\partial_l\boldsymbol{\nu}_a+\nu_{ab}\boldsymbol{\gamma}^b_l-\boldsymbol{\gamma}^a_r\partial_l
\mathbf{u}_r
\end{equation}
which is equation (12.3.13) in \cite{Er2001}. Here $\frac{D}{dt}$ acts on the one-form $\boldsymbol{\gamma}\in\Omega^1(\mathcal{D},\mathbb{R}^3)$ as $\frac{D}{dt}\boldsymbol{\gamma}:=\frac{\partial}{\partial
t}\boldsymbol{\gamma}+\boldsymbol{\pounds}_\mathbf{u}\boldsymbol{\gamma}$.

\medskip  

\noindent \textbf{Lagrangian and Hamiltonian formulation of micropolar liquid
crystals.} We now show that the system of equations
\eqref{Micropolar_L_C_Eringen} and \eqref{gamma_evolution} can be obtained by
affine Euler-Poincar\'e and affine Lie-Poisson reduction. As in the case of
micropolar fluids, the symmetry group is the semidirect product
$\operatorname{Diff}(\mathcal{D})\,\circledS\,\mathcal{F}(\mathcal{D},SO(3))$.
The advected quantities are the \textit{mass density} $\rho$, the
\textit{microinertia} tensor $j\in\mathcal{F}(\mathcal{D},Sym(3))$, and the
\textit{wryness tensor} $\gamma\in\Omega^1(\mathcal{D},\mathfrak{so}(3))$. The
action of the symmetry group on the variables $\rho$ and $j$ is \textit{linear}
and is the same as for micropolar fluids. The action of $( \eta , \chi)$ on the
wryness tensor $\gamma$ is 
\textit{affine} and is given by
\[
\gamma\mapsto\chi^{-1}(\eta^*\gamma)\chi+\chi^{-1}T\chi.
\]
The Lagrangian is the same as that for micropolar fluids, except for the fact
that the free energy depends also on the wryness tensor. We thus have
\begin{equation}\label{micropolar_L_C_Lagrangian}
l(\mathbf{u},\boldsymbol{\nu},\rho,j,\gamma)=\frac{1}{2}\int_\mathcal{D}\rho\|\mathbf{u}\|^2\mu+\frac{1}{2}\int_\mathcal{D}\rho(j\boldsymbol{\nu}\!\cdot\!\boldsymbol{\nu})\mu-\int_\mathcal{D}\rho\Psi(\rho^{-1},j,\gamma)\mu.
\end{equation}
The computation of the affine Euler-Poincar\'e equations is similar to that for
micropolar fluids, except for the equation associated to the variable
$\boldsymbol{\nu}$ for which we give some details below.

Using the Lagrangian \eqref{micropolar_L_C_Lagrangian} we find that the
evolution for $\boldsymbol{\nu}$ is
\[
\rho\frac{1}{2}\frac{D}{dt}\widehat{j\boldsymbol{\nu}}=2\left(j\frac{\partial\Psi}{\partial
j}\right)_A-\operatorname{div}\left(\rho\frac{\partial\Psi}{\partial\gamma}\right)-\left[\gamma_i,\rho\frac{\partial\Psi}{\partial\gamma_i}\right],
\]
where the index $A$ denotes the antisymmetric part of the matrix. According to
our conventions, $ \frac{D}{dt}\widehat{j \boldsymbol{\nu}}$ on the right hand
side means that $ \frac{D}{dt}$ is applied to every entry of the matrix valued
function $ \widehat{j \boldsymbol{\nu}}: \mathcal{D} \rightarrow 
\mathfrak{so}(3)$. Using the
equality
\[
\widehat{\frac{\partial\Psi}{\partial\boldsymbol{\gamma}_i}}=2\frac{\partial\Psi}{\partial\gamma_i},
\]
we obtain
\[
\rho\frac{D}{dt}(j\boldsymbol{\nu})=4\widetilde{\left(j\frac{\partial\Psi}{\partial
j}\right)_A}-\operatorname{div}\left(\rho\frac{\partial\Psi}{\partial\boldsymbol{\gamma}}\right)-\boldsymbol{\gamma}_i\times\rho\frac{\partial\Psi}{\partial\boldsymbol{\gamma}_i},
\]
where $\widetilde{\;} $ denotes the inverse of $\widehat{\;}$. We now use the
axiom of objectivity to simplify this expression.

\begin{lemma}\label{objectivity_micropolar_L_C} Suppose that the free energy
$\Psi$ verifies the axiom of objectivity, that is,
\[
\Psi(\rho^{-1},A^{-1}iA,A^{-1}\boldsymbol{\gamma}
A)=\Psi(\rho^{-1},i,\boldsymbol{\gamma}),\;\text{for all $A\in SO(3)$ $($or
$O(3)$$)$.}
\]
Then
\begin{equation}\label{objectivity_condition}
2\left(j\frac{\partial \Psi}{\partial
j}\right)_A=\left(\left(\frac{\partial\Psi}{\partial\boldsymbol{\gamma}}\right)^T\boldsymbol{\gamma}-\boldsymbol{\gamma}\left(\frac{\partial\Psi}{\partial\boldsymbol{\gamma}}\right)^T\right)_A.
\end{equation}
\end{lemma}
\textbf{Proof.} Consider a curve $A(t)\in SO(3)$ such that $A(0) = I_3$ and
$\dot A(0)=\xi\in\mathfrak{so}(3)$. Differentiating the equality
$\Psi(A(t)^{-1}jA(t),A(t)^{-1}\boldsymbol{\gamma}A(t))=\Psi(j,\boldsymbol{\gamma})$
at $t=0$, we obtain the condition
\[
\mathbf{D}\Psi(j,\boldsymbol{\gamma})(j\xi-\xi
j,\boldsymbol{\gamma}\xi-\xi\boldsymbol{\gamma})=0,\;\text{for all $\xi\in
\mathfrak{so}(3)$.}
\]
Using the equalities
\begin{align*}
\mathbf{D}\Psi(j,\boldsymbol{\gamma})(j\xi-\xi
j,&\boldsymbol{\gamma}\xi-\xi\boldsymbol{\gamma})=\operatorname{Tr}\left(\frac{\partial
\Psi}{\partial j}(j\xi-\xi
j)\right)+\operatorname{Tr}\left(\left(\frac{\partial\Psi}{\partial\boldsymbol{\gamma}}\right)^T(\boldsymbol{\gamma}\xi-\xi\boldsymbol{\gamma})\right)\\
&=\operatorname{Tr}\left(\left(\frac{\partial \Psi}{\partial
j}j-j\frac{\partial \Psi}{\partial
j}+\left(\frac{\partial\Psi}{\partial\boldsymbol{\gamma}}\right)^T\boldsymbol{\gamma}-\boldsymbol{\gamma}\left(\frac{\partial\Psi}{\partial\boldsymbol{\gamma}}\right)^T\right)\xi\right),
\end{align*}
and the identity $2\left(j\frac{\partial \Psi}{\partial
j}\right)_A=j\frac{\partial \Psi}{\partial j}-\frac{\partial \Psi}{\partial
j}j$, we obtain the result.$\qquad\blacksquare$

\medskip

Note that equation \eqref{objectivity_condition}, can be rewritten in
$\mathbb{R}^3$ as
\[
4\widetilde{\left(j\frac{\partial \Psi}{\partial
j}\right)_A}=\boldsymbol{\gamma}^a\times\frac{\partial\Psi}{\partial\boldsymbol{\gamma}^a}+\boldsymbol{\gamma}_k\times\frac{\partial\Psi}{\partial\boldsymbol{\gamma}_k}.
\]
Using these results for the equation for $\boldsymbol{\nu}$, the affine
Euler-Poincar\'e equations associated to the Lagrangian
\eqref{micropolar_L_C_Lagrangian} read

\begin{equation}\label{micropolar_L_C}
\left\lbrace
\begin{array}{ll}
\vspace{0.2cm}\displaystyle\rho\left(\frac{\partial}{\partial t}
\mathbf{u}+\nabla_\mathbf{u}\mathbf{u}\right)=\operatorname{grad}\frac{\partial
\Psi}{\partial\rho^{-1}}-\partial_k\left(\rho\frac{\partial\Psi}{\partial\boldsymbol{\gamma}^a_k}\boldsymbol{\gamma}^a\right),\\
\vspace{0.2cm}\displaystyle j\frac{D}{d
t}\boldsymbol{\nu}-(j\boldsymbol{\nu})\times\boldsymbol{\nu}=-\frac{1}{\rho}\operatorname{div}\left(\rho\frac{\partial\Psi}{\partial\boldsymbol{\gamma}}\right)+\boldsymbol{\gamma}^a\times\frac{\partial\Psi}{\partial\boldsymbol{\gamma}^a},\\
\vspace{0.2cm}\displaystyle\frac{\partial}{\partial
t}\rho+\operatorname{div}(\rho \mathbf{u})=0,\qquad\,\,\,\displaystyle
\frac{D}{d t}j+[j,\hat{\boldsymbol{\nu}}]=0,\\
\vspace{0.2cm}\displaystyle\frac{\partial}{\partial
t}\gamma+\boldsymbol{\pounds}_\mathbf{u}\gamma+\mathbf{d}^\gamma
\hat{\boldsymbol{\nu}}=0.
\end{array} \right.
\end{equation}
Recall that $ \mathbf{d} ^\gamma  \hat {\boldsymbol{\nu} } = \mathbf{d} \hat{ \boldsymbol{\nu}} + [\gamma, \hat{\boldsymbol{\nu}}]$. Thus, we have recovered equations \eqref{Micropolar_L_C_Eringen} for
non-dissipative micropolar liquid crystals, together with the equation
\eqref{gamma_evolution}, up to a change of variables $\gamma\mapsto-\gamma$.

\medskip

Consider the right-invariant Lagrangian
$L_{(\rho_0,j_0,\gamma_0)}$ induced on the tangent bundle\\ $T\left[\operatorname{Diff}(\mathcal{D})\,\circledS\,\mathcal{F}(\mathcal{D},SO(3))\right]$
by the Lagrangian \eqref{micropolar_L_C_Lagrangian}. A curve
$(\eta,\chi)\in
\operatorname{Diff}(\mathcal{D})\,\circledS\,\mathcal{F}(\mathcal{D},SO(3))$ is
a solution of the Euler-Lagrange equations associated to
$L_{(\rho_0,j_0,\gamma_0)}$ if and only if the curve
\[
(\mathbf{u},\hat{\boldsymbol{\nu}}):=(\dot\eta\circ\eta^{-1},\dot\chi\chi^{-1}\circ\eta^{-1})\in\mathfrak{X}(\mathcal{D})\,\circledS\,\mathcal{F}(\mathcal{D},SO(3))
\]
is a solution of the equations \eqref{micropolar_L_C} with initial conditions
$(\rho_0,j_0,\gamma_0)$. The evolution of the mass density $\rho$, the
microinertia $j$, and the wryness tensor $\gamma$ is given by
\[
\rho=J(\eta^{-1})(\rho_0 \circ\eta^{-1}),\quad j=\left(\chi
j_0\chi^{-1}\right)\circ\eta^{-1},\quad\text{and}\quad\gamma=\eta_*\left(\chi\gamma_0\chi^{-1}+\chi
T\chi^{-1}\right).
\]
If the initial value $\gamma_0$ is zero, then the evolution of $\gamma$ is
given
by
\[
\gamma=\eta_*\left(\chi T\chi^{-1}\right).
\]
This relation is usually taken as a definition of $\gamma$ when using equation
\eqref{Micropolar_L_C_Eringen}. We consider $\gamma$ as an independent variable
and therefore the system \eqref{micropolar_L_C} contains an evolution equation
for $\gamma$.

The Legendre transformation and the Hamiltonian formulation can be carried out
as in the case of micropolar fluids. The affine Lie-Poisson bracket consists of
the sum of the Lie-Poisson bracket 
\eqref{Poisson_bracket_micropolar} with the term
\[
\int_\mathcal{D}\left[\left(\mathbf{d}^\gamma\frac{\delta
f}{\delta\kappa}+{\boldsymbol{\pounds}}_{\frac{\delta
f}{\delta\mathbf{m}}}\gamma\right)\cdot\frac{\delta
g}{\delta\gamma}-\left(\mathbf{d}^\gamma\frac{\delta
g}{\delta\kappa}+{\boldsymbol{\pounds}}_{\frac{\delta
g}{\delta\mathbf{m}}}\gamma\right)\cdot\frac{\delta f}{\delta\gamma}\right]\mu
\]
due to the presence of the variable $\gamma$.
\medskip

The Kelvin-Noether circulation theorem applied to micropolar liquid crystals
yields the relation
\[
\frac{d}{dt}\oint_{c_t}\mathbf{u}^\flat=\oint_{c_t}\frac{\partial\Psi}{\partial
i}\!\cdot\!\mathbf{d}i+\frac{\partial\Psi}{\partial
\gamma}\!\cdot\!\mathbf{i}_{\_\,}\mathbf{d}\gamma-\frac{1}{\rho}\operatorname{div}\left(\rho\frac{\partial\Psi}{\partial\gamma}\right)\!\cdot\!\gamma.
\]
The $\gamma$-circulation yields the relation
\[
\frac{d}{dt}\oint_{c_t}\boldsymbol{\gamma}=\oint_{c_t}\boldsymbol{\nu}\times\boldsymbol{\gamma}
\]
in $\mathbb{R}^3$.

\medskip  

\noindent \textbf{Remark.} According to equation (12.9.1) in \cite{Er2001}, a
liquid crystal flow is called \textit{smectic} if the constraint
\[
\operatorname{Tr}(\boldsymbol{\gamma}) = \sum_{i=1}^3 \boldsymbol{\gamma}^i_i
=0
\]
is satisfied. However, note that  this constraint is not preserved by the evolution
\[
\gamma=\eta_*\left(\chi\gamma_0\chi^{-1}+\chi T\chi^{-1}\right)
\]
in general. This is consistent with the fact that 
the last equation in \eqref{micropolar_L_C}, which can be written equivalently in vectorial form as 
\[
\frac{\partial \boldsymbol{\gamma}}{\partial t} + 
{\boldsymbol{\pounds}}_{\mathbf{u}}  \boldsymbol{\gamma} + \mathbf{d}
\boldsymbol{\nu} + \boldsymbol{\gamma} \times \boldsymbol{\nu} = 0,
\]
does \textit{not\/} imply that if the
initial condition for $\boldsymbol{\gamma} $ has trace zero then
$\operatorname{Tr} \boldsymbol{\gamma} = 0 $ for all time.


\subsubsection{Microstretch theory of polymeric liquid crystals}
\label{polymeric_microstretch}

This approach is based on the equations for microstretch continua given in
\eqref{Eringen_microstretch_continua}.
The difference from the microstretch fluid treated before is that for polymeric
liquid crystals the free energy $\Psi$ depends also the \textit{wryness tensor}
$\gamma\in\Omega^1(\mathcal{D},\mathfrak{so}(3))$ and on the
\textit{microstrain} $e\in\Omega^1(\mathcal{D})$. The \textit{constitutive
equations} in the \textit{non-dissipative} case are given by
\begin{align*}
t_{kl}&=\frac{\partial\Psi}{\partial\rho^{-1}}\delta_{kl}-\rho\frac{\partial\Psi}{\partial\boldsymbol{\gamma}^a_k}\boldsymbol{\gamma}^a_l-\rho\frac{\partial\Psi}{\partial
e_k}e_l,\qquad
m_{kl}=\rho\frac{\partial\Psi}{\partial\boldsymbol{\gamma}^l_k},\\
m_k&=\frac{\partial\Psi}{\partial e_k},\qquad\text{and}\qquad
s-t=2\rho\left(\frac{\partial\Psi}{\partial
j_{kl}}j_{kl}+\frac{\partial\Psi}{\partial j_0}j_0\right)
\end{align*}
see equations (16.3.13) and (16.3.13) in \cite{Er2001}, where
\[
\Psi=\Psi(\rho^{-1},j,\gamma,e): \mathbb{R}\times
Sym(3)\times\mathfrak{gl}(3)\times\mathbb{R}^3\rightarrow\mathbb{R}
\]
denotes the \textit{free energy}. The axiom of objectivity requires that
\[
\Psi(\rho^{-1},A^{-1}jA,A^{-1}\boldsymbol{\gamma}
A,A^{-1}e)=\Psi(\rho^{-1},j,\boldsymbol{\gamma},e),
\]
for all $A\in O(3)$ (for nematics and nonchiral smectics), or for all $A\in
SO(3)$ (for cholesterics and chiral smectics). See paragraphs 16.4, 16.6 and
16.7 in \cite{Er2001} for the choice of the free energy for nematic, smectic,
and cholesteric polymers respectively.

Assuming that $f_l=0$ and $l_k=0$, the equation
\eqref{Eringen_microstretch_continua} for microstretch continua become
\begin{equation}\label{Microstretch_L_C_Eringen}
\left\lbrace
\begin{array}{l}\vspace{0.2cm}\displaystyle\rho\frac{D}{dt}\mathbf{u}_l=\partial_l\frac{\partial\Psi}{\partial\rho^{-1}}-\partial_k\left(\rho\frac{\partial\Psi}{\partial\boldsymbol{\gamma}^a_k}\boldsymbol{\gamma}^a_l+\rho\frac{\partial\Psi}{\partial
e_k}e_l\right),\\
\vspace{0.2cm}\displaystyle
\rho\sigma=\partial_k\left(\rho\frac{\partial\Psi}{\partial
e_k}\right)-2\rho\left(\frac{\partial\Psi}{\partial
j_{kl}}j_{kl}+\frac{\partial\Psi}{\partial j_0}j_0\right),\\
\vspace{0.2cm}\displaystyle\rho\sigma_l=\partial_k\left(\rho\frac{\partial\Psi}{\partial\boldsymbol{\gamma}^l_k}\right)-\varepsilon_{lmn}\rho\left(\frac{\partial\Psi}{\partial\boldsymbol{\gamma}^a_m}\boldsymbol{\gamma}^a_n+\frac{\partial\Psi}{\partial
e_m}e_n\right),\\
\vspace{0.2cm}\displaystyle\frac{D}{dt}\rho+\rho\operatorname{div}\mathbf{u}=0,\qquad\displaystyle
\frac{D}{dt}j_{kl}+(\varepsilon_{kpr}j_{lp}+\varepsilon_{lpr}j_{kp})\boldsymbol{\nu}_r=0.\end{array}\right.
\end{equation}
These are the equations for \textit{non-dissipative microstretch polymeric
liquid crystals} as studied in section 16 of \cite{Er2001}.

To these equations one adds the evolution of $\boldsymbol{\gamma} \in  \Omega ^1
( \mathcal{D} , \mathbb{R} ^3)$ and
$e \in  \Omega ^1 ( \mathcal{D} )$ given by 
\[
\frac{D}{dt}\boldsymbol{\gamma}^a_l =
\partial_l\boldsymbol{\nu}_a+\widehat{\boldsymbol{\nu}}_{ab}\boldsymbol{\gamma}^b_l-\boldsymbol{\gamma}^a_r\partial_lu_r
\]
and
\[
\frac{D}{dt}e_k=\nu_{0,k}+e_i\mathbf{u}_{i,k}
\]
which is equation (16.3.8) in \cite{Er2001}.
Like in equation \eqref{gamma_evolution}, $ D /dt $ acts on the
one-forms $\boldsymbol{\gamma}$ and $e$ as $D/dt=\partial/\partial
t+\boldsymbol{\pounds}_\mathbf{u}$.

\medskip  

\noindent \textbf{Lagrangian and Hamiltonian formulation of polymeric liquid
crystals.}
We now show that the equation \eqref{Microstretch_L_C_Eringen} can be obtained
by affine Euler-Poincar\'e and affine Lie-Poisson reduction. As in the case of
microstretch fluids, the symmetry group is the semidirect product
$\operatorname{Diff}(\mathcal{D})\,\circledS\,\mathcal{F}(\mathcal{D},CSO(3))$.
The advected quantities are the \textit{mass density} $\rho$, the
\textit{microinertia} tensor $i\in\mathcal{F}(\mathcal{D},Sym(3))$, the
\textit{wryness tensor} $\gamma\in\Omega^1(\mathcal{D},\mathfrak{so}(3))$, and
the \textit{microstrain} $e\in\Omega^1(\mathcal{D})$. The action of the
symmetry
group on the variables $\rho$ and $i$ is \textit{linear} and is the same as for
microstretch fluids. The action of $( \eta , \chi) \in
\operatorname{Diff}(\mathcal{D})\,\circledS\,\mathcal{F}(\mathcal{D},CSO(3))$ on
the wryness tensor $\gamma$ is 
\textit{affine} and is given by
\begin{equation}\label{action_gamma}
\gamma\mapsto\overline{\chi}^{-1}(\eta^*\gamma)\overline{\chi}+\overline{\chi}^{-1}T\overline{\chi},
\end{equation}
where $\overline{\chi}\in\mathcal{F}(\mathcal{D},SO(3))$ is defined by the
equality
\[
\chi=\operatorname{det}(\chi)^{1/3}\overline{\chi},
\]
and is the \textit{microrotation part} of
$\chi\in\mathcal{F}(\mathcal{D},CSO(3))$. The determinant
$\operatorname{det}(\chi)^{1/3}$ can be seen as the \textit{microstretch part}
of $\chi$. The action of $( \eta , \chi)$ on the microstrain $e$ is also
\textit{affine} and is given by
\begin{equation}\label{action_e}
e\mapsto
\eta^*e+\frac{1}{3}\operatorname{det}(\chi)^{-1}\mathbf{d}\left(\operatorname{det}(\chi)
\right).
\end{equation}
We now explain why these affine actions are natural. The variables $e$ and
$\gamma$ can be seen as the symmetric and antisymmetric part of a
\textit{strain
tensor} $\zeta\in\Omega^1(\mathcal{D},\mathfrak{cso}(3))$. More precisely, we
have
\[
\zeta=e I_3+\gamma,
\]
where $3e:=\operatorname{Tr}(\zeta)$ and $\gamma:=\zeta_A$. The affine action
of $\operatorname{Diff}(\mathcal{D})\,\circledS\,\mathcal{F}(\mathcal{D},CSO(3))$
is the natural action of the automorphism group onto the connections of the
trivial principal bundle $CSO(3)\times\mathcal{D} \rightarrow  \mathcal{D}$, as
defined in the general theory (see \eqref{affine_representation_gamma}), that
is,
\[
\zeta\mapsto \chi^{-1}\eta^*\zeta\chi+\chi^{-1}T\chi.
\]
By taking the trace of this action we find the affine action \eqref{action_e},
whereas the antisymmetric part gives the affine action \eqref{action_gamma}.

We now give the associated right infinitesimal actions and diamond operations.
We have
\begin{align*}
\gamma\mathbf{u}&=\boldsymbol{\pounds}_\mathbf{u}\gamma,\quad
\gamma\nu=[\gamma,\widehat{\boldsymbol{\nu}}],\quad
w\diamond_1\gamma=(\operatorname{div}w)\!\cdot\!\gamma-w\!\cdot\!\mathbf{i}_{\_\,}\mathbf{d}\gamma,\quad
w\diamond_2\gamma=[\gamma_i,w_i]\\
e\mathbf{u}&=\boldsymbol{\pounds}_\mathbf{u}e,\quad e\nu=0,\quad
f\diamond_1
e=(\operatorname{div}f)\!\cdot\!e-f\!\cdot\!\mathbf{i}_{\_\,}\mathbf{d}e,\quad
f\diamond_2 e=0.
\end{align*}
Relative to the two group one-cocycles
\[
C_1(\chi)=\overline{\chi}^{-1}T\overline{\chi}\quad\text{and}\quad
C_2(\chi)=\frac{1}{3}\operatorname{det}(\chi)^{-1}\mathbf{d}\operatorname{det}(\chi)
\]
we have
\begin{align*}
\mathbf{d}C_1(\nu)&=\mathbf{d}\widehat{\boldsymbol{\nu}},\quad\mathbf{d}C_1^T(w)=-\operatorname{div}w\\
\mathbf{d}C_2(\nu)&=\mathbf{d}\nu_0,\quad\mathbf{d}C_2^T(f)=-\frac{1}{3}(\operatorname{div}f)I_3.
\end{align*}
The Lagrangian is the same as that for microstretch fluids, except for the fact
that the free energy depends also on the wryness tensor $\gamma$ and on the
microstrain $e$. We thus have
\begin{equation}\label{microstretch_L_C_Lagrangian}
l(\mathbf{u},\nu,\rho,i,\gamma,e)=\frac{1}{2}\int_\mathcal{D}\rho\|\mathbf{u}\|^2\mu+\frac{1}{2}\int_\mathcal{D}\rho(i\nu\!\cdot\!\nu)\mu-\int_\mathcal{D}\rho\Psi(\rho^{-1},i,\gamma,e)\mu,
\end{equation}
where $i\nu\!\cdot\!\nu=\operatorname{Tr}((i\nu)^T\nu)$. 
The computation of the associated Euler-Poincar\'e equations involves the
following generalization of Lemma \ref{objectivity_micropolar_L_C}

\begin{lemma} Suppose that the free energy $\Psi$ verifies the axiom of
objectivity, that is,
\[
\Psi(\rho^{-1},A^{-1}iA,A^{-1}\boldsymbol{\gamma}
A,A^{-1}e)=\Psi(\rho^{-1},i,\boldsymbol{\gamma},e),\;\text{for all $A\in SO(3)$
$($or $O(3)$$)$.}
\]
Then
\begin{equation}\label{objectivity_condition_polymeric}
2\left(j\frac{\partial \Psi}{\partial
j}\right)_A=\left(\left(\frac{\partial\Psi}{\partial\boldsymbol{\gamma}}\right)^T\boldsymbol{\gamma}-\boldsymbol{\gamma}\left(\frac{\partial\Psi}{\partial\boldsymbol{\gamma}}\right)^T-e\otimes\frac{\partial\Psi}{\partial
e}\right)_A.
\end{equation}
\end{lemma}

Using this lemma, a direct (but long) verification shows that the affine
Euler-Poincar\'e equations associated to the Lagrangian
\eqref{microstretch_L_C_Lagrangian} are
\begin{equation}\label{microstretch_L_C}
\left\lbrace
\begin{array}{ll}
\vspace{0.2cm}\displaystyle\rho\left(\frac{\partial}{\partial t}
\mathbf{u}+\nabla_\mathbf{u}\mathbf{u}\right)=\operatorname{grad}\frac{\partial
\Psi}{\partial\rho^{-1}}-\partial_k\left(\rho\frac{\partial\Psi}{\partial\boldsymbol{\gamma}^a_k}\boldsymbol{\gamma}^a+\rho\frac{\partial\Psi}{\partial
e_k}e\right),\\
\vspace{0.2cm}\displaystyle\frac{j_0}{2}\left(\frac{D}{d
t}\nu_0-\nu_0^2\right)+(j\boldsymbol{\nu})\!\cdot\!\boldsymbol{\nu}=2\left(j_0\frac{\partial
\Psi}{\partial j_0}+j\!\cdot\!\frac{\partial \Psi}{\partial
j}\right)-\frac{1}{\rho}\operatorname{div}\left(\rho\frac{\partial
\Psi}{\partial e}\right),\\
\vspace{0.2cm}\displaystyle j\frac{D}{d
t}\boldsymbol{\nu}-2\nu_0j\boldsymbol{\nu}-(j\boldsymbol{\nu})\times\boldsymbol{\nu}=-\frac{1}{\rho}\operatorname{div}\left(\rho\frac{\partial\Psi}{\partial\boldsymbol{\gamma}}\right)+\boldsymbol{\gamma}^a\times\frac{\partial\Psi}{\partial\boldsymbol{\gamma}^a}+e\times\frac{\partial\Psi}{\partial
e},\\
\vspace{0.2cm}\displaystyle\frac{\partial}{\partial
t}\rho+\operatorname{div}(\rho \mathbf{u})=0,\qquad\,\,\,\displaystyle
\frac{D}{d t}j+2\nu_0 j+[j,\widehat{\boldsymbol{\nu}}]=0,\\
\vspace{0.2cm}\displaystyle\frac{\partial}{\partial
t}\gamma+\boldsymbol{\pounds}_\mathbf{u}\gamma+[\gamma,\widehat{\boldsymbol{\nu}}]+\mathbf{d}\widehat{\boldsymbol{\nu}}=0,\qquad\,\,\,\displaystyle
\frac{\partial}{\partial
t}e+\boldsymbol{\pounds}_\mathbf{u}e+\mathbf{d}\nu_0=0.
\end{array} \right.
\end{equation}
Thus, we have recovered equations \eqref{Microstretch_L_C_Eringen} for
\textit{non-dissipative polymeric liquid crystals} together with the
conservation laws for $\gamma$ and $e$, up to a change of variables $\nu\mapsto
-\nu^T$ and $\gamma\mapsto-\gamma$.
Recall that the microstretch rate $\nu_0\in\mathcal{F}(\mathcal{D})$ and the
microrotation rate $\boldsymbol{\nu}\in\mathcal{F}(\mathcal{D},\mathbb{R}^3)$
are constructed from the variable
$\nu\in\mathcal{F}(\mathcal{D},\mathfrak{cso}(3))$ through the decomposition
\[
\nu=\nu_0 I_3+\widehat{\boldsymbol{\nu}}.
\]

\medskip

As before, the Lagrangian reduction can be carried out, by starting with the
right-invariant Lagrangian
$L_{(\rho_0,i_0,\gamma_0,e_0)}:T\left[\operatorname{Diff}(\mathcal{D})\,\circledS\,\mathcal{F}(\mathcal{D},CSO(3))\right]\rightarrow\mathbb{R}$
induced by the Lagrangian \eqref{microstretch_L_C_Lagrangian}. The evolution of
the linear advected quantities $\rho$ and $i$ is the same as in the case of
microstretch fluids. The evolution of the affine advected quantities is given
by
\[
\gamma=\eta_*\left(\overline{\chi}\gamma_0\overline{\chi}^{-1}+\overline{\chi}
T\overline{\chi}^{-1}\right)\quad\text{and}\quad
e=\eta_*\left(e_0+\frac{1}{3}\operatorname{det}(\chi)\mathbf{d}\left(\operatorname{det}(\chi)^{-1}\right)\right).
\]
If the initial values are zero, then the evolution of $\gamma$ and $e$ is given
by
\[
\gamma=\eta_*\left(\overline{\chi}
T\overline{\chi}^{-1}\right)\quad\text{and}\quad
e=\frac{1}{3}\eta_*\Big(\operatorname{det}(\chi)\mathbf{d}
\left( \operatorname{det}(\chi)^{-1}\right) \Big).
\]
\cite{Er2001} takes these relations as definitions of $\gamma$ and $e$.

\medskip

The Legendre transformation and the Hamiltonian formulation can be carried out
as in the case of microstretch fluids. The affine Lie-Poisson bracket consists
of the sum of the Lie-Poisson bracket 
\eqref{Poisson_bracket_micromorphic} with the two terms
\begin{align*}
&\int_\mathcal{D}\left[\left(\mathbf{d}^\gamma\left(\frac{\delta
f}{\delta\kappa}\right)_A+{\boldsymbol{\pounds}}_{\frac{\delta
f}{\delta\mathbf{m}}}\gamma\right)\cdot\frac{\delta
g}{\delta\gamma}-\left(\mathbf{d}^\gamma\left(\frac{\delta
g}{\delta\kappa}\right)_A+{\boldsymbol{\pounds}}_{\frac{\delta
g}{\delta\mathbf{m}}}\gamma\right)\cdot\frac{\delta
f}{\delta\gamma}\right]\mu\\
&\quad+\int_\mathcal{D}\left[\left(\mathbf{d}\left(\frac{\delta
f}{\delta\kappa}\right)_0+{\boldsymbol{\pounds}}_{\frac{\delta
f}{\delta\mathbf{m}}}e\right)\frac{\delta g}{\delta
e}-\left(\mathbf{d}\left(\frac{\delta
g}{\delta\kappa}\right)_0+{\boldsymbol{\pounds}}_{\frac{\delta
g}{\delta\mathbf{m}}}e\right)\frac{\delta f}{\delta e}\right]\mu
\end{align*}
due to the presence of the variables $\gamma$ and $e$.

\medskip

The Kelvin-Noether circulation theorem applied to polymeric liquid crystals
yields the relation
\[
\frac{d}{dt}\oint_{c_t}\mathbf{u}^\flat=\oint_{c_t}\frac{\partial\Psi}{\partial
i}\!\cdot\!\mathbf{d}i+\frac{\partial\Psi}{\partial
\gamma}\!\cdot\!\mathbf{i}_{\_\,}\mathbf{d}\gamma-\frac{1}{\rho}\operatorname{div}\left(\rho\frac{\partial\Psi}{\partial\gamma}\right)\!\cdot\!\gamma+\frac{\partial\Psi}{\partial
e}\!\cdot\!\mathbf{i}_{\_\,}\mathbf{d}e-\frac{1}{\rho}\operatorname{div}\left(\rho\frac{\partial\Psi}{\partial
e}\right)e.
\]
The $\gamma$-circulation yields the relations
\[
\frac{d}{dt}\oint_{c_t}\boldsymbol{\gamma}=\oint_{c_t}\boldsymbol{\nu}\times\boldsymbol{\gamma}
\]
in $\mathbb{R}^3$, and
\[
\frac{d}{dt}\oint_{c_t}e=0.
\]


\subsubsection{Ordered micropolar theory}

This approach is developed in \cite{LhRe2004}. It is based on the micropolar
theory and use the Oseen-Z\"ocher-Frank free energy. As we will see, it
gives a direct generalization of the Ericksen-Leslie equations.

Lhuillier and Rey consider the general equations for micropolar continua
\eqref{Eringen_micropolar_continua}, together with the \textit{constitutive
relations} in the \textit{non-dissipative case},
\[
t_{kl}=\frac{\partial F}{\partial\rho^{-1}}\delta_{kl}-\rho\frac{\partial
F}{\partial\mathbf{n}^a_{,\,k}}\mathbf{n}^a_{,\,l}\quad\text{and}\quad
m_{kl}=\varepsilon_{lab}\mathbf{n}_a\rho\frac{\partial
F}{\partial\mathbf{n}^b_{,\,k}};
\]
see equations $(10)$ in \cite{LhRe2004}. Note that here we adapted these
relations to the compressible case. Thus, equations
\eqref{Eringen_micropolar_continua} read
\begin{equation}\label{Micropolar_director_Lhuillier_Rey}
\left\lbrace
\begin{array}{l}\vspace{0.2cm}\displaystyle\rho\frac{D}{dt}\mathbf{u}_l=\partial_l\frac{\partial
F}{\partial\rho^{-1}}-\partial_k\left(\rho\frac{\partial
F}{\partial\mathbf{n}^a_{,\,k}}\mathbf{n}^a_{,\,l}\right),\\
\vspace{0.2cm}\displaystyle\rho\sigma_l=\partial_k\left(\varepsilon_{lab}\mathbf{n}_a\rho\frac{\partial
F}{\partial\mathbf{n}^b_{,\,k}}\right)-\varepsilon_{lmn}\rho\frac{\partial
F}{\partial\mathbf{n}^a_{,\,m}}\mathbf{n}^a_{,\,n},\\
\vspace{0.2cm}\displaystyle\frac{D}{dt}\rho+\rho\operatorname{div}\mathbf{u}=0,\qquad\displaystyle
\frac{D}{dt}j_{kl}+(\varepsilon_{kpr}j_{lp}+\varepsilon_{lpr}j_{kp})\boldsymbol{\nu}_r=0.\end{array}\right.
\end{equation}
To these equations one needs to add the evolution for $\mathbf{n}$
\[
\frac{D}{dt}\mathbf{n}=\boldsymbol{\nu}\times\mathbf{n},
\]
which is equation (9) in \cite{LhRe2004}. Recall that $\sigma = \frac{D}{dt} (j
\boldsymbol{ \nu}) $.

\medskip  

\noindent \textbf{Lagrangian and Hamiltonian formulation of ordered micropolar theory
of liquid crystals.}  We now show that the equations
\eqref{Micropolar_director_Lhuillier_Rey} can be obtained by Euler-Poincar\'e
and Lie-Poisson reduction. As in the case of micropolar fluids, the symmetry
group is the semidirect product
$\operatorname{Diff}(\mathcal{D})\,\circledS\,\mathcal{F}(\mathcal{D},SO(3))$.
The advected quantities are the \textit{mass density} $\rho$, the
\textit{microinertia} tensor $j\in\mathcal{F}(\mathcal{D},Sym(3))$, and the
\textit{director} $\mathbf{n}\in\mathcal{F}(\mathcal{D},\mathbb{R}^3)$. The
representation of the symmetry group on the variables $\rho$ and $j$ is the
same
as for micropolar fluids. The representation on the director is the same as
that
for the director theory.

The Lagrangian is the same as that for micropolar fluids, except the fact that
it involves the elastic free energy $F(\rho^{-1},\mathbf{n},\nabla\mathbf{n})$,
which is usually taken to be the Oseen-Z\"ocher-Frank free energy. We thus
obtain
\[
l(\mathbf{u},\boldsymbol{\nu},\rho,j,\mathbf{n})=\frac{1}{2}\int_\mathcal{D}\rho\|\mathbf{u}\|^2\mu+\frac{1}{2}\int_\mathcal{D}\rho(j\boldsymbol{\nu}\!\cdot\!\boldsymbol{\nu})\mu-\int_\mathcal{D}\rho
F(\rho^{-1},\mathbf{n},\nabla\mathbf{n})\mu.
\]
The associated Euler-Poincar\'e equations are
\begin{equation}\label{EP_ordered_micropolar}
\left\{
\begin{array}{ll}
\vspace{0.2cm}\displaystyle\rho\left(\frac{\partial}{\partial
t}\mathbf{u}+\nabla_\mathbf{u}\mathbf{u}\right)=\operatorname{grad}\frac{\partial
F}{\partial \rho^{-1}}-\partial_i\left(\rho\frac{\partial
F}{\partial\mathbf{n}_{,i}}\!\cdot\!\nabla\mathbf{n}\right),\\
\vspace{0.2cm}\displaystyle \rho
\frac{D}{dt}(j\boldsymbol{\nu})=\mathbf{h}\times\mathbf{n}.
\end{array}
\right.
\end{equation}
The advection equations are
\begin{equation}\label{advection_ordered_micropolar}
\left\{
\begin{array}{ll}
\vspace{0.2cm}\displaystyle\frac{\partial}{\partial
t}\rho+\operatorname{div}(\rho\mathbf{u})=0,\\
\vspace{0.2cm}\displaystyle\frac{D}{dt}j+[j,\hat{\boldsymbol{\nu}}]=0,\\
\vspace{0.2cm}\displaystyle\frac{D}{dt}\mathbf{n}=\boldsymbol{\nu}\times\mathbf{n}.
\end{array}
\right.
\end{equation}
The evolution of the advected quantities is given by
\[
\rho=J(\eta^{-1})(\rho_0\circ\eta^{-1}),\quad j=\left(\chi
j_0\chi^{-1}\right)\circ\eta^{-1}
\quad\text{and}\quad\mathbf{n}=\left(\chi\mathbf{n}_0\right)\circ\eta^{-1}.
\]
Thus we have recovered equations \eqref{Micropolar_director_Lhuillier_Rey}
together with the evolution of the director. One just needs to prove that the
equations for the variable $\boldsymbol{\nu}$ are equivalent in
\eqref{EP_ordered_micropolar} and \eqref{Micropolar_director_Lhuillier_Rey}. As
shown in the following lemma, this is a consequence of the axiom of
objectivity.

\begin{lemma} Suppose that the free energy $F$ verifies the property
\[
F(\rho^{-1},A^{-1}\mathbf{n},A^{-1}\nabla\mathbf{n}A)=F(\rho^{-1},\mathbf{n},\nabla\mathbf{n}),\quad\text{for
all $A\in SO(3)$ $($or $O(3)$$)$.}
\]
Then the matrix
\[
\left(\frac{\partial
F}{\partial\nabla\mathbf{n}}\right)^T(\nabla\mathbf{n})-(\nabla\mathbf{n})\left(\frac{\partial
F}{\partial\nabla\mathbf{n}}\right)^T-\mathbf{n}\left(\frac{\partial
F}{\partial\mathbf{n}}\right)^T
\]
is symmetric.
\end{lemma}

Note that in this expression the last term denotes the multiplication of a column vector with a row vector and the result of this operation is a $ 3 \times  3 $ matrix whose $ (i,j) $ entry is $ \mathbf{n} _i \frac{ \partial F}{ \partial \mathbf{n} _j } $.

Using this lemma, we obtain the equality
\begin{equation}\label{useful_equality}
\partial_k\left(\varepsilon_{lab}\mathbf{n}_a\rho\frac{\partial
F}{\partial\mathbf{n}^b_{,\,k}}\right)-\varepsilon_{lmn}\rho\frac{\partial
F}{\partial\mathbf{n}^a_{,\,m}}\mathbf{n}^a_{,\,n}=(\mathbf{h}\times\mathbf{n})_l,
\end{equation}
therefore, the equations associated to the variable $\boldsymbol{\nu}$ in
\eqref{Micropolar_director_Lhuillier_Rey} and \eqref{EP_ordered_micropolar} are
equivalent.

The Hamiltonian and Lie-Poisson bracket can be computed as in the preceding
examples. The Kelvin-Noether circulation theorem has the same form as that of
the Ericksen-Leslie equations, namely
\[
\frac{d}{dt}\oint_{c_t}\mathbf{u}^\flat=\oint_{c_t}\frac{1}{\rho}\nabla\mathbf{n}^T\!\cdot\!\mathbf{h}.
\]


\subsubsection{Comparison of the three theories}

In this subsection we summarize the known relationships between the three theories for liquid crystals presented in this paper. We shall prove that the director theory of Ericksen-Leslie is a particular case of the 
ordered micropolar theory of Lhuillier-Rey. Therefore, one needs to compare the latter with the micropolar theory of Eringen. As will be shown, these two theories, while close, do not seem to be equivalent.

\begin{theorem} The Ericksen-Leslie equations are a particular case of the equations given by the ordered micropolar theory.
\end{theorem}
\textbf{Proof.} As we have seen in Theorem \ref{equivalence}, if $\|\mathbf{n}\|=1$ and
$\boldsymbol{\nu}=\mathbf{n}\times\frac{D}{dt}\mathbf{n}$, then the
Ericksen-Leslie equations \eqref{E_L} are equivalent to the equations
\eqref{EP_n_liquid_crystals} and \eqref{advection_liquid_crystals}. It turns
out
that these equations are a particular case of the equations
\eqref{EP_ordered_micropolar} and \eqref{advection_ordered_micropolar} given by
the ordered micropolar theory. To see this, it suffices to assume that the
microinertia $j$ is constant and given by $j=JI_3$, where $J$ is the
microinertia constant appearing in the director theory, and $I_3$ is the
identity $3\times 3$ matrix.$\qquad\blacksquare$

\medskip

Thus, \textit{the ordered micropolar theory is a
generalization of the director theory, which takes into account the variation
of microinertia.} We now compare the ordered director theory with the micropolar theory of Eringen for a particular choice of the initial condition for the microinertia $ j $. This choice imposes the condition that $ j $ is a moment of inertia. This will use some technical lemmas.

\begin{lemma}\label{lemma_1} Let $j$ and $\mathbf{n}$ be solutions of the
equations \eqref{EP_ordered_micropolar} and
\eqref{advection_ordered_micropolar}. Let $j_0$ and $\mathbf{n}_0$ be the
initial values and suppose that
\[
j_0=J(I_3-\mathbf{n}_0\otimes\mathbf{n}_0),
\]
where $J$ is a scalar constant. Then
\[
j=J(I_3-\mathbf{n}\otimes\mathbf{n})
\]
for all time.
\end{lemma}
\textbf{Proof.} From the Lagrangian formulation, we know that the evolution of
$j$ and $\mathbf{n}$ is
\[
\mathbf{n}=(\chi\mathbf{n}_0)\circ\eta^{-1}\quad\text{and}\quad j=\left(\chi
j_0\chi^{-1}\right)\circ\eta^{-1}.
\]
Since
$\chi(\mathbf{n}\otimes\mathbf{n})\chi^{-1}=(\chi\mathbf{n})\otimes(\chi\mathbf{n})$
for all $\chi\in SO(3)$, we obtain 
\[
j=\left(\chi
j_0\chi^{-1}\right)\circ\eta^{-1}=J(I_3-(\chi\mathbf{n}_0)\otimes(\chi\mathbf{n}_0))=J(I_3-\mathbf{n}\otimes\mathbf{n}).\qquad\blacksquare
\]

\medskip

\begin{lemma}\label{lemma_2} Let $\boldsymbol{\nu}$, $j$, and $\mathbf{n}$ be
solutions of the equations \eqref{EP_ordered_micropolar} and
\eqref{advection_ordered_micropolar}. Define
\[
\boldsymbol{\gamma}:=(\nabla\mathbf{n})\times\mathbf{n}\in\Omega^1(\mathcal{D},\mathbb{R}^3),
\]
that is, $ \boldsymbol{\gamma} (\mathbf{v}) = [( \nabla \mathbf{n} ) \mathbf{v}
] \times \mathbf{n} \in  \mathcal{F}( \mathcal{D} , \mathbb{R} ^3 )$ for every $
\mathbf{v} \in  \mathfrak{X} ( \mathcal{D} ) $.
Then $\boldsymbol{\gamma}$ verifies the equation
\[
\frac{\partial}{\partial
t}\boldsymbol{\gamma}+\boldsymbol{\pounds}_\mathbf{u}\boldsymbol{\gamma}+\boldsymbol{\gamma}\times\boldsymbol{\nu}+\frac{j}{J}\mathbf{d}\boldsymbol{\nu}=0.
\]
\end{lemma}
\textbf{Proof.} Using the equation
\[
\frac{D}{dt}\mathbf{n}=\boldsymbol{\nu}\times\mathbf{n},
\]
and $\|\mathbf{n}\|=1$, we compute
\begin{align*}\frac{D}{dt}\boldsymbol{\gamma}&=\nabla
\left(\frac{D}{dt}\mathbf{n}\right)\times\mathbf{n}+\nabla\mathbf{n}\times
\frac{D}{dt}\mathbf{n}\\
&=(\mathbf{d} \boldsymbol{\nu}\times\mathbf{n})\times\mathbf{n}+(\boldsymbol{\nu}\times\nabla\mathbf{n})\times\mathbf{n}+\nabla\mathbf{n}\times(\boldsymbol{\nu}\times\mathbf{n})\\
&=(\mathbf{n}\!\cdot\!\mathbf{d}\boldsymbol{\nu})\mathbf{n}-\mathbf{d}\boldsymbol{\nu}+(\mathbf{n}\!\cdot\!\boldsymbol{\nu})\nabla\mathbf{n}-(\nabla\mathbf{n}\!\cdot\!\boldsymbol{\nu})\mathbf{n}\\
&=\boldsymbol{\nu}\times(\nabla\mathbf{n}\times\mathbf{n})-(\mathbf{d}\boldsymbol{\nu}-(\mathbf{n}\!\cdot\!\mathbf{d}\boldsymbol{\nu})\mathbf{n})\\
&=\boldsymbol{\nu}\times\boldsymbol{\gamma}-(I_3-\mathbf{n}\otimes\mathbf{n})\mathbf{d}\boldsymbol{\nu}.\qquad\blacksquare\end{align*}

\begin{theorem} 
\label{th:comparison}
Suppose (see the following lemma) that the free energy $F$ can
be written in terms of
\[
j:=J(I_3-\mathbf{n}\otimes\mathbf{n})\quad\text{and}\quad
\boldsymbol{\gamma}:=\nabla\mathbf{n}\times\mathbf{n},
\]
that is, there exists a function $\Psi :\mathbb{R}\times
Sym(3)\times\mathfrak{gl}(3)\rightarrow\mathbb{R}$ such that
\[
F(\rho^{-1},\mathbf{n},\nabla\mathbf{n})=\Psi(\rho^{-1},j,\boldsymbol{\gamma}).
\]
Let $(\mathbf{u},\boldsymbol{\nu},\rho,j,\mathbf{n})$ be a solution of the
equations \eqref{EP_ordered_micropolar} and
\eqref{advection_ordered_micropolar}, and suppose that the initial conditions
verify $j_0=J(I_3-\mathbf{n}_0\otimes\mathbf{n}_0)$. Then
$(\mathbf{u},\boldsymbol{\nu},\rho,j,\boldsymbol{\gamma})$ is a solution of the
system

\begin{equation}\label{Eringen_like}
\left\lbrace
\begin{array}{ll}
\vspace{0.2cm}\displaystyle\rho\left(\frac{\partial}{\partial t}
\mathbf{u}+\nabla_\mathbf{u}\mathbf{u}\right)=\operatorname{grad}\frac{\partial
\Psi}{\partial\rho^{-1}}-\partial_k\left(\rho\frac{\partial\Psi}{\partial\boldsymbol{\gamma}^a_k}\boldsymbol{\gamma}^a\right),\\
\vspace{0.2cm}\displaystyle j\frac{D}{d
t}\boldsymbol{\nu}-(j\boldsymbol{\nu})\times\boldsymbol{\nu}=-\frac{1}{\rho}\operatorname{div}\left(\rho\frac{j}{J}\frac{\partial\Psi}{\partial\boldsymbol{\gamma}}\right)+\boldsymbol{\gamma}^a\times\frac{\partial\Psi}{\partial\boldsymbol{\gamma}^a},\\
\vspace{0.2cm}\displaystyle\frac{\partial}{\partial
t}\rho+\operatorname{div}(\rho \mathbf{u})=0,\qquad\,\,\,\displaystyle
\frac{D}{d t}j+[j,\hat{\boldsymbol{\nu}}]=0,\\
\vspace{0.2cm}\displaystyle\frac{\partial}{\partial
t}\boldsymbol{\gamma}+\boldsymbol{\pounds}_\mathbf{u}\boldsymbol{\gamma}+\boldsymbol{\gamma}\times\boldsymbol{\nu}+\frac{j}{J}\mathbf{d}\boldsymbol{\nu}=0.
\end{array} \right.
\end{equation}
This system is very close but distinct from the system \eqref{micropolar_L_C}
studied by Eringen. The difference is due to the presence of the factor
$\frac{j}{J}$ in the second and last equations.
\end{theorem}
\textbf{Proof.} Let $(\mathbf{u},\boldsymbol{\nu},\rho,j,\mathbf{n})$ be a
solution of the equations \eqref{EP_ordered_micropolar} and
\eqref{advection_ordered_micropolar}. The equations associated to the
conservation of mass and microinertia in \eqref{Eringen_like} are clearly
verified since they are identical. Using that
$j_0=J(I_3-\mathbf{n}_0\otimes\mathbf{n}_0)$, we obtain
$j=J(I_3-\mathbf{n}\otimes\mathbf{n})$, by lemma \ref{lemma_1}. Thus by Lemma
\ref{lemma_2}, the last equation in \eqref{Eringen_like} is verified.

Using that the relation $\boldsymbol{\gamma}=\nabla\mathbf{n}\times\mathbf{n}$
reads $\boldsymbol{\gamma}^a_i=\varepsilon^a_{bc}\mathbf{n}^b_{,i}\mathbf{n}^c$
in coordinates, we obtain
\[
\frac{\partial F}{\partial \mathbf{n}^b_{,j}}=\frac{\partial
\Psi}{\partial\boldsymbol{\gamma}^a_i}\frac{\partial
\boldsymbol{\gamma}^a_i}{\partial \mathbf{n}^b_{,j}}=\frac{\partial
\Psi}{\partial\boldsymbol{\gamma}^a_j}\varepsilon^a_{bc}\mathbf{n}^c.
\]
This shows that
\[
\partial_k\left(\rho\frac{\partial
F}{\partial\mathbf{n}^a_{,\,k}}\mathbf{n}^a_{,\,l}\right)=\partial_k\left(\rho\frac{\partial\Psi}{\partial\boldsymbol{\gamma}^a_k}\boldsymbol{\gamma}^a_l\right).
\]
Thus, the first equation is verified. We also have
\begin{align*}
\varepsilon_{lab}\mathbf{n}^a\frac{\partial
F}{\partial\mathbf{n}^b_{,\,k}}&=\varepsilon_{lab}\mathbf{n}^a\frac{\partial
\Psi}{\partial\boldsymbol{\gamma}^d_k}\varepsilon^d_{bc}\mathbf{n}^c=\mathbf{n}^a\mathbf{n}^l\frac{\partial
\Psi}{\partial\boldsymbol{\gamma}^a_k}-\frac{\partial
\Psi}{\partial\boldsymbol{\gamma}^l_k}\\
&=-(I_3-\mathbf{n}\otimes\mathbf{n})_{la}\frac{\partial
\Psi}{\partial\boldsymbol{\gamma}^a_k}=-\left(\frac{j}{J}\frac{\partial
\Psi}{\partial\boldsymbol{\gamma}}\right)_{lk}.
\end{align*}
Using this equality and the relation \eqref{useful_equality}, the second
equation of \eqref{EP_ordered_micropolar} reads
\begin{align*}
\rho\frac{D}{dt}(j\boldsymbol{\nu})_l&=\partial_k\left(\varepsilon_{lab}\mathbf{n}^a\rho\frac{\partial
F}{\partial\mathbf{n}^b_{,\,k}}\right)-\varepsilon_{lmn}\rho\frac{\partial
F}{\partial\mathbf{n}^a_{,\,m}}\mathbf{n}^a_{,\,n}\\
&=-\partial_k\left(\rho\left(\frac{j}{J}\frac{\partial
\Psi}{\partial\boldsymbol{\gamma}}\right)_{lk}\right)-\varepsilon_{lmn}\rho\frac{\partial\Psi}{\partial\boldsymbol{\gamma}^a_m}\boldsymbol{\gamma}^a_n.
\end{align*}
This can be written as
\[
\rho\frac{D}{dt}(j\boldsymbol{\nu})=-\operatorname{div}\left(\rho\frac{j}{J}\frac{\partial
\Psi}{\partial\boldsymbol{\gamma}}\right)+\rho\boldsymbol{\gamma}^a\times\frac{\partial\Psi}{\partial\boldsymbol{\gamma}^a},
\]
which is exactly the second equation in
\eqref{Eringen_like}.$\qquad\blacksquare$
\medskip 

No Lie-Poisson or Euler-Poincar\'e interpretation of the system \eqref{Eringen_like} is known. This system would coincide with 
Eringen's system \eqref{micropolar_L_C}, if $j/J $ could be taken
equal to one. This, however, cannot be achieved since $j $ is
an advected variable; $J $ is a constant. We interpret Theorem
\ref{th:comparison} as saying that the the Lhuillier-Rey and
Eringen formulations are very close but not equivalent.

\medskip

We now show that the hypothesis made on the free energy $F$ is verified in the
case of the Oseen-Z\"ocher-Frank free energy.

\begin{lemma} Let $\mathbf{n}$ a unit vector. Define
\[
\boldsymbol{\gamma}:=\nabla\mathbf{n}\times\mathbf{n}\quad\text{and}\quad
j=I_3-\mathbf{n}\otimes\mathbf{n}.
\]
Then
\begin{align*}
\operatorname{Tr}(\boldsymbol{\gamma})&=\mathbf{n}\!\cdot\!\operatorname{curl}\mathbf{n},\\
\operatorname{Tr}(\boldsymbol{\gamma}^T\boldsymbol{\gamma})&=\operatorname{Tr}(\nabla\mathbf{n}^T\nabla\mathbf{n})=(\operatorname{div}\mathbf{n})^2+(\mathbf{n}\cdot\operatorname{curl}\mathbf{n})^2+\|\mathbf{n}\times\operatorname{curl}\mathbf{n}\|^2\\
&\qquad\qquad\qquad\qquad\qquad+\operatorname{div}((\mathbf{n}\cdot\nabla)\mathbf{n}-\mathbf{n}\operatorname{div}\mathbf{n}),\\
\operatorname{Tr}(\boldsymbol{\gamma}^T\boldsymbol{\gamma}(\mathbf{n}\otimes\mathbf{n}))&=\|\mathbf{n}\times\operatorname{curl}\mathbf{n}\|^2.
\end{align*}
Thus,
\[
\operatorname{Tr}(\boldsymbol{\gamma}^T\boldsymbol{\gamma}j)=(\operatorname{div}\mathbf{n})^2+(\mathbf{n}\cdot\operatorname{curl}\mathbf{n})^2+\operatorname{div}((\mathbf{n}\cdot\nabla)\mathbf{n}-\mathbf{n}\operatorname{div}\mathbf{n}).
\]
\end{lemma}
\textbf{Proof.} Note that the relation
$\boldsymbol{\gamma}=\nabla\mathbf{n}\times\mathbf{n}$ reads
\[
\boldsymbol{\gamma}_i=\partial_i\mathbf{n}\times\mathbf{n}=
\left[\begin{array}{c}n^3\partial_in^2-n^2\partial_in^3\\
n^1\partial_in^3-n^3\partial_in^1\\
n^2\partial_in^1-n^1\partial_in^2\end{array}\right].
\]
Thus, we obtain
\begin{align*}
\mathbf{n}\cdot\operatorname{curl}\mathbf{n}&=n^1(\partial_2n^3-\partial_3n^2)+n^2(\partial_3n^1-\partial_1n^3)+n^3(\partial_1n^2-\partial_2n^1)\\
&=(n^3\partial_1n^2-n^2\partial_1n^3)+(n^1\partial_2n^3-n^3\partial_2n^1)+(n^2\partial_3n^1-n^1\partial_3n^2)\\
&=\boldsymbol{\gamma}_1^1+\boldsymbol{\gamma}^2_2+\boldsymbol{\gamma}^3_3
=\operatorname{Tr}(\boldsymbol{\gamma}).
\end{align*}
This shows the first equality.

In order to check the second assertion, we compute
$\boldsymbol{\gamma}^T\boldsymbol{\gamma}$ in terms of $\mathbf{n}$. We have
\[
\boldsymbol{\gamma}^T(\mathbf{u})\cdot \mathbf{v}
=\mathbf{u}\cdot\boldsymbol{\gamma}(\mathbf{v})
=\mathbf{u}\cdot(\nabla\mathbf{n}(\mathbf{v})\times\mathbf{n})
=\nabla\mathbf{n}(\mathbf{v})\cdot(\mathbf{n}\times\mathbf{u})
=\mathbf{v}\cdot\nabla\mathbf{n}^T(\mathbf{n}\times\mathbf{u}),
\]
therefore, we have
$\boldsymbol{\gamma}^T(\mathbf{u})=\nabla\mathbf{n}^T(\mathbf{n}\times\mathbf{u})$
and we can compute
\begin{align*}
\boldsymbol{\gamma}^T\boldsymbol{\gamma}(\mathbf{u})
&=\nabla\mathbf{n}^T[\mathbf{n}\times(\boldsymbol{\gamma}(\mathbf{u}))]
=\nabla\mathbf{n}^T[\mathbf{n}\times(\nabla\mathbf{n}(\mathbf{u})\times\mathbf{n})]\\
&=\nabla\mathbf{n}^T[(\mathbf{n}\cdot\mathbf{n})\nabla\mathbf{n}(\mathbf{u}))-(\mathbf{n}\cdot\nabla\mathbf{n}(\mathbf{u}))\mathbf{n}]=\nabla\mathbf{n}^T\nabla\mathbf{n}(\mathbf{u}).
\end{align*}
We have obtained the equality
$\boldsymbol{\gamma}^T\boldsymbol{\gamma}=\nabla\mathbf{n}^T\nabla\mathbf{n}$.
The formula
\[
\operatorname{Tr}(\nabla\mathbf{n}^T\nabla\mathbf{n})=(\operatorname{div}\mathbf{n})^2+(\mathbf{n}\cdot\operatorname{curl}\mathbf{n})^2+\|\mathbf{n}\times\operatorname{curl}\mathbf{n}\|^2+\operatorname{div}((\mathbf{n}\cdot\nabla)\mathbf{n}-\mathbf{n}\operatorname{div}\mathbf{n})
\]
can be checked directly, see for example Lemma 3.3 in \cite{Virga1994}. This
proves the second equality.

The third assertion follows from the equalities
\begin{align*}
\operatorname{Tr}\left(\boldsymbol{\gamma}^T\boldsymbol{\gamma}(\mathbf{n}\otimes\mathbf{n})\right)&
=\operatorname{Tr}\left((\boldsymbol{\gamma}\mathbf{n})^T\boldsymbol{\gamma}\mathbf{n}\right)
=\|\boldsymbol{\gamma}\mathbf{n}\|^2
=\|(\nabla\mathbf{n})\mathbf{n}\times\mathbf{n}\|^2\\
&=\|(\nabla\mathbf{n})\mathbf{n}\|^2=\|\mathbf{n}\times\operatorname{rot}\mathbf{n}\|^2,
\end{align*}
where the last one follows from the identity
$(\nabla\mathbf{n})\mathbf{n}=-\mathbf{n}\times\operatorname{rot}\mathbf{n}$;
see Lemma 3.3 in \cite{Virga1994}.$\qquad\blacksquare$

\medskip

As a consequence of this lemma, the expressions associated to chirality, twist,
splay,  and bend appearing in the Oseen-Z\"ocher-Frank free energy can be
expressed in terms of the variables $j$ and $\boldsymbol{\boldsymbol{\gamma}}$.
We have
\begin{align*}
\mathbf{n}\!\cdot\!\operatorname{curl}\mathbf{n}&=\operatorname{Tr}(\boldsymbol{\boldsymbol{\gamma}}),\qquad
(\mathbf{n}\!\cdot\!\operatorname{curl}\mathbf{n})^2=\operatorname{Tr}(\boldsymbol{\gamma})^2,\\
(\operatorname{div}\mathbf{n})^2&=\operatorname{Tr}(\boldsymbol{\gamma}^T\boldsymbol{\gamma}j)-\operatorname{Tr}(\boldsymbol{\gamma})^2,\quad\text{modulo
a divergence},\\
\|\mathbf{n}\times\operatorname{curl}\mathbf{n}\|^2&=\operatorname{Tr}(\boldsymbol{\gamma}^T\boldsymbol{\gamma})-\operatorname{Tr}(\boldsymbol{\gamma}^T\boldsymbol{\gamma}j).
\end{align*}
Thus, in terms of $j$ and $\boldsymbol{\gamma}$, the Oseen-Z\"ocher-Frank
free energy reads
\begin{align*}
\rho\Psi(\rho^{-1},j,\boldsymbol{\gamma})=&K_2\operatorname{Tr}(\boldsymbol{\gamma})+\frac{1}{2}K_{11}\left(\operatorname{Tr}(\boldsymbol{\gamma}^T\boldsymbol{\gamma}j)-\operatorname{Tr}(\boldsymbol{\gamma})^2\right)+\frac{1}{2}K_{22}\operatorname{Tr}(\boldsymbol{\gamma})^2\\
&+\frac{1}{2}K_{33}\left(\operatorname{Tr}(\boldsymbol{\gamma}^T\boldsymbol{\gamma})-\operatorname{Tr}(\boldsymbol{\gamma}^T\boldsymbol{\gamma}j)\right),
\end{align*}
up to the addition of a divergence. This functional clearly satisfies the
axiom of objectivity.

\subsubsection{Remark on the use of other groups}

From a mathematical point of view, the previous approaches generalize to any order parameter Lie group $\mathcal{O}\subset GL(3)^+$.

\medskip  

\noindent \textbf{$\gamma$-theory.} In the case of Eringen's theory, it suffices to consider the group $\operatorname{Diff}(\mathcal{D})\,\circledS\,\mathcal{F}(\mathcal{D},\mathcal{O})$ acting on the advected quantities
\[
(\rho,i,\zeta)\in \mathcal{F}(\mathcal{D}) \times Sym(3) \times \Omega^1(\mathcal{D},\mathfrak{o})
\]
as
\[
\rho\mapsto J\eta(\rho\circ\eta),\quad i\mapsto \chi^T(i\circ\eta)\chi, \quad\zeta\mapsto \operatorname{Ad}_{\chi^{-1}}\eta^*\zeta+\chi^{-1}T\chi.
\]
If $\mathcal{O}=SO(3)$ then $\zeta=\gamma$ is the \textit{wryness tensor}, and we recover the theory of micropolar liquid crystals (see \S\ref{subsubsec:micropolar}). If $\mathcal{O}=CSO(3)$ then $\zeta=(\gamma,e)$, where $\gamma$ is the \textit{wryness tensor} and $e$ is the \textit{microstrain} and we recover the microstretch theory of polymeric liquid crystals (see \S\ref{polymeric_microstretch}). In general, we obtain a theory of ``$\mathcal{O}$-\textit{liquid crystals}" whose Lagrangian given by
\[
l(\mathbf{u},\nu,\rho,i,\zeta)=\frac{1}{2}\int_\mathcal{D}\rho\|\mathbf{u}\|^2\mu+\frac{1}{2}\int_\mathcal{D}\rho (i\nu\!\cdot\!\nu)\mu-\int_\mathcal{D}\rho\Psi(\rho^{-1},i,\zeta)\mu,
\]
and the variable $\zeta$ is interpreted as a connection on the trivial $\mathcal{O}$-principal bundle over $\mathcal{D}$. The associated affine Euler-Poincar\'e equations \eqref{EPComplexFluid_covariant_form} are
\begin{equation}\label{general_liquid_crystals}
\left\lbrace
\begin{array}{ll}
\vspace{0.2cm}\displaystyle\rho\left(\frac{\partial}{\partial t}
\mathbf{u}+\nabla_\mathbf{u}\mathbf{u}\right)=\operatorname{grad}\frac{\partial
\Psi}{\partial\rho^{-1}}-\partial_k\left(\rho\frac{\partial\Psi}{\partial\zeta_k^a}\zeta^a\right),\\
\vspace{0.2cm}\displaystyle P\left(i\left(\frac{\partial }{\partial t}\nu+\mathbf{d}\nu(\mathbf{u})-\nu\nu-2\frac{\partial\Psi}{\partial i} \right)\right)=0,\\
\vspace{0.2cm}\displaystyle\frac{\partial}{\partial
t}\rho+\operatorname{div}(\rho \mathbf{u})=0,\qquad\,\,\,\displaystyle
\frac{\partial}{\partial t}i+\mathbf{d}i(\mathbf{u})+\nu^Ti+i\nu=0,\\
\vspace{0.2cm}\displaystyle\frac{\partial}{\partial
t}\zeta+\pounds_{\mathbf{u}}\zeta+[\zeta,\nu]+\mathbf{d}\nu=0,
\end{array} \right.
\end{equation}
where $P:\mathfrak{gl}(3)\rightarrow\mathfrak{o}$ denotes the orthogonal projection onto the Lie algebra $\mathfrak{o}$, associated to the inner product
\[
a\!\cdot\!b=\operatorname{Tr}(a^Tb),\quad a,b\in\mathfrak{o}.
\]

\medskip  

\noindent \textbf{$\mathbf{n}$-theory.} Recall that in the case of the Ericksen-Leslie and Lhuillier-Rey theories the director is a map $\mathbf{n}:\mathcal{D}\rightarrow\mathbb{R}^3$, on which $\operatorname{Diff}(\mathcal{D})\,\circledS\,\mathcal{F}(\mathcal{D},SO(3))$ acts linearly as
\begin{equation}\label{representation_SO(3)}
\mathbf{n}\mapsto \chi^{-1}(\mathbf{n}\circ\eta).
\end{equation}
This representation can be generalized to other groups in two ways. First the action \eqref{representation_SO(3)} clearly still makes sense for any matrix Lie group $\mathcal{O}$. In this case, the advection equation for $\mathbf{n}$ reads
\[
\frac{D}{dt}\mathbf{n}=\nu\mathbf{n},
\]
where $\nu\in\mathcal{F}(\mathcal{D},\mathfrak{o})$, and the variable $\mathbf{n}$ evolves by 
\[
\mathbf{n}=(\chi\mathbf{n}_0)\circ\eta^{-1}.
\]
Note that, in general, the norm $\|\mathbf{n}(x)\|$ is not constant. For example, if $\mathcal{O}=SO(K)$, for $K$ positive definite, then $\mathbf{n}$ describes an ellipsoid. As in the case of microfluids, the typical cases to consider are
\[
K_1=\left[\begin{array}{ccc} \;1\;&0\;&0\;\\ \;0\;&1\;&0\;\\ \;0\;&0\;&\varepsilon\;\end{array}\right]\quad\text{or}\quad K_2=\left[\begin{array}{ccc} \;\varepsilon\;&0\;&0\;\\ \;0\;&\varepsilon\;&0\;\\ \;0\;&0\;&1\;\end{array}\right],\quad 0<\epsilon\ll 1.
\]
The group $SO(K_1)$ should be useful for the description of smectic liquid crystals.
If $\mathcal{O}=CSO(3)$, then the norm $\|\mathbf{n}\|$ evolves as
\[
\|\mathbf{n}(\eta(x))\|=\operatorname{det}(\chi(x))^{1/3}\|\mathbf{n}_0(x)\|.
\]
Therefore, the norm depends on the ``stretch part" $\operatorname{det}(\chi)^{1/3}$ of $\chi\in CSO(3)$. More generally, we can use the group $CSO(K)$ to model the theory of \textit{aniso\-tropic microstretch liquid crystals}. This constitutes a first way of generalizing the representation \eqref{representation_SO(3)} to other groups.

A second way to generalize the representation \eqref{representation_SO(3)} consists in replacing the director $\mathbf{n}:\mathcal{D}\rightarrow \mathbb{R}^3$ by a Lie algebra valued variable $n:\mathcal{D}\rightarrow \mathfrak{o}$ on which the group $\operatorname{Diff}(\mathcal{D})\,\circledS\,\mathcal{F}(\mathcal{D},\mathcal{O})$ acts as
\begin{equation}\label{action_Ad_n}
n\mapsto \operatorname{Ad}_{\chi^{-1}}(n\circ\eta).
\end{equation}
In this case, the advection equation reads
\[
\frac{D}{dt} n=\operatorname{ad}_\nu n,
\]
and the variable $n$ evolves as
\[
n=\left(\operatorname{Ad}_\chi n_0\right)\circ\eta^{-1}.
\]
Remark that when $\mathcal{O}=SO(3)$, both approaches coincide because the birth representation and the adjoint representation of $SO(3) $ on $\mathbb{R}^3$ are identical. When $\mathcal{O}=CSO(3)$, (microstretch case) then we can write $n=\widehat{\mathbf{n}}+m I_3$ and the action \eqref{action_Ad_n} reads
\[
(\mathbf{n},m)\mapsto (\overline{\chi}^{-1}(\mathbf{n}\circ\eta),m\circ\eta),
\]
where
\[
\overline{\chi}:=\frac{1}{\operatorname{det}(\chi)^{1/3}}\chi:\mathcal{D}\rightarrow SO(3),
\]
and the evolutions are given by
\[
\mathbf{n}=\left(\overline{\chi}\,\mathbf{n}_0\right)\circ\eta^{-1}\quad\text{and}\quad m=m_0\circ\eta^{-1}.
\]
Therefore $\mathbf{n}$ can be seen as a director, since $\|\mathbf{n}(x)\|$ is constant in time. Since the evolution of the variable $m$ does not depend on the micromotion and the evolution of $\mathbf{n}$ depend only on the $SO(3)$ part of the micromotion $\chi\in CSO(3)$, this approach cannot be used for the description of microstretch liquid crystals. This shows that the first generalization seems physically more interesting.

\medskip  

\noindent \textbf{Acknowledgments.} The authors acknowledge the partial support of the
Swiss National Science Foundation. Our special thanks go to Darryl Holm for his
invaluable explanations of the physical phenomena described in the examples
treated in this paper and for pointing out that a general abstract theory that
would encompass all these examples was lacking. His patience and numerous
clarifications were crucial in our understanding of these models.




\end{document}